\documentclass[12pt, a4paper]{book}

\usepackage{vuwthesis}
\usepackage{amsmath}
\usepackage{amsthm}
\usepackage{amssymb}
\usepackage{graphicx}
\usepackage{rotating}
\usepackage{slashbox}

\def\d{{\mathrm{d}}}
\def\signum{\ensuremath{\mathrm{signum}}}
\newcommand{\diff}[1]{\ensuremath{\mathrm{d}{#1}}}
\newcommand{\dx}[1]{\diff{#1}}
\newcommand{\dr}{\ensuremath{\frac{\mathrm{d}\phantom{r}}{\dx{r}}}}
\newcommand{\eqprime}[1]{\tag{\ref{#1}$^\prime$}}

\newcommand{\dTwo}{\ensuremath{\delta_2}}

\newcommand{\orho}{\ensuremath{\overline{\rho}}}
\newcommand{\orr}{\ensuremath{\overline{r}}}

\def\lint{\hbox{\Large $\displaystyle\int$}} %needs \usepackage{amssymb}
\newtheorem{theorem}{Theorem}

\newtheorem{defn}{Definition}

\newcommand{\pacs}[1]{\par \footnotesize \indent PACS numbers: #1 \par}
\newcommand{\keywords}[1]{\par \footnotesize \indent Keywords: #1\par}

\newcommand{\paperauthor}[1]{\centerline{\Large #1}\bigskip}
\newcommand{\publishnotice}[1]{\par\bigskip\centerline{\bf #1}\bigskip}

\begin{document}

\frontmatter
% Book style knows about front matter
% Report style doesn't so you need to set roman numbering etc yourself :-(

%%%%%%%%%%%%%%%%%%%%%%%%%%%%%%%%%%%%%%%%%%%%%%%%%%%%%%%

%\title{An investigation of highly compact objects and black holes in Einstein's General Relativity}

%\title{Through the looking glass: Windows to quantum gravity}

%\title{The bloody thesis}

%\title{One monkey, one typewriter and a very limited time constraint\\
%\vskip 1cm
%\small{(It ain't Shakespear, baby!)}}
\title{Much ado about nothing\\
\vskip 1cm
\small{A treatise on empty and not so empty spacetimes}}
\author{Damien James Martin}

\subject{Mathematics}
\abstract{In this thesis three separate problems relevant to general
relativity are considered. Methods for algorithmically producing all the
solutions of isotropic fluid spheres have been developed over the last five
years. A different and somewhat simpler algorithm is discussed here, as well
as algorithms for anisotropic fluid spheres. The second and third problems are
somewhat more speculative in nature and address the nature of black hole
entropy. Specifically, the second problem looks at the genericity of the
so-called quasinormal mode conjecture introduced by Hod, while the third
problem looks at the near-horizon structure of a black hole in hope of gaining
an understanding of why so many different approaches yield the same entropy.
A method of finding the asymptotic QNM structure is found based on the Born
series, and serious problems for the QNM conjecture are discussed. The work in
this thesis does not completely discount the possibility that the QNM
conjecture is true. New results released weeks before this thesis was finished showed that the QNM conjecture was flawed. Finally, the near-horizon structure of a black hole is
found to be very restricted, adding credence to the ideas put forward by
Carlip and Solodukhin that the black hole entropy is related to an inherited
symmetry from the classical theory.}
% Books don't normally have abstracts, and this is a bit of a hack

% Uncomment the appropriate degree
%\phd
\mscthesisonly
%\mscwithhonours
%\mscbothparts
% \otherdegree{DEGREE OR DIPLOMA NAME}

\maketitle

\chapter{Acknowledgements}

I have many people that I would like to thank for enabling me to write this thesis. There are many things inside it that can still be improved, and almost all of the results discussed within can be extended. More could have been done, and the fact that it was not is my own fault. I still wish to thank this moment to thank the people that have supported me in this work, even though at times I am sure they wondered why they bothered!

\paragraph{}
I would like to thank the School of Chemical and Physical Sciences for their hospitality for the duration of this degree, as pressure was placed on the mathematics department for office space. I was always sure that I could turn to members of the staff for guidance without being shunned as a ``mathematician''! In particular, John Lekner was always willing and interested in my work.

\paragraph{}
As for the actual guidance of my work, I could not have asked for a better supervisor. Matt ``Just do it'' Visser was not only able to guide me on the intricacies of black hole physics, and not only be able to repeat the Nike slogan at very frequent intervals, but also kept on at me persistently to ensure that things got done properly and by the deadline. Even aided by Silke Wienfurtner, this was a mammoth task for him to undertake and I appreciate the efforts made, even if it did not seem like it at the time.

\paragraph{}
I would like to thank many of the people that I met and talked to me at GR17 in Dublin. While Dublin was too late to seriously affect this thesis, I gained many valuable insights. In particular I would like to thank Steven Carlip for clarifying aspects of the near horizon conformal symmetries, John Baez for pointing out that the Hod conjecture was actually false and Donald Marolf for presenting an alternative view on black hole holography. On the more personal side of things, I would like to thank Francis and Sue Cook for many pleasant evenings and home-cooked meals. 

\paragraph{}
Finally I would like to extend my gratitude to Shoko Jin for being there to listen to me and support me at the difficult stages of this thesis, and putting up with me at my most irrational\footnote{Between any two rational moments lie an infinite number of irrational ones}. I am not sure this thesis would have even been finished without the support she showed.

\tableofcontents

%%%%%%%%%%%%%%%%%%%%%%%%%%%%%%%%%%%%%%%%%%%%%%%%%%%%%%%

% book style knows about mainmatter
% if you are using report style you will have to rest page numbering etc.
\chapter{Introduction}
This thesis looks at three problems in general relativity:
\begin{itemize}
\item The construction of ``physically reasonable'' static spherically symmetric
solutions of Einstein's field equations.
\item The determination of the asymptotic spectrum of quasinormal modes of
dirty black holes.
\item Investigation of a conformal symmetry at the event horizon of arbitrary
stationary black holes.
\end{itemize}
The first problem is the odd man out. It looks at the bounds that
general relativity alone can place on a star in equilibrium without imposing an
equation of state. It is shown that no statement about a maximum mass can be
made, although various statements about how compact an object is, \emph{can} be
made. It is also shown how to take the gravity profile of an isotropic fluid
(that is, one where the tangential and radial pressures are equal) and from
this deduce the pressure profile, the density profile and (hence trivially) the
mass profile. Bounds are also discussed that ensure physically reasonable
\emph{anisotropic fluids}. Three separate algorithms are presented for the
anisotropic case, but all require the specification of a measure of the
anisotropy -- a hard thing to determine either theoretically or
observationally!

\paragraph{}
The other two problems are both ``windows to quantum gravity''; they look at
the properties of \emph{classical} black holes in general relativity and see
what this can tell us about \emph{quantum} black holes and ultimately quantum
gravity. There are now many approaches on how to quantise gravity (particularly
``toy models'' in 2+1 dimensions) and they all seem to reproduce the
semi-classical formulae for black hole entropy and temperature. We are left with
three possibilities:
\begin{itemize}
\item Black hole entropy and temperature are not generic. It is a coincidence
that every proposed model so far of quantum gravity reproduces these results.
\item Black hole entropy and temperature are not generic. Loop quantum gravity,
strings and other models of quantum gravity are somehow the ``same'' theory
described differently (M-theory).
\item Black hole entropy and temperature \emph{are} generic. Something that all
the approaches take for granted (such as diffeomorphism invariance) ultimately
requires these results. As a result calculations of a particular model will give
these results and hence cannot be used to discriminate between theories.
\end{itemize}
The view taken in this thesis is that the semi-classical results are generic.
To state this precisely one needs to be a lot more careful -- extending
Newtonian gravity to the quantum realm certainly would not reproduce black hole
horizons, let alone black hole entropy or temperature. A less extreme example
would be to investigate the statistics of horizon states as carried out by
Polychronakos \cite{Polychronakos:2003} where the entropy was found to be
proportional to the area provided that the individual states where
\emph{somewhat} distinguishable. Nevertheless it is believed that under suitable
restrictions that the black hole entropy and temperature are generic in the
sense that a large class of wildly different theories of quantum gravity will
give results compatible with the semi-classical calculation.

\begin{figure}[t]
\begin{center}
\begin{minipage}{0.4\textwidth}
\begin{center}
\includegraphics[width=0.85\textwidth]{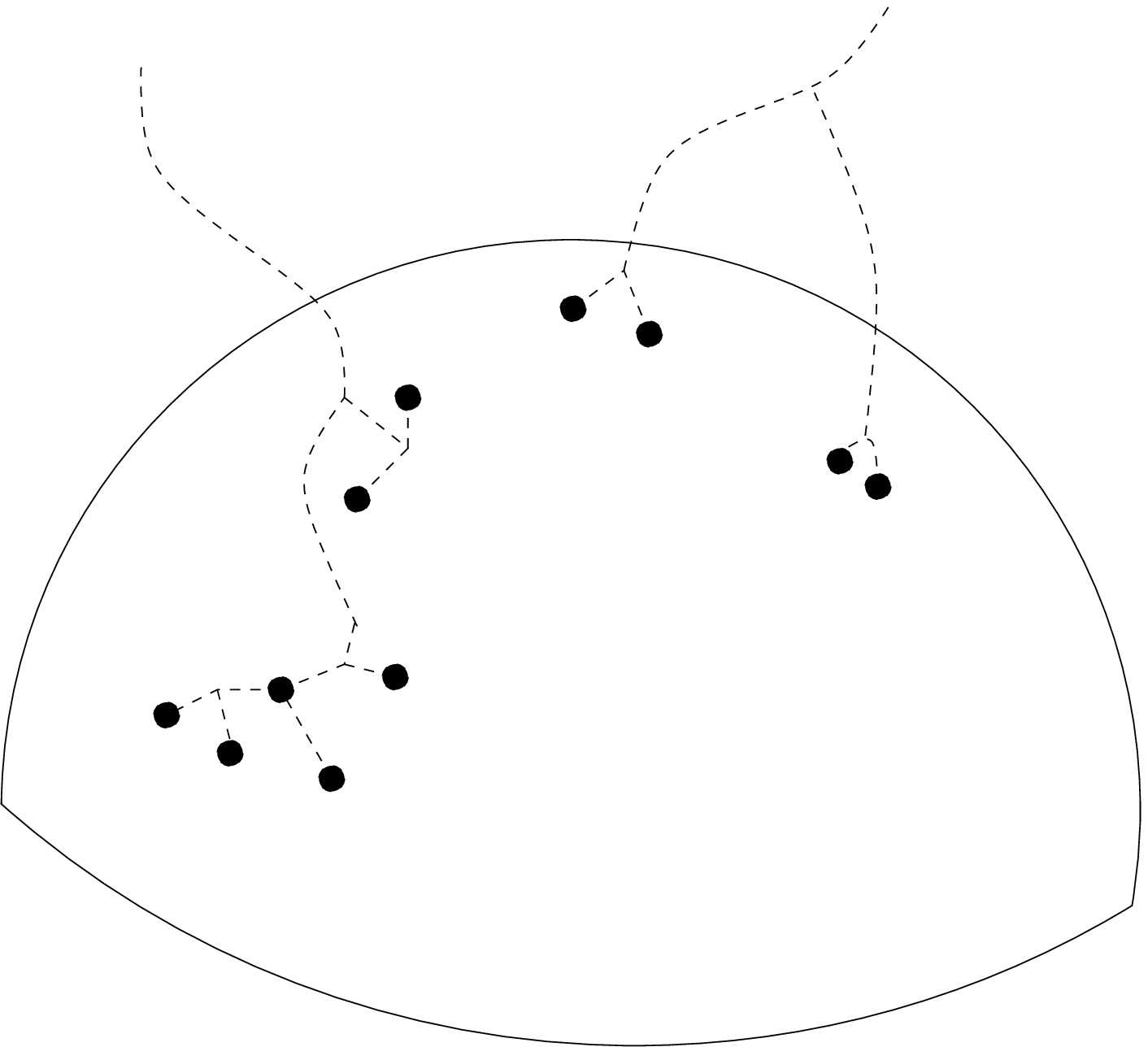}
\end{center}
\end{minipage}
\begin{minipage}{0.4\textwidth}
\begin{center}
\includegraphics[width=0.85\textwidth]{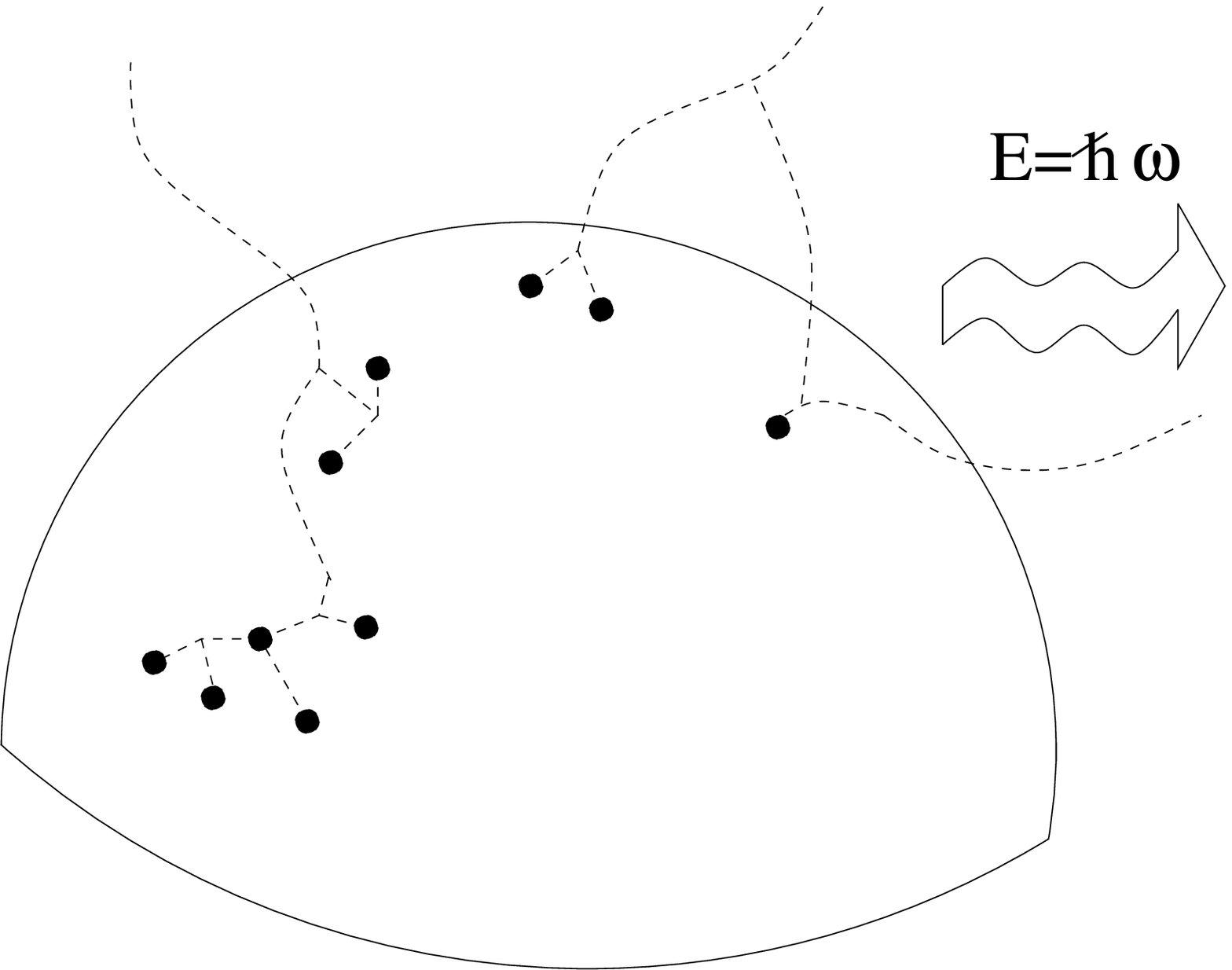}
\end{center}
\end{minipage}
\end{center}
\begin{small}
\caption{{\footnotesize The area of a black hole horizon is calculated in LQG by the spin links intersecting the horizon, indicated by black dots. On the right a link breaks off the horizon, and goes off toward infinity. Hod's conjecture is that highly damped modes have energies $\sim\hbar \omega$, and hence relate this loss of energy to the area of horizon lost}} \label{fig:horizon_links}
\end{small}
\end{figure}

\paragraph{}
The quasinormal mode conjecture, or Hod conjecture, speculated that we could find information out about details of quantum gravity by looking at the resonances of a black hole relaxing back into a stationary state after a perturbation. These resonances, or quasinormal modes (QNMs), dictate the late time response of a black hole to an external perturbation. The QNM are harmonic functions with complex frequency, so as well as oscillating they are exponentially damped. This damping occurs because energy is lost both toward infinity and over the black hole horizon. The QNM conjecture related properties of these purely classical QNM to an ambiguity in loop quantum gravity (LQG). 

\paragraph{}
To understand how QNM may be related to LQG, an understanding of how LQG calculates the area of a black hole is needed. We construct a trivalent network, called a spin network, that underlies the geometry of space. Each link on the network is labelled by a representation of the gauge group of LQG, usually taken to be $SU(2)$. In the case of $SU(2)$ the representations are the same as the representations of particles of spin $j$, hence the name spin network. The area of a black hole horizon is calculated as a function of the links that intersect the horizon (see the left hand side of figure \ref{fig:horizon_links}). If a particle is emitted from the horizon, as on the right hand side of figure \ref{fig:horizon_links}, then the area decreases. But a decrease in area implies a decrease in mass, the exact amount of mass decrease depends on the angular momentum of the link carried away and the surface gravity of the original black hole. The mass $m$, or energy, that the spin link carries off gives rise to a frequency $\omega = m c^2/\hbar$. As the spin links can only by assigned half integer spins $j$, the possible values for $\omega$ for a Schwarzschild black hole are restricted to
\begin{equation}
\omega = \frac{c^3}{8\pi GM}\;\ln(2j + 1).
\end{equation}
On the other hand, the QNMs of a Schwarzschild black hole appeared from numerical calculations to have the asymptotic form
\begin{equation}
\omega_{\textrm{QNM}} = \frac{c^3 \;\ln 3}{8 \pi GM}  + \frac{c^3}{4 \pi GM}\left(n + \frac{1}{2}\right)i \quad\quad(\textrm{as }n\rightarrow\infty).\label{eq:intro_qnm_asymptotic}
\end{equation}
If one assumed that Hawking radiation was due to primarily spin 1 links, then the (real) frequency of a QNM would act like the emission of a particle with energy $E=\hbar \omega$! To get the Schwarzschild mass-area relationship correct required the value of the Barbero--Immirzi parameter $\gamma$, an arbitrary parameter in LQG, to be fixed to a specific value. The Barbero--Immirzi parameter chooses the valid sector of LQG, and so cannot take different values for different problems. So it seemed that assuming that treating QNMs as the emission of particles allowed LQG to be specified unambiguously. An excited flurry of work began to verify if \eqref{eq:intro_qnm_asymptotic} could be established analytically, not that it was just numerically close to the ``$\ln 3$'' form. It was recently shown that the real part of the quasinormal modes was not of the ``$\ln 3$'' form by Domagala and Lewandowski \cite{Domagala:2004}, and hence the QNM conjecture did not hold for the Schwarzschild black hole, let alone more complicated examples! However, the work we did in chapter \ref{chap:QNM_intro} and chapter \ref{chap:qnm_find} was on the basis of extending the QNM conjecture to other black hole spacetimes, as it was thought to hold at least on Schwarzschild spacetime.

\paragraph{}
The final problem discussed in the thesis looks at why different theories give the same result for the black hole entropy. It may be that the entropy calculation only checks that the theory of gravity can be expressed as a 2D conformal field theory on the horizon. The conformal symmetry in 2D allows an almost complete determination of the density of states, and hence the entropy. There is speculation, initiated by Carlip and Solodukhin, that classical general relativity can be expressed as a conformal field theory on a 2D section near the horizon. As a Killing horizon is a null surface, a Killing horizon cannot be modelled by a massive field theory, hence breaking conformal invariance. The research carried out by Dr.~Visser, Dr.~Medved and myself is much more modest than constructing a full conformal field theory modelling classical general relativity. Instead, we focus on the near horizon geometry of a black hole under the restriction that each of the curvature invariants remains finite.

\section*{Structure of the thesis}
When writing, choices must be made about the point at which you refer to papers
for background and what you include. This thesis has been written with the goal
of being accessible to people with a basic background in quantum field theory
and general relativity, and much of the background material that could have been
referenced has been included to enable this thesis to show why the topics included
are of general interest. Although complete separation between research and
literature review is not possible, I have attempted to use even chapters as an
introduction and motivation while expanding on original research in odd
chapters. The research chapters give more explicit calculations and details to the papers included in the 
appendices. 

\paragraph{}
This thesis is made up of five chapters and seven appendices. In the version of this thesis that was 
handed in, the appendices are full copies of the papers published, and were produced as a
collaboration with Dr.~Matt Visser and Dr.~Allan Joseph Medved. Because these papers are available
on the arXiv, this is inappropriate for the thesis on the arXiv. Instead the appendices have been truncated to just the abstract, journal and arXiv references so that people reading this can pursue the papers they are interested in. The only other differences between the final version and this one are this paragraph and parts of the conclusions section.
\section*{Conventions}
The original source is unknown, but it has often been said that the beauty of
standards is that there are so many to choose from. In this section I outline
the basic conventions that I use throughout the thesis.

\subsubsection*{Signature}
The signature of spacetime is taken to be $(- + + +)$ throughout.

\subsubsection*{Differential geometry}
The Riemann tensor is defined to be 
\begin{equation}
R_{abc}{}^{d}\omega_d = \left(\nabla_a\nabla_b - \nabla_b\nabla_a\right)\omega_c = \left[\nabla_a,\nabla_b\right]\omega_c
\end{equation}
for any one-form $\omega_c$. The Ricci tensor is defined to be the contraction
over the first and third indexes:
\begin{equation}
R_{ab} = R^{c}{}_{acb}.
\end{equation}
These conventions agree with the definitions given in chapter 3 of Wald \cite{book:Wald}.

\paragraph{}
The Ricci rotation coefficients are defined to be
\begin{equation}
\omega_{\mu\nu a} = e_{(\mu)}{}^b\nabla_a e_{(\nu)b}, 
\end{equation}
where $e_{(\alpha)}$ are the basis vectors of a tetrad. The Greek letters are
used to emphasise the first two indexes are labels and the last index is a
coordinate. Technically I should write $(\omega_{\mu\nu})_{a}$, as there are 6
one-forms $\omega_{\mu\nu}$ each with four components. The brackets are overly
cumbersome and have been dropped. This \emph{differs} from the convention used
by Wald. 

\paragraph{}
Similarly, the \emph{spin coefficients} are given by
\begin{equation}
\omega_{\mu\nu\hat{a}} = \omega_{\mu\nu b} e^{(b)}{}_a.
\end{equation}

\subsubsection*{Symmetrisation}
\begin{itemize}
\item The Lie bracket is defined in the standard way. For example, for vector fields the Lie
bracket is identical to the commutator:
\begin{equation}
[\zeta,\psi]^a = \zeta^b \partial _b \psi^a - \psi^b\partial_b \zeta^a,
\end{equation}
\item Symmetrisation and antisymmetrisation on indexes include a normalisation
factor:
\begin{align}
A_{a_1 a_2 \ldots a_n [b_1 \ldots b_m]} &= \frac{1}{m!} \sum_{\sigma \in S_m}
\signum(\sigma) A_{a_1 a_2 \ldots a_n b_{\sigma(1)} \ldots b_{\sigma(m)}},\\
A_{a_1 a_2 \ldots a_n (b_1 \ldots b_m)} &= \frac{1}{m!} \sum_{\sigma \in S_m}
 A_{a_1 a_2 \ldots a_n b_{\sigma(1)} \ldots b_{\sigma(m)}}.
\end{align}
$S_m$ is the group of permutations on $m$ objects, and $\signum(\sigma)$ is $1$ ($-1$) if the
permutation is even (odd). Any indices that lie between vertical bars are
exempt from (anti)symmetrisation. For example:
\begin{subequations}
\begin{align}
A_{(ab)} &= \frac{1}{2}(A_{ab} + A_{ba})\\
A_{[ab]} &= \frac{1}{2}(A_{ab} - A_{ba})\\
B_{[a|b|c]} &= \frac{1}{2}(B_{abc} - B_{cba}).
\end{align}
\end{subequations}
\end{itemize}
The nature of symmetrisation is quite different and there is no possibility
of confusion. While it would be slightly more consistent to adopt a
normalisation for Lie brackets or drop normalisation for symmetrisation, a
higher value has been placed on the ability to compare directly with the
literature.

\subsubsection*{Variable definitions}
In this section I try and establish the notation that will be used throughout the thesis. This table is not meant to be exhaustive, but is meant to cover common definitions that occur throughout portions of the thesis. These are ordered roughly by order of appearance in the thesis, but some exceptions are made to keep related variables together.

\begin{center}
\begin{tabular}{|cc|p{10cm}|}
\hline
Symbol & Chapters & Meaning\\
\hline
$T_{\mu\nu}$ &All& The stress-energy tensor.\\
$M$ & All & The total mass of a star or black hole, as measured from infinity.\\
$m(r)$ & 1--2 & The total mass which is less than a coordinate distance $r$ from the origin. Also known as the Hawking quasi-local mass. Note the consistency condition $M = m(\infty).$\\
$R$ & All & The coordinate distance that marks the edge of a star \emph{or} the event horizon of a black hole \emph{or} the Ricci scalar.\\
$r$ & All & A radial coordinate. Note that this coordinate is \emph{not} the radial distance, except in flat space. If there is an origin, $r$ is chosen so that the area of a set of points equidistant from the origin on a constant time surface has area $4\pi r^2$.\\
$\rho$ & All & The energy density as measured by a particular observer.   \\
\hline
$a$ & 1--2 & The specific angular momentum $a = J/M$.\\
$P$ & 1--2 & The pressure (isotropic spacetimes only).\\
$P_r$ & 1--2 & The radial pressure (spherically symmetric spacetimes only).\\
$P_t$ & 1--2 & The tangential pressure (spherically symmetric spacetimes only).\\
\hline
$A$ & All & The area of a black hole horizon or a general area.\\
$A_n$ & 2--3 & The eigenvalues of the area operator. The $n$ acts as an index for different eigenvalues, and depending on the theory may mean a collection of indices.\\
$a_n$ & 2--3 & The contribution to the area by a single spin link.\\
$\omega_n$ & 2--3 & The (complex) angular frequency of the quasi-normal mode.\\
\hline
$N$ & 4--5 & The lapse function.\\
$n$ & 4--5 & The direction that is normal (and not tangent) to the horizon. $n=0$ on the horizon.\\
$\varphi$ & 4--5 & A direction that is axisymmetric and is a free coordinate on the horizon.\\
$z$ & 4--5 & The other coordinate that becomes a free coordinate on the horizon. No particular symmetry.\\
\hline
\end{tabular}
\end{center}

\subsubsection*{Energy conditions}
The energy conditions adopted in this thesis are summarised in table
\ref{table:energy_conditions}, and discussed in more detail below:

\begin{itemize}
\item The null energy condition (NEC):\\
For all future pointing null vectors $k^a$
\begin{equation}
T_{ab}k^a k^b \geq 0.
\end{equation}
In terms of the principal pressures this implies that
\begin{equation}
\rho + p_i \geq 0, \quad\quad i=1,2,3.
\end{equation}
\item The weak energy condition (WEC):\\
Traditionally stated, the WEC states that for all timelike
vectors $t^a$ that
\begin{equation}
T_{ab}t^a t^b \geq 0.\label{eq:WEC}
\end{equation}
Any (unit) timelike vector can be a tangent to an observer's world line. The
WEC condition states that the energy density measured by any observer is
non-negative. Note that we do not need to assume the existence of principal
pressures to make this statement.\\
The WEC $\Rightarrow$ NEC as $T_{ab}$ is a multilinear map and hence
continuous in both arguments. As any future directed timelike vector satisfies
\eqref{eq:WEC} any null vector will as well.\footnote{This argument can be
made because we are comparing all vectors in a particular tangent space, and
are then unaffected by singularities in other points in spacetime. Continuity
arguments would not hold when studying global structure, such as extending the
chronological future of a point to the causal future.}
\item The dominant energy condition (DEC):
The DEC assumes that the WEC holds, and that for all future directed timelike
vectors $t^a$ that $-T_{ab} t^{b}$ is a future directed non-spacelike vector.
This ensures that the net energy flow does not exceed the speed of light. By
construction DEC $\Rightarrow$ WEC. In terms of the principal pressures:
\begin{equation}
\rho \geq |p_i|,\quad\quad(i=1,2,3) 
\end{equation}
\item The strong energy condition (SEC):
The strong energy condition states that for all timelike
vectors $t^a$ that
\begin{equation}
T_{ab}t^a t^b \geq -\frac{1}{2}T,
\end{equation}
which by the same continuity argument used in the WEC case. In terms of
principal pressures we have
\begin{equation}
\rho + \sum_{i=1}^{3} p_i \geq 0,\quad\quad \rho + p_i \geq 0,
\quad\quad(i=1,2,3).
\end{equation}
Note that the SEC implies the NEC, it \emph{does not} imply the WEC. For
example, matter with a negative energy density but sufficiently high pressures
could satisfy the SEC but would violate the WEC.
\end{itemize}

\begin{table}[t]
\begin{center}
\begin{footnotesize}
\begin{tabular}{|l@{\,}lp{0.3\textwidth}p{0.4\textwidth}|}
\hline
Energy condition & &Definition & Meaning (for any observer)\\
\hline 
Null Energy Condition &(NEC)& $T_{ab}k^a k^b \geq 0$ & ---\\
Weak Energy Condition &(WEC)& $T_{ab} t^a t^b \geq 0$ & Non-negative energy
densities.\\
Strong Energy Condition &(SEC)& $T_{ab} t^a t^b \geq -T/2`$ &---\\
Dominant Energy &(DEC)& 1. WEC and & Energy densities non-negative\\
Condition&& 2. $-T^{a}{}_{b} t^b$ is a future
directed non-spacelike vector&and the net flow of energy is less than the speed
of light.\\
\hline
\end{tabular}
\end{footnotesize}
\caption{A summary of the energy conditions. In the above $k^a$ is any null
vector and $t^a$ is any timelike or null vector. The last column gives a simple
physical interpretation for generic matter where
applicable.}\label{table:energy_conditions}
\end{center}
\end{table}
Hawking's area theorem for black hole
horizon relies on the NEC, and hence evaporation of a black hole must violate
the NEC. Violating the NEC implies violating the DEC, SEC and WEC as well
\cite{book:Wald}. In fact, it is well known that quantum fields violate many
of the traditional energy conditions. In chapter \ref{chap:bounds} it is
assumed that for bulk classical matter that violations of the energy
conditions are insignificant. For a contrary point of view, see the essay by
Barcel\'{o} and Visser \cite{twilight}. Subsequent chapters look at
near-horizon structure and Hawking radiation, so the energy conditions are not
applicable.

\paragraph{}
While the energy conditions are used to bound the types of matter we have
available, the theorems in general relativity (such as the singularity
theorems) usually rely on statements about the geometry. By assuming the
Einstein equations the energy conditions can be transformed into statements
about the geometries we are interested in. The NEC becomes the condition the
$R_{ab} k^a k^b = G_{ab} k^ak^b \geq 0$, the null Ricci convergence condition.
The WEC becomes the Einstein convergence condition $G_{ab}t^at^b$ and the SEC
becomes the Ricci convergence condition $R_{ab}t^at^b$.

\paragraph{}
The definition of the energy conditions used conforms with Hawking and Ellis
\cite{Hawk74}.

\mainmatter

%%%%%%%%%%%%%%%%%%%%%%%%%%%%%%%%%%%%%%%%%%%%%%%%%%%%%%%

% individual chapters included here
\chapter{Constructing perfect fluid spheres}\label{chap:bounds}

\section{Introduction}
The most basic model for a star is one in \emph{hydrostatic equilibrium}, a
condition which connects the rate of change of pressure throughout the star
with the density profile. When supplemented with a relation between $\rho$ and
$P$ (a so-called \emph{equation of state}) we can then integrate out the
equations and find both the density and pressure profile. The simplest example
of this method is to assume a fluid with $P = \rho^\gamma$, an example which is
treated in most textbooks \cite{Glen2000,book:Weinberg}. The
disadvantage
of such a method is that we need to know not only general relativity but also a
great deal of nuclear physics at extremely high densities to get realistic
profiles out for compact objects such as neutron stars. If more than one type
of matter is present or different phases exist then the ``equation of state''
used to integrate the density profile is really an equation that ties the
density at a specific radius to the pressure. Hence the function $P(\rho)$ may
not be invertible. Even in the cases that such a relation is invertible it will
generally be very different from the equations of state of the matter making up
the star.\footnote{An extreme example is a constant density star. By combining
different phases of matter it is possible to have a fluid of constant density,
but it would still be compressible (i.e. at other densities the relation $\rho
= \hbox{Const}$ would not be valid), whereas if the relation
$\rho=\hbox{Const}$ was assumed to hold true at all pressures then we would
have a incompressible star and infinite speed of sound. \cite{MTW}} Beyond the
purely technical problems of finding equilibrium pressure and density relations
there is also the problem that we do not have direct experimental evidence for
how matter behaves under such pressures. This is reflected in the difficulty of
finding a strict mass limit for neutron stars and other superdense objects such
as quark stars \cite{Glen2000}.

\paragraph{}
An alternative \emph{algorithmic} approach can be used to generate any
spherically symmetric static solution of Einstein's equation without recourse
to the details of the material that makes it up. For \emph{Newtonian} gravity
the condition of hydrostatic equilibrium is quite simple:
\begin{equation}
\frac{\d P_r}{\d r} = -g(r)\rho(r) = -\frac{m(r)\rho(r)}{r^2} \label{eq:Hydrostatic}
\end{equation}
where $g(r)$ is the local acceleration due to gravity, $m(r)$ is the mass
contained within a radius $r$ and $\rho(r)$ is the density. The mass within a
radius $r$ is simply related to the density:
\begin{equation*}
m(r) = 4\pi\int_0^{\bar{r}}\bar{r}^2 \rho(\bar{r})\,\d \bar{r}.
\end{equation*}
If we tried to use $\rho(r)$ as a known function we have a second order
differential equation to solve:
\begin{equation*}
\frac{\d^2 P_r}{\d r^2} +\left(\frac{2}{r}-\frac{1}{\rho}\frac{\d \rho}{\d r}\right)\frac{\d P_r}{\d r} + 4\pi \rho^2 = 0.
\end{equation*}
We are relatively fortunate in this case, as the differential equation has a
nice solution, albeit in terms of an integral of the density:
\begin{equation}
P_r(r) = \int\frac{C_1 - 4\pi\int r^2\rho(r)\,\d r}{r^2}\rho(r)\,\d r + C_2.\label{eq:NewtonPressure}
\end{equation}
This is what is meant by an \emph{algorithmic procedure}: given an arbitrary
function (in this case $\rho(r)$) we can then use that to generate an exact
solution for the pressure profile \eqref{eq:NewtonPressure} and mass
distribution $m(r)$ (trivial). Because the pressure does not act as a source
in Newtonian gravity we have no way of knowing what the tangential pressure is
doing without some assumption of the nature of matter making up the star.

\paragraph{}
The above algorithm is effective and simple and it will allow us to find the
(radial) pressure profile for \emph{any} time-independent spherically symmetric
Newtonian fluid with no external forces. By allowing arbitrary functions for
$\rho(r)$ we could have solutions that are clearly unphysical, such as any
solution with $\rho(r)$ negative. The advantage of this particular algorithm is
that ensuring that $\rho(r)$ is positive is trivial, as this function is freely
specifiable. The only non-trivial constraint that we can sensibly impose on the
\emph{Newtonian} case is that the pressure remains finite everywhere. That is,
we demand:
\begin{itemize}
\item The exact solution can be integrated out for radial pressure.
\item Any reasonable conditions ($\rho \geq 0$, $P_r$ finite) are easily imposed.
\item The ``generating function'' is not an esoteric function but has direct physical relevance.
\end{itemize}
Unfortunately attempts to generalise this equation to relativistic stars fail
as the generalisation of \eqref{eq:NewtonPressure} is a \emph{non-linear}
second order differential equation. The non-linearity arises for a simple
physical reason; now the pressure also acts as a source of gravity as well.

\paragraph{}
In contrast, let us return to \eqref{eq:Hydrostatic} and now let $g(r)$ be the
freely specifiable function. Simple algebra allows us to find $m(r)$, and
differentiation allows us to find $\rho(r)$ and then $P_r(r)$:
\begin{eqnarray*}
m(r) = r^2 g(r) &=& 4\pi \int_0^r \bar{r}^2 \rho(\bar{r})\,\d\bar{r}\\
 \Rightarrow \rho(r) &=& \frac{2rg(r) + r^2 \frac{\d g}{\d r}}{4\pi r^2}\\
\Rightarrow P_r(r) &=& -\frac{1}{2\pi}\int\frac{g(r)^2}{r}\,\d r - \frac{1}{8\pi}g(r)^2 + P_{r}(0).
\end{eqnarray*}
The differential equation \eqref{eq:Hydrostatic} has roughly the same form for
relativistic stars, but now pressure and density are sources. Despite being
non-linear, it is still a first order differential equation and specifying the
function $g(r)$ allows one to construct the density and pressure profiles. In
both the Newtonian and relativistic cases we would like to ensure that:
\begin{itemize}
\item The energy density is non-negative, and
\item The pressure is finite and non-negative.
\end{itemize}
It is comparatively more difficult to ensure the non-negativity of energy and
pressure in the second algorithm, as it involves choosing a function $g$ and
then computing the density. In the relativistic case we also want to ensure:
\begin{itemize}
\item The compactness $\chi(r) = 2m(r)/r$ is always less than one (no black
holes), and
\item That the various energy conditions are satisfied.
\end{itemize}
These conditions have no sensible analogue for perfect fluid spheres in Newtonian
gravity.

\paragraph{}
The rest of this chapter is devoted to properties of perfect fluid spheres in
general relativity. First an algorithm for perfect fluid spheres analogous to
the one presented here for Newtonian stars is developed. It is then interesting
to ask questions like
\begin{itemize}
\item Can we place bounds on the generating functions $g(r)$ so that we get reasonable solutions?
\item Can we bound the internal compactness $\chi(r)$?
\item Can we find more general bounds than the Buchdahl-Bondi bound $\chi = \chi(R) < 8/9$?
\end{itemize}

\paragraph{}
A few comments are in order on why we have restricted ourselves to looking at
time independent non-rotating fluids, as most stars are radiating (destroying
the time independent condition\footnote{For all reasonable systems involving
finite energy.}) and rotating (which leads to non-static metrics and equatorial
bulges breaking spherical symmetry). 
The effect of radiation can be treated as a perturbation in almost all of the
astrophysically relevant examples, and on \emph{small} timescales is swamped
by the effect of rotation. The problems facing an algorithmic procedure for a
stationary rotating star or planet are pragmatic ones. The first is that there
is no equivalent of Birkhoff's theorem for rotating systems, so we have no
reason for thinking that the exterior solutions to rotating bodies are
described by the Kerr metric \cite[particularly \S 7.1,\S 12.3]{book:Wald}. That does
not stop us from restricting our
attention to only those stationary solutions that \emph{do} join onto the Kerr
metric as a first attack on the problem. However, despite numerous attempts no
solutions of the Einstein equation in \emph{reasonable} matter have been found
that join smoothly onto the Kerr vacuum, although  various
perturbation techniques have been developed
\cite{Hartle:1967,Wiltshire:2003}. Thus the idea of being able to generate all
physical solutions to describe rotating bodies ``algorithmically'' would be
far-fetched at the current time, as not even a single explicit solution is
known. To give a rough estimate of the order of the effects being neglected, a
table of values has been constructed (table \ref{table:aRM}). In a
asymptotically flat, spherically
symmetric system Birkhoff's theorem is the statement that the solution in
vacuum is a piece of the Schwarzschild solution (see \S\ref{sec:birkhoff} for
a proof). The number of degrees of freedom inside matter is also tightly
constrained, making stationary spherically symmetric metrics suitable for solving
algorithmically. We shall see that the algorithm becomes complicated even under
these restrictions, and that a useful algorithm only exists for matter models
that are also isotropic.

\begin{table}[th]
\begin{footnotesize}
\begin{tabular}{ccccccc}
\hline
Name& Mass&Mean radius&$\Delta R/R_{\textrm{eq}}$& $M/R$& $a/R$ & $a/M$\\
\hline
% Earth mass: 6 x 10^{24} kg, radius: 6x10^{6} meters
Earth & 4 mm & $6.4 \times 10^{3}$ km&$3 \times 10^{-6}$& $7\times 10^{-10}$&$3\times 10^{-7}$&429\\
Jupiter&1.4 m&71500 km&$7\times 10^{-2}$ &$2\times 10^{-8}$&$1\times 10^{-5}$&524\\
Sun& $1.5$ km&  695 500 km&$9.1\times 10^{-6}$& $2\times 10^{-6}$& $5\times 10^{-7}$&$0.234$\\
Neutron star&\multicolumn{6}{c}{$M \sim 1.4 M_\odot$, $R \sim 20$ km, period $\sim 1$ s}\\
&2.1 km& 20 km &$\ll 1$& 0.105&$2\times 10^{-4}$& $1.9\times 10^{-3}$\\
\hline
\end{tabular}
\end{footnotesize}
\caption{{\footnotesize A table showing the typical size of rotation effects on astrophysical
bodies in units in which $c = G = 1$. Notice that for bodies with low compactness ($2M/R$) it is possible for
the parameter $a/M \gg 1$. The oblateness, defined as the difference between the
polar and equatorial radius over the equatorial radius, is also given to
indicate departure from spherical symmetry.}}\label{table:aRM}
\end{table}

\paragraph{}
As seen in table \ref{table:aRM}, the specific angular momentum may exceed the
mass of the star or planet, unlike a black hole. An extreme example of such a
violation comes about from considering a planet orbiting a
star. The (orbital) angular momentum can be approximated by Newtonian gravity:
\begin{equation}
L = m_p r^2 \omega = m_p \sqrt{M_{\odot}r},
\end{equation}
where $m_p$ is the mass of the planet and $M_\odot$ is the mass of the star,
and $m_p \ll M_{\odot}$. The specific angular momentum $a$ is then
\begin{equation}
a = \frac{L}{M_\odot+m_p} = \frac{m_p}{M_\odot + m_p} \sqrt{M_{\odot} r}
\approx \frac{m_p}{M_{\odot}} \sqrt{M_{\odot} r}.
\end{equation}
Expressing this in terms of the Schwarzschild radius of the planet and star
gives
\begin{equation}
a = \frac{R_{sp}}{R_{s\odot}}\sqrt{\frac{R_{s\odot} r}{2}},
\end{equation}
or that the ratio of specific angular momentum to mass is given by
\begin{equation}
\frac{a}{M_{\textrm{tot}}} \approx \frac{a}{M_\odot} 
=\frac{R_{sp}}{R_{s\odot}}\sqrt{\frac{2r}{R_{s\odot}}}.
\end{equation}
The orbital radius $r$ is of the order of astronomical units, making this
ratio incredibly large. Taking Jupiter and the Sun gives
\begin{equation}
\left(\frac{a}{M}\right)_{\textrm{Jupiter + Sun}} \approx 21.
\end{equation}
Looking at the ratio $a/M$ for planets alone typically yields enormous values.
\paragraph{}

\section{Setting up the problem}
\subsection{Spherical symmetry}\label{sec:birkhoff}
We start with a spherically symmetric metric and try and see what constraints
this places on the geometry. The presentation given here closely follows
Misner, Thorne and Wheeler \cite{MTW} (\S 23.3). By \emph{spherical symmetry} it is meant that the group of isometries of
spacetime possess $SO(3)$ as a subgroup. It is also assumed that for every
point $x$ except the origin that the action of this subgroup is a
two-dimensional surface, denoted $S(x)$. Naturally the origin, if one exists,
will be mapped to a zero dimensional surface (namely the origin itself for any
rotation). For nonstandard topologies (such as wormholes) it is not necessary
for an origin to exist. Each of these two dimensional surfaces can be given an
induced metric of a sphere
\begin{equation}
\d s^2|_{S(x)} = R(x)^2\d\Omega^2.
\end{equation}
where $\d\Omega^2 = \d \theta^2 + \sin^2\theta\d\phi^2$ is the metric on a
(unit) two-sphere and $R(x)$ is a ``scaling'' function that depends on the
point $x$ of the $r-t$ submanifold. The spheres are a submanifold and are also
\emph{maximally symmetric} implying that we can always separate the coordinates
on a submanifold
\begin{align}
\d s^2&=g_{00}(r,t)\d t^2+ 2g_{01}(r,t)\,\d r \d t + g_{11}(r,t)\,\d r^2 + R(r,t)^2\,\d\Omega^2\\
&=g_{00}\left(\d t + \frac{g_{01}}{g_{00}}\d r\right)^2 + \left(g_{11}-\left(\frac{g_{01}^2}{g_{00}}\right)\right)\d r^2 + R(r,t)^2\,\d\Omega^2
\end{align}
In Misner et al \cite{MTW} they carry out this construction in considerably
more detail but at the cost of assuming that the spacetime is stationary. The
time coordinate can be redefined as $\bar{t}$ in the following way to ensure
the cross-term vanishes. Introduce the (as yet undetermined) function
$f(\bar{t},r)$:
\begin{align}
t &= \bar{t} + f(\bar{t},r)\\ 
\d t &= \left(1+\frac{\partial f}{\partial t}\right)\d\bar{t} + \frac{\partial f}{\partial r}\d r
\end{align}
We then ensure that the cross term vanishes by defining $f(\bar{t},r)$ as
\begin{equation}
f(\bar{t},r) = -\int\frac{g_{01}(t,r)}{g_{00}(t,r)}\,\d r
\end{equation}
We have now got a definition of $\bar{t}$ in terms of $t$, up to an arbitrary constant. The metric now takes the desired form:
\begin{align}
\Rightarrow \d s^2 &= g_{00} \d \bar{t}^2 + \left(g_{11}-\frac{g_{01}^2}{g_{00}}\right)\d r^2 + R(r,t)^2\,\d\Omega^2.
\end{align} 
Finally the radial coordinate is redefined so that $R(x)=\bar{r}$. This means
we can now write the metric in the form (dropping the bars):
\begin{equation}
\d s^2 = -\zeta(r,t)^2\,\dx{t}^2 + \eta(r,t)^2\,\dx{r}^2 +
r^2\,\dx{\Omega}^2\label{eq:spherically_symmetric_metric}
\end{equation}
for some functions $\eta$ and $\zeta$. The squares of the functions have been
introduced to ensure that spacetime has a Lorentzian structure.

\paragraph{}
Mathematically there are lots of things that can go wrong with this
construction globally. The ``timeshift'' function $f(\bar{t},r)$ or the
``radial function'' $R(x)$ may not be invertible, but the implicit function
theorem (almost) guarantees that \emph{locally} we can put the metric in this
form. Notice that the only condition so far is spherical symmetry, the field
equations have not yet been imposed. Doing so in vacuum leads us to Birkhoff's
theorem:
\vfill

\newpage
\begin{theorem}
(Birkhoff's theorem) In a spherically symmetric spacetime the \emph{only}
solution to the Einstein equations (without cosmological constant) in vacuum is
a piece of the Schwarzschild solution.\label{thm:birkoff}
\end{theorem}
\begin{proof}
The field equations state that the Einstein tensor vanishes identically in
vacuum. Calculating the Einstein tensor from the metric in an orthonormal frame
gives the following equations:
\begin{align}
G_{\hat{t}\hat{t}} &= \frac{\left(2 r\frac{\partial \eta}{\partial r} +\eta^3 - \eta \right)}{r^2\eta^3} = 0\\
G_{\hat{r}\hat{r}} &= \frac{\zeta\eta^2-\zeta-2r\frac{\partial\zeta}{\partial r}}{\eta^2\zeta r^2} = 0\\
G_{\hat{t}\hat{r}} &= \frac{2}{\zeta \eta^2 r}\frac{\partial \eta}{\partial t} = 0\\
G_{\hat{\theta}\hat{\theta}} &= 0\\
G_{\hat{\phi}\hat{\phi}} &= G_{\hat{\theta}\hat{\theta}} ,
\end{align}
where the last expression follows from spherical symmetry, not from the requirement that the Einstein tensor vanishes. The angular expressions have not (yet) been given as they are incredibly ugly in their current form; huge simplifications are obtained by solving the other equations first. As a point of notation the Einstein tensor has been evaluated in an orthonormal frame, which will be denoted here and throughout by placing ``hats'' on the indices. 

\paragraph{}
The $G_{\hat{t}\hat{r}}$ equation tells us that the function $\eta$ is independent of time. The $G_{\hat{t}\hat{t}}$ equation can then be explicitly solved for $\eta(r)$:
\begin{equation*}
\eta(r) = \pm \sqrt{\frac{r}{r - 2M}}
\end{equation*}
where $M$ is currently just a (fortuitously named) constant of integration. Substituting $\eta(r)$ back into the $G_{\hat{r}\hat{r}}$ equation gives
\begin{equation*}
\zeta(r,t)^2 = \zeta(r)^2 = A\left(1-\frac{2M}{r}\right)
\end{equation*}
where $A$ is another constant of integration. 

\paragraph{}
The next logical step is introduce the angular components of the Einstein tensor and check that they do indeed vanish:
\begin{align*}
G_{\hat{\theta}\hat{\theta}} &= \frac{\frac{\partial \eta}{\partial r} \zeta - \frac{\partial\zeta}{\partial r} \eta}{r\eta^3\zeta} + \frac{\frac{\partial\eta}{\partial r}\frac{\partial\zeta}{\partial r} - \frac{\partial^2 \zeta}{\partial r^2}\eta}{ \eta^3\zeta}\\
&= 0;
\end{align*}
as it had to be, by virtue of the Bianchi identity. By rescaling the time coordinate the metric can be put into standard Schwarzschild form:
\begin{equation}
\d s^2 = -\left(1-\frac{2M}{r}\right)\,\d t^2 + \frac{1}{1-\frac{2M}{r}}\,\d r^2 + r^2\,\d\Omega^2.
\end{equation}
\end{proof}

\paragraph{}
It should be noted that before applying the Einstein equation we were able to
transform into coordinates that ensured the metric decomposed into spacelike
and timelike parts i.e. $g_{ti} = 0$, $i=r,\theta,\phi$. Hence if the spacetime
is \emph{stationary} (i.e. there is a timelike Killing vector) and spherically
symmetric automatically imply that it is \emph{static} (i.e. that there is a
spacelike hypersurface orthogonal to the orbits of the timelike Killing vector
field). An example of a stationary but non-static case is the Kerr black hole,
as the rotation rate is not changing but the surfaces orthogonal to the
timelike Killing field are not spacelike. In looking at stars we shall always
be looking at stationary cases (and hence static by spherical symmetry) so no
use will be made of this fact in this thesis.

\paragraph{}
The final comment about Birkhoff's theorem is that we know that spherically symmetric spacetimes do not radiate -- even if the entire spacetime is not static! More specifically the vacuum region is forced to be static even if the matter regions are undergoing (spherically symmetric) pulsations.

\subsection{Spherically symmetric matter}

So far we have only looked at the field equations in vacuum. The full Einstein field equations are
\begin{equation}
G_{\mu\nu} = 8\pi T_{\mu\nu}
\end{equation}
where $T_{\mu\nu}$ is the energy-momentum tensor. Performing the necessary incantations of the previous section we take the metric to be of the form
\begin{equation}
\d s^2 = -\zeta(r)^2\d t^2 + \frac{1}{1-2m(r)/r}\,\d r^2 + r^2\,\d\Omega^2.
\label{eq:1.1.metric}
\end{equation}
I am putting in an additional assumption now that the spacetime is stationary,
which is \emph{not} guaranteed inside matter. I have also replaced $\eta(r)$
with the function $m(r)$, a label that will be justified as we look at the field
equations. Calculating the Einstein tensor directly from the metric
\eqref{eq:1.1.metric} in an orthonormal frame to find (where primes denote
differentiation with respect to $r$):
\begin{subequations}\label{eq_group:1.1.field_eq}
\begin{align}
 G_{\hat{t}\hat{t}} &= 8\pi\rho = \frac{2}{r^2}\frac{\dx{m}}{\dx{r}}\label{eq:1.1.field_eq:t}\\
G_{\hat{r}\hat{r}} &= 8\pi P_r = \frac{2}{r\zeta}\left(1-\frac{2m}{r}\right)\zeta^\prime - \frac{2m}{r^3}\label{eq:1.1.field_eq:r}\\
G_{\hat{\theta}\hat{\theta}} &= 8\pi P_t = \frac{1}{\zeta}\sqrt{1-\frac{2m}{r}}\left(\zeta^\prime\sqrt{1-\frac{2m}{r}}\right)^\prime+\frac{\zeta^\prime}{r\zeta}\left(1-\frac{2m}{r}\right)-\frac{1}{r}\left(\frac{m}{r}\right)^\prime\label{eq:1.1.field_eq:theta}\\
G_{\hat{\phi}\hat{\phi}} &= G_{\hat{\theta}\hat{\theta}}\label{eq:1.1.field_eq:phi},
\end{align}
\end{subequations}
where we have used the Einstein field equations only to relate $G$ to the
principal pressures. Again we see that \eqref{eq:1.1.field_eq:phi} holds and is
a direct consequence of spherical symmetry. The label
$m(r)$ is justified as the ``mass inside radius $r$'' by integrating from an
arbitrary but constant $r_0$ to $r$\footnote{The mass ``inside'' a radius $r$
only makes sense if we can define a point $r=0$. In cases that we cannot
define $r=0$ (wormhole throats, for example) it would be better to say the
mass between $r_0$ and $r$).}
\eqref{eq:1.1.field_eq:t}
\begin{equation*}
m(r) = 4\pi\int_{r_0}^r \bar{r}^2\rho(\bar{r})\,\d \bar{r} + m_0,
\end{equation*}
which is the \emph{Hawking quasi-local mass} in spherical symmetry. This is
precisely the equation you would expect for a spherically symmetric mass
\emph{in a Euclidean space}. In a curved space one would anticipate that the
volume form would have to be replaced by $\sqrt{r(r-2m(r))}\,\dx{r}$. This would
define the \emph{proper mass} \cite{book:Wald}, but as gravity itself gravitates
we also have a ``gravitational mass''. In spherically symmetric spacetime the
gravitational mass \emph{exactly} compensates for the smaller measure being used
leading to the na\"{\i}ve formula one would expect in flat space. This result is
\emph{not} generally true, and unless the spacetime is asymptotically simple there
does not appear to be a meaningful way of defining a mass. The asymptotically \emph{flat} case is discussed in chapter 11 of Wald's book \cite{book:Wald}. Note that if we require the geometry to be regular at $r=0$ (which
includes the case where we do not want a black hole) then $m_0=0$. For future reference the local acceleration is given:
\begin{theorem}
The local acceleration measured by a (timelike) observer in a static spherically
symmetric spacetime is purely radial with magnitude $a(r) =
\sqrt{1-2m(r)/r}\;\zeta^\prime(r)/\zeta(r)$.
\end{theorem}
\begin{proof}
We use the fact that the spacetime is static, so that we have a Killing vector
field $k_a = \partial / \partial t$. Let us have an observer who is ``at rest'', meaning that
the four velocity is parallel to the Killing field (i.e. $(r,\theta,\phi)$ is
constant). We choose to normalising the four velocity by
\begin{equation}
g_{a b}V^a V^b = -1\qquad \Rightarrow\qquad V^a = \left(\frac{1}{\zeta}, 0, 0,
0\right),
\end{equation}
which is possible as the motion of the observer is timelike. These conditions imply that $V^a = (\zeta^{-1}, 0, 0, 0) = \zeta^{-1}\; k^a$.

\paragraph{}
We can then calculate the four acceleration as follows:
\begin{align}
a^a &= V^b\nabla_b V^a\\
    &= \frac{1}{\zeta}k^b \nabla_b \left(\frac{1}{\zeta} k^a\right)\\
    &= \frac{1}{\zeta}\left(k^b\nabla_b \frac{1}{\zeta}\right)k^a + \frac{1}{\zeta^2}k^b\nabla_b k^a\\
&= \frac{1}{\zeta}k^b\frac{\partial \zeta^{-1}}{\partial x^b} k^a+ \frac{1}{\zeta^2}k^b\nabla_b k^a.
\end{align}
In this case we know that $\zeta$ is a function of $r$ alone, so $\nabla\zeta^{-1}$ will be pointing in the purely radial direction. As the Killing field $k^a$ is pointing along (and indeed defining!) the timelike direction, the first term must be zero. 
\begin{align*}
\Rightarrow a^a &= \frac{1}{\zeta^2}k^b(\nabla_b k_c)g^{ac} \quad\quad\textrm{(as we have a \emph{metric} tensor)}\\
&=-\frac{1}{\zeta^2}k^b (\nabla_c k_b)g^{ac}\quad\quad\textrm{(Killing's equations)}\\
&= -\frac{1}{2\zeta^2}(\nabla_c k^b k_b) g^{ac} = -\frac{1}{2\zeta^2}(\nabla_c \zeta^2) g^{ac}\\
&= -g^{ac}\frac{\nabla_c\zeta}{\zeta},
\end{align*}
which is zero unless $a=r$. For the radial acceleration we have
\begin{equation}
a^r = g^{rr}\frac{\zeta^\prime}{\zeta}
\end{equation}
which completes the proof.
\end{proof}
Instead of using the acceleration due to gravity directly it is often more convenient to express things in terms of $g(r)$:
\begin{defn}
Let $g(r)$ be related to the $g_{tt}$ component of the metric as follows:
\begin{eqnarray}
g(r) &\equiv& \frac{1}{2}\frac{\d\phantom{r}}{\d r}\ln\left(-g_{tt}\right)\\
&=& \frac{\zeta(r)^\prime}{\zeta(r)}\\
&=& a^rg_{rr}\quad\quad\quad(\textrm{no sum})
\end{eqnarray}
\end{defn}
Alternatively we can write the $g_{tt}$ component of the metric in terms of
$g(r)$ by a simple rearrangement of the definition:
\begin{equation}
g_{tt} = -\exp\left(2\int^r g(\bar{r})\;\d \bar{r}\right).
\end{equation}

\paragraph{}
We can eliminate $\zeta(r)$ from the Einstein tensor in favour of the $g(r)$, a
quantity that has a direct and simple physical interpretation. The
$G_{\hat{r}\hat{r}}$ equation can be rearranged to give
\begin{equation}
g(r) = \frac{m(r) + 4\pi r^3 P_r(r)}{r(r-2m(r))}. \label{eq:1.1.define_g}
\end{equation}
The $G_{\hat{\theta}\hat{\theta}}$ equation takes a little more work. By
substituting $\zeta(r)^\prime = \zeta(r) g(r)$ we find
\begin{align}
8\pi P_t &= \frac{1}{\zeta}\sqrt{1-\frac{2m}{r}}\left((g\zeta)^\prime\sqrt{1-\frac{2m}{r}} - \frac{\zeta g}{\sqrt{1-2m/r}}\left(\frac{m}{r}\right)^\prime\right) \nonumber\\&\quad + \frac{g}{r}\left(1-\frac{2m}{r}\right) - \frac{1}{r}\left(\frac{m}{r}\right)^\prime\nonumber\\
&=\left(\frac{1}{\zeta}(g^\prime\zeta + g\zeta^\prime) + \frac{g}{r}\right)\left(1-\frac{2m}{r}\right) - \left(g+\frac{1}{r}\right)\left(\frac{m}{r}\right)^\prime\nonumber\\
&= \left(g^\prime + g^2 + \frac{g}{r}\right)\left(1-\frac{2m}{r}\right) -  \left(g+\frac{1}{r}\right)\left(\frac{m}{r}\right)^\prime\label{eq:1.1.Pt_and_g}.
\end{align} 
The algorithm for isotropic stars is obtained by taking $P_t = P_r$. When
coupled with the equation of hydrostatic equilibrium (discussed below) the
solution can be found in terms of $g(r)$ quite simply.

\subsection{Hydrostatic equilibrium}
\begin{figure}[tb]
\begin{center}
\includegraphics[width=0.6\textwidth]{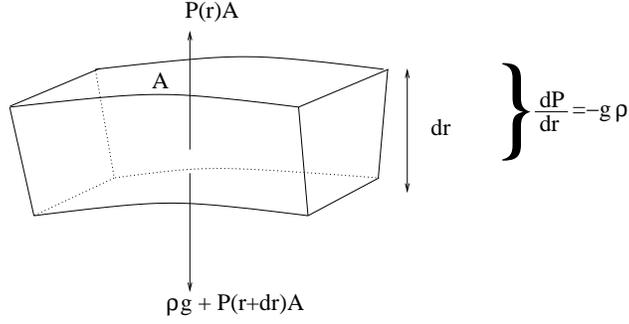}
\caption{The forces acting on a fluid parcel element subject to Newtonian gravity. The tangential forces are ignored as they do not play a role in the equilibrium condition. Here $P$ denotes radial pressure and $A$ denotes the area of the fluid parcel element}\label{fig:NewtonHydrostatic}
\end{center}
\end{figure}
For Newtonian self-gravitating fluids the equations of hydrostatic equilibrium
are obtained easily and are physically motivated (see figure
\ref{fig:NewtonHydrostatic}). For anisotropic relativistic fluids a nicely
motivated approach does not appear to be possible. We start with the expression
for $g(r)$ \eqref{eq:1.1.define_g} and differentiate:
\begin{eqnarray}
g^\prime &=& \frac{m^\prime + 4\pi (r^3P_r)^\prime}{r(r-2m)} - 2(r-m-m^\prime r)\frac{m + 4\pi r^3 P_r}{r^2(r-2m)^2}\\
&=& \frac{m^\prime}{r(r-2m)} + 4\pi\frac{(r^3P_r)^\prime}{r(r-2m)} - 2(r-m-m^\prime r)\frac{g}{r(r-2m)}\\
&=& \frac{1}{r(r-2m)}\bigg\{(1+2gr)m^\prime + 12\pi r^2 P_r + 4\pi r^3P_r^\prime \nonumber\\
&&\quad\quad- 2(r-m)g\bigg\}\label{eq:g_prime0}.
\end{eqnarray}
We can solve the $G_{\hat{\theta}\hat{\theta}}$ equation \eqref{eq:1.1.Pt_and_g}
to find an expression for $g^{\prime}$ as well:
\begin{equation}
g^\prime = \frac{(8\pi P_t)r}{r-2m} + \frac{gr+1}{r-2m}\left(\frac{m}{r}\right)^\prime - g^2 - \frac{g}{r}\label{eq:g_prime}.
\end{equation}
Eliminating $g^\prime$ we solve \eqref{eq:g_prime0} and \eqref{eq:g_prime} for
$P_r^\prime$:
\begin{align}
%4\pi r^3 P_r^\prime &=& g^\prime r(r-2m) - m^\prime(1+2gr) - 12\pi r^2P_r + 2(r-m)g\nonumber\\
4\pi r^3 P_r^\prime &= 8\pi P_t r^2 + (1+gr)\left(\left(\frac{m}{r}\right)^\prime r - m^\prime\right) - m^\prime gr \\
&\quad\quad - g^2r(r-2m) + gr -12\pi r^2 P_r.\nonumber
\end{align}
The derivatives of $m(r)$ are eliminated via the $G_{\hat{t}\hat{t}}$ equation
\eqref{eq:1.1.field_eq:t}:
\begin{align}
4\pi r^3 P_r^\prime =& 8\pi r^2(P_t - P_r) - \frac{m(1+gr)}{r} -4\pi r^3 g\rho\\
&\quad\quad  - g^2r(r-2m) + gr - 4\pi r^2 P_r\nonumber.
\end{align}
 We expand all but one of the factors of $g(r)$ and
collect terms:
\begin{align}
4\pi r^3 P^\prime_r =& -4\pi r^2 P_r - 4\pi r^3 g\rho + 8\pi r^2(P_t - P_r)\\
&\quad\quad +\frac{1}{r(r-2m)}\bigg\{(m+4\pi r^3P_r)r - (m+4\pi r^3 P_r)^2\nonumber\\
&\quad\quad - m(r-m+4\pi r^3P_r)\bigg\}\nonumber.
\end{align}
Some remarkable cancellations in the bracket lead to
\begin{align}
%&=& -4\pi r^2 P_r -4\pi r^3 g \rho+ 8\pi r^2(P_t - P_r) + \frac{1}{r(r-2m)}\left((m+4\pi r^3P_r)((r-m)-4\pi r^3 P_r)- m((r-m)+4\pi r^3P_r)\right)\\
4\pi r^3 P^\prime_r=&-4\pi r^2 P_r -4\pi r^3 g \rho+ 8\pi r^2(P_t - P_r) \\
&\quad\quad + \frac{4\pi r^3 P_r}{r(r-2m)}\left[ 
(r-2m)-(m+4\pi r^3P_r)\right]\nonumber,
\end{align}
or, in its final form
\begin{equation}
\frac{\d P_r}{\d r} = -g (\rho + P_r) + \frac{2(P_t-P_r)}{r}. 
\label{eq:TOVaniso}
\end{equation}
This equation is known as the anisotropic Tolman Oppenheimer Volkoff (TOV)
equation.  In the case of an isotropic fluid we have the more familiar TOV
equation.
\begin{equation}
\frac{\d P_r}{\d r} = \frac{\d P}{\d r} = -g(\rho + P)\label{eq:TOVeq}.
\end{equation}  

%\paragraph{}
%Finally we introduce a measure of anisotropy. There are many ways that one could go about defining such a measure, and the moment we introduce the most ``obvious'' one from the field equations:
%\begin{equation}
%G_{\hat{r}\hat{r}} - G_{\hat{\theta}\hat{\theta}} = 8\pi(P_{r} - P_{t}) = -\frac{r}{\zeta}\sqrt{1-\frac{2m}{r}}\left(\frac{1}{r}\sqrt{1-\frac{2m}{r}}\zeta^\prime\right)^\prime + r\left(\frac{m}{r^3}\right)^\prime.\label{eq:1.1.aniso}
%\end{equation}

\section{Algorithmic construction}
\subsection{Isotropic fluid}
The most difficult thing about the algorithmic constructions is deciding which
functions we are going to eliminate and which we are going to keep. We have a
plethora of functions to choose from : $P \equiv P_r = P_t$, $g$, $\rho$, $m$
and $\zeta$. By referring to \eqref{eq:spherically_symmetric_metric}, we  see
that spherical symmetry allows us to reduce our problem to at most two free
functions . As we are interested in \emph{isotropic fluids} the extra condition
$P_r = P_t$ restricts us to one arbitrary function, the so-called ``generating
function''.

\paragraph{}
While in principle we have one freely specifiable function we are not guaranteed
at this stage that we will be able to find a closed form solution for the
resulting differential equation. If we were to take the mass as a ``generating
function'' then finding $\rho$ is trivial:
\begin{equation}
\rho = \frac{1}{4\pi r^2}\frac{\d m(r)}{\d r}.
\end{equation}
Then we try and construct the pressure profile:
\begin{eqnarray}
\frac{\d P}{\d r} &=& -g(\rho + P)\\
&=& -\left(\frac{m + 4\pi r^3 P}{r(r-2m)}\right)(\rho + P)\\
&=& -\frac{m\rho}{r(r-2m)} - \left(\frac{m + 4\pi r^3\rho}{r(r-2m)}\right)P - \frac{4\pi r^3}{r(r-2m)} P^2.
\end{eqnarray}
This is a non-linear differential equation although it will in general have a
solution, and for some choices $m(r)$ there will be a closed form solution,
there is no closed form solution for \emph{arbitrary} functions $m(r)$. The
culprit is the last term which makes the equations non-linear. Note that in the
Newtonian case the first term would not have $m$ in the denominator and the
other two terms would not exist at all; hence in cases where the star is not too
compact one can neglect the last term, integrate the resulting equation and have
an approximate solution significantly better than the Newtonian solution. This
is not of much use to us as we are looking for exact solutions and bounding the
compactness by the condition that the pressure remains finite; in such limiting
cases the last term becomes dominant rather than negligible!

\paragraph{}
Attempts to generate solutions from the density $\rho(r)$ leads to a very
similar conclusion. To eliminate $m(r)$ we need to solve \eqref{eq:1.1.define_g}
and differentiate now leading to a \emph{second-order} non-linear differential
equation. After these attempts it may be a surprise that any closed form
algorithm exists at all.

\paragraph{}
To actually construct the algorithm we start with the isotropic version of
\eqref{eq:g_prime}:
\begin{equation}
g^\prime = \frac{8\pi P r}{r-2m} + \frac{gr+1}{r-2m}\left(\frac{m}{r}\right)^\prime - g^2 - \frac{g}{r}.
\end{equation}
We choose to write our function in terms of mass and the gravity profiles. We
eliminate $P$ by using the $G_{\hat{r}\hat{r}}$ Einstein equation
\eqref{eq:1.1.field_eq:r}:
\begin{equation}
g^\prime = \frac{g}{r}-\frac{2m}{r^2(r-2m)} + \frac{gr+1}{r-2m}\left(\frac{m}{r}\right)^\prime - g^2.
\end{equation}
This is a Ricatti differential equation in $g$, so again we cannot specify
$m(r)$ and find a general closed form solution. However, when considered as a
differential equation in $m(r)$ we see that it is actually a linear differential
equation:
\begin{equation}
m^\prime = \frac{r^2}{1+gr}\left(g^\prime +g^2 - \frac{g}{r}\right) + \frac{m}{r(1+rg)}\left(3(1+gr)  -2r^2(g^\prime  + g^2)\right).
\end{equation}
As a first order linear differential equation we are able to integrate it directly by introducing an appropriate integrating factor. 

\paragraph{}
The integrating factor (denoted by $\lambda$) can be found by solving the homogeneous differential equation:
\begin{eqnarray}
\lambda^\prime &=& \lambda\left(\frac{3}{r}  -2r\frac{(g^\prime  + g^2)}{1+rg}\right).
\end{eqnarray}
The solution takes the rather simple final form:
\begin{eqnarray}
\lambda &=&  \exp\left(\int\frac{3}{r} -  2\frac{r(g^\prime  + g^2)}{1+rg}\,\d r\right)\\
&=& \exp\left(3\ln |r| - 2\int\frac{g^\prime r + g^2r}{1+rg}\,\d r\right)\\
&=& r^3 \exp\left(-2\int \frac{(gr)^\prime - g + g^2r}{1+rg}\,\d r\right)\\
&=& r^3 \exp\left(-2\ln |1+ rg| +2\int g\frac{1-gr}{1+gr}\,\d r\right)\\
&=& \frac{r^3}{(1+rg)^2}\exp\left(2\int_{r_0}^r g\frac{1-gr}{1+gr}\,\d r\right),
\end{eqnarray}
where in the last integral we have explicitly introduced the constant of
integration $r_0$. It is now a simple matter to obtain a closed form solution
for $m(r)$ (with the constant of integration absorbed into the \emph{indefinite}
integral):
\begin{eqnarray}
m(r) &=& \frac{r^3}{(1+rg)^2}\exp\left(2\int_{r_0}^r g\frac{1-gr}{1+gr}\,\d r\right)\\
&&\quad\quad\times\int \frac{r}{(1+gr)\lambda}\left((gr)^\prime - 2g +g^2r \right) \,\d r\nonumber\\
&=& \frac{r^3}{(1+rg)^2}e^{2\vartheta(r)}\int \frac{1+rg}{r^2}e^{-2\vartheta(r)}\left((gr)^\prime - 2g +g^2r \right) \,\d r\label{eq:algo0},
\end{eqnarray}
where we have introduced the function
\begin{equation}
\vartheta(r) = \int^r_{r_0} g(\bar{r})\frac{1-g(\bar{r})\bar{r}}{1+g(\bar{r})\bar{r}}\,\d\bar{r}.
\end{equation}
Formally we have done what we have set out to achieve; given a generating
function $g(r)$ we can construct the mass profile and from the
$G_{\hat{r}\hat{r}}$ equation we have the pressure profile.

\paragraph{}
We can improve it by eliminating the function $g^\prime$ by an integration by
parts. Performing integration by parts on the offending term:
\begin{align}
\int\frac{1+gr}{r^2}(gr)^\prime e^{-2\vartheta(r)}\,\d r&= \int (gr)^\prime
\frac{e^{-2\vartheta(r)}}{r^2}\,\d r + \frac{1}{2}\int (g^2 r^2)^\prime\frac{e^{-2\vartheta(r)}}{r^2}\,\d r\\
&\hspace{-22ex}= gr \frac{e^{-2\vartheta(r)}}{r^2}+ \frac{1}{2} (g^2
r^2)\frac{e^{-2\vartheta(r)}}{r^2} - \int gr\left(1 + \frac{gr}{2}\right) \left(\frac{e^{-2\vartheta(r)}}{r^2}\right)^\prime\,\d r\\
&\hspace{-22ex}= \frac{g}{r}e^{-2\vartheta(r)}\left(1 + \frac{gr}{2}\right)
+ \int \frac{g}{r}\left(1 + \frac{gr}{2}\right) \left(\frac{2}{r} + 2g\frac{1-gr}{1+gr}\right)e^{-2\vartheta(r)}\,\d r\\
&\hspace{-22ex}= \frac{g}{r}e^{-2\vartheta(r)}\left(1 + \frac{gr}{2}\right)
+ \int \frac{g}{r^2}\frac{2 + gr}{1+gr} \left(1-gr\right)^2e^{-2\vartheta(r)}\,\d r.
\end{align}
After a little bit of algebra we find that $m(r)$ is given by
\begin{equation}
m(r) = \frac{gr^2}{(1+rg)^2}\left(1 + \frac{gr}{2}\right) + \frac{r^3}{(1+rg)^2}e^{2\vartheta}\int \frac{2g^2(g^2 r^2 - 3)}{r(1+rg)}e^{-2\vartheta(r)}\,\d r\label{eq:algo1},
\end{equation}
where there is still a constant of integration to be determined.

\paragraph{}
For stars with finite central pressure the integrand in \eqref{eq:algo1} is well behaved. Sufficiently close to the centre of the star we can make the following approximations:
\begin{equation}
m(r) \approx \frac{4}{3}\pi r^3 \rho_c,\quad\quad P(r) \approx P_c
\end{equation}
which implies
\begin{equation}
g(r) \approx\frac{4}{3}\pi r \frac{\rho_c + 3 P_c}{1-8\pi r^2 \rho_c/3}\approx \frac{4}{3}\pi r(\rho_c + 3P_c)\label{eq:approx_g_small_r}
\end{equation}
i.e. the gravity in well behaved stars is linear at small $r$, making the
integrand well behaved.
\paragraph{}
Another ``well behaved'' algorithm comes from considering a different integration by parts in \eqref{eq:algo0}:
\begin{eqnarray}
&&\int \frac{1+rg}{r^2}\left((gr)^\prime - 2g +g^2r \right)e^{-2\vartheta} \,\d r\\
&&\quad\quad = \int (1+rg)\left(\frac{g^\prime}{r} - \frac{g}{r^2} +\frac{g^2}{r} \right)e^{-2\vartheta} \,\d r\\
&&\quad\quad = \int \left(\left(\frac{g}{r}\right)^\prime + gg^\prime - \frac{g^2}{r} + (1+rg)\frac{g^2}{r} \right)e^{-2\vartheta} \,\d r\\
&&\quad\quad = \int \left(\left(\frac{g}{r}\right)^\prime + \frac{1}{2}(g^2)^\prime  + g^3 \right)e^{-2\vartheta} \,\d r\\
&&\quad\quad = \frac{g}{r}e^{-2\vartheta} + \frac{1}{2}g^2e^{-2\vartheta}\\
&&\quad\quad\quad\quad\quad +2\int\left[ g\left(\frac{1-gr}{1+gr}\right)\left(\frac{g}{r}+ \frac{1}{2}g^2\right)e^{-2\vartheta} + \frac{g^3}{2} e^{-2\vartheta}\right]\,\d r\nonumber\\
&&\quad\quad = \frac{g}{r}e^{-2\vartheta}\left(1 + \frac{gr}{2}\right) +2\int \frac{g^2}{r}\left(\frac{1}{1+gr}\right)\bigg\{\left(1+ \frac{gr}{2}\right)(1-rg)\nonumber\\
&&\quad\quad\quad\quad\quad\quad\quad\quad\quad\quad\quad\quad  
+ \frac{gr}{2}(1+gr)\bigg\}e^{-2\vartheta}\,\d r\\
&&\quad\quad = \frac{g}{r}e^{-2\vartheta}\left(1 + \frac{gr}{2}\right) +2\int \frac{g^2}{r(1+rg)}e^{-2\vartheta}\,\d r,
\end{eqnarray}
which leads to the algorithm \cite{Martin:2003jc}:
\begin{align}
m(r) &= \frac{gr^2}{(1+rg)^2}\left(1 + \frac{gr}{2}\right) +\frac{2r^3}{(1+rg)^2}e^{2\vartheta} \int \frac{g^2}{r(1+rg)}e^{-2\vartheta}\,\d r\\
&= \frac{gr^2}{(1+rg)^2}\left(1 + \frac{gr}{2}\right) +\frac{2r^3 e^{2\vartheta}}{(1+rg)^2}\left(C_{r_0} +\int_{r_0}^r \frac{g^2 e^{-2\vartheta}}{r(1+rg)}\,\d r\right),
\label{eq:ALGO2}
\end{align}
where the constant of integration has been put in explicitly. This algorithm is slightly better than \eqref{eq:algo1} insofar as $g$ only enters as a quadratic rather than a quartic. To compute the pressure we simply use the $G_{\hat{r}\hat{r}}$ equation:
\begin{equation}
p(r) = {1\over8\pi[1+r\;g(r)]^2} 
\left[
- g^2 - 2[1+2rg] e^{2\vartheta}
\left\{ C_{r_0} 
+ \int {2g^2\over r(1+rg)} \; e^{-2\vartheta} \; \d r
\right\}
\right].
\end{equation}

\subsection{Anisotropic fluid}
We now consider the more general case of an anisotropic fluid: the radial
pressure $P_r$ is not required to be the same as the tangential pressure $P_t$.
This case is a little more artificial as we now have the freedom of \emph{both}
arbitrary functions of $r$ allowed in the metric by the constraints of a static
spherically symmetric spacetime. This tends to make the results of this section
somewhat ``artificial'' and so the derivations are done in less detail.

\paragraph{}
We are able to tell about the gross features of anisotropy by looking back at
the anisotropic TOV equation \eqref{eq:TOVaniso}:
\begin{equation}
\frac{\d P_r}{\d r} = -g (\rho + P_r) + \frac{2(P_t-P_r)}{r}. 
\end{equation}
The first term is non-positive by the NEC ($|P_r| < \rho$,
$\rho>0$).\footnote{We cannot assume that $P_r \geq 0$. For example, we can have
an empty solid sphere immersed in an atmosphere. The radial pressure is pushing
inward ($P_r < 0$), but the rigidity of the
surface creates a tangential pressure which can support the object against
collapse.} The radial pressure is zero at the surface of the star by definition,
and in the isotropic case the derivative shows that $P_r$ comes to zero from
above so $P_r \geq 0$. The more negative $\d P_r/\d r$ is the larger the central
pressure, and hence we have:
\begin{itemize}
\item $P_t > P_r$: 
One would expect this condition to hold in \emph{rigid solids}. The tangential
pressure would help the solid retain its shape. Thus less radial pressure
would be required to stop collapse. The fall off of $P_r$ as the surface is
approached will be gentler, or equivalently the central pressure will be
lower, than an isotropic star of the same radius and mass. We can see this
mathematically by looking at \eqref{eq:TOVaniso} and seeing that the second term
is positive.\\
One may have $P_t$ slightly larger than $P_r$  in ``rocky'' planets such as the Earth. 
Astrophysically compact objects such as neutron stars and white dwarfs may form crusts that would allow the tangential pressure to exceed the radial pressure. The most extreme example of this effect would be a Dyson sphere, which relies on the tangential pressure over the surface of the sphere for its stability. As the central pressure is \emph{lowered} these stars are able to support a higher maximum mass than their isotropic counterparts.
\item $0 < P_t < P_r$:
This condition would occur in fluids with \emph{weak} surface tension; gravity
is still needed to hold the fluid together. Most ``normal'' stars would satisfy
this condition, although the effect will typically be negligible especially
compared to rotation. As a consequence of the surface tension the radial
pressure profile must be steeper and hence the central pressure is greater than
its isotropic counterpart.
\item $P_t < 0$:
This is the case for fluids with strong surface tension, or so-called
\emph{self-bound matter} (see chapters 8 and 12 of \cite{Glen2000}). These are
forms of matter that do not require the gravitational force to be bound, such as
small water droplets. The radial pressure in such cases does not have to go
smoothly to zero at the surface. The radial pressure profile will be pushed up
further in comparison to the isotropic case, so the central pressure will also
be higher.\\
We have experimental evidence of objects satisfying this condition but they are
all small such as raindrops. There is some speculation that hybrid or quark
stars may satisfy this condition; this is an assertion that the nuclear force is capable of binding the
star even if gravitational forces were absent. However there is no observational
evidence of any astrophysically relevant body satisfying this condition.
\end{itemize}
These cases show the qualitative difference that introducing anisotropy makes.
One needs to be careful as it is possible to construct solutions where neither
the radial pressure or the tangential pressure is dominant throughout the star.
The requirement of spherical symmetry requires that $P_r = P_t = P_c$ at the
centre of the star as well, so in particular we cannot have a static solution
where $P_t < 0$ all the way through the star.

\paragraph{}
Although the effect of anisotropy will typically be much less than that of rotation in stars, for completeness the problem is addressed. Three separate algorithms are discussed:

\begin{itemize}
\item By using $g(r)$ and the \emph{anisotropy function} $(P_r - P_t)$ to generate the mass and pressure profiles.
\item By using $g(r)$ and the tangential pressure to generate new solutions.
\item By using $g(r)$ and $\delta(r) = P_t / P_r$.
\end{itemize}

\subsection{The anisotropy function solution}
This solution is perhaps the most ``direct'' attack as we simply integrate the
first order linear differential equation. With the integrating factors somewhat
simplified we obtain
\begin{eqnarray}
m(r) &=& \frac{8\pi r^3}{(1+rg(r))^2}e^{2\vartheta(r)}\int \frac{1+rg(r)}{r}(P_r - P_t)e^{-2\vartheta(r)}\,\dx{r}\label{eq:afs}\\
&&\quad\quad + \bigg\{g(r)r^2\frac{(1+\frac{1}{2}rg(r))}{(1+rg(r))^2}\nonumber\\&&\quad\quad + 
\frac{\exp(2\vartheta(r))r^3}{(1+g(r)r)^2}\left(C + 2\int \frac{g(r)^2}{r(1+rg(r))}\exp(-2\vartheta(r))\right)\bigg\}\nonumber\\
&=& m_{\textrm{aniso}} + m_{\textrm{iso}}.
\end{eqnarray}
In the last line we have written the mass as the sum of \emph{an} isotropic mass
and an anisotropic mass. The isotropic mass is defined to be the mass of a star
with generating function $g(r)$ but with $P_r = P_t$ throughout. The anisotropic
mass is defined as whatever is left over. By comparison with our algorithm for
isotropic fluids \eqref{eq:ALGO2} we see that the first term in \eqref{eq:afs}
is the anisotropic mass and the second term is the isotropic mass.

\paragraph{}
An alternative definition of an isotropic mass
would be the mass profile that generated $P_r$, under the constraint that $P_t =
P_r$. While this is a legitimate construction [although not able to be solved in
closed form in general] it is distinct from the one I have chosen to use. It is
important to note that the pressure profile corresponding to the isotropic
mass profile the way I have defined it will \emph{not} give the same $P_r$.

\paragraph{}
Some general comments about this solution:
\begin{itemize}
\item It is very easy to find the isotropic limit as this function decomposes
into an isotropic and anisotropic part.
\item Care must be used when saying the ``isotropic part''. What I am asserting
is that this is the mass generated by an isotropic fluid sphere by the function
$g(r)$. I am \emph{not} asserting that $P_{r}(\hbox{isotropic}) =
P_{r}(\hbox{anisotropic})$.
\item The function $(P_r - P_t)$ is a difficult thing to determine
experimentally. Theoretically the best limit I can place on it is $2\rho$ by the
DEC.
\item As our boundary condition we can choose $\lim_{r\rightarrow
0}\frac{m(r)}{r^3} = \frac{4}{3}\pi\rho_c$. We also choose the integrals to have
lower limits of integration at zero, allowing evaluation of the constant of
integration:
\begin{equation}
\lim \frac{m(r)}{r^3}= \frac{4}{3}\pi\rho_c = \lim_{r\rightarrow 0}\frac{g(r)}{r} + C.
\end{equation}
With the expansion of $g$ for small $r$ given by \eqref{eq:approx_g_small_r} we find that $C=-4\pi p_c$. The advantage of this boundary condition is we do not need to know anything about the anisotropy to calculate it. (For a look at other types of boundary conditions on \emph{isotropic} stars see appendix \ref{app:bounds_paper}).
\item The positivity of the mass and the isotropic mass allows us to
(trivially) state
\begin{equation}
-m_{\textrm{aniso}}\leq m_{\textrm{iso}}
\end{equation}
for realistic stellar models. Note that this is not guaranteed by the
algorithm, only by our prejudice for what constitutes physically reasonable
solutions. 
\end{itemize}

\subsection{The tangential pressure solution}
Using the $G_{\hat{r}\hat{r}}$ equation in terms of $g$ we have
\begin{equation}
\frac{8\pi P_rr^2}{1+rg(r)} = \frac{2rg(r)}{1+rg(r)}-\frac{2m(r)}{r}\left(\frac{1+2g(r)r}{1+g(r)r}\right).
\end{equation}
Using this we now have
\begin{equation}
m(r) = \frac{r^3}{I(r)}\left(\int\frac{g(r)^2r + g^\prime(r)r - g(r) -8\pi P_t r}{r^2(1+rg(r))}I(r)\,\dx{r}+C_2\right)
\end{equation}
where the integrating factor $I(r)$ is given by the rather unsavoury expression
\begin{equation}
I(r) = \exp\left(2\int\frac{2rg(r) + 1 + r^2g(r)^2 + g^\prime(r)r^2}{r(1+g(r)r)}\,\dx{r}\right)
\end{equation}
An integration by parts simplifies the integrating factor considerably:
\begin{equation}
I(r) = r^2(1+rg(r))^2\exp\left(2\int\frac{g(r)^2r}{1+rg(r)}\,\dx{r}\right).
\end{equation}
Yet another integration by parts will get rid of the derivative of $g(r)$ in the main expression, as well as some algebra and actually doing one of the integrals (!) we finally obtain:
\begin{eqnarray}
m(r) &=& \frac{r}{2} + \frac{r}{(1+rg(r))^2}\exp\left(-2\int\frac{g(r)^2r}{1+rg(r)}\,\dx{r}\right)\\
&&\times\left\{C_3 - 8\pi\int
P_t(1+g(r)r)r\exp\left(2\int\frac{g(r)^2r}{1+rg(r)}\,\dx{r}\right)\right\}\nonumber
\end{eqnarray}
\begin{itemize}
\item This solution is fairly ``tidy''
\item Finding isotropic solutions is messy (eg. substitute for $P_t$, use the $G_{\hat{r}\hat{r}}$ equation to find $P_r$ and then use this for new $P_t$. Rinse, lather, repeat.)
\item Is $P_t$ easily measurable? Can it be estimated realistically? This is a
difficult question, although I suspect that the answer is no.
\item To guarantee that $2m(r)/r < 1$ the term in curly braces must be negative.
\item Boundary condition at $r=0$: $m(r) \approx \frac{4\pi}{3}\rho_c r^3$ implies that $C_3 = -\frac{1}{2}$. (All top level integrals run from zero to r, nested run from zero to integration variable).
\item Setting the lower bound of integration of the integrating factor at infinity and using the boundary condition that $m(r \geq R) = M$ and $P(r \geq R) = 0$ gives 
\begin{align}
M &= \frac{r}{2} + C_3\frac{r(r-2M)^2}{(r-M)^2}\exp\left(2\int_r^\infty\frac{M^2}{\bar{r}(\bar{r}-2M) (\bar{r}-M)}\,\dx{\bar{r}}\right)\\
&= \frac{r}{2} + C_3\frac{r(r-2M)^2}{(r-M)^2}\left(\frac{(r-M)^2}{r(r-2M)}\right)\\
&= \frac{r}{2} + C_3(r-2M) 
\end{align}
which gives the simple condition $C_3 = -\frac{1}{2}$. 
\item An easy inequality comes from requiring that $2m(r)/r < 1$ throughout the fluid:
\begin{equation}
\left(C_3 - 8\pi\int P_t(1+g(r)r)r\exp\left(2\int\frac{g(r)^2r}{1+rg(r)}\,\dx{r}\right)\right) < 0,
\end{equation}
It is less useful than one might hope as it involves an integration.
\end{itemize}

\subsection{The $\delta$ solution}
Consider the equation
\begin{eqnarray}
m^\prime(r) &=& \frac{2g(r)r(1-\delta)}{1+g(r)r} + \frac{(g(r)^2r + g^\prime(r)r - g(r))r}{1+g(r)r}\\
&&\quad\quad + m\left(\frac{3}{r} - 2r\frac{g(r)^2 + g^\prime(r)}{1+g(r)r} - \frac{2}{r}\frac{(1+2g(r)r)}{1+g(r)r}(1-\delta)\right)\nonumber\\
&=& \frac{(1-2\delta)g(r) + g(r)^2r + g^\prime(r)r}{1+g(r)r}r \\
&&\quad\quad + m\left(\frac{3}{r} - \frac{2}{r}\left(\frac{g(r)^2r^2 + (g(r)r^2)^\prime + (1-\delta)-2\delta g(r)r}{1+g(r)r}\right)\right)\nonumber
\end{eqnarray}
with $\delta(r) = P_t/P_r$.

\paragraph{}
The integrating factor can be written down in closed form, but is long. The
function $f$ is introduced to shorten the notation
\begin{equation}
\exp\left(2\int f(r)\,\dx{r}\right) \equiv \exp\left(2\int\left[g(r)\frac{g(r)r+1-2\delta(r)}{1+g(r)r}+\frac{1-\delta(r)}{r(1+g(r)r)}\right] \,\dx{r}\right).
\end{equation}
The solution for $m(r)$ is given by
\begin{equation}
m(r) = \frac{g(r)r^2(1+r g(r)/2)}{(1+g(r)r)^2} + \frac{r^3e^{-2\int f\,\d r}}{(1+g(r)r)^2}\left(\int\frac{g(r)^2(1+\delta(r))}{r(1+g(r)r)}e^{2\int f\,\dx{r}} + C_4\right).
\end{equation}
This is an algorithm, although not easy to use or find properties of. In
particular, the most outstanding features of this algorithm are:
\begin{itemize}
\item Problems at the surface as, although $P_r$ vanishes $P_t$
typically will not, implying that $\delta(R)=\infty$.
\item The integrating factor is rather cumbersome as it involves
both $g$ and $\delta$.
\item However, by letting $\delta(r)=1$ identically we easily recover the
isotropic formula.
\end{itemize}

\subsection{Conclusions}\label{sec:algo_conclusions}
Algorithms for generating the solutions for both the isotropic and anisotropic
cases have been developed. In the isotropic case we are able to get a solution
by supplying $g(r)$, while in the anisotropic case we have to supplement this
with an additional function. If anything these algorithms work \emph{too} well,
as we can easily pick functions that lead to unphysical results. Specifically we
would like to ensure
\begin{itemize}
\item The energy density $\rho$ is non-negative
\item The dominant energy conditions ($\rho > |P_r|$ and $\rho > |P_t|$) are satisfied.
\item The central pressure is finite.
\item That the compactness $\chi(r)<1$ so that a black hole is not sitting inside our star.
\end{itemize}
With the exception of ensuring that the central pressure is finite, it is
difficult to see how these conditions could be satisfied without picking $g(r)$,
calculating the mass, density and pressure profiles and then explicitly checking
that all of the above conditions are satisfied. In the next section we start
looking at bounds which restrict the nature of ``physical'' solutions. The
manner in which the bounds are obtained are reasonably general but have the
disadvantage of not constructing valid solutions that saturate the bounds, or
indeed demonstrate that the bounds can be saturated at all! Notice that we have
to invoke energy conditions in the case $P_t > P_r$ as an arbitrarily low
central pressure could otherwise be obtained by letting $P_t$ become high
enough.

\paragraph{}
It is interesting to note is that it is commonly claimed that the TOV equation
\eqref{eq:TOVeq} tells us that pressure contributes as a \emph{source} of
gravity, and so the pressure is higher than in Newtonian stars of the same
density profile. This ``regeneration of pressure'' effect leads to a point where
the gravitational attraction of the pressure is greater than the repulsion,
hence the star is forced to collapse on itself. It is claimed that this
mechanism is responsible for the Buchdahl--Bondi limit to be discussed in more
detail in the next section. This argument is elegant and persuasive, but a
glance at \eqref{eq:TOVaniso} shows us that it is seriously misleading. If the
radial and the tangential pressures are allowed to be different we see that it
is the radial pressure that acts as a source and that the tangential pressure
(if positive) actually \emph{reduces} the effect of gravity. We have also
constructed an algorithm that uses $P_t$ as free data, hence seeing pressure as
contributing as an attractive source of the gravitational field is too
na\"{\i}ve. 

\section{Compactness bounds}

It has been well known for years that the compactness of a static 
spherically symmetric fluid $\chi(R) \equiv 2M/R$ is bounded above by $8/9$, a
result known as the Buchdahl--Bondi bound \cite{Bondi,Buchdahl}. The result is
significant as it shows that a star cannot get arbitrarily close to forming a
black hole ($\chi = 1$). No such bound exists in the Newtonian case; although
the concept of a black hole does not exist\footnote{There are ``dark stars'' in
Newtonian relativity where the escape velocity is at least the speed of light.
However these stars are qualitatively very different; as the star can be a
fluid free of singularities. While the light cannot escape to infinity, it will
be able to propagate out to some finite distance $r_c$. Observers with $r<r_c$
will see a perfectly normal star, and may be unaware that it is ``dark'' to
distant observers.} we can still require that the central pressure remains
finite. The condition for hydrostatic equilibrium in a Newtonian star is
\begin{equation}
\frac{\d P}{\d r} = -g(r)\rho(r),\quad\quad g(r) = \frac{m(r)}{r^2} = \frac{4}{3}\pi r \bar{\rho},\qquad\textrm{(Newtonian)}
\end{equation}
which is a linear differential equation for $P(r)$. If the density is finite
throughout the star then the average density and $g(r)$ are also finite, hence
for stars that have a finite radius $R$ will also have finite central pressure.
We can add a \emph{partial} general relativistic correction onto $g(r)$:
\begin{equation}
\frac{\d P}{\d r} = -g(r)\rho(r),\quad\quad g(r) = \frac{m(r) + 4\pi r^3 P(r)}{r^2(1-2m/r)}, \qquad\textrm{(Post-Newtonian)}
\end{equation}
which still leaves a linear partial differential equation for $P(r)$. The
solution to this equation is 
\begin{eqnarray}
P(r) &=& \exp\left(-4\pi\int^r \frac{r\rho(r)}{1-2m(r)/r}\,\d r\right)\\
&&\quad\times\left\{C_0 -\int_0^r \frac{m(\bar{r})\rho(\bar{r})}{\bar{r}(\bar{r}-2m(\bar{r}))}\exp\left(4\pi\int^{\bar{r}} \frac{x\rho(x)}{1-2m(x)/x}\,\d x\right)\,\d\bar{r}\right\}\nonumber.
\end{eqnarray}
It is now obvious that we cannot have $\chi(r) = 1$ anywhere in the star, as the
integrands will blow up. A calculation of a constant density star shows that the compactness can get arbitrarily close to one. The condition of constant density gives us the density and mass profiles easily:
\begin{equation}
\rho = \rho_0, \quad\quad m(r) = \frac{4}{3} \pi r^3 \rho_0.
\end{equation}
The integrating factor can be expressed in terms of elementary functions:
\begin{eqnarray}
\exp\left(\int^r \frac{4 \pi r\rho(r)}{1-2m(r)/r}\,\d r\right) &=& 
\exp\left(4\pi \int^r \frac{r\rho_0}{1-\frac{8}{3}\pi r^2\rho_0}\right)\\
&=& \exp\left(\frac{3}{4}\int^{r=\sqrt{3\chi/8\pi \rho_0}}\frac{1}{1-\chi}\,\d\chi\right)\\
&=& \left(1 - \chi(r)\right)^{3/4}\\
&=& \left(1 - \frac{8}{3}\pi r^2 \rho_0\right)^{3/4},
\end{eqnarray}
which implies
\begin{eqnarray}
P(r) &=& \left(1 - \chi(r)\right)^{-3/4}\left\{C_0 - \frac{4\pi(\rho_0)^2}{3}\int_0^r r \left(1-\frac{8}{3}\pi r^2 \rho_0\right)^{-1/4}\,\d r\right\}\nonumber\\
&=& \left(1 - \chi(r)\right)^{-3/4}\left\{C_0 - \frac{4\pi(\rho_0)^2}{3}\int_0^{\chi(r)}  \left(1-\chi\right)^{-1/4}\,\frac{3\d \chi}{16 \pi \rho_0}\right\}\\
&=&  \left(1 - \chi(r)\right)^{-3/4}\left\{C_0 - \frac{\rho_0}{4}\int_0^{\chi(r)}  \left(1-\chi\right)^{-1/4}\,\d \chi\right\}\\
&=& C_0(1-\chi(r))^{3/4} -  \frac{\rho_0}{3}\\
&=& C_0\left(1-\frac{8}{3}\pi \rho_0 r^2\right)^{3/4} -  \frac{\rho_0}{3} ,
\end{eqnarray}
where the last equation makes reminds us that we cannot choose $\chi(r)$ arbitrarily as we are looking at a constant density solution. At the surface of the star $P = 0$ and we have:
\begin{equation}
P(R) = 0 =\left(P_c + \frac{\rho_0}{3}\right) (1-\chi)^{3/4} - \frac{\rho_0}{3}
\end{equation}
which implies that the compactness $\chi = \chi(R)$ is given by
\begin{equation}
\chi = 1-\left(\frac{\rho_0}{3P_c +\rho_0}\right)^{4/3}
\end{equation}
which shows that we may have the compactness as close to one as desired by allowing $P_c/\rho_0$ to become large (but still finite).

\paragraph{}
When we look at the full TOV equation with pressure as a ``source'' we obtain a non-linear differential equation. We will find that in the case of isotropic stars with a non-increasing density profile that the central pressure will blow up as the Buchdahl--Bondi limit is approached. This effect is sometimes called regeneration of pressure, although this term has to be used carefully (cf \S \ref{sec:algo_conclusions}).

\subsection{Generalising Buchdahl--Bondi: $P_r \geq P_t$}

In this section we consider a star subject to two restrictions:
\begin{itemize}
\item The radial pressure $P_r$ is never less than the tangential pressure $P_t$.
\item The \emph{averaged density} defined by
\begin{equation}
\bar{\rho} = \frac{3m(r)}{4\pi r^3}
\end{equation}
is a non-increasing function of $r$.
\end{itemize}
The original derivation of the Buchdahl--Bondi limit stated that $\chi < 8/9$ for isotropic fluids with non-increasing density profiles. That derivation is quite complicated and difficult to extend to more general results. More modern proofs, such as the proof given in Wald \cite{book:Wald}, are very easy to generalise to $P_r \geq P_t$. As a smaller tangential pressure means that more radial pressure is required to achieve hydrostatic equilibrium, one would expect on physical grounds that the Buchdahl--Bondi bound would hold in this case as well. 

\paragraph{}
Rearranging expressions \eqref{eq:1.1.field_eq:r} and \eqref{eq:1.1.field_eq:theta} and using the assumption that $P_r \geq P_t$ we have:
\begin{align}
8\pi(P_{r} - P_{t})&= -\frac{r}{\zeta}\sqrt{1-\frac{2m}{r}}\left(\frac{1}{r}\sqrt{1-\frac{2m}{r}}\zeta^\prime\right)^\prime + r\left(\frac{m}{r^3}\right)^\prime\geq 0.
\end{align}
The first term is bounded above:
\begin{align}
r\left(\frac{m}{r^3}\right)^\prime &\geq \frac{r}{\zeta}\sqrt{1-\frac{2m}{r}}\left(\frac{1}{r}\sqrt{1-\frac{2m}{r}}\zeta^\prime\right)^\prime\label{eq:bb0}
\end{align}
But here the left hand side is proportional to $\d\bar{\rho}/\d r$ which is non-positive by hypothesis. The bound \eqref{eq:bb0} requires that the rhs is non-positive. But all the quantities outside of the derivative are positive, so we have the result that the derivative must be non-positive:
\begin{align}
\dr\left(\frac{1}{r}\sqrt{1-\frac{2m}{r}}\frac{\dx{\zeta}}{\dx{r}}\right) &\leq 0.\label{eq:1.1.pr_bigger_pt_inequality}
\end{align}
As the quantity inside the brackets decreases monotonically with $r$, it is bounded below by its value at the surface of the star, $r=R$.\footnote{In fact we know that if $r_1 < r_2$ then $( )_1 > ( )_2$. But at the surface we know the geometry joins smoothly onto the Schwarzschild spacetime and can evaluate one of these bounds.} i.e.
\begin{align}
\frac{1}{R}\sqrt{1-\frac{2M}{R}}\frac{\dx{\zeta}}{\dx{r}}\bigg\vert_{r=R}&\leq\frac{1}{r}\sqrt{1-\frac{2m}{r}}\frac{\dx{\zeta}}{\dx{r}}\bigg\vert_{r=r}.
\end{align}
For all $r\geq R$ Birkhoff's theorem tells us that the geometry is (locally) the Schwarzschild solution. The requirement that the first derivative is continuous allows us to evaluate the left hand side. After a rearrangement we obtain
\begin{align}
\frac{\dx{\zeta}}{\dx{r}}\bigg\vert_{r=r} &\geq \frac{M}{R^3}\frac{r}{\sqrt{1-\frac{2m}{r}}}\label{eq:1.2.diff_eq_zeta}\\
\zeta\vert_{r=R} - \zeta\vert_{r=0} &\geq \frac{M}{R^3}\int_0^R\frac{r}{\sqrt{1-\frac{2m}{r}}}\dx{r}\\
\zeta(0) &\leq \sqrt{1-\frac{2M}{R}} - \frac{M}{R^3}\int_0^R\frac{r}{\sqrt{1-\frac{2m}{r}}}\dx{r}\label{eq:1.2.bound_zeta}.
\end{align}
While we cannot perform the final integral as we do not know $m(r)$, we \emph{can} bound as we know the average density is non-increasing: 
\begin{equation}
m(r)\geq\frac{M}{R^3}r^3
\end{equation}
which allows us to bound the integral in \eqref{eq:1.2.bound_zeta}
\begin{equation}
\int_0^R\frac{r}{\sqrt{1-\frac{2m}{r}}}\dx{r} \geq \int_0^R\frac{r}{\sqrt{1-\frac{2M}{R^3}r^2}}\dx{r}.
\end{equation}
The final integral can now be performed easily
\begin{equation}
0\leq\zeta(0)\leq\frac{3}{2}\sqrt{1-\frac{2M}{R}} - \frac{1}{2}.
\end{equation}
The lack of horizons can only be satisfied if $\zeta > 0$ throughout the spacetime, which is impossible if $\chi = 2M/R > 8/9$. When we consider the places where \emph{inequalities} entered the argument we see that there are three places: initially $P_r \geq P_t$ which is saturated by an isotropic fluid, $\d \bar{\rho}/\d r \leq 0$ in obtaining \eqref{eq:1.1.pr_bigger_pt_inequality} which is saturated by $\bar{\rho}^\prime = 0$ and the bound on the mass function $m(r) \geq Mr^3/R^3$ which is also saturated by $\bar{\rho}^\prime = 0$. Thus the constant density or \emph{interior Schwarzschild} solution saturates all these bounds and in this case we have
\begin{equation}
\zeta(0) = \frac{3}{2}\sqrt{1-\frac{2M}{R}} - \frac{1}{2}
\end{equation}
demonstrating that $\chi < 8/9$ is the best limit that we can achieve \emph{without invoking any energy conditions}.

\paragraph{}
To extend this result, we can integrate \eqref{eq:1.2.diff_eq_zeta} from $r$ to $R$. This will give us information about $\zeta(r)$ anywhere inside the star. Repeating the above calculations we find:
\begin{align}
\zeta(r) &\leq \sqrt{1-\frac{2M}{R}} - \frac{M}{R^3}\int_r^R\frac{r}{\sqrt{1-\frac{2m}{r}}}\dx{r}\\
&\leq \sqrt{1-\frac{2M}{R}} + \frac{1}{4}\left(2\sqrt{1-\frac{2M}{R}u}\bigg|^{u=1}_{u=(r/R)^2}\right)\\
&= \frac{1}{2}\left(3\sqrt{1-\frac{2M}{R}} - \sqrt{1-\frac{2M}{R}\left(\frac{r}{R}\right)^2}\right).\label{eq:zeta_bound}
\end{align}
But the RHS is just $\zeta(r)$ for a constant density star with $\bar{\rho} = 3M/4\pi R^2$. Denoting the constant density solution by $\zeta^*(r)$ we rewrite \eqref{eq:zeta_bound} as
\begin{equation}
\zeta(r) \leq \zeta^*(r)\label{eq:zeta_bound2}
\end{equation}
for stars with non-increasing average density. This was done in \cite{Visser2003a}, see appendix \ref{app:bounds_paper}. A more significant generalisation would be to treat stars that have $P_r \leq P_t$ everywhere, as is done in the \S \ref{sec:Pr_less_Pt}.

\subsubsection{Improved bounds with the DEC.}
From equation \eqref{eq:1.2.diff_eq_zeta} and equation \eqref{eq:zeta_bound2} we have
\begin{equation}
g(r) = \frac{\zeta^\prime}{\zeta}\geq \frac{M}{R^3}\frac{r}{\zeta^{*}(r)\sqrt{1-2m/r}} \geq  \frac{M}{R^3}\frac{r}{\zeta^{*}(r)\sqrt{1-\chi (r/R)^2}} = \frac{\zeta^{*\prime}}{\zeta} = g^*(r).\label{eq:g_inequality}
\end{equation}
Hence throughout the star the local acceleration due to gravity is always greater or equal to that of the constant density (interior Schwarzschild) solution. We can then use the field equation \eqref{eq:1.1.field_eq:r} to show that
\begin{equation}
\lim_{r\rightarrow 0}8\pi P_r(r) = \lim_{r\rightarrow 0}\left[\frac{2}{r}g(r) - \frac{2m(r)}{r^3}\right] = \lim_{r\rightarrow 0}\left(\frac{2}{r}g(r)\right) - \frac{8\pi}{3} \rho_c.
\end{equation}
We note that we have $g(r)\geq g^*(r)$ and that $\rho_c \geq \rho^*$.\footnote{This last inequality follows from the fact that at the centre $\rho_c = \bar{\rho_c}$ while $\rho^* = \bar{\rho}(R)$. We then apply the condition that average density is decreasing:  $\bar{\rho}(r) \geq \rho^*$.} Hence we have
\begin{equation}
\lim_{r\rightarrow 0}8\pi P_r(r) = \lim_{r\rightarrow 0}(\frac{2}{r}g(r)) - \frac{8\pi}{3} \rho_c \geq \lim_{r\rightarrow 0}(\frac{2}{r}g^*(r)) - \frac{8\pi}{3} \rho_c
\end{equation}
or using \eqref{eq:approx_g_small_r} to expand $g$ and $g^*$:
\begin{equation}
P_r(0) \geq P_r^*(0) + \frac{1}{3}\left(\rho^* - \rho_c\right).
\end{equation}
This is actually a \emph{very} weak bound. As shown in either appendix B or at the end of section \ref{sec:thin_shell} we can conclude that
\begin{equation}
P_r(0) \geq P_r^{*}(0).
\end{equation}
Note that we do \emph{not} have $P_r(r) \geq P_r^{*}(r)$ throughout the star, see appendix B for more information concerning this point. However, we see that if the interior Schwarzschild solution violates the dominant energy condition (DEC) $P_r^*(0) > \rho$ then all other decreasing density solutions will violate it too. The pressure profile for the interior Schwarzschild solution is given by
\begin{equation}
P^*_r(r) = \rho^* \frac{\sqrt{1-\chi(r/R)^2} - \sqrt{1-\chi}}{3\sqrt{1-\chi}-\sqrt{1-\chi(r/R)^2}}\label{eq:DEC_buchdahl_lmt}
\end{equation}
and in particular the central pressure is given by
\begin{equation}
P^*_r(0) = \rho^*\frac{1 - \sqrt{1-\chi}}{3\sqrt{1-\chi}-1}.
\end{equation}
As the pressure decreases throughout the star, the maximum value is obtained at the centre. The dominant energy condition will be satisfied \emph{in the Schwarzschild interior solution} iff
\begin{equation}
\frac{1 - \sqrt{1-\chi}}{3\sqrt{1-\chi}-1} \leq 1,
\end{equation}
or in terms of a compactness, $\chi \leq 3/4$. This is not a maximum for all decreasing density stars as $\rho_c \geq \rho^*$. In general we need to satisfy
\begin{equation}
\frac{P^*}{\rho_c} \leq 1.
\end{equation}
Using this condition and \eqref{eq:DEC_buchdahl_lmt} we have the result
\begin{equation}
\chi \leq \frac{4[\rho_c/\rho^*](2[\rho_c/\rho^*] + 1]}{(3[\rho_c / \rho^*] + 1)^2},\quad \frac{\rho_c}{\rho^*}\geq 1.
\end{equation}
Note that as the central pressure gets higher the Buchdahl--Bondi limit is obtained. This time we have only established that $\chi \leq 8/9$ is a bound for a solution that obeys the DEC, we have not provided an example which saturates the bound. With more work, better bounds may be available.

\subsection{Bounds on compactness if $P_r \leq P_t$}\label{sec:Pr_less_Pt}

When $P_r$ is less than $P_t$ the tangential pressure is doing more
to stop gravitational collapse than the radial pressure. Hence the required
radial pressure is \emph{less} than the isotropic case, and we would expect
that even stars that have a non-increasing average density profile should be
able to violate the Buchdahl--Bondi bound.  By inspection of
\eqref{eq:TOVaniso} it is clear that we should be able to get the compactness
as close to one as desired if we impose no upper limit on $P_t$. One limit we
can place on the compactness will come about from insisting that the dominant
energy condition holds: $|P_t| \leq \rho$ \cite[we are explicitly dealing with a
type I stress-energy tensor of \S 4.3]{Hawk74}. The bounds become
difficult to saturate because the requirement of spherical symmetry means that
$P_t|_{r=0} = P_r|_{r=0}$ and the condition $P_t < \rho$ is in general
difficult to satisfy. We make an attempt in \ref{sec:aniso_const_density}.

\paragraph{}
To find more general bounds we look at the two measures of anisotropy:
\begin{itemize}
\item Bounding the fractional change in the anisotropy, smoothed out by the energy density
\begin{equation}
%\delta_{\textrm{frac}}(\epsilon) 
\delta_{P_r}(\epsilon):= \sup\left[\frac{P_t + \epsilon\rho}{P_r+\epsilon\rho}\right] - 1 = 
\sup\left[\frac{P_t - P_r}{P_r + \epsilon\rho}\right].
\end{equation}
In the case that $\epsilon$ is zero this reduces to the fractional change in anisotropy.
\item Bounding the anisotropy as a fraction of the energy density
\begin{equation}
\delta_{\rho} := \sup\left[\frac{P_t - P_r}{\rho}\right]
\end{equation}
\end{itemize}
Notice that because of spherical symmetry, we have $P_r = P_t$ at $r=0$, implying that both $\delta_\rho$ and $\delta_{P_r}(\epsilon)$ are non-negative.

\subsubsection*{Bounding by $\delta_{P_r}(\epsilon)$}
If $\delta_{P_r}(\epsilon)$ exists and is finite, we know that
\begin{equation}
\delta_{P_r}(\epsilon)\geq \frac{P_t - P_r}{P_r + \epsilon \rho} = \frac{G_{\hat{\vartheta}\hat{\vartheta}} - G_{\hat{r}\hat{r}}}{G_{\hat{r}\hat{r}} + \epsilon G_{\hat{t}\hat{t}}}, 
\end{equation}
as $\delta_{P_r}(\epsilon)$ is the supremum. We can use the field equations \eqref{eq_group:1.1.field_eq} to find an expression bounding $m$:
\begin{align}
8\pi(P_t - P_r) &\leq (G_{\hat{r}\hat{r}} + \epsilon G_{\hat{t}\hat{t}})\delta_{P_r}(\epsilon).
\end{align}
Expanding out the pressures on the LHS:
\begin{align}
\frac{r}{\zeta}\sqrt{1-\frac{2m}{r}}\left(\frac{1}{r}\sqrt{1-\frac{2m}{r}}\zeta^\prime\right)^\prime& - r\left(\frac{m}{r^3}\right)^\prime\leq \left(G_{\hat{r}\hat{r}} + \frac{2\epsilon}{r^2}m^\prime\right)\delta_{P_r}(\epsilon)
\end{align}
which implies
\begin{align}
2\delta_{P_r}(\epsilon)\left(\epsilon\frac{m^\prime}{r^2} - \frac{m}{r^3}\right) + r\left(\frac{m}{r^3}\right)^\prime &\geq \frac{r}{\zeta}\sqrt{1-\frac{2m}{r}}\Bigg\{\left(\frac{1}{r}\sqrt{1-\frac{2m}{r}}\zeta^\prime\right)^\prime\label{eq:1.boundAniso:1} \\
&\quad\quad\quad\quad - \frac{2\delta_{P_r}(\epsilon)}{r^2}\sqrt{1-\frac{2m}{r}}\zeta^\prime\Bigg\}.\nonumber
\end{align}
The goal is now to attack each side independently, and make the LHS proportional to the average density. We are then able to get bounds for stars that have a non-increasing average density profile. The LHS is easy, if somewhat ugly:
\begin{equation}
2\delta_{P_r}(\epsilon)\left(\epsilon\frac{m^\prime}{r^2} - \frac{m}{r^3}\right) + r\left(\frac{m}{r^3}\right)^\prime = 2\delta_{P_r}(\epsilon)\;\epsilon\;r^{-2+\frac{1}{\epsilon}}\left(\frac{m}{r^{1/\epsilon}}\right)^\prime + r\left(\frac{m}{r^3}\right)^\prime.
\end{equation}
While this is not \emph{proportional} to the derivative of the average density if $\epsilon \neq 1/3$ it is easy to show via the product rule that if the average density is decreasing then $m/r^{3+n}$ is decreasing for positive $n$:
\begin{equation}
\left(\frac{m}{r^{3+n}}\right)^\prime = \left(\frac{m}{r^3}\right)^\prime\frac{1}{r^n} - \left(\frac{m}{r^3}\right)\frac{n}{r^{n+1}}.\label{eq:1.density_prod_rule}
\end{equation}
Hence we can guarantee this term is negative if $0< \epsilon \leq 1/3$ and if the average density is decreasing. 

\paragraph{}
The RHS is a lot more messier, but it is fairly easy to show that
\begin{align}
&\frac{r}{\zeta}\sqrt{1-\frac{2m}{r}}\left(\left(\frac{1}{r}\sqrt{1-\frac{2m}{r}}\zeta^\prime\right)^\prime - \frac{2\delta_{P_r}(\epsilon)}{r^2}\sqrt{1-\frac{2m}{r}}\zeta^\prime\right)\\ 
&\quad\quad\quad\quad=\frac{r^{1+2\delta_{P_r}(\epsilon)}}{\zeta}\sqrt{1-\frac{2m}{r}}\left(r^{-(1+2\delta_{P_r}(\epsilon))}\sqrt{1-\frac{2m}{r}}\zeta^\prime\right)^\prime\nonumber.
\end{align}
By taking $\epsilon \leq 1/3$ we can make \eqref{eq:1.boundAniso:1} into the much nicer statement that
\begin{equation}
\eqprime{eq:1.boundAniso:1}
\left(r^{-1-2\delta_{P_r}(\epsilon)}\sqrt{1-\frac{2m}{r}}\zeta^\prime\right)^\prime \leq 0.
\end{equation}
The procedure is now very much the same as the $P_r \geq P_t$ case already discussed. By joining continuously onto the Schwarzschild metric we can obtain
\begin{align}
0&\geq \frac{1}{r^{1+2\delta_{P_r}(\epsilon)}}\sqrt{1-\frac{2m}{r}}\zeta^\prime\bigg\vert^R_{\bar{r}} 
\end{align}
implying that
\begin{align}
\frac{1}{\bar{r}}\sqrt{1-\frac{2m}{\bar{r}}}\zeta^\prime(\bar{r}) &\geq \frac{M}{R^{3+2\delta_{P_r}(\epsilon)}} = \frac{\chi}{2R^{2+2\delta_{P_r}(\epsilon)}}\\
\zeta^\prime(\bar{r}) &\geq \frac{\chi}{2R^{2+2\delta_{P_r}(\epsilon)}}\frac{\bar{r}^{1+2\delta_{P_r}(\epsilon)}}{\sqrt{1-2m/\bar{r}}}\\
\zeta(x) &\leq \zeta(R) - \frac{\chi}{2R^{2+2\delta_{P_r}(\epsilon)}}\int_x^R\,\frac{\bar{r}^{1+2\delta_{P_r}(\epsilon)}}{\sqrt{1-2m/\bar{r}}}\,\d \bar{r}\\
&= \sqrt{1-\chi} - \frac{\chi}{2R^{2+2\delta_{P_r}(\epsilon)}}\int_x^R\,\frac{\bar{r}^{1+2\delta_{P_r}(\epsilon)}}{\sqrt{1-2m/\bar{r}}}\,\d \bar{r}\label{eq:1:anisoBound:2}.
\end{align}
As the density is decreasing we can again appeal to the fact that $m(r) \geq M\frac{r^3}{R^3}$ to be able to analytically bound this integral. Doing this we obtain
\begin{align}
\zeta(x)&\leq \sqrt{1-\chi} - \frac{\chi}{4(1+\delta_{P_r}(\epsilon))}\left({}_2F_1\left(\frac{1}{2},1+\delta;2+\delta;\chi\right)\right.\nonumber \\
&\quad \left.- \left(\frac{x}{R}\right)^{2+2\delta_{P_r}(\epsilon)}{}_2F_1\left(\frac{1}{2},1+\delta;2+\delta;\left[\frac{x}{R}\right]^2\chi\right)\right).
\end{align}  
Here ${}_2F_1$ is the Gauss hypergeometric function which takes on reasonably simple form when $\delta_{P_r}(\epsilon)$ is an even integer. I am also going to restrict attention to the central value, in which case
\begin{equation}
0\leq\zeta(0)\leq \sqrt{1-\chi} - \frac{\chi}{4(1+\delta_{P_r}(\epsilon))}{}_2F_1\left(\frac{1}{2},1+\delta;2+\delta;\chi\right)\label{eq:1.anisoBound:3}.
\end{equation}

\subsubsection*{Bounding by $\delta_\rho$}
As $\delta_\rho$ is non-negative, we can use the Einstein equations to give:
\begin{align}
r\dr\left(\frac{1}{r}\sqrt{1-\frac{2m}{r}}\frac{\dx{\zeta}}{\dx{r}}\right)&\leq \delta_\rho G_{\hat{t}\hat{t}} + r\dr\left(\frac{m}{r^3}\right)\\
&\leq (2\delta_\rho + 1)r^{3/(2\delta_\rho + 1)-2}\bigg(\frac{\dx{m}}{\dx{r}}r^{-3/(2\dTwo + 1)}\nonumber\\
& - \frac{3mr^{-3/(2\delta_\rho + 1) - 1}}{2\delta_\rho + 1}\bigg)\\
&\hspace{-10ex}\leq \frac{1}{\eta}r^{3\eta-2}\dr\left(\frac{m}{r^{3\eta}}\right),\quad\quad\eta\equiv(2\delta_\rho + 1)^{-1}\leq 1.
\end{align}
Because in this case we know that the $r$ exponent on the ``density-like'' term is less than three we cannot apply the simple result of \eqref{eq:1.density_prod_rule} to deduce that the LHS is decreasing. We can continue as the expression on the RHS is still simple enough to integrate by parts:
\begin{align}
\left.\frac{1}{r}\sqrt{1-\frac{2m}{r}}\frac{\dx{\zeta}}{\dx{r}}\right|^{R}_{r}&\leq\frac{1}{\eta}\int_r^R r^{3(\eta-1)}\dr\left(\frac{m}{r^{3\eta}}\right)\,\d r\\
&\leq \left.\frac{m}{r^3}\right|^R_r + 3(1-\eta)\int_r^R \frac{m}{r^4}\,\dx{r}.
\end{align}
We can then find the bound for an arbitrary position throughout the star:
\begin{align}
\frac{1}{r}\sqrt{1-\frac{2m}{r}}\frac{\dx{\zeta}}{\dx{r}} &\geq \frac{1}{\eta}\left(\frac{m}{r^3} - \frac{M}{R^3}(1-\eta)\right)-\frac{3(1-\eta)}{\eta}\int_r^R\frac{m}{r^4}\,\dx{r}.
\end{align}
It does not seem clear from this that a simple bound can be placed on the compactness with knowledge of $\delta_\rho$, making it significantly less useful than the previous case.

\subsubsection{Hypergeometric functions}

So far we have obtained the bound \eqref{eq:1.anisoBound:3} but this bound is not written in terms of elementary functions. The hypergeometric equation is defined to be
\begin{align}
{}_2F_1(a,b;c;x) &= 1 + \frac{ab}{c}x + \frac{a(a+1)b(b+1)}{2c(c+1)}x^2\\
&\quad\quad + \frac{a(a+1)(a+2)b(b+1)(b+2)}{6c(c+1)(c+2)}x^3 + \ldots\nonumber\\
&=1+\sum_{i=1}^\infty \,\frac{(a)(a+1)\ldots(a+i-1)(b)\ldots(b+i-1)}{c(c+1)\ldots(c+i-1)i!}x^i .
\end{align}
It is immediately apparent that the hypergeometric function is symmetric in its first two parameters $a$ and $b$. The cases when either of these parameters are integers lead to tractable closed expressions for the Gauss hypergeometric function in terms of more elementary functions. In particular we have
\begin{align}
{}_2F_1\left(\frac{1}{2},1;2;\chi\right) &= \frac{2}{1+\sqrt{1-\chi}}\\
{}_2F_1\left(\frac{1}{2},2;3;\chi\right) &= \frac{4}{3\chi^2}(2-(2+\chi)\sqrt{1-\chi})\\
{}_2F_1\left(\frac{1}{2},3;4;\chi\right) &= \frac{2}{5\chi^3}(8 - (3\chi^2 + 4\chi + 8)\sqrt{1-\chi})\\
{}_2F_1\left(\frac{1}{2},4;5;\chi\right) &= \frac{1}{35\chi^4}(128 - (40\chi^3 + 48\chi^2 + 64\chi + 128)\sqrt{1-\chi}).
\end{align}
These relations can be found by asking Maple, looking them up in mathematical tables, or using the relation
\begin{equation}
{}_2F_1(a,b;c;\chi) = (1-\chi)^{c-a-b}\,{}_2F_1(c-a,c-b;c;\chi),
\end{equation}
and then using the power series expansion in the new hypergeometric. We can also find the limiting behaviour directly from the power series:
\begin{align}
\lim_{n\rightarrow\infty}{}_2F_1\left(\frac{1}{2},1+n;2+n;\chi\right) &\approx 1 + \frac{1}{2}x + \frac{\frac{1}{2}\frac{3}{2}}{2!}x^2 +\ldots\\
&= \frac{1}{\sqrt{1-\chi}}.
\end{align}

\begin{figure}[tp]
\begin{center}
\includegraphics[width=0.5\textwidth, angle=-90]{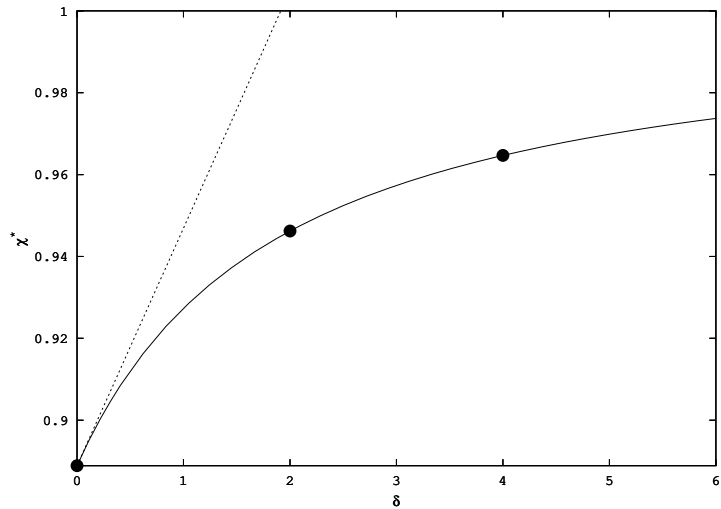}
\caption{The bound on the compactness $\chi_*$ as a function of the anisotropy parameter.}\label{fig:1.bound_chi:1}
\end{center}
\end{figure}

\paragraph{}
These results can be put back into \eqref{eq:1.anisoBound:3} giving us bounds on the compactness. Let $\chi_*$ denote the value of $\chi$ that saturates \eqref{eq:1.anisoBound:3} in the sense that
\begin{equation}
\sqrt{1-\chi_*} - \frac{\chi_*}{4(1+\delta_{P_r}(\epsilon))}\,{}_2F_1\left(\frac{1}{2},1+\delta_{P_r}(\epsilon);2+\delta_{P_r}(\epsilon);\chi_*\right) = 0.\label{eq:chiStar}
\end{equation}
We can then find the bound on the compactness exactly for these few cases:
\begin{align}
\delta_{P_r}(\epsilon) = 0 &\Rightarrow \chi_{*} = \frac{8}{9};\\
\delta_{P_r}(\epsilon) = 1 &\Rightarrow \chi_{*} = \frac{3+\sqrt{105}}{14};\\
\delta_{P_r}(\epsilon) = 2 &\Rightarrow \chi_* = 1-\bigg(\frac{(4235+110\sqrt{1111})^{1/3}}{33} \\
&\quad\quad\quad\quad\quad\quad+ \frac{5}{(4235+110\sqrt{1111})^{1/3}} - \frac{2}{3}\bigg)^2;\nonumber\\
\delta_{P_r}(\epsilon) = \infty &\Rightarrow \chi_* = 1.
\end{align}
A plot showing the points $\delta_{P_r}(\epsilon) = 0,1,2$ as well a numerical solution of \eqref{eq:chiStar} and its tangent are shown in figure \ref{fig:1.bound_chi:1}. Notice that if the anisotropy is unbounded then we cannot say anything about a star that is not a black hole, and this bound suggests that we can get arbitrarily close. Note that we do not currently have a way of obtaining a solution that saturates these bounds, so better bounds may exist.

\subsubsection{An attempt at saturation}\label{sec:aniso_const_density}
We see how close we can get to saturation of our bounds by looking again at a constant density star; this saturates the non-increasing condition. We would also like to let $P_t = \rho_0$, saturating the limit set by the dominant energy condition. This saturation would cause difficulties with the requirement $P_r = P_t$ at the origin set by spherical symmetry, so we assume that the form of $P_t$ is
\begin{equation}
P_t(r) = \begin{cases} \rho_0& r\geq r_\star\\
P_r& r< r_\star
\end{cases}.
\end{equation}
Note that the while the radial pressure profile may not be discontinuous there is no such requirement on the \emph{tangential} pressure profile. Thus we can get arbitrarily close to saturating both our bounds.

\paragraph{}
We begin with the anisotropic TOV equation \eqref{eq:TOVaniso} in the region $r \geq \epsilon$:
\begin{eqnarray}
\frac{\d P_r}{\d r} &=& -g(\rho_0 + P_r) + \frac{2}{r}(\rho_0-P_r)\\
&=& \left(\frac{2}{r}-g\right)\rho_0 - \left(g+\frac{2}{r}\right)P_r\label{eq:dPdr_max_aniso}.
\end{eqnarray}
The solution to this equation is given in terms of hypergeometric functions, but is not easy to interpret directly. We can learn quite a bit from investigating particular solutions. But first we rewrite \eqref{eq:dPdr_max_aniso} in terms of the compactness:
\begin{align}
\frac{\d P_r}{\d r} &= \left(\frac{2}{r}-\frac{(m(r) + 4\pi r^3 P_r)}{r(r-2m(r))}\right)\rho_0 - \left(\frac{m(r) + 4\pi r^3 P_r}{r(r-2m(r))}+\frac{2}{r}\right)P_r\\
&= \left(\frac{2}{r}-\frac{(Mr^3/R^3 + 4\pi r^3 P_r)}{r(r-2M(r/R)^3)}\right)\left(\frac{3\chi}{8\pi R^2}\right)\nonumber\\
&\quad\quad\quad\quad - \left(\frac{M(r/R)^3 + 4\pi r^3 P_r}{r(r-2M(r/R)^3)}+\frac{2}{r}\right)P_r\\
&= \left(\frac{2}{r}-\frac{r(\chi/R^2 + 8\pi  P_r)}{2(1-\chi r^2/R^2)}\right)\left(\frac{3\chi}{8\pi R^2}\right)\nonumber\\
&\quad\quad\quad\quad - \left(\frac{r(\chi/R^2 + 8\pi  P_r)}{2(1-\chi r^2/R^2)}+\frac{2}{r}\right)P_r\label{eq:dPdr_terms_chi}.
\end{align}
At the surface of the star we know that $P_r = 0$. We see that
\begin{eqnarray}
\frac{\d P_r}{\d r}\bigg|_{r=R} &=& \left(\frac{2}{R}-\frac{\chi}{2R(1-\chi)}\right)\left(\frac{3\chi}{8\pi R^2}\right)\nonumber\\
&=&\begin{cases} 
\geq 0&\textrm{ if }\chi \leq 4/5\\
< 0&\textrm{ if }\chi > 4/5
\end{cases}.
\end{eqnarray}
Hence if $\chi$ is less than $4/5$ the gradient is positive and the radial pressure near the surface is \emph{negative}. Gravity is not strong enough to hold the star together under the very strong imposed tangential pressures and the radial pressure must ``help''. But then the second term in \eqref{eq:dPdr_terms_chi} is strictly positive, the only negative contribution begin the $r\chi/R^2$ piece in the numerator of the first term, which is maximised when $r=R$. Thus the pressure gradient is always positive and hence as we follow the profile \emph{inward} the radial pressure gets more severely negative.

\begin{figure}[t]
\begin{center}
\includegraphics[width=0.5\textwidth, angle=-90]{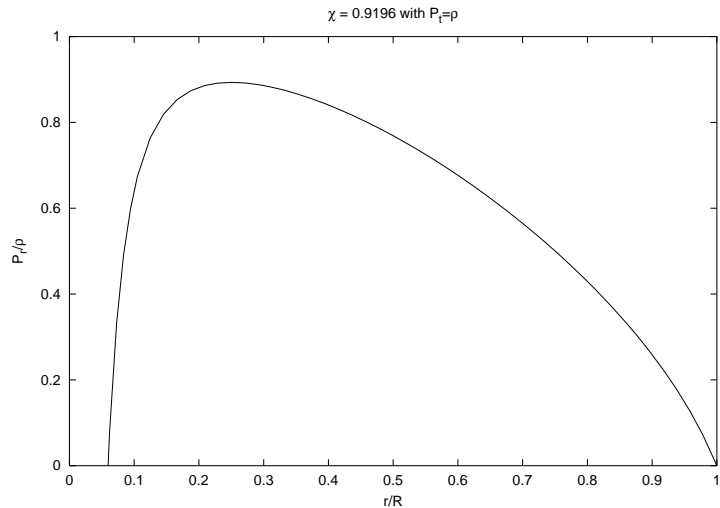}
\caption{{\footnotesize The radial pressure of a star with compactness $\chi = 0.9196$. The second intercept is at $r_\star \approx 0.07$ and a constant density solution with compactness $\chi_\star = 0.9196(0.07)^2 \approx 0.005$ can be used in the region $r < r_\star$.}}\label{fig:near_saturate}
\end{center}
\end{figure}

\paragraph{}
If $\chi>4/5$ then the radial pressure near the surface is positive and decreases to zero. For $\chi \lesssim 0.92$ the pressure eventually goes back to zero and then becomes negative, diverging as we approach the centre. For $\chi \gtrsim 0.92$ the radial pressure is positive throughout, but diverges off to positive infinity as the centre is approached. To construct a solution that obeys the energy condition, one approach would be:
\begin{itemize}
\item For $0.8 < \chi < 0.92$ we have two places where $P_r(r)=0$, denoted by $r=r_\star$ (the interior solution) and $r=R$ (the surface).
\item We use the solution $P_t = \rho$ in the region $r_\star \leq r \leq R$.
\item For $r \leq r_\star$ we can generate an isotropic solution for a constant density star. It is equivalent to considering a constant density star of density $\rho_0$ and radius $r_\star$ as $P_r(r_\star)=0$ and we only require one initial condition for the solution. The compactness of this isotropic star would be
\begin{equation}
\chi_\star = \frac{2M_\star}{r_\star} = \frac{8}{3}\pi \rho_0 r_\star^2 = \chi\left(\frac{r_\star}{R}\right)^2
\end{equation}
\end{itemize}
This gives a legitimate solution provided that $\chi_\star < 8/9$. A solution is shown for $\chi = 0.9196$ in figure \ref{fig:near_saturate}. 

%\paragraph{}
%How close is this to saturating our previous limit? 

\subsection{Theoretical limit on compactness: $\bar{\rho}^\prime\leq 0$}
We have shown that as $\delta_{P_r}(\epsilon)$ approaches infinity that $\chi_\star$ approaches one \eqref{eq:chiStar}. We have also shown that an explicit example that obeys the energy conditions with $\chi = 0.9196$ in the last section. The question now is if the dominant energy condition places any bound on $\delta_{P_r}(\epsilon)$, hence ultimately a bound on $\chi_\star$. Note that it may not be possible to construct a stellar model with $\chi = \chi_\star(\sup[\delta_{P_r}(\epsilon)])$, but at least we know that $\chi_\star(\sup[\delta_{P_r}(\epsilon)])$ will serve as an upper bound.

\paragraph{}
From the definition of $\delta_{P_r}(\epsilon)$ we have
\begin{equation*}
\delta_{P_r}(\epsilon) = \sup\left[\frac{P_t - P_r}{P_r + \epsilon \rho}\right]
\end{equation*}
where 
\begin{itemize}
\item $0 < \epsilon \leq 1/3$  (Inequality condition)
\item $|P_r| \leq \rho$, $|P_t| \leq \rho$   (Dominant energy condition)
\item $\bar{\rho}^\prime \leq 0$ (Decreasing average density condition)
\item $\delta_{P_r}(\epsilon) \geq 0$  (Spherical symmetry requires $P_r(0) = P_t(0)$)
\end{itemize}
As an extremely crude estimate, we may use the dominant energy condition to give us
\begin{equation*}
\delta_{P_r}(\epsilon) \leq \sup\left[\frac{|P_t| + |P_r|}{|P_r + \epsilon\rho|}\right] \leq 2\sup\left[\frac{\rho}{|P_r + \epsilon\rho|}\right].
\end{equation*}
The denominator blows up for particular values of $\epsilon$ if $P_r$ is negative somewhere in the star. Provided that $P_r \geq -\rho/3$ we can state that
\begin{equation}
\inf_{\epsilon \in (0,1/3]} \delta_{P_r}(\epsilon) < \infty,
\end{equation}
and hence we can find some bound on the compactness. If we add the stronger hypothesis that $P_r \geq 0$ we have 
\begin{equation}
\inf_{\epsilon \in (0,1/3]} \delta_{P_r}(\epsilon) \leq \inf_{\epsilon \in (0,1/3]} \frac{\rho - 0}{0+\epsilon\rho} = 3.
\end{equation}
This allows us to conclude that
\begin{equation}
\chi \leq \chi_\star(3) \approx 0.974.
\end{equation}
It should be noted that to get this result we still have the restriction that the average density is decreasing outward and that the radial pressure is always positive. It should also be noted that our attempt with the constant density solution got us nowhere near this bound, suggesting that it is probably very weak.

% Cut and paste from here on in
%
\section{Pressure inside the star}
\subsection{An upper bound on pressure if $P_r\geq P_t$ and $\bar{\rho}^\prime\leq 0$}
We can obtain both an upper and lower bound on the pressure inside at a radius $r$. As we are assuming that $P_r$ is greater than $P_t$ throughout the star and that the average density is decreasing we can use \eqref{eq:1.1.pr_bigger_pt_inequality}
\begin{equation*}
\dr\left(\frac{1}{r}\sqrt{1-\frac{2M}{R}}\frac{\dx{\zeta}}{\dx{r}}\right)\leq 0
\end{equation*}
and integrating first from $r$ to $r_0$ ($r<r_0$) we obtain
\begin{align}
\frac{1}{r_0}\sqrt{1-\frac{2m_0}{r_0}}\frac{\dx{\zeta}}{\dx{r}}\bigg\vert_{r=r_0}&\equiv K_0\leq \frac{1}{r}\sqrt{1-\frac{2m}{r}}\frac{\dx{\zeta}}{\dx{r}}\big\vert_{r=r}
\end{align}
which in turn implies
\begin{align}
\frac{\dx{\zeta}}{\dx{r}} &\geq \frac{K_0r}{\sqrt{1-2m/r}}\\
\zeta(r_0) - \zeta(0)&\geq K_0\int_0^{r_0}\frac{r}{\sqrt{1-2m/r}}\,\dx{r}\\
&\geq K_0\int_0^{r_0}\frac{r}{\sqrt{1-2Mr^2/R^3}}\,\dx{r}\\
&\geq K_0\frac{r_0^3}{2m_0}\left(1-\sqrt{1-\frac{2m_0}{r_0}}\right).
\end{align}
Which gives a bound on $\zeta$ at the centre of the star:
\begin{align}
\zeta(0) &\leq \zeta(r_0)\left(1- \frac{1}{r_0}\sqrt{1-\frac{2m_0}{r_0}}\frac{\zeta(r_0)^\prime}{\zeta(r_0)} \left(1-\sqrt{1-\frac{2m_0}{r_0}}\right)\right).
\end{align}
The notation is by and large self explanatory, $m_0\equiv m(r_0)$ where $r_0$ is the arbitrary (but fixed!) radius. From \eqref{eq:1.1.define_g} we know
\begin{equation}
g(r_0)= \frac{\zeta(r_0)^\prime}{\zeta(r_0)} = \frac{m_0+4\pi r_0^3P_{r0}}{r_0(r_0-2m_0)}.
\end{equation}
Quite a but of algebra later we find that
\begin{equation}
P(r_0) \equiv P_0 \leq \frac{1}{4\pi r^2_0}\left(\sqrt{1-\chi_0} + 1 - \frac{3}{2}\chi_0\right),
\end{equation}
so we find that the pressure is bounded \emph{above}. On first sight this is not incredibly surprising; we have said that we have a given mass at a given radius, if the pressure became too large gravity would not be strong enough to hold the star together violating the stationary assumption we have already made. What \emph{is} surprising is that this bound decreases as the compactness $\chi(r_0)\equiv 2m_0/r_0$ increases and becomes zero as $\chi_0\rightarrow 8/9$. Physically the pressure \emph{increases} as the mass of the star increases -- what gives?

\paragraph{}
We need to understand what this inequality is actually trying to tell us. If the pressure exceeds this value then we are going to have an infinite pressure by the time we reach the centre of the star. As $\chi \rightarrow 8/9$ then any radial pressure will result in infinite pressure at $r=0$, hence why the bound is saturating. But we will also have to ensure that the pressure is large enough so that the star does not collapse, which would result in a lower bound. This bound does not, however give us any information about the central pressure.

\subsection{More bounds with the TOV equation}
We can use the TOV equation  now to find a lower bound on pressure mentioned in the previous section. This bound is more intuitive, as it comes about from requiring that the radial pressure is able to support the star rather than what pressure is required to stop a metric component changing sign. Of course, when the intersection of these bounds is empty we cannot hope to find solutions!

\begin{align}
\frac{\dx{P}}{\dx{r}} &= -\frac{m+4\pi r^3P}{r^2(1-2m/r)}(\rho + P)\\
&\leq -\frac{m\rho}{r^2(1-2m/r)} = -\frac{mm^\prime}{4\pi r^4(1-2m/r)},
\end{align}
therefore
\begin{align}
P(R)-P(0) &=-P(0) \leq -\frac{1}{4\pi}\int_0^R \frac{mm^\prime}{r^4}\sum_{i=0}^\infty\left(\frac{2m}{r}\right)^i\,\,\dx{r},
\end{align}
which implies
\begin{align}
P(0)&\geq \frac{1}{4\pi}\sum_i\,2^i\,\int_0^M\frac{m^{i+1}m^\prime}{r^{4+i}}\,\dx{r}\\
&\geq \frac{1}{4\pi}\sum_i\,2^i\,\left(\frac{1}{i+2}\frac{m^{i+2}}{r^{4+i}}\bigg\vert_{m=0}^{m=M} + \frac{4+i}{2+i}\int_0^R\,\frac{m^{i+2}}{r^{i+5}}\,\dx{r}\right)\\
&\geq \frac{1}{4\pi}\sum_i\,2^i\,\left(\frac{1}{i+2}\frac{M^{i+2}}{R^{4+i}}\right)\\
&\geq \frac{1}{4\pi}\sum_{j=2}^{\infty}\,\frac{1}{4R^2j}\left(\frac{2M}{R}\right)^j\\
&\geq -\frac{1}{16\pi R^2}\left(\ln\left(1-\chi\right) + \chi\right).
\end{align}
This holds for any spherically symmetric star, subject only to the constraints that the pressure and density are positive. We have not made any assumptions about how the matter is distributed throughout the star. In the limit that the compactness approaches one, the pressure in the centre of the star becomes infinite to support the mass of the star from gravitational collapse. 

\section{Variations on a theme}
In deriving the generalised Buchdahl-Bondi bound on the compactness we assumed that the density was non-increasing throughout the star so that we could bound an integral and that the pressure was isotropic. This bound happens to be saturated by the Schwarzschild interior solution giving an upper limit on the compactness of $8/9$.

\paragraph{}
By allowing an anisotropic pressure but still insisting that the density decreased as we went through the star it was found that the compactness could get arbitrarily close to one. It should be noted that in saying this I am assuming the existence of solutions of Einstein's equations that have constant $\delta_{P_r}(\epsilon)$ except for an arbitrarily small region near the origin, as $P_r = P_t$ at the origin is required by spherical symmetry. By imposing the DEC we could claim that $\chi_*\leq 0.974$.

\paragraph{}
Kovetz \cite{KovetzII} concentrated his attention on isotropic fluids but no longer insisted that the average density decreased throughout the star. He looked at what would happen to the pressure profile as mass was moved around a perfect fluid without changing its total mass. This idea allows a comparison of the pressure profiles of two perfect fluid solutions with different density profiles. Do do this, Kovetz divides a fluid solution into ``thin shells'' and mass is taken from one shell at radius $b+\delta b$ and redistributed on a shell at radius $a+\delta a$, where $a<b$. In the new solution the \emph{average} density behaves like
\begin{equation*}
\bar{\rho}_{\textrm{new}} = \bar{\rho} + \delta \bar{\rho},\quad \delta\bar{\rho}\begin{cases}
\geq 0&\text{if $a\leq r \leq b$}\\
=0& \text{otherwise}
\end{cases}
\end{equation*}
and from this Kovetz derives that the central pressure increases. Note that the total mass of the star remains unchanged so the variation in the (non-averaged) \emph{density} cannot be non-negative or non-positive throughout the star. It is then argued that by successive use of thin shells that if we have any two solutions with $\bar{\rho_1} \geq \bar{\rho_2}$ everywhere but the same mass then the central pressure in star 1 is greater than the central pressure in star 2. One can strengthen this result: if $\bar{\rho}_1 \geq \bar{\rho}_2$ everywhere then the central pressures obey the relation $P_{c,1} \geq P_{c,2}$. To do this we shall need to present Kovetz's argument in a slightly different way.

\paragraph{}
Here the argument is presented in terms of a variation rather than a ``thin shell'' and Kovetz's result is re-derived. The variational argument also allows for the change in average density to be non-monotonic, although no simple \emph{general} statement about the central pressure can be given in this case. A simple argument for obtaining the central pressure increase result is also presented. 

\subsection{A simple derivation for a ``thin shell''}\label{sec:thin_shell}
As we are moving mass \emph{inward} $m(r)$ remains unchanged until we reach $r=b$, as $m(r)$ is the mass within a radius $r$ and does not depend on its distribution. A more formal derivation would be to use $\delta \bar{\rho}$ is zero until $r=b$. The density $\rho$ does not change \emph{except} on the thin shells themselves, so it is certainly the same for $r > b$. Then the (isotropic) TOV equation for $r>b$
\begin{equation*}
\frac{\d P_r}{\d r} = -\frac{m(r) + 4\pi r^3 P_r}{r(r-2m)}(P_r + \rho),
\end{equation*}
where there are \emph{no} variations, so the rate of change of $P_r$ is the same for $r>b$. As this is a first order differential equation it needs to be supplemented by an initial condition, a suitable one being that the radial pressure vanishes on the surface. But this is the same boundary condition for both solutions, so $P_r$ is \emph{identical} for $r>b$.

\paragraph{}
For definiteness's sake let us say that we are moving matter inward so $m_{\text{new}}(r) \geq m_{\text{old}}(r)$. Then
\begin{align}
g(r) &= \frac{m_{\text{new}} + 4\pi r^3 P_{r,\text{new}}}{r(r-2m_{\text{new}})}\\
&\geq \frac{m_{\text{old}} + 4\pi r^3 P_{r,\text{new}}}{r(r-2m_{\text{old}})}.
\end{align}
This translates into the following bound on the radial pressure gradient:
\begin{align}
\frac{\d P_{r,\text{new}}}{\d r} &\leq -\frac{m_{\text{old}} + 4\pi r^3 P_{r,\text{new}}}{r(r-2m_{\text{old}})}(P_{r,\text{new}} + \rho).
\end{align}
No effort has been made to distinguish the new and the old densities as they differ only on arbitrarily thin shells. Hence:
\begin{align}
\frac{\d P_{r,\text{new}}}{\d r} &\leq \frac{\d P_{r,\text{old}}}{\d r}
\end{align}
and in particular
\begin{align}
P_{r,\text{new}}(b) - P_{r,\text{new}}(r) &\leq P_{r,\text{old}}(b) - P_{r,\text{old}}(r).
\end{align}
But the pressure profiles are identical for $r\geq b$. Hence we have
\begin{equation}
P_{r,\text{new}}(r) \geq P_{r,\text{old}}(r)
\end{equation}
with equality holding for all $r\geq b$. To summarise: as matter is moved inward from a thin shell at $r=b$ that 
\begin{itemize}
\item For $r\geq b$ the pressure is unchanged.
\item For $r \leq b$ the new pressure (and in particular the central pressure) is greater than or equal to the previous pressure.
\end{itemize}

\subsection{General variations}

We now generalise the results from Kovetz's thin shells. In Kovetz's original paper one mass profile was turned into another by moving thin shells inward \cite{KovetzII}. As a result more care was needed with the matching conditions so that one could find a variation in the pressure for each shell, and then integrate over the shells. Here we choose a slightly different approach, we allow for general \emph{small} variations (but do not restrict ourselves to monotonic variations) of two solutions.

\paragraph{}
We will write
\begin{equation}
m_n(r) = m(r) + \delta m(r)
\end{equation}
to mean the mass contained within a radius $r$ after moving matter around is the same as that in our initial star plus some variation $\delta m(r)$. So far this is standard notation and we will prefix ``$\delta$'' in front of other quantities as well -- no confusion should result. 
%\centerline{\sf *** Find a ``home'' for this section ***}
%\paragraph{}
%To begin, we cast $g(r)$ into a form that is easier to manipulate:
%\begin{align}
%g(r) &\equiv \frac{\zeta^\prime}{\zeta}\\
%&= \frac{4\pi r(P + \rho)}{1-8\pi\orho\, r^2/3} + \frac{1}{2}\left(\ln(1-\frac{2m}{r})\right)^\prime\\
%&= \frac{4\pi r^3 P + m}{r^2(1-8\pi\orho\, r^2/3)}\label{eq:g_inMass}\\
%&= \frac{4\pi r}{1-8\pi\orho\, r^2/3}\left(P + \frac{\orho}{3}\right)\label{eq:g_inRho}
%\end{align}
%\centerline{\sf ****************************}

\paragraph{}
Variations in $g(r)$ takes the following form:
\begin{align}
\delta g &= \frac{\partial g}{\partial P}\delta P + \frac{\partial g}{\partial m}\delta m\nonumber\\
&=\frac{4\pi r}{1-2m/r}\,\delta P + \frac{1+8\pi r^2 P}{r^2(1-2m/r)^2}\delta m\label{eq:vary_g}.
\end{align}
We remove references to the pressure (but not the \emph{variation} in pressure) by expressing $P=P_r$ in terms of $g(r)$ using the $G_{\hat{r}\hat{r}}$ equation \eqref{eq:1.1.field_eq:r}
\begin{equation}
\delta g(r) = \frac{4\pi r}{1-2m/r}\,\delta P + \frac{1+2g(r)r}{r^2(1-2m/r)}\,\delta m\label{eq:vary_g2}.
\end{equation}
The variation in the pressure is the expression that we are really interested in. By taking the first variations of the TOV equation:
\begin{align}
\delta P^\prime &= -(P+\rho)\,\delta g - g \delta P - g\delta\rho\\
&=-\left(\frac{4\pi r}{1-2m/r}\left(P+\rho\right)\right)\delta P - \frac{(2gr + 1)}{r^2(1-2m/r)}\left(P+\rho\right)\delta m - g\, \delta P - g\,\delta\rho \label{eq:preK12}.
\end{align}
We could eliminate pressure by substituting again from \eqref{eq:1.1.field_eq:r} but it is easier to note
\begin{align}
g(r) &= \frac{m(r) + 4\pi r^3 P_r(r)}{r^2(1-2m(r)/r)}\\
 & = \frac{4\pi rP_r(r)}{1-2m(r)/r} + \frac{1}{1-2m(r)/r}\left(\frac{m(r)}{r^2} - \frac{m(r)^\prime}{r} + \frac{m(r)^\prime}{r}\right)\\
&= \frac{4\pi r^2 P_r(r) + m(r)^\prime}{r(1-2m(r)/r)} - \frac{1}{1-2m(r)/r}\left( \frac{m(r)^\prime}{r} - \frac{m(r)}{r^2}\right)\\
&= \frac{4\pi r(P_r + \rho)}{1-2m(r)/r} - \frac{1}{1-2m(r)/r}\left(\frac{m(r)}{r}\right)^\prime.
\end{align}
We can drop the subscript ``r'' on the pressure as we are dealing with isotropic fluids. This allows elimination of $P$ in \eqref{eq:preK12}
\begin{align}
\delta P^\prime &= -\left(2g-\frac{1}{2}\left(\ln(1-2m/r)\right)^\prime\right)\,\delta P \nonumber\\
&\quad\quad - \frac{(1+2rg)}{4\pi r^3}\left(g + \frac{(m/r)^\prime}{1-2m/r}\right)\,\delta m - g\,\delta \rho\label{eq:K12}.
\end{align}
This equation is a first order differential equation in $\delta P$, so first we find the solution to the homogeneous equation to obtain the integrating factor. Setting $\delta\rho$ and $\delta m$ to zero in \eqref{eq:K12} leaves us with
\begin{align}
\delta P_H^\prime &= -\left(2g - \frac{1}{2}\left(\ln(1-2m/r)\right)^\prime\right)\delta P_H\\
&= \left(\ln\left(\frac{\sqrt{1-2m/r}}{\zeta^2}\right)\right)^\prime\delta P_H.
\end{align}
The homogeneous solution can be found by direct integration:
\begin{align}
\delta P_H(r) &= \frac{C\sqrt{1-2m/r}}{\zeta(r)^2}.
\end{align}
This allows us to get the general solution in the standard way:
\begin{align}
\delta P(r) &= -\delta P_H \, \int_{r_0}^r \frac{\cdots \delta\rho + \cdots\delta m}{\delta P_H}\nonumber\\
&=\frac{\sqrt{1-2m/r}}{\zeta(r)^2}\lint_r^{r_0}\,\frac{\zeta(\orr)^2}{\sqrt{1-2m/\orr}}\Bigg\{g(\orr)\,\delta\rho(\orr)\nonumber\\
&\quad + \frac{1+2g(\orr)\orr}{4\pi \orr^3}\left(g(\orr) + \frac{(m/r)^\prime}{1-2m/\orr}\right)\,\delta m\Bigg\}\,\d \orr.
%\orr^2(1-2m/\orr)}\left(P+\rho\right)\delta m\right)\,\dx{\orr}\nonumber
\end{align}
Here $r_0$ is a constant of integration to be fixed by a boundary condition. By moving mass around it is possible that we change the location of the surface of the star. To avoid ambiguity I shall use $R$ to denote the maximum radius of the two solutions. To evaluate $r_0$ note that $\delta P(R)$ must be zero for both solutions as it is at the surface of one solution and either at the surface of the other solution or in vacuum, implying that $r_0 = R$ is a suitable boundary condition. 

\begin{align}
\delta P(r) &= \frac{\sqrt{1-2m/r}}{\zeta(r)^2}\lint_r^{R}\,\frac{\zeta(\orr)^2}{\sqrt{1-2m/\orr}}\Bigg\{g(\orr)\,\delta\rho(\orr) \nonumber\\
&\quad+ \frac{1+2g(\orr)\orr}{4\pi \orr^3}\left(g(\orr) + \frac{(m/r)^\prime}{1-2m/\orr}\right)\delta m\Bigg\}\,\dx{\orr}.
\end{align}
Let us take the first term in the integrand and integrate by parts:
\begin{align}
\int_r^R \frac{\zeta(\orr)^2 g(\orr)}{\sqrt{1-2m/\orr}}\,\delta\rho(\orr)\dx{\orr} &= \frac{1}{4\pi}\int_r^R \frac{\zeta(\orr)^2 g(\orr)}{\orr^2\sqrt{1-2m/\orr}}\,\left(\delta m(\orr)\right)^\prime\dx{\orr}\\
&= -\frac{1}{4\pi}\left(\frac{\zeta(r)^2 g(r)}{r^2\sqrt{1-2m/r}}\delta m(r) \right.\nonumber\\
&\phantom{mespace}\left.+ \int_r^R \left(\frac{\zeta(\orr)^2 g(\orr)}{\orr^2\sqrt{1-2m/\orr}}\right)^\prime\,\delta m \, \dx{\orr}\right).
\end{align}
Hence 
\begin{align}
\delta P(r) &= -\frac{g(r)\delta m(r)}{4\pi r^2} + \frac{\sqrt{1-2m/r}}{4\pi\zeta(r)^2}\lint_r^{R} \Bigg\{\frac{\zeta(\orr)^2(1+2g(\orr)\orr)}{\orr^3\sqrt{1-2m/r}}\nonumber\\
&\times\left(g(\orr)+\frac{(m/r)^\prime}{1-2m/\orr}\right) -\left(\frac{\zeta(\orr)^2 g(\orr)}{\orr^2\sqrt{1-2m/\orr}}\right)^\prime\Bigg\}\,\delta m(\orr) \, \dx{\orr} \label{eq:the_integral}.
\end{align}

\paragraph{}
The resulting integral is now in a form that we can work with, as we know the sign of $\delta m$ and can try to simplify the bracketed expression. Let us start with the differentiation and recall that $\zeta^\prime = g\,\zeta$:
\begin{align}
(\zeta^2)^\prime &= 2 \zeta \zeta^\prime = 2\zeta^2\, g.
\end{align}
We can rewrite the final term in \eqref{eq:the_integral} as
\begin{align} 
\left(\frac{\zeta^2 g}{r^2\sqrt{1-2m/r}}\right)^\prime &= 
\frac{2\zeta^2 g^2 + \zeta^2 g^\prime}{r^2\sqrt{1-2m/r}} - \frac{2\zeta^2g}{r^3\sqrt{1-2m/r}} \nonumber\\
&\quad\quad + \left(\frac{m}{r}\right)^\prime\frac{\zeta^2 g}{r^2(1-2m/r)^{3/2}}\\
&=\frac{\zeta^2}{r^2\sqrt{1-2m/r}}\left(2 g^2 + g^\prime - 2g/r + \left(\frac{m}{r}\right)^\prime\frac{g}{1-2m/r}\right)\\
&=\frac{\zeta^2}{r^2\sqrt{1-2m/r}}\left(2 g^2 + g^\prime - 2g/r - \frac{g}{2}\left(\ln(1-\frac{2m}{r})\right)^\prime \right).
\end{align}
As the pressure is isotropic we can use \eqref{eq:1.1.Pt_and_g} to relate $P$ and $g^\prime$. 
\begin{align}
2g^2 + g^\prime - \frac{2g}{r} &= \left(g^2 + g^\prime + \frac{g}{r}\right) + g^2 - \frac{3g}{r}\\
&= \frac{8\pi P + (g + 1/r)(m/r)^\prime}{1-2m/r} + g^2 - \frac{3g}{r}\\
&= \frac{2g(1-2m/r) - 2m/r^2 + (gr + 1)(m/r)^\prime}{r(1-2m/r)} + g^2 - \frac{3g}{r}
\end{align}
\begin{align}
&= -\frac{g}{r} + g^2 + g\frac{\left(\frac{m}{r}\right)^\prime}{1-2m/r} + \frac{r^2\left(\frac{m}{r}\right)^\prime - 2m}{r^3(1-2m/r)}\\
&= -\frac{g}{r} + g^2 + \left(\frac{m}{r}\right)^\prime \frac{g}{1-2m/r} + \frac{4\pi(\rho - \orho)}{1-2m/r}
\end{align}
or in terms of the original derivative:
\begin{equation}
\left(\frac{\zeta^2 g}{r^2\sqrt{1-2m/r}}\right)^\prime = 
\frac{\zeta^2}{r^2\sqrt{1-2m/r}}\left(g^2  + \left(\frac{m}{r}\right)^\prime\frac{2g}{1-2m/r} + \frac{4\pi\left(\rho - \orho\right)}{1-2m/r} - \frac{g}{r} \right).
\end{equation}
Now we can go back to \eqref{eq:the_integral}
\begin{align}
&\delta P(r) +\frac{g(r)}{4\pi r^2}\delta m(r) = \frac{\sqrt{1-2m/r}}{4\pi\zeta(r)^2}\int_r^{R} \left(\frac{\zeta(\orr)^2(1+2g(\orr)\orr)}{\orr^3 \sqrt{1-2m/\orr}}\left(g(\orr)+\frac{(m/\orr)^\prime}{1-2m/\orr}\right)\nonumber\right.\\
&\phantom{mespace} \left. - \left(\frac{\zeta(\orr)^2 g(\orr)}{\orr^2\sqrt{1-2m/\orr}}\right)^\prime\right)\,\delta m(\orr) \, \dx{\orr}. 
\end{align}
Expanding out the derivative in the final term gives
\begin{align}
&\delta P(r) +\frac{g(r)}{4\pi r^2}\delta m(r) = \frac{\sqrt{1-2m/r}}{4\pi\zeta(r)^2}\int_r^{R} 
   \frac{\zeta(\orr)^2}{\orr^2\sqrt{1-2m/\orr}}\nonumber\\
&\phantom{mespace}\times \left[\frac{(1+2g(\orr) \orr)}{\orr}\left(g(\orr) + \frac{(m/\orr)^\prime}{1-2m/\orr}\right) \right.\nonumber\\
&\phantom{mespace}\left. - \left(g^2 + \frac{2g(\frac{m}{\orr})^\prime}{1-2m/\orr} + \frac{4\pi(\rho-\orho)}{1-2m/\orr} + \frac{g}{\orr}\right)\right]\delta m(\orr)\,\dx{\orr}.
\end{align}
Writing out the terms and cancelling allows for considerable simplification
\begin{align}
&\delta P(r) +\frac{g(r)}{4\pi r^2}\delta m(r)\nonumber\\
&= \frac{\sqrt{1-2m/r}}{4\pi\zeta(r)^2}\int_r^{R} 
   \frac{\zeta(\orr)^2}{\orr^2\sqrt{1-2m/\orr}}\left(g(\orr)^2 - \frac{4\pi(\rho-\orho)}{1-2m/\orr} + \frac{(m/r)^\prime}{\orr(1-2m/\orr)}\right)\,\delta m(\orr)\,\d \orr\nonumber\\
&= \frac{\sqrt{1-2m/r}}{4\pi\zeta(r)^2}\int_r^{R} 
   \frac{\zeta(\orr)^2}{\orr^2\sqrt{1-2m/\orr}}\left(g(\orr)^2 - \frac{4\pi(\rho-\orho)}{1-2m/\orr} + \frac{4\pi(\rho-\orho/3)}{1-2m/\orr}\right)\,\delta m(\orr)\,\d \orr\nonumber\\
&= \frac{\sqrt{1-2m/r}}{4\pi\zeta(r)^2}\int_r^{R} 
   \frac{\zeta(\orr)^2}{\orr^2\sqrt{1-2m/\orr}}\left(g(\orr)^2 + \frac{8\pi\orho}{3(1-2m/\orr)}\right)\,\delta m(\orr)\,\d \orr.
\end{align}
Now we impose the condition that mass is moved inward or outward by requiring that $\delta m(r)$ be non-negative or non-positive respectively. Looking at the case where $\delta m(r) \geq 0$ we obtain:
\begin{align}
\delta P(r) +  \frac{g(r)}{4\pi r^2}\delta m(r) &= \frac{\sqrt{1-2m/r}}{4\pi\zeta(r)^2}\int_r^{R} \frac{\zeta(\orr)^2}{\orr^2\sqrt{1-2m/\orr}}\nonumber\\
&\quad\quad\times\left(g^2 + \frac{2m(\orr)/\orr^3}{1-2m(\orr)/\orr} \right)\delta m(\orr)\,\dx{\orr}\label{eq:varyP}\\
&\geq 0,
\end{align}
as all terms in the integrand are positive. In particular, as $\delta m(0) = 0$ (both solutions have no mass at $r=0$) we obtain Kovetz's result
\begin{equation}
\delta P_c \geq 0.
\end{equation}

\subsection{Variation in $g(r)$}
We can use the explicit expression for a variation in pressure express the change in $g(r)$ by referring back to \eqref{eq:vary_g}:
\begin{align}
\delta g&=\frac{4\pi r}{1-2m/r}\,\delta P + \frac{1+2g(r) r}{r^2(1-2m/r)}\delta m\\
&= -\frac{g(r)r}{r^2(1-2m/r)}\delta m(r) + \frac{r}{\sqrt{1-2m/r}\zeta(r)^2}\int_r^{R} \frac{\zeta(\orr)^2}{\orr^2\sqrt{1-2m/\orr}}\bigg\{g^2 \nonumber\\
&\phantom{mespace}+ \frac{8\pi\orho}{3(1-2m/\orr)} \bigg\}\delta m(\orr)\,\dx{\orr} + \frac{1+2 g(r) r}{r^2(1-2m/r)}\delta m(r)\\
&= \frac{1 +  g(r)r}{r^2(1-2m/r)}\delta m(r) + \frac{r}{\sqrt{1-2m/r}\zeta(r)^2}\int_r^{R} \frac{\zeta(\orr)^2}{\orr^2\sqrt{1-2m/\orr}}\bigg\{g^2 \nonumber\\
&\phantom{mespace}+ \frac{8\pi\orho}{3(1-2m/\orr)} \bigg\}\delta m(\orr)\,\d \orr.
\end{align}
It is clear that if $\delta m$ is everywhere greater (less) than zero then $\delta g$ would also be greater (less) than zero. i.e. the local acceleration due to gravity increases as energy is moved toward the centre of the star. Note that we have no requirement that the average density decreases. If the average density does decrease as we go out, then we know $\delta m(r) = m(r) - M(r/R)^3 \geq 0$ and we recover \eqref{eq:g_inequality} as a special case. The integral formulae for $\delta P$ and $\delta g$ are of course still valid if $\delta m$ is not non-negative or non-positive corresponding to mass being moved around but not purely inward or outward. However there is very little one can say about this general situation.

\section{Conclusions}

Even within the confines of a spherically symmetric spacetime there is a lot of freedom for generating solutions. It is appropriate to reiterate a comment made at the beginning of Chapter 5 of \cite{Hawk74}; that one can take any metric and define the energy momentum tensor so that the Einstein equations are satisfied. To restrict our attention to \emph{interesting} solutions requires some prejudice on our part as to the nature of possible stress tensors. The attitude adopted in this chapter has been to ignore specific models of matter such as an equation of state but to make assumptions about the nature of any kind of matter. As noted in \cite{twilight} even the mild assumptions made in this chapter may be fundamentally flawed. It must also be admitted that such processes may become astrophysically relevant, as violation of the null energy condition is required for Hawking radiation to take place.\footnote{The area increase theorem requires the null convergence condition: $R_{ab} W^a W^b \geq 0$ for any null vector $W$. This is implied by Einstein equations ($R_{ab} \propto T_{ab}$) and the null energy condition ($T_{ab} W^a W^b \geq 0$ for null $W^a$). Hawking radiation requires a decrease in mass (and hence area) so must violate the NEC \cite{Hawk74}.} It is still hoped that the energy conditions describe bulk matter well, and outside of measurements of the cosmological constant no violations of any of the energy conditions have yet been found experimentally \emph{in bulk matter}\footnote{The Casimir effect is one example of violations in the vacuum for small systems}.

\paragraph{}
The goal of this chapter was to find some way of describing ``reasonable'' spherically symmetric solutions. The isotropic solutions were able to be generated by specification of $g(r)$, the local acceleration due to gravity. By the conditions of isotropy and spherical symmetry we only have one free function in the metric and the sceptical reader may ask why our method is of interest: after all does it not just formalise the idea of taking an arbitrary function, getting a spherically symmetric metric, generating a solution and then defining the energy-momentum tensor to be whatever comes out the other side in exactly the way mentioned in Hawking and Ellis? While in essence this is what happens we have the additional nice properties that
\begin{itemize}
\item The gravity profile is physically measurable.
\item A class of physically unreasonable solutions can be discarded by requiring that $g(r) \geq 0$ without discarding any physically reasonable solutions.
\item Another class of physically unreasonable solutions can be discarded by requiring that $\lim_{r\rightarrow 0} g(r)/r = (P_c + \rho_c/3)$.
\item The solutions generated are guaranteed to be isotropic: $P_r = P_t$. This is a good description of most stars throughout the bulk.
\end{itemize}
The condition for a finite central pressure comes about from a single integral:
\begin{equation}
\int_0^R \frac{g(r)^2}{r(1+r\;g(r))}\exp\left(-2\int_0^r g(\bar{r})\frac{1-\bar{r}\;g(\bar{r})}{1+\bar{r}\;g(\bar{r})}\,\d \bar{r}\right)\,\d r < \infty\label{eq:finite_central_P}
\end{equation}
and as the pressure profile decreases throughout an \emph{isotropic} star we will then be guaranteed of a finite pressure everywhere. We can actually get a ``quick and dirty'' test of generating functions $g(r)$ for finite pressure. We start by noting that
\begin{equation}
-g(r) < g(r) \frac{1 - r\;g(r)}{1 + r\;g(r)} < g(r).
\end{equation}
Using the lower bound, we have that \eqref{eq:finite_central_P} is satisfied if
\begin{eqnarray}
\int_0^R \frac{g(r)^2}{r(1+r\;g(r))}\exp\left(+2\int_0^r g(\bar{r})\,\d\bar{r}\right)\,\d r &<& \infty
\end{eqnarray}
In contrast, the condition that $\rho \geq 0$ and $\rho \geq P$ are both exceedingly difficult to enforce. It is an unanswered question what the most general functions are that satisfy these constraints. It is unclear if such a project would even have a succinct answer. 

\paragraph{}
To this end other approaches have been developed, bounding the behaviour of the compactness throughout the star. The famous Buchdahl--Bondi result $\chi < 8/9$ is re-derived for stars that have a decreasing average density. To satisfy the DEC as well, we are able to show that the Schwarzschild interior solution requires that $\chi < 3/4$. A bound is established that limits to $8/9$ in the limit of the central density going to infinity for a star in which the average density decreases outward. However, no solutions that conform to the DEC have yet been found, and it is possible that better bounds exist. It is an unfortunate side effect of not imposing an equation of state that we are not able to obtain a maximum mass, only bound the compactness. One can see this by looking at the algorithmic solution for the mass \eqref{eq:ALGO2}:
\begin{equation}
m(r) = \frac{gr^2}{(1+rg)^2}\left(1 + \frac{gr}{2}\right) +\frac{2r^3}{(1+rg)^2}e^{2\vartheta}\left(C_{r_0} +\int_{r_0}^r \frac{g^2}{r(1+rg)}e^{-2\vartheta}\,\d r\nonumber\right)
\end{equation}
We change to the dimensionless variables:
\begin{eqnarray*}
x &=& \frac{r}{R}\\
G(x) &=& g(r)R = g(Rx)R\\
\bar{m}(x) &=& \frac{m(r)}{M} = \frac{m(Rx)}{M}
\end{eqnarray*}
We begin by transforming $\vartheta(r)$:
\begin{eqnarray*}
\vartheta(r) &=& \int_{r_0}^r g(\bar{r}) \frac{1 - \bar{r}\;g(\bar{r})}{1+g(\bar{r})\bar{r}}\,\d \bar{r}\\
&=& \int_{x_0}^x \frac{G(\bar{x})}{R}\frac{1 - \bar{x}\;G(\bar{x})}{1+\bar{x}\;G(\bar{x})}\,\d(R\bar{x})\\
&:=& \vartheta(x)
\end{eqnarray*}
Hence our equation changes to
\begin{eqnarray}
M\bar{m}(x) &=& \frac{GRx^2}{(1+xG)^2}\left(1 + \frac{Gx}{2}\right) \\
&&\quad\quad\quad+\frac{2Rx^3}{(1+Gx)^2}e^{2\vartheta}\left(C_{r_0}R^2 +\int_{x_0}^x \frac{G^2}{x(1+Gx)}e^{-2\vartheta}\,\d x\nonumber\right)\nonumber\\
\bar{m}(x) &=& \frac{R}{M}\left\{\frac{Gx^2}{(1+xG)^2}\left(1 + \frac{Gx}{2}\right)\right.\\
&&\left.\quad\quad\quad +\frac{2x^3}{(1+Gx)^2}e^{2\vartheta}\left(C_{r_0}R^2 +\int_{x_0}^x \frac{G^2}{x(1+Gx)}e^{-2\vartheta}\,\d x\nonumber\right)\right\}\nonumber
\end{eqnarray}
The only term inside the braces that contains a mention of scale is an arbitrary constant of integration, and it becomes clear that we may scale our solutions to arbitrarily high masses. This is unfortunate, as masses are much easier to infer from the nearby orbiting bodies; the compactness requires the additional information about the size of the body. Hence we see immediately that general relativity does not impose a ``maximum mass'' of a star, and to find such information we do need to supplement our work with an equation of state.

\paragraph{}
In the anisotropic case one has a lot more freedom. In this case the number of important results seem rather limited. While one can construct an algorithm in this case as well it is far from clear that it is useful. The compactness bounds are harder to state as well: if one is prepared to ignore the DEC then it seems that the compactness can be arbitrarily close to one. If the DEC is taken into account \emph{and} the radial pressure is assumed positive throughout \emph{and} the average density decreases throughout the star then it has been shown that $\chi \leq 0.974$. An explicit example is constructed with $\chi = 0.9196$, and it is difficult to see how to improve on it, although not all constraints have been saturated in obtaining this profile.

\paragraph{}
Perhaps the most important thing that we learn from anisotropy is to take the ``regeneration of pressure'' argument with a pinch of salt.

\def\et{\ensuremath{e_{(0)}}}
\def\er{\ensuremath{e_{(1)}}}
\def\eth{\ensuremath{e_{(2)}}}
\def\ephi{\ensuremath{e_{(3)}}}

\chapter{Quasinormal modes, dirty black holes and Zerilli's equation}
\label{chap:QNM_intro}
\section{Introduction}
Consider a stationary geometry such as a star or black hole in equilibrium
and perturb it, say by infalling matter or a particle with a small impact
parameter.  For small perturbations we can linearise the geometry about the
stable background solution, and the result is a dampened simple
harmonic system. Hence the solutions are superpositions of the form:
\begin{align}
g_{ab} &\sim g_{ab}^{(0)}+\delta g_{ab}\;\Re\left\{\exp(-i\omega_n t)\right\}\label{eq:chap2_omega_defined}\\ 
&= g_{ab}^{(0)}+ (\delta g_{ab})\exp(\Im(\omega_n) t)\cos(\Re(\omega_n) t),\quad\quad(\Im(\omega_n) < 0),
\end{align}
where $g^{(0)}_{ab}$ represents the value of $g_{ab}$ in the unperturbed
geometry. The ``resonant'' frequencies $\omega_n$ are referred to as the
\emph{quasinormal modes} (QNMs) or \emph{ringing modes}.

\paragraph{}
In numerical relativity it is expected that one can model systems at late times
accurately by using superpositions of a finite number of quasinormal modes over
a spatially bounded region. Hence gravitational wave detectors are ``tuned'' to
pick up frequencies of the order of the real part of the QNM with the smallest imaginary 
part, as they will decay the slowest. These are referred to in the literature as the 
``lowest lying QNMs''. In
looking at stellar structure, QNMs also determine the stability of the system
under small perturbations. More detail can be found in the review article
\cite{Kokkotas} which discusses these applications. For all of these
applications it is unsurprising that it is the slowly damped modes that are
relevant.

\paragraph{}
In contrast, Hod \cite{HodBohr} showed that if we take the real part of the
\emph{highly damped modes} of a Schwarzschild black hole to be a ``transition
frequency'', then it follows that the area of the black hole is quantised and
some information on the area spectrum is available. Hod also showed that this
argument could be used to determine the \emph{Barbero--Immirzi} parameter $\gamma$, an
arbitrary parameter in \emph{loop quantum gravity} (LQG), making the theory
essentially unique. For this conjecture to have validity we must find the same
value of $\gamma$ for any black hole, not just a Schwarzschild black hole. The
next two chapters are devoted to looking at the QNMs of spherically symmetric
black holes surrounded by matter to see if the claim made by Hod is generic.

\subsubsection{Hod's QNM conjecture}
An investigation of the \emph{highly damped} QNMs led to a proposal by Hod
\cite{HodBohr}: that the QNMs of a black hole given by \emph{classical} general
relativity tells us something about \emph{quantum} gravity. The asymptotic
spectrum for the Schwarzschild black hole was numerically found to obey
\cite{Andersson:1993,Nollert:1985ev}:
\begin{equation}
\omega_n = \frac{c^3\ln 3}{8\pi G M} - \frac{ic^3}{4GM}\left(n+\frac{1}{2}\right), \quad\quad \textrm{as }n\rightarrow\infty.\label{eq:qnm_asym}
\end{equation} 
Recently this asymptotic form has been ``proven'' analytically by Motl and Neitzke
\cite{Motl,MotlNeitzke}. Hod's observation was that we could take the energy of
a quanta emitted by a black hole to be:
\begin{equation}
\Delta E = c^2 \Delta M = \hbar \;\Re(\omega_n) = \frac{\hbar c^3\ln 3}{8 \pi GM} =  \frac{c^6\ln 3}{8\pi G^2 M}\ell_P^2 \label{eq:deltaE_2_ReO}
\end{equation}
via a ``correspondence principle'', where $\Delta M$ is the mass lost by the black hole and
$\ell_P$ is the Planck length. But the horizon's area is determined by the mass
with
\begin{equation}
A = 16 \pi\left(\frac{GM}{c^2}\right)^2
\end{equation}
which would imply that emission of such a quanta would be accompanied by a
decrease in area:
\begin{equation}
\Delta A = 32 \pi\frac{ G^2 M \Delta M}{c^4}.\label{eq:deltaA_deltaM}
\end{equation} 
Taking the change in mass from \eqref{eq:deltaE_2_ReO} shows that the change in area is
\begin{equation}
\Delta A= 4\ell_P^2\ln 3. \label{eq:change_in_area}
\end{equation}
This suggests that the area of a black hole is quantised, and has an
equi-spaced spectrum! More precisely, there is
  an area operator and its eigenvalues are equally spaced. \footnote{It is seen in 
  \S\ref{sec:states_discussion} that
  this conclusion is not automatic. It is possible to obtain a non-equi-spaced
  spectrum provided different assumptions are made. At the moment the arguments
  are heuristic only to provide motivation for the QNM conjecture.} The entropy
of the black hole is given by the Bekenstein-Hawking formula:
\begin{equation}
S_{\mathrm{BH}}= \frac{A}{4\ell_P^2} = \frac{N\Delta A}{4\ell_P^2} = \ln\left( 3^N\right),
\end{equation}
which has led to speculation that the entropy of a black hole is due to a
``condensation'' of 3-state systems at the horizon (such as spin-1 links of a
spin network).

\paragraph{}
This leads us to the \emph{QNM conjecture}:
\begin{quote}
  The QNM conjecture is that the real part of the asymptotic form of the QNM of
  a black hole leads to a ``transition frequency''. This transition frequency
  is related to the \emph{dominant} change in the area spectrum. For a generic black hole:
\begin{subequations}
  \begin{equation}
    \Delta A = \frac{\partial A}{\partial M}\; \frac{\hbar \Re(\omega_n)}{c^2}.
  \end{equation}
  For the \emph{specific} example of a Schwarzschild black hole that motivated this
  conjecture we have (from equations \eqref{eq:deltaE_2_ReO} and
  \eqref{eq:deltaA_deltaM})
\begin{equation}
\Delta A = 32\pi \frac{G^2 \hbar}{c^6}M \;\Re(\omega_n) 
               = \frac{32\pi GM}{c^3}\;\ell_P^2\Re(\omega_n). 
\end{equation}
\end{subequations}
\end{quote}

\subsubsection{Loop quantum gravity and the QNM conjecture}
The QNM mode conjecture is not a theory of quantum gravity, but an observation.
It says nothing about where the entropy comes from and does not give a
formalism for performing calculations. All that the QNM conjecture does is tie
the QNM spectrum to the area spectrum of the black hole. LQG is a specific theory of
quantum gravity, with one adjustable parameter $\gamma$ called the
Barbero--Immirzi parameter. By making LQG compatible with the QNM conjecture,
we arrive at a unique prediction for $\gamma$. This requires us to run the
previous argument \emph{backward}: LQG supplies the area spectrum, up to an
arbitrary constant $\gamma$. The QNM conjecture fixes the absolute spacing in
the spectrum and hence predicts a value for $\gamma$.

\paragraph{} 
An area spectrum $\{ A_n \}$ in LQG is given by \cite{Ashtekar:2001}:
\begin{equation}
A_n = 8\pi \gamma \ell_P^2\sum_{2j=0}^\infty n_{j} \sqrt{j(j + 1)},\label{eq:std_lqg_area}
\end{equation}
where $n_j$ is the number of spin-links with spin $j$. If we assume that almost
all of the spin-links that pass through the horizon of a black hole have the
same spin value $j_{\textrm{dom}}$ then the area is
\begin{equation}
A_n \approx 8\pi \gamma \ell_P^2 N \sqrt{j_{\textrm{dom}}(j_{\textrm{dom}} +
1)}\label{eq:num_states}.
\end{equation}
A spin $j$ link on a spin network has $2j+1$ possible values. Hence we have a
possible $(2j + 1)^N$ states on the horizon itself, or an entropy of
\begin{equation}
S \approx N\ln(2j_{\textrm{dom}}+1) = \frac{\ln(2j_{\textrm{dom}}+1)}{8\pi \gamma
\ell_P^2\sqrt{j_{\textrm{dom}}(j_{\textrm{dom}} + 1)}}\;A_n
\end{equation}
where the number of states $N$ has been approximated via \eqref{eq:num_states}.
This can be made compatible with the Bekenstein-Hawking entropy formula provided we choose
\begin{equation}
 \gamma = \frac{\ln(2j_{\textrm{dom}}+1)}{2\pi\sqrt{j_{\textrm{dom}}(j_{\textrm{dom}} + 1)}}\,\,.\label{eq:gamma_entropy}
\end{equation}
There is some theoretical prejudice for $j_{\textrm{dom}}=1/2$, as fermions
would fit neatly into such a theory. Instead, we now repeat Hod's argument: a
loss of a spin $j_{\textrm{dom}}$ link would emit energy $\hbar\Re(\omega)$.
This would imply a change in area is given by \eqref{eq:num_states}
\begin{equation}
\Delta A = 8\pi \gamma \ell_P^2 \sqrt{j_{\textrm{dom}}(j_{\textrm{dom}} + 1)}.\label{eq:area_j}
\end{equation}
Inverting the argument in \eqref{eq:deltaE_2_ReO} implies that the ``transition
frequency'' should be given by
\begin{equation}
\Re(\omega) = \frac{\Delta M c^2}{\hbar}= \frac{\Delta A c^6}{32 \pi \hbar G^2 M} = \frac{c^3 \Delta A}{32 \pi G M \ell_P^2}.
\end{equation}
Substituting in for the area from \eqref{eq:area_j}:
\begin{equation}
\Re(\omega_n)= \frac{8\pi \gamma \ell_P^2 
\sqrt{j_{\textrm{dom}}(j_{\textrm{dom}}+ 1)}}{32 \pi GM \ell_P^2}c^3 
=\frac{\gamma\sqrt{j_{\textrm{dom}}(j_{\textrm{dom}} + 1)}}{4 GM}.
\end{equation}
Equation \eqref{eq:gamma_entropy} constrains $\gamma$ in a 
LQG model to recover the correct expression for the entropy. We then find
that
\begin{equation} 
\Re(\omega_n) = \frac{c^3\ln(2j_\textrm{dom} + 1)}{8\pi GM}.
\end{equation}
This is the real part of the asymptotic spectrum \eqref{eq:qnm_asym} if we
choose $j_{\textrm{dom}} = 1$! Hence the QNM conjecture and LQG are not only
compatible, but forcing such compatibility uniquely determines $\gamma$.

\subsubsection{Issues with the QNM conjecture}
While the QNM conjecture is interesting, it relies on a numerical coincidence
and it leaves many questions unanswered. We have seen that the QNM conjecture relies on
each spin link of spin $j$ adding the same amount to the area (at least asymptotically), it
does not rely on there being an equi-spaced area spectrum for different values of $j$. Indeed,
the standard LQG spectrum provides an explicit counter-example! Provided that one spin dominates
and each of those dominating spin links providing an equal amount to the area allows the 
black hole to decrease its area in multiples of a fixed amount, and satisfy the
QNM conjecture. However, the prediction that the black hole loses area (and
hence mass squared, in the Schwarzschild case) in constant discrete amounts is
at odds with the emitted radiation (Hawking radiation) being purely blackbody.
The prediction that the black hole should be a condensate of links with spin 1
also raises the question of how fermions are incorporated into LQG. Why does this
correspondence only occur for the rapidly decaying modes (i.e.  large $n$)? These
are interesting questions and will be addressed in \S \ref{sec:qnm_background}.

\paragraph{}
The research in the next two chapters is more modest. All of the foregoing
considerations were focused on a Schwarzschild black hole -- that is a spherically symmetric black hole in an otherwise empty spacetime -- but if the
QNM correspondence is fundamental it should apply to \emph{all} black hole
spacetimes. As the general black hole entropy is formulated in terms of the
surface gravity $\kappa$, the QNMs should also rely on $\kappa$. This is
significant as $\kappa$ is not completely determined by the black hole, but is
affected by (non-black hole) matter outside the horizon. Hence we introduce the
notion of a \emph{dirty black hole} as a black hole surrounded by matter, and
investigate if the QNM conjecture is still valid. In terms of LQG specifically,
we know that $\gamma$ is a parameter that can only be fixed once; that is, two
different geometries should not give different values for $\gamma$. If one
enforces LQG \emph{only} to be consistent with the Bekenstein--Hawking entropy
then different geometries do give the same value for $\gamma$
\eqref{eq:gamma_entropy}, although $j_{\textrm{dom}}$ is still undetermined
(see \S B of \cite{Ashtekar:2001}). Our task will be to see if the QNM
conjecture will predict $j_{\textrm{dom}} = 1$ for all static, spherically
symmetric black hole geometries.

\paragraph{}
The next section of this thesis outlines in more detail some of the arguments
for and against the QNM conjecture. Readers familiar with the arguments or
wanting to see concrete results would be best to look at \S
\ref{sec:intro_qnm}, where QNMs are described as a scattering problem, and then
at \S\ref{sec:wave_eqn} where the wave equations for dirty black hole
spacetimes are derived. In chapter \ref{chap:qnm_find} the different methods of
finding quasinormal modes are discussed.

\section{The first window: Alice's curiosity}\label{sec:qnm_background}
\subsection{Black hole entropy}
It has long been known that one can associate a temperature and entropy with
black holes, as was initially advocated by Bekenstein \cite{Bekenstein:1973}.
One argument is that black holes have many features reminiscent of
thermodynamic quantities. In particular, the \emph{area} of a black hole
horizon plays an important role, as can be seen by considering the following
results:
\begin{itemize}
\item \emph{Reversible} processes leave the area unchanged, a result due to
Christo\-doulou \cite{Christodoulou:1970}.
\item The area theorem, proved by Hawking in 1971, showed that subject to the
  null energy condition the area of a black hole never decreases.
  \cite{Hawking:1971,Hawk74}.
\end{itemize}
The area theorem trivially contains Christodoulou's result, as the initial and
final areas are the same for a reversible process and the area never decreases.
Despite this redundancy, Christodoulou's result nicely emphasises the idea that
the area of a black hole is an adiabatic invariant. Stationary black holes can
be described by a limited number of parameters -- mass, charge and angular
momentum -- and all of these can be changed by reversible processes
\cite{Christodoulou:1970}. Hence if a black hole has entropy, it is plausible
that:
\begin{equation}
S_{\textrm{BH}} = S(A).
\end{equation}
At this stage we only know that $S(A)$ is a monotonically increasing function
of area. To obtain the functional form of the entropy, Bekenstein argued that
truly reversible processes were impossible: black holes
had to increase their area and that the amount of increase was independent of
the parameters of the black hole! To see this requires Christodoulou's result
that the increase in the area of a horizon, when a black hole captures a
particle of mass $m$ with a turning point at proper distance $b$ away from the
horizon, is given by \cite{Christodoulou:1970,HodBohr}:
\begin{equation}
(\Delta A)_{\textrm{min}} = 8\pi G\;\frac{m b}{c^2}.
\end{equation}
The process is then reversible if and only if $b=0$; that is, the turning point
occurs \emph{on} the horizon. Bekenstein argued that because of the uncertainty
principle, the turning point for a real particle could not be determined
exactly. Taking a particle that \emph{classically} had a turning point at the
horizon would be subject to some uncertainty in both the momentum and position,
leading to
\begin{align}
(\Delta A)_{\textrm{min}} &= \frac{8\pi G}{c^2} \frac{\sqrt{m^2 c^4 + (\delta P)^2 c^2}}{c^2}\;\delta x \geq \frac{8\pi G}{c^2} \delta P\;\delta x\\
&\geq 4\pi \frac{ G \hbar}{c^3} = 4 \pi \ell_P^2.
\end{align}
This minimum area increase is completely independent of the characteristics of
the black hole! This led Bekenstein to the idea that the entropy was
proportional to the area and that the area was discrete with \emph{equi-spaced
  eigenvalues}.

\paragraph{}
Bekenstein's original suggestion of an entropy associated with black holes was
motivated by a very different consideration: ensuring that the validity of the
second law of thermodynamics was not violated. In fact the suggestion pre-dated
both the area theorem and Christodoulou's result! If black holes did not have
an entropy, then by lowering objects into a black hole we could ``lose'' the
entropy associated with that object and hence violate the second law of
thermodynamics! However, while this argument suggests that an entropy is
associated with a black hole, it does not tell us what form it takes. The
result that the entropy is proportional to the \emph{area} leads to an
interesting problem -- that the entropy of an object is bounded by the amount
of area it would have if made into a black hole. This is the idea behind the
\emph{holographic principle}. Unfortunately, na\"{\i}ve attempts to create a
bound on the amount of entropy an object can have that is compatible with the
holographic principle are plagued with difficulties, and it is an ongoing task
to find a satisfactory statement. However, if we insist that the total
entropy, or so-called \emph{generalised entropy},
never decreases, we will require some sort of holographic principle.

\paragraph{}
Notice that neither of these arguments suggest any sort of ``microstates'' of a
black hole that the entropy is associated with. This is another unresolved
problem in black hole thermodynamics which will be re-addressed in chapter
\ref{chap:conformal}.

%\subsection{Hawking radiation}
%\label{sec:hawking_radiation}
%\input{Hawking}
%\begin{itemize}
%\item BH Temperature: mention frequency mixing, Unruh effect. Actual illustrate the ``eternal BH'' case %and mention Matsubara frequencies and black body spectrum for \emph{a single black hole}
%\item Mention first law of BHT: not limited to the Schwarzschild case.
%\item Mention good semiclassical support for black hole temperature and entropy.
%\end{itemize}

\subsection{Discussion concerning the QNM conjecture}\label{sec:states_discussion}
There are many unanswered questions about the QNM conjecture that leave it
somewhat unsatisfactory. The most important such question is whether or not the
conjecture is generic, or just a numerical coincidence for the Schwarzschild
black hole.  Even if the result turns out to be generic, we still have to
explain how the discrete spectrum from the QNM gives rise to blackbody Hawking
radiation. We also have to fit fermions into a theory that seems to be
dominated by states with a \emph{three fold} area degeneracy\footnote{These are
  ``spin-1 links'' in LQG, but the QNM conjecture allows for any three
  microstate subsystem.}. Many of these questions have been raised by
Polychronakos \cite{Polychronakos:2003}.

\paragraph{}
Let us first introduce the issue of an the logical relationship between the QNM
conjecture, LQG and quantum gravity. The QNM conjecture is non-generic as it
\emph{requires} the area of a black hole horizon to be quantised. This
quantisation is not required by a quantum theory of gravity; there exist
``perfectly good'' string theories that do not quantise area.\footnote{Here
  perfectly good string theories means that string theories that do not
  quantise area suffer no more technical problems than those that do.} In LQG
the basic variables are holonomies which describe how to parallel transport
spinors, and the fundamental excitations of the geometry is analogous to
electric field lines, or quantum vortices. In the classical limit the
geometries approach the standard area (see \cite{Ashtekar:2001} for a review,
or \cite{RovelliBook,Rovelli:1995ge,Rovelli:1995ac} for calculations of the
area spectrum and the relationship of these ``vortices'' to spin network
states).  Each vortex line penetrating a surface then contributes to its area.
The vortex lines are formally analogous to spin network states and can be
labelled by the half-integers. Although the QNM conjecture is most often used
to determine the Barbero-Immirzi parameter in LQG, it should be noted that the
conjecture and LQG are logically independent. In particular:
\begin{itemize}
\item If the QNM conjecture is false then it does not tell us \emph{anything}
  about the nature of quantum gravity, and in particular about LQG. 
\item If LQG is false, then we have discarded a specific model that quantises
  area on the horizon. The QNM conjecture could still be compatible with
  another theory of quantum gravity that quantises area.
\item If the area is not quantised, then of course LQG and the QNM conjecture
  are both false.
\end{itemize}

\paragraph{}
Once we have assumed that the area is quantised, we must decide on an
appropriate area spectrum. As seen in the introduction, the QNM conjecture
implies that (for a non-rotating black hole) the emission of any quanta gives the
same change in area. In a generic spectrum this would imply that one particular
transition dominates. In an equi-spaced area spectrum transitions would could
occur between the $n$ and the $n-c$ states for any $n$ and some constant $c$. In
principle it is also possible for different transitions in a non-equi-spaced
spectrum to produce the same change in area via selection rules. In LQG there
are two spectra that are frequently discussed. The first is the standard
spectrum (c.f. \eqref{eq:std_lqg_area}):
\begin{subequations}
\begin{equation}
a_i = 8\pi \gamma\ell_{P}^2 \sqrt{\frac{i}{2}\left(\frac{i}{2}+1\right)},\label{eq:lqg_std}
\end{equation}
where $a_i$ is the area from a spin link with spin $i/2$. The other comes from
applying ``radiative corrections'' on \eqref{eq:std_lqg_area} (see
\cite{Alekseev:2000hf})
\begin{equation}
a_i = 4\pi\gamma\ell_{P}^2(i+1).\label{eq:lqg_radiative}
\end{equation}
\end{subequations}

\paragraph{}
One of the arguments presented in Polychronakos's paper
\cite{Polychronakos:2003} was that fermions could be included in LQG, while
still having the dominant contribution to the area due to links with spin one,
thus remaining compatible with the QNM conjecture. More precisely, a theory of
LQG with gauge group $SU(2)$ is still compatible with the QNM conjecture.
Polychronakos demonstrated that the dominance of spin one links could occur
due to thermodynamic suppression of spin half particles.

\paragraph{}
To see how this comes about, consider a horizon of total area $A$. Denote the
number of spin links that pass through the horizon with spin $i/2$ by $n_i$.
Then it follows that
\begin{equation}
A = \sum_{i=0}^{\infty} n_i a_i \label{eq:partition_of_A},
\end{equation}
where $a_i$ is the area added by each link with spin $i/2$. The details of the
area spectrum are left open at this point in the argument, allowing us to
compare \eqref{eq:lqg_std} to \eqref{eq:lqg_radiative} later on.\footnote{The
  most general spectrum would be a function of all the $n_i$, and would not
  require that each spin link of the same value would add the same amount of
  area. In the case where the horizon area is some arbitrary function we cannot
  relate the QNM conjecture and LQG. However, the purpose of this argument is
  simply to show that with the inclusion of fermions, LQG and the QNM conjecture are not
  mutually inconsistent.} Each set $\{n_i\}$ satisfying
\eqref{eq:partition_of_A} is called a partition of $A$. A link with spin $i/2$
has a degeneracy given by
\begin{equation}
g_i = i+1.
\end{equation}
Here it is assumed that the area added by a link is given by the total spin $j$
and all of the $2j+1$ internal states are independent.

\paragraph{}
Now a partition function for the \emph{area} is constructed
\cite{Polychronakos:2003,Gour:2004gk}
\begin{equation}
Z(\beta) = \sum_{\{m_i\}} \prod_{i=0}^\infty (i+1)^{m_i}\exp(-\beta m_i a_i).
\end{equation}
Explicitly we are summing over all sequences of positive integers -- \emph{not}
only partitions of the area $A$. Here we are looking at a grand canonical
ensemble, rather than a micro-canonical ensemble. Note the absence of
factorials, implying that the spin-links with the same value of $j$ are
\emph{not} indistinguishable.  Note also that this distinguishability (or partial
distinguishability \cite{Polychronakos:2003}) is an assumption that allows
fermions to be incorporated, and has not been ``proven''. For standard
thermodynamic arguments to be valid it is necessary to assume that the black
hole is enclosed and that the total area of the system is conserved. As pointed
out by Gour and Suneeta \cite{Gour:2004gk} there is no natural analogue of a heat
bath in this system.  Instead we rather recklessly assume that standard
thermodynamic arguments hold.\footnote{Note that the assumption that ``area
thermodynamics'' is valid is weak, but this has \emph{nothing} to do with
standard black hole thermodynamics which are provable statements in classical
general relativity.}

\paragraph{}
The rest of the argument is nicely given by Gour and Suneeta, given the
assumption that we can use thermodynamic arguments. If we have $\beta$ less than
a critical value $\beta_c$ then higher spins dominate the partition function and
the partition function does not converge. In particular, the contribution from a
given spin $i/2$ will not converge unless $\beta > \beta_i$, where $\beta_i$ is
given by
\begin{equation}
\beta_i = \frac{\ln(g_i)}{a_i} = \frac{\ln(i+1)}{a_i}.\label{eq:beta_i}
\end{equation}
To see this, note that the sequences of positive integers has the natural
numbers as a subsequence. Thus the partition function has the geometric series
as a subseries, and \eqref{eq:beta_i} ensures converges of this particular
subsequence. To ensure all the subsequences converge it is necessary that
\begin{equation}
\beta > \beta_c = \max_{i}\beta_i.
\end{equation}
Provided $\beta > \beta_c$ we have
\begin{equation}
Z(\beta) = \prod_{i=0}^\infty \frac{1}{1-(i+1)\exp(\beta a_i)} .
\end{equation}

\paragraph{}
As the area is the analogue of the energy for this ``thermodynamic'' system we
have
\begin{align}
\langle A\rangle &= -\frac{\d \ln(Z)}{\d \beta} 
                                =\sum_{i=0}^\infty\frac{(i+1)a_i\exp(-\beta a_i)}{1- (i+1)\exp(-\beta a_i)}.
\end{align}
The numerator vanishes rapidly, provided that the denominator is well behaved.
Thus large areas are possible if $\beta \approx \beta_c$. 

\paragraph{}
Nothing more can be said until an area spectrum has been assumed. Using the
standard area spectrum \eqref{eq:lqg_std} we have
\begin{subequations}
\begin{align}
\lim_{j\rightarrow 0}\beta_j &= 0\\
\beta_{1} &= \frac{\ln 2}{4\pi \gamma \ell_P^2 \sqrt{3}} = \frac{0.0318}{\gamma \ell_P^2}\\
\beta_{2} &= \frac{\ln 3}{8\pi \gamma \ell_P^2 \sqrt{2}} =   \frac{0.0309}{\gamma \ell_P^2}\\
\beta_{3} &= \frac{\ln 4}{4\pi \gamma \ell_P^2 \sqrt{15}}= \frac{0.0285}{\gamma \ell_P^2}.
\end{align}
\end{subequations}
All subsequent spins have a lower $\beta_i$, thus the standard spectrum predicts
domination by spin $1/2$ particles. The prediction for $\gamma$ does \emph{not}
agree with the QNM conjecture.

\paragraph{}
However, if one uses the equi-spaced spectrum \eqref{eq:lqg_radiative}:
\begin{subequations}
\begin{align}
\beta_{0}   &= \frac{\ln 1}{4\pi \ell_P^2\gamma} = 0\\
\beta_{1} &= \frac{\ln 2}{8\pi \gamma\ell_P^2} =  \frac{0.0276}{\gamma\ell_P^2}\\
\beta_{2} &= \frac{\ln 3}{12\pi \gamma\ell_P^2}=\frac{0.0291}{\gamma\ell_P^2}\\
\beta_{3} &= \frac{\ln 4}{16\pi\gamma} = \frac{0.0276}{\gamma\ell_P^2},
\end{align}
\end{subequations}
with all subsequent $\beta_i$s lower. Then the dominant contribution is from
$i=2$, or spin links with spin 1. Note that the fermions are still described by
the theory but are suppressed. 

\paragraph{}
Although there are a number of aspects of this treatment that are at best
unsatisfactory, it provides a concrete example of how LQG with fermions and the
QNM conjecture can be compatible. Before taking the argument too seriously,
it would need to explain why the area can be
treated as a conserved quantity. More seriously, objects that are not black
holes tend to have \emph{larger} areas than a black hole of the same mass;
so the above argument would suggest that everything macroscopic was a
collection of spin-1 links!

\paragraph{}
Polychronakos has also looked at the possibility that black hole states are
fully distinguishable. Assuming the QNM conjecture, the Bekenstein-Hawking
entropy cannot be (exactly) recovered in LQG without changing the gauge group
to $SO(3)$ and hence excluding coupling to fermions. 

\paragraph{}
A separate issue facing the QNM conjecture is Hawking radiation. While the
conjecture is not sensitive to the details of the area spectrum it does tell us
that the radiation from the black hole will be a discrete spectrum rather than
a continuous (blackbody) spectrum predicted by Hawking.  Polychronakos
\cite{Polychronakos:2003} states that this may be analogous to an atom in
thermal equilibrium with a heat bath which would radiate its own line spectrum
if placed in isolation. The problem with this analogy is that blackbody
radiation of atoms is derived when many ``independent oscillators'' are
present, whereas black hole temperature calculations assume the metric of a
single black hole. Only at high frequencies does the Hawking radiation ``look''
like a black body spectrum.

\paragraph{}
The condition that the QNM conjecture is generic is the most serious and also
the hardest to check. The goal of the next two chapters is to investigate dirty black hole 
spacetimes, and try to determine if the ``correspondence principle'' that applies to the Schwarzschild 
black hole applies to these more general black hole spacetimes.

\section{An introduction to quasinormal modes}\label{sec:intro_qnm}
The idea of a quasinormal mode was presented very briefly in the introduction
of this chapter. Here, we extend that idea, and show how QNMs are
defined and what they are used for. Of direct relevance to the investigation of
the QNM of black holes will be the interpretation of QNM as a scattering
problem in quantum mechanics. This motivates the ``monodromy'' approach of Motl
and Neitzke \cite{MotlNeitzke} and the ``Born approximation'' of Visser, Medved
and Martin \cite{Medved:2003rg}. In addition, a brief overview will be given on
how quasinormal modes are used in other areas of astrophysics.

\subsection{Defining quasinormal modes}\label{sec:def_QNM}
The following is a modified discussion from Kokkotas and Schmidt
\cite{Kokkotas}. Let us consider a wave equation with a potential function
that vanishes provided that $ |x|> x_0$:
\begin{equation}
\frac{\partial^2 \chi(t,x)}{\partial t^2} - \frac{\partial^2 \chi(t,x)}{\partial x^2} + V(x) \chi(t,x) = 0\label{eq:wave_eqn0}.
\end{equation}
When \emph{normal modes} are studied we restrict attention to systems of finite
energy. In this case the operator $\hat{H}$, defined by
\begin{equation}
\hat{H} = -\frac{\partial^2\phantom{x}}{\partial x^2} + \hat{V}(x),
\end{equation}
is self-adjoint and has a continuous spectrum if the domain of $x$ and $t$ is
infinite. Because of the continuous spectrum, we no longer have a 
square-integrable eigenfunction \cite{Kokkotas}. By introducing distributions
we can still write superpositions of solutions as new solutions.

\paragraph{}
Normal modes are useful in studying steady oscillations of a system and
non-localised finite energy solutions. They are not suitable for defining the
evolution of data of compact support or systems that radiate energy off to
infinity. Notice that these problems only occur in infinite systems where the
spectrum is continuous. Even finite systems can be dissipative, and here normal
modes do not work either. In modelling perturbations, we want to treat some
initial perturbation that is generically of compact support and that will
radiate away to infinity; for this we use \emph{quasinormal modes}. To get
decay of the solutions we require that operator $H$ is no longer self-adjoint,
so that we may have complex eigenvalues. Hence we cannot look at the space of
finite energy solutions! The physical reason for this is that the wave equation
is reversible, so if we have a finite energy solution at $t=0$ that decays at
late times it must \emph{grow} at early times. This is not a problem physically
as something would cause the initial perturbation; it would not be a decay of
an infinite ``perturbation'' at an infinite time in the past!

\paragraph{}
We are assuming that the potential vanishes if $|x| > x_0$, and is always
positive. Then the solution $\chi(t,x)$ is always bounded above. Define the
\emph{Laplace transform} of $\chi(t,x)$:
\begin{equation}
\chi_L(s,x) = \int_0^\infty \chi(t,x)\exp(-s t)\,\d t
\end{equation}
which makes sense for all $\Re(s) > 0$. The wave equation \eqref{eq:wave_eqn0} becomes
\begin{equation}
s^2 \chi_L - \frac{\partial^2 \chi_L}{\partial x^2} + V\chi_L = s\chi(0,x) + \frac{\partial\chi}{\partial t}(0,x)\label{eq:laplace_transform}
\end{equation}
where the right hand side contains all the initial data. One approach to
solving such a problem is to use \emph{Green's functions}. To find a Green's
function one looks at solutions that give a ``delta function'' contribution. A
well known method is to find two linearly independent solutions to the
\emph{homogeneous equation} $f_+$ and $f_-$. A Green's function can then be
explicitly found \cite{book:Bence}:
\begin{equation}
G(s,x_1,x_2) = \frac{1}{W(s)}\times\begin{cases}
f_-(s,x_1)\;f_+(s,x_2)&(x_1 < x_2)\\
f_-(s,x_2)\;f_+(s,x_1)&(x_1 > x_2)
\end{cases}
\end{equation}
where $W(s)$ is the Wronskian of the two solutions:
\begin{equation}
W(s) = \left| \begin{array}{cc}
f_-& f_+\\
\frac{\partial f_-}{\partial x}& \frac{\partial f_+}{\partial x}
\end{array}\right|.
\end{equation}
After finding a Green's function, the original differential equation is solved,
albeit in integral form:
\begin{equation}
\chi(t,x) = \int_{-\infty}^\infty G(s,x_1,x_2)\left\{s\chi(0,x) + \frac{\partial\chi}{\partial t}(0,x)\right\}\,\d s.
\end{equation}
In selecting $f_+$ and $f_-$ it is convenient to pick them such that they are
bounded as $x$ becomes large. As they are solutions of the homogeneous equation
and the potential is assumed to vanish, we have
\begin{equation}
f = \exp(\mp sx), \quad(|x| > x_0)
\end{equation}
Hence we pick $f_+$ so that it decays for large $x$ and $f_-$ so that it decays at large negative $x$:
\begin{equation}
f_{\pm}(x) = \exp(\mp s x), \quad(\pm x > x_0) \label{eq:decay_relations}.
\end{equation}
Although the large distance behaviour of $f_+$ is $\exp(-sx)$, toward small
values of $x$, $f_+$ is represented by a linear combination of exponential
terms. This is similar to the tracking of WKB solutions used in the monodromy
approach to QNM.

\paragraph{}
We finally have enough background to define the QNM themselves! Suppose that there is a (complex) frequency $\omega_n$ such that:
\begin{equation}
f_+(x) = c(s) f_-(x)
\end{equation}
where $c(s)$ is a complex number. This would imply that $f_+(x)$ tends to one
at large positive $x$, and to $c(s)$ at large negative $x$. In particular the
Wronskian of the two solutions vanishes and the Green's function becomes
singular. The fact that the solution is bounded means that the Green's function
exists and is analytic for $\Re(s)>0$, so that any poles if they occur must
occur for $\Re(s)<0$. We call such values of $s$ \emph{quasinormal} and they
form a countable set for positive potentials of compact support
\cite{Bachelot:1993dp,Kokkotas}, hence we label them $s_n$.

\begin{figure}[t]
\begin{center}
\includegraphics[width = 0.64\textwidth]{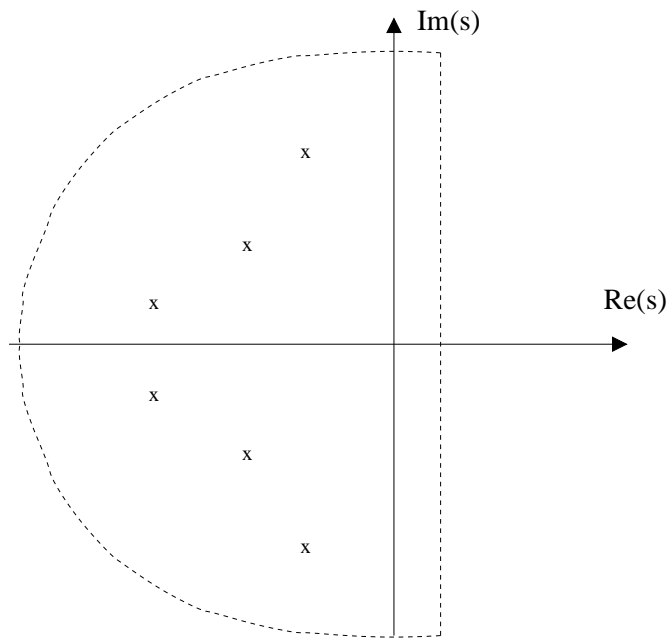}
\caption{Taking the inverse Laplace transform. The poles in the Green's function are denoted by ``x'' and the contour of integration is the dotted line.}
\label{fig:greens_function}
\end{center}
\end{figure}

\paragraph{}
We can then take the inverse Laplace transform or \emph{Bromwich integral} \cite{book:Bence}:
\begin{equation}
\chi(t,x) = \frac{1}{2\pi i }\int_{-\infty}^\infty \chi_L(a+ib,x)\exp((a+ib
)t)\,\d b
\end{equation}
where $a>0$ for convergence reasons. By closing the contour as shown in figure
\ref{fig:greens_function} and using the Cauchy residue theorem, we see that it
is the poles of the Green's function that contribute to the line integral. The
 solution in a finite region of space $(x_0 < x < x_1)$ and at late times
($t > t_1$) can be approximated by the finite sum:
\begin{equation}
\chi_N(t,x) = \sum_{n=1}^N d_n \exp(s_n t) f_+(s_n, x) \label{eq:perturb_approx}
\end{equation}
where the $s_n$ are quasinormal frequencies ordered by their real parts: $\Re(x_n) \leq \Re(s_{n+1})$. The coefficients $d_n$ depend only on the initial data.

\paragraph{}
So how good an approximation is \eqref{eq:perturb_approx}? Bacholet and
Motet-Bacholet \cite{Bachelot:1993dp} showed that for positive potential of
compact support,
\begin{equation}
\left|\chi(t,x) - \chi_N(t,x)\right| \leq C(x_0,x_1;t_1;\epsilon)\exp\left((-|\Re(s_{N+1})| + \epsilon)t\right)\label{eq:error}
\end{equation}
where $\epsilon$ is greater than zero. At late times the approximation
\eqref{eq:perturb_approx} becomes very close to the exact numerical solutions,
making QNM a very effective way of studying perturbations.

\paragraph{}
In models of black holes the ``potential function'' $V(x)$ does not have
compact support and this introduces additional complications. The functions
$f_+$ and $f_-$ may not exhibit exponential decay as large values of $x$ are
approached. The longer tails of the ``quasinormal eigenfunctions'' means that
the decay of the error \eqref{eq:error} is also less than exponential. As
pointed out by Kokkotas and Schmidt \cite{Kokkotas}, if one restricts attention
to a finite interval of time $t_0 < t < t_1$ then the wave equation cannot
depend on the values of the potential for large $x$ and exponential decay of
the error is recovered.

\subsection{Relating QNM to scattering}
The solution $\psi(x)$ to the  time independent Schr\"odinger equation
\begin{equation}
-\frac{\hbar^2}{2m}\frac{\partial^2 \psi}{\partial x^2} + V(x)\psi = E\psi\label{eq:schro}
\end{equation}
will also solve the \emph{homogeneous} Laplace transform of the wave equation
\begin{equation}
-\frac{\partial^2 \psi}{\partial x^2} + V(x)\psi = -s^2 \psi
\end{equation}
if we identify $E=-s^2$ and $m=\sqrt{\hbar}/2$.  Under this identification, the
two systems are equivalent. From the previous discussion we know that QNMs are
restricted to $\Re(s) < 0$ if $V(x)$ is positive and of compact support. In
terms of $E$, this means that any value of $E$ is allowed except for values
lying on the negative real axis.\footnote{Such energies are a requirement for
  bound states. Note that a negative energy eigenvalue does not necessarily
  imply a bound state as the state must also be normalisable to qualify as a
  bound state.}

\paragraph{}
We can show that negative real $E$ is not allowed directly from
\eqref{eq:schro}. Let us multiply \eqref{eq:schro} by the complex conjugate
$\psi^*$ and integrate over the real line:
\begin{align}
-\int_{-\infty}^{\infty}\psi^*\frac{\partial^2 \psi}{\partial x^2}\d x + \int_{-\infty}^{\infty} V(x)\psi^* \psi\,\d x &= E\int_{-\infty}^{\infty}\psi^* \psi\,\d x = E
\end{align}
where the wavefunction is taken to be normalised. Applying integration by parts to the first term gives
\begin{align}
\int_{-\infty}^{\infty} \psi^*\frac{\partial^2 \psi}{\partial x^2}\d x &=
\psi^*\frac{\partial \psi}{\partial x}\Bigg|_{-\infty}^{\infty} - \int_{-\infty}^{\infty} \frac{\partial \psi^*}{\partial x}\frac{\partial \psi}{\partial x}\d x\\
&= - \int_{-\infty}^{\infty} \left|\frac{\partial \psi}{\partial x}\right|^2\d x.
\end{align}
Hence the Schr\"{o}dinger equation becomes
\begin{align}
\int_{-\infty}^{\infty} \left|\frac{\partial \psi}{\partial x}\right|^2\d x + \int_{-\infty}^{\infty} V(x)|\psi|^2\,\d x &= E < 0.
\end{align}
Positivity of the first term implies that
\begin{align}
\int_{-\infty}^{\infty} V(x)|\psi|^2\,\d x < 0
\end{align}
which contradicts the assumption that $V(x)$ is never negative. Note that values
of $E$ on the negative real axis are permitted if $V(x) < 0$ anywhere, although
these values can only be associated with finite energy on a point spectrum. If
a negative eigenvalue $E_n$ \emph{does} exist, then solutions can grow
exponentially in time. 

\begin{figure}[t]
\begin{minipage}{0.4\textwidth}
\begin{center}
\includegraphics[width=\textwidth]{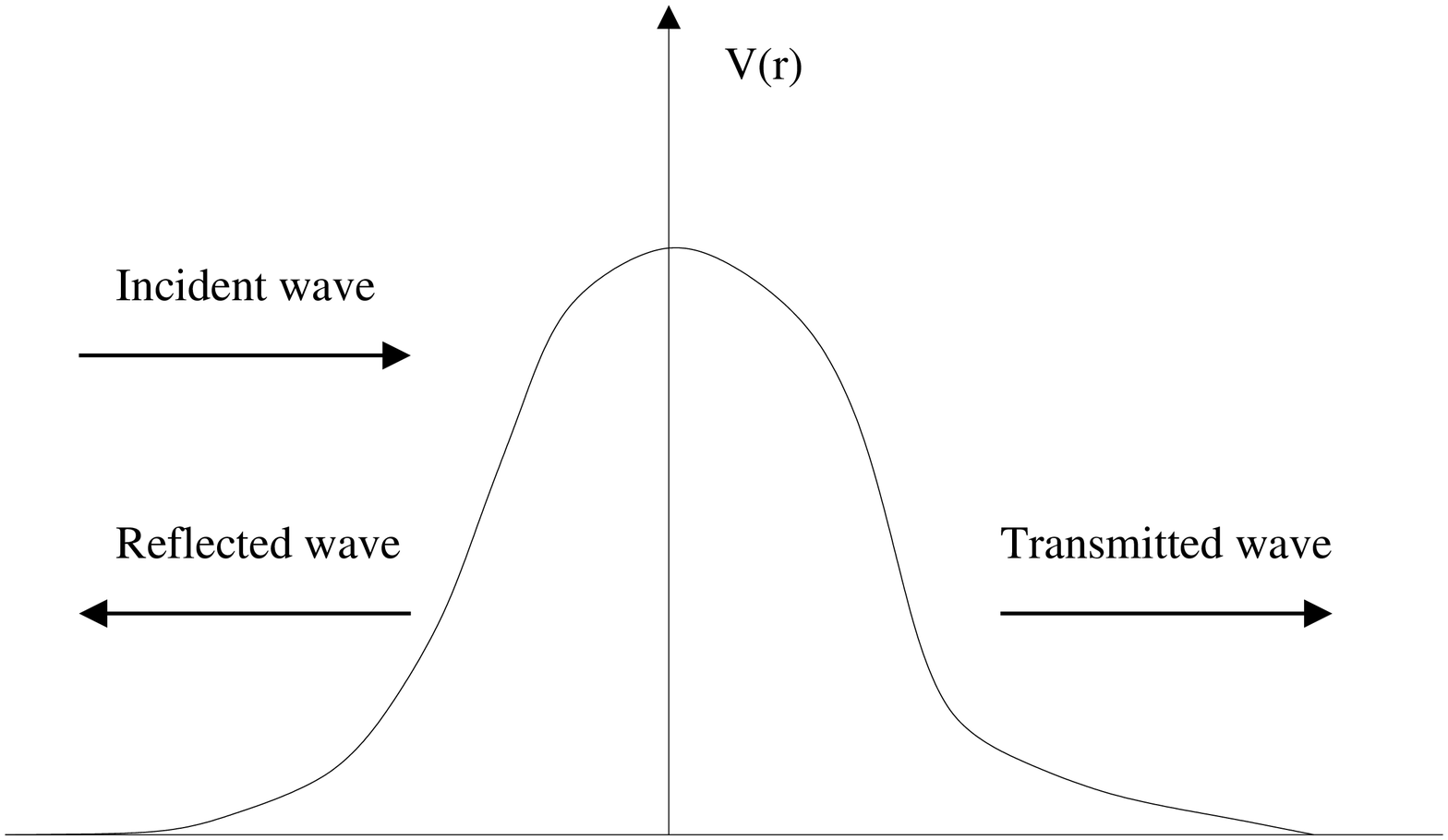}
\end{center}
\end{minipage}
\hspace{\stretch{1}}
\begin{minipage}{0.4\textwidth}
\begin{center}
\includegraphics[width=\textwidth]{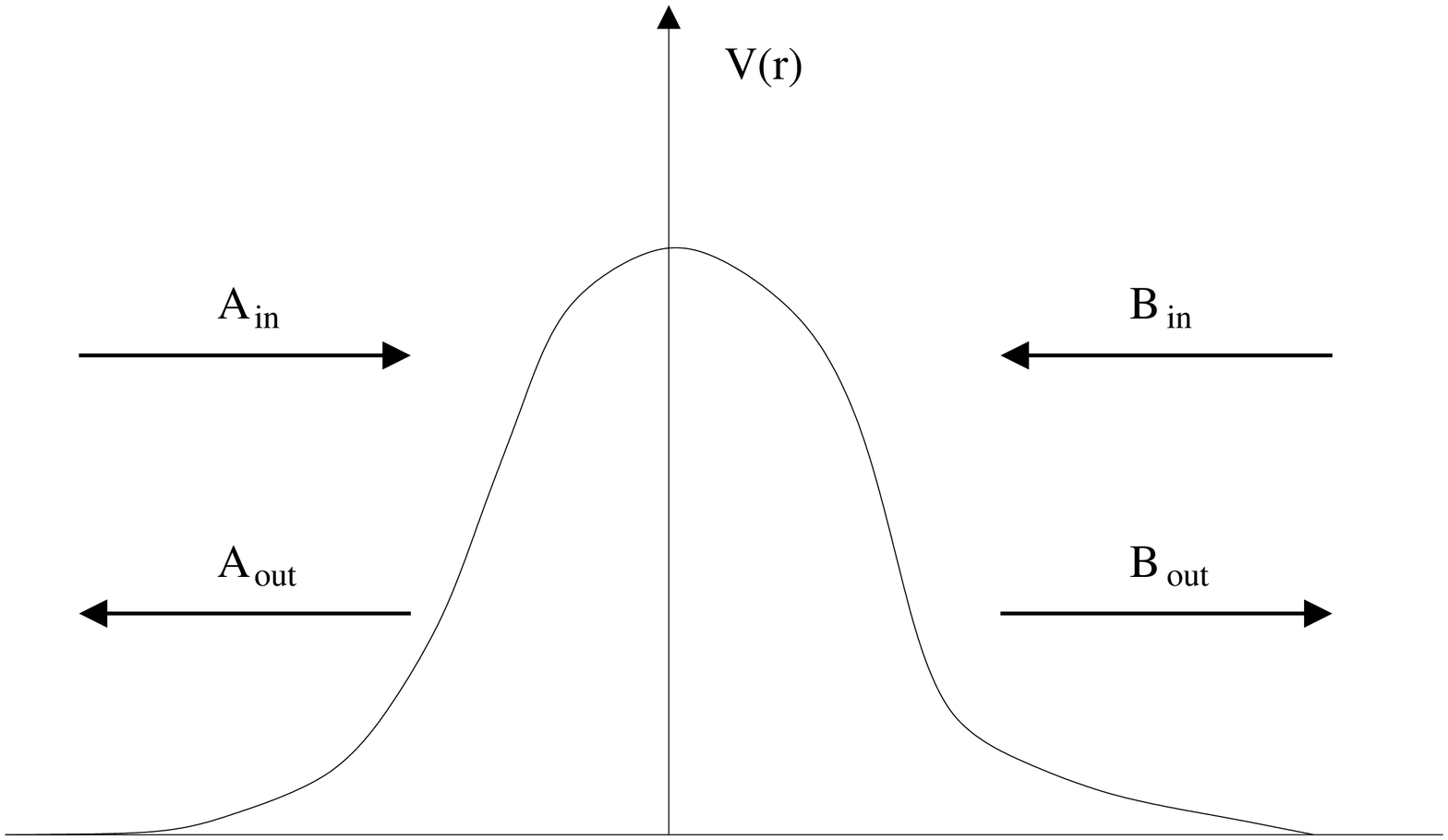}
\end{center}
\end{minipage}
\begin{footnotesize}
\caption[Scattering off a potential]{If particles are only incident from one side, then a clear separation between the reflected and transmitted particles can be made (left hand figure). If we have particles incident from \emph{both} sides, then the ingoing amplitudes are still incident waves ($A_{\textrm{in}}$ and $B_{\textrm{in}}$), but the outgoing waves are a superposition of transmitted and reflected waves. }\label{fig:scattering}
\end{footnotesize}
\end{figure}
\paragraph{}
As the wave equation is a linear differential equation in $\psi$ (or $\chi$)
then the solutions linearly superpose. We can place purely left travelling
waves off at infinity, by interacting with a potential these will turn into a
mixture of transmitted and reflected \emph{amplitudes}. As the wave equation
(or Schr\"odinger's equation) is second order we have two linearly independent
solutions. Let us choose one solution that is left travelling when $V(x) = 0$
denoted $\psi_L$, the other purely right travelling denoted $\psi_R$. Then we have in
general
\begin{equation}
\psi(x) = C_1 \psi_L(x) + C_2 \psi_R(x) 
\end{equation}
in regions where the potential vanishes. Let us analyse waves coming in from
the left only and hitting a potential.  Toward the left the solution will be a
superposition of \emph{reflected} waves and the incident wave. The boundary
condition of no incoming radiation from the right implies that the solution
will be purely right travelling. Only one boundary condition has been fixed so
far, let us fix the incoming amplitude as $A_{\textrm{in}}$. Let us assume that
the potential vanishes once $|x|> x_0$.  Then the solution can be found:
\begin{align}
\psi(x) = 
\begin{cases}
A_{\textrm{in}} \psi_L(x) + A_{\textrm{out}} \psi_R(x)&x
< -x_0\\
B_{\textrm{out}} \psi_R(x) &x > \phantom{-}x_0
\end{cases}.
\end{align}
This is not equivalent to solving the problem as it does not tell us about the solution \emph{inside} the potential. It is often the case that we are only interested in what happens far away from the potential. We can find the behaviour a large distance from the potential by finding the so-called \emph{reflection} and \emph{transmission} amplitudes.
\begin{equation}
\begin{array}{l}
A_{\textrm{out}} = R_A\; A_{\textrm{in}}\\
B_{\textrm{out}} = T_A\; A_{\textrm{in}}
\end{array}
\quad\quad \textrm{(Incoming from the left only).}
\end{equation}
If the ``energy'' is real (or $s$ is purely imaginary) then we have the ``conservation law'':
\begin{equation}
|R_A|^2 + |T_A|^2 = 1.
\end{equation}
For obvious reasons, $R_A$ and $T_A$ are called the (left) reflection and transmission amplitudes. Analogous relations exist if there are only incident waves from the right:
\begin{equation}
\begin{array}{l}
A_{\textrm{out}} = R_B\; B_{\textrm{in}}\\
B_{\textrm{out}} = T_B\; B_{\textrm{in}}
\end{array}
\quad\quad \textrm{(Incoming from the right only)}.
\end{equation}
It is more difficult to give the interpretation of reflection and transmission as any outgoing radiation will be a superposition of reflected and transmitted parts. However, the principle of superposition makes finding the solution very easy:
\begin{equation}
\left(\begin{array}{c}
A_{\textrm{out}}\\
B_{\textrm{out}}
\end{array}\right)
=
\left(\begin{array}{cc}
R_A& T_B\\
T_A& R_B\\
\end{array}\right)
\left(\begin{array}{c}
A_{\textrm{in}}\\
B_{\textrm{in}}
\end{array}\right).
\end{equation}
The matrix here is the familiar \emph{S-matrix} from quantum mechanics.

\paragraph{}
We know that as $x$ becomes large (small) $\psi(x)\rightarrow f_+(x)$ ($\psi\rightarrow f_-(x)$) as it is the well-behaved (decaying) solution in this regime. At a quasinormal mode the functions $f_+(x)$ and $f_-(x)$ become linearly dependent, and we can have
\begin{equation}
\psi(x) \propto f_+(x)
\end{equation}
with well behaved solutions at both large positive and negative $x$. But these solutions are \emph{purely outgoing}: $A_{\textrm{in}} = B_{\textrm{in}} = 0$! In terms of the S-matrix we have
\begin{equation}
\left(\begin{array}{c}
A_{\textrm{out}}\\
B_{\textrm{out}}
\end{array}\right)
=
\left(\begin{array}{cc}
R_A& T_B\\
T_A& R_B\\
\end{array}\right)
\left(\begin{array}{c}
0\\
0
\end{array}\right).
\end{equation}
Hence at a QNM the S-matrix becomes singular! We can also tell that QNM modes occur only for purely outgoing waves at infinity and the horizon.

\paragraph{}
This similarity with scattering problems in quantum mechanics means that we can try and adapt techniques in quantum mechanics to finding QNM. The techniques that are particularly useful are modified WKB approaches and Born approximation approaches as they are particularly sensitive to poles in the S-matrix, and hence in the Green's function.

\paragraph{}
Some care is warranted as it not only QNMs that correspond to poles in the S-matrix. Bound states of potentials where nothing propagates out to infinity also cause poles in the S matrix. For purely positive potentials this is not an issue, but bound states can develop if the potential is negative anywhere. We have also made a lot of use of the fact that the potential vanishes, so that we can talk about ``out'' states without having to be too careful about whether or not a sensible meaning can be given. The cases dealt with in black hole physics are \emph{not} of compact support, but decay exponentially. It is assumed that ``outgoing'' states still make sense.

%\subsubsection{Stability}
%Discuss: 
%         Stability
%        Introduce the polar equations and Zerelli equations for AC. Schild
%         Isospectral
%         A brief discourse on Kerr and Kerr-Newman 
%         Methods of calculating
\section{Wave equations for dirty black hole\\ spacetimes}\label{sec:wave_eqn}

Instead of a restricting attention to a Schwarzschild black hole, the spacetime
is chosen to be a \emph{dirty black hole}:
\begin{equation}
\d s^2 = -\exp(-2\Phi(r))\left(1-\frac{b(r)}{r}\right)\d t^2 + \frac{1}{1-b/r}\d r^2 + r^2 \d\Omega^2. \label{eq:dbh_metric}
\end{equation}
This black hole spacetime is manifestly spherically symmetric and static,
however it is not required that the exterior region is in vacuum. A quick
computation shows that the energy density is given by
\begin{equation}
\rho = \frac{1}{8\pi r^3}\frac{\d b(r)}{\d r}.
\end{equation}
This metric is capable of describing the special cases often discussed in the
literature, such as Reissner--Nordstr\"om and Schwarzschild-de Sitter, as well as
more generic black holes.

\paragraph{}
The restriction to spherical symmetry is one that is used to make the
calculations more tractable. The condition that the metric is stationary (i.e.
time independent) is a reasonable one as it is difficult to define QNM in a
generic time dependent background. The additional restriction that the spacetime
is also \emph{static} is not necessary but simplifies calculations. In
particular, we know that $g_{rr}$ and $g_{tt}$ must go to infinity and zero
respectively at the same locations to preserve the signature of the metric. As
the spacetime is stationary, the place where $g_{tt}$ vanishes is an infinite
redshift surface as seen from infinity. Thus the spacetime \eqref{eq:dbh_metric}
has a horizon where
\begin{equation}
b(r) = r.
\end{equation}
Note that while $g_{tt}$ and $1/g_{rr}$ must share the same zeros, they may be
different in other locations. The function $\Phi(r)$ allows for this
\cite{Visser:1992qh}. In the special cases of Schwarzschild, Ressnier--Nordstr\"om
and Schwarzschild--de Sitter $\Phi = 0$ ($g_{tt} = 1/g_{rr}$), but this is not
required by the conditions that the spacetime is spherically symmetric and
static. So that $g_{tt}$ and $1/g_{rr}$ still preserve the same zeros $\Phi$
will be assumed to be finite at the location of any horizons.

\subsection{The generalised tortoise coordinate $r_\star$}
When defining wave equations, it is useful to make the $t$--$r$ piece of the
metric
\emph{conformally flat}:
\begin{align}
\d s^2 &= \exp(-2\Phi(r))\left(1-\frac{b}{r}\right)\left(-\d t^2 + \frac{\exp(2\Phi(r))}{(1-b/r)^2}\d r^2\right) + r^2\d\Omega^2\\
&= \exp(-2\Phi(r))(1-b/r)\left(-\d t^2 + \d r_{\star}^2\right) + r^2\d\Omega^2.
\end{align}
The last equation introduces the tortoise coordinate $r_\star$ up to an arbitrary
constant by
\begin{equation}
\frac{\d r_\star}{\d r} = \frac{\exp(\Phi(r))}{1-b/r}.\label{eq:tortoise}
\end{equation}

\paragraph{}
It is intuitively clear that the ``flattening'' of the space to produce
$r_\star$ will mean $r_\star$ will approach infinity as a horizon is approached.
To prove this, expand $b(r)$ about a horizon $r_0$:
\begin{equation}
b(r) = r_0 + \left.\frac{\d b}{\d r}\right|_{r_0}(r-r_0) + 
       \frac{1}{2!}\left.\frac{\d^2 b}{\d r^2}\right|_{r_0}(r-r_0)^2 + \ldots
\end{equation}
where $b(r_0) = r_0$ has been used. This implies that
\begin{align}
\frac{\d r_\star}{\d r} &= \frac{ r \exp(\Phi(r))}{r-b}\\
r_\star &\approx \int \frac{r\exp(\Phi(r))}{(r-r_0)(1 - b^\prime)}\,\d r\\
&\approx \frac{\exp(\Phi(r_0))}{1-b^\prime}\int \frac{r}{r-r_0} \,\d r\\
&= \frac{\exp(\Phi(r_0))}{1-b^\prime}\left(r-r_0+r_0\ln\left|r-r_0\right|\right).\label{eq:approach_horizon}
\end{align}
The special case $b^\prime = 1$, corresponding to an extremal horizon, has been
excluded and is discussed further in \cite{Visser:1992qh}. As $r\rightarrow r_0$
the generalised tortoise coordinate tends to infinity.

\paragraph{}
Near a horizon the $r_\star$ and $r$ are related to the surface gravity of the
horizon. For a spherically symmetric static black hole the surface gravity is
given by
\begin{equation}
\kappa = \frac{1}{2}\left|\frac{\d g_{tt}}{\d r}\right|_{r_0}.
\end{equation}
The surface gravity will be discussed more carefully in chapter \ref{chap:conformal}. In a dirty black hole spacetime we have
\begin{equation}
\kappa = \frac{\exp(-2\Phi(r_0))}{2r_0}\left|(1-b^\prime)\right| 
\end{equation}
The equation \eqref{eq:approach_horizon} then becomes
\begin{align}
r_\star &\approx \pm \frac{1}{2\kappa \exp(\Phi)}\left(\frac{r-r_0}{r_0}+\ln\left|r-r_0\right|\right)\\
&\approx \pm \frac{1}{2\kappa \exp(\Phi)}\ln|r-r_0|.
\end{align}
Inverting this relation we have
\begin{equation}
r-r_0 \approx \exp(\pm 2\kappa \exp(\Phi(r_0)) r_\star) \label{eq:close2horz}
\end{equation}
as the horizon is approached.

\paragraph{}
The potentials that describe dirty black hole spacetimes generally have
polynomial falloff as a horizon is approached. As seen from
\eqref{eq:close2horz} this implies that the falloff in $r_\star$ will be
exponential. If the region of interest lies between the outermost horizon and
infinity, however, then polynomial decay occurs for \emph{both} $r$ and
$r_\star$ as infinity is approached.\footnote{As the outermost horizon is
  approached the decay is still exponential in $r_\star$.}

\paragraph{}
The next task is to find the wave equations for different types of fields. In slightly more detail, the function that plays the role of the potential relies not only on the spacetime, but on the type of field under consideration and its spin. In practice, for the static case we are perturbing the geometry but requiring that the linearised Ricci tensor is unchanged. Thus the QNMs of a field are determined by how that field contributes to the stress energy tensor. However, in the literature there are comments about the potential for spin $j$. More appropriately, these are the gravitational, electromagnetic and scalar potentials. In vacuum spacetimes the difference between spin $j$ and a specific field is not great, as the field is not in the spacetime and at linear order basic fields look the same. We are also spoilt in vacuum spacetimes by the fact that different polarisations of the same field are isospectral, so while different polarisations are technically associated with different potentials the spectra are the same and people sloppily talk about ``the'' potential for a spin $j$ field. In the non-vacuum case different polarisations are not necessarily isospectral, for example Victor Cardoso showed in his PhD thesis \cite{Cardoso:2004pj} that the two gravitational wave polarisations are not isospectral. The motto is that the effective potential in a generic spacetime needs to consider the polarisation and the specifics of the field. 

\paragraph{}
Although I disagree with the interpretation of the ``potential of spin $j$'', it is a commonly used potential in the literature. For this reason, I repeat the standard formula below. The spin $j$ effective potential in a Schwarzschild spacetime is claimed to be \cite{Kokkotas}
\begin{equation}
V_{\ell}(r) = \left(1-\frac{2M}{r}\right)\left[\frac{\ell(\ell+1)}{r^2} + \frac{2M}{r^3}(1-j^2)\right].
\end{equation}
\subsection{Scalar wave equations}
Even in the case of \emph{massless} scalar waves propagating on a background
there is some cause for ambiguity. The simplest generalisation of the wave
equation for scalar particles is
\begin{equation}
\Box \phi - m^2\phi + \left(\zeta-\frac{1}{6}\right)\phi= g^{ab}\nabla_a \nabla_b \phi- m^2\phi + \left(\zeta-\frac{1}{6}\right)R\phi = 0.\label{eq:KG}
\end{equation}
Here $m$ is the mass of the excitations of the field and $\zeta$ is the \emph{conformal coupling}. If $\zeta$ is zero and $m=0$ then \eqref{eq:KG} is conformally invariant. The case $\zeta = 1/6$ eliminates the final term and the resulting equation is called the minimally coupled Klein-Gordon equation. Note that the mass automatically breaks conformal invariance, so for massive fields $\zeta = 1/6$ is almost universally used. However, the following is no more complicated by keeping $\zeta$ general.

\paragraph{}
Let us rewrite the term involving the Laplacian. We have
\begin{align}
\Box \phi &= g^{ab} \nabla_a \nabla_b \phi\\
&= g^{ab} \nabla_a \phi_{,b}
\end{align}
where $\phi_{,b}$ is a one-form. By exploiting the metric connexion we have
\begin{align}
\Box \phi &= \nabla_a \phi^{,a}\\
&= \partial_a \phi^{,a} + \Gamma^{a}_{ab}\,\phi^{,b}.
\end{align}
We also exploit the well known formula \cite{book:Wald} 
\begin{equation}
\Gamma^{a}_{ab} = \partial_b \ln(\sqrt{|g|})
\end{equation}
to obtain
\begin{align}
\Box\phi &= \partial_a \phi^{,a} + \phi^{,b}\partial_b \ln(\sqrt{|g|})\\
&= \partial_a \phi^{,a} + \phi^{,a}\partial_a \ln(\sqrt{|g|})\\
&= \frac{1}{\sqrt{|g|}}\partial_a\left(\phi^{a}\sqrt{|g|}\right)\\
&= \frac{1}{\sqrt{|g|}}\partial_a\left(g^{ab}\sqrt{|g|}\phi_{,b}\right).\label{eq:boxphi1}
\end{align} 

\paragraph{}
We now have to apply \eqref{eq:boxphi1} to the dirty black hole spacetime:
\begin{align}
\sqrt{|g|} &= \exp(-\Phi(r)) r^2 \sin\theta.
\end{align}
The diagonal metric simplifies terms immensely:
\begin{align}
\left(g^{ab}\sqrt{|g|}\phi_{,b}\right)_{,a} &= \sum_{a} \left(g^{aa}\sqrt{|g|}\phi_{,a}\right)_{,a}\\
%&= \sum_{a} \partial_a\left(g^{aa}e^{-\Phi(r)}r^2\sin\theta\;\partial_a \phi\right)\\
&= e^{-\Phi(r)}r^2\sin\theta\;\left\{ g^{tt} \partial_t^2 \phi + g^{\varphi\varphi}\partial_{\varphi}^2\phi\right\} \nonumber\\
&\quad+ e^{-\Phi(r)}r^2 g^{\theta\theta} \partial_\theta \sin\theta \partial_\theta\phi + \sin{\theta}\partial_r\left(g^{rr}e^{-\Phi(r)}r^2\partial_r \phi\right)\\
&= \sqrt{|g|}\left\{ g^{tt} \partial_t^2 \phi + \frac{1}{r^2\sin^2\theta}\partial_{\varphi}^2\phi\right\} + \frac{\sqrt{|g|}}{\sin\theta} \frac{1}{r^2} \partial_\theta \sin\theta \partial_\theta\phi\nonumber\\
&\quad + \sin{\theta}\partial_r\left(g^{rr}e^{-\Phi(r)}r^2\partial_r \phi\right)\\
&= \sqrt{|g|} g^{tt} \partial_t^2 \phi + \sin{\theta}\partial_r\left(g^{rr}e^{-\Phi(r)}r^2\partial_r \phi\right)\nonumber\\
&\quad\quad + \frac{\sqrt{|g|}}{r^2}\left(\frac{1}{\sin\theta}\partial_\theta \sin\theta\partial_\theta + \frac{1}{\sin^2 \theta}\partial^2_\varphi\right)\phi.
\end{align}
The final term in parenthesis is the Laplacian on the unit sphere! Let us
denote
\begin{equation}
\nabla_2^2 = \frac{1}{\sin\theta}\partial_\theta \sin\theta\partial_\theta + \frac{1}{\sin^2 \theta}\partial^2_\varphi.
\end{equation}
Our four dimensional Laplacian then takes the form
\begin{align}
\Box \phi &= \frac{1}{\sqrt{|g|}}\left(g^{ab}\sqrt{|g|}\phi_{,b}\right)_{,a}\\
&= g^{tt} \partial_t^2 \phi + \frac{e^{\Phi(r)}}{r^2}\partial_r\left(g^{rr}e^{-\Phi(r)}r^2\partial_r \phi\right) + \frac{1}{r^2}\nabla_2^2\phi
\end{align}
We now separate this equation into angular and non-angular parts by making the ansatz
\begin{equation}
\phi(r,\theta,\varphi,t) = r^2 \rho(r,t) Y(\theta,\varphi)
\end{equation}
The Klein-Gordon equation \eqref{eq:KG} separates:
\begin{equation}
\frac{r^2 g^{tt}}{\rho}\partial_t^2 \rho + \frac{e^{\Phi(r)}}{\rho r^2}\partial_r\left(g^{rr}e^{-\Phi(r)}r^2\partial_r r^2\rho\right) - \left[m^2 + (\zeta + 1/6)R\right]r^2 = -\frac{\nabla_2^2 Y}{Y}.
\end{equation}
It is worthwhile to note that $m$ is independent of $\theta$ and $\varphi$, as it is a scalar in a spherically symmetric geometry. In the case of a conformal field we are using $m^2 = R/6$ which may depend on $r$.

\paragraph{}
To find regular solutions of the angular equation we require that \cite{book:Bence}
\begin{equation}
\nabla_2^2 Y = -\ell(\ell + 1) Y.
\end{equation}
Hence the angular pieces are the Legendre polynomials. The non-angular pieces are less trivial:
\begin{align}
&g^{tt}\partial_t^2 r^2 \rho + \frac{e^{\Phi(r)}}{r^2}\partial_r\left(g^{rr}e^{-\Phi(r)}r^2\partial_r (r^2\rho)\right) - \frac{\ell(\ell + 1)}{r^2} (r^2\rho)
\nonumber\\
&\quad\quad - \left[m^2 + (\zeta + 1/6)R\right](r^2\rho)  = 0.\label{eq:scalar_prewave_eqn}
\end{align}

\paragraph{}
The equation \eqref{eq:scalar_prewave_eqn} differs from a simple wave equation
in the appearance of first derivative terms. We may simplify the term with radial derivatives in \eqref{eq:scalar_prewave_eqn} by transforming to the generalised tortoise coordinate \eqref{eq:tortoise}:
\begin{align}
\partial_r g^{rr}e^{-\Phi(r)}r^2\partial_r&=
\partial_r r^2(1-b/r)e^{-\Phi(r)}\partial_r\\
&= \partial_r r^2 \frac{\d r}{\d r_\star} \frac{\partial}{\partial r} = \partial_r r^2 \frac{\partial}{\partial r_\star}.
%&= 2r \frac{\partial }{\partial r_\star} + r^2 \partial_r \partial_{r_\star}\\
%&= 2r \partial_{r_\star} + \exp(-\Phi(r))(1-b/r) r^2 \partial_{r_\star}^2
\end{align}
To eliminate the first derivative, define the function $P(r,t)$ via
\begin{equation}
r^2 \rho(r,t) = r^s P(r,t). \label{eq:ansatz_anderung}
\end{equation}
The relevant term is then
\begin{align}
 \partial_r g^{rr}e^{-\Phi(r)}r^2\partial_r (r^2 \rho)&= 
\partial_r g^{rr}e^{-\Phi(r)}r^2\partial_r(r^s P)\\
&= \partial_r r^2 \frac{\partial}{\partial r_\star} r^s P\\
&= \partial_r \left(r^{2+s} \frac{\partial P}{\partial r_\star} + s r^{s+1} \frac{\d r}{\d r_\star} P\right)\\
&= \frac{\d r_\star}{\d r} \Bigg\{(2+s)r^{1+s}\frac{\d r}{\d r_\star} \frac{\partial P}{\partial r_\star} + r^{2+s} \frac{\partial^2 P}{\partial r_\star^2}\nonumber\\  
&\quad + s(s+1)r^{s} \left(\frac{\d r}{\d r_\star} \right)^2 P + sr^{s+1} \frac{\d\phantom{r_\star} }{\d r_\star}\left( \frac{\d r}{\d r_\star}\right)P\nonumber\\
&\quad + 
s r^{s+1} \frac{\d r}{\d r_\star} \frac{\partial P}{\partial r_\star}\bigg\}.
\end{align}
Expanding out the derivative and collecting terms gives
\begin{align}
\partial_r g^{rr}e^{-\Phi(r)}r^2\partial_r (r^2 \rho) &= 2(1+s)r^{1+s}\frac{\partial P}{\partial r_\star} + r^{2+s}\frac{\d r_\star}{\d r} \frac{\partial^2 P}{\partial r_\star^2}\nonumber\\  
&\quad + s(s+1)r^{s} \frac{\d r}{\d r_\star} P + sr^{s+1} \frac{\d\phantom{r}}{\d r}\left( \frac{\d r}{\d r_\star}\right)P.
\end{align}
The unwanted first derivative is eliminated by the choice $s=-1$. The radial term simplifies dramatically:
\begin{align}
\partial_r g^{rr}e^{-\Phi(r)}r^2\partial_r (r^2 \rho) &= r\frac{\d r_\star}{\d r} \frac{\partial^2 P}{\partial r_\star^2} - \frac{\d\phantom{r}}{\d r}\left( \frac{\d r}{\d r_\star}\right)P\\
&=\frac{r e^{\Phi(r)}}{(1-b/r)} \frac{\partial^2 P}{\partial r_\star^2} - \frac{\d\phantom{r}}{\d r}\left( e^{-\Phi(r)}(1-b/r)\right)P\\
&= \frac{r e^{\Phi(r)}}{(1-b/r)} \frac{\partial^2 P}{\partial r_\star^2} + \left( \Phi^\prime (1-b/r) + (b/r)^\prime\right)e^{-\Phi(r)}P.
\end{align}
In terms of the tortoise coordinate and the \emph{Ansatzs\"{a}nderung} \eqref{eq:ansatz_anderung}, equation \eqref{eq:scalar_prewave_eqn} becomes
\begin{align}
&-\frac{e^{2\Phi(r)}}{1-b/r}\partial_t^2 r^{-1} P + \frac{1}{r^2}\left(\frac{r\exp(2\Phi(r))}{(1-b/r)} \frac{\partial^2 P}{\partial r_\star^2} + \left( \Phi^\prime (1-b/r) + (b/r)^\prime\right)P\right)\nonumber\\
&\quad\quad\quad\quad - \left[m^2+(\zeta+1/6)R\right](r^{-1} P) - \frac{\ell(\ell + 1)}{r^2} (r^{-1} P) = 0.
\end{align}
By multiplying out by the coefficient of the time derivatives we get the equation for a scalar field in a dirty black hole in the form of a wave equation:
\begin{align}
\frac{\partial^2 P}{\partial t^2} - \frac{\partial^2 P}{\partial r_\star^2} + 
V(r)P = 0.\label{eq:scalar_wave_eqn}
\end{align}
The ``potential'' $V(r)$ is given by
\begin{equation}
V(r) = \left[\frac{\ell(\ell + 1)}{r^2} + m^2 +\zeta + \frac{1}{6} -\left(\frac{\Phi^\prime (1-b/r) + (b/r)^\prime}{r}\right)\right] (1-b/r)\;e^{-2\Phi(r)}.\label{eq:scalar_potential}
\end{equation}

\paragraph{}
Notice that the differential equation is expressed in terms of $r_\star$, while
the potential is expressed as a simple function of $r$. The problem of finding
QNMs of scalar fields reduces to finding the purely outgoing harmonic solutions to \eqref{eq:scalar_potential}.

\subsection{Gravitational perturbations -- a wave equation}\label{sec:wave_eqn_gravity}

The gravitational case is somewhat more complicated. By virtue of spherical symmetry we can take any perturbation to be axisymmetric (see \S 4.1 of \cite{Book:ChandrasekharI}). Let us take the metric in the general axisymmetric form:
\begin{align}
\d s^2 &= -\exp(2[\nu^{0}(r)+\delta_t(r,t)])\d t^2 + \exp(2[\mu^{0}_r(r)+\delta_r(r,t)])\d r^2\nonumber\\
&\quad + \exp(2[\mu^{0}_\theta(r)+\delta_\theta(r,t)])\d \theta^2 + \exp(2[\mu^{0}_\varphi(r,\theta)+\delta_\varphi(r,\theta,t)])\nonumber\\
&\quad\times\left(\d \varphi - q_r(r,\theta,t)\d r - q_\theta(r,\theta,t) - q_t(r,\theta,t)\right)^2.\label{eq:general_axisymmetric}
\end{align}
The functions $\nu^{0}(r)$, $\mu^{0}_r(r)$, $\mu^{0}_\theta(r)$ and $\mu^{0}_{\varphi}(r,\theta)$ are arbitrary at this point and represent the background spacetime. A dirty black hole spacetime is recovered when we take
\begin{subequations}
\begin{align}
\label{eq:QNM_intro_coord_numu}
\exp(2\nu^{0}(r)) &= \exp(-2\Phi(r))(1-b/r)\\
\exp(2\mu^{0}_r(r))&= \frac{1}{1-b/r}\\
\exp(2\mu^{0}_\theta(r)) &= r^2\\
\exp(2\mu^{0}_\varphi(r,\theta)) &= r^2 \sin^2\theta.
\end{align}
\end{subequations}
The functions $\delta_i$ and $q_i$ are then considered to be perturbations on
the background spacetime. Gravitational wave equations are found by insisting
that the perturbations be ``massless''; more precisely we insist that the
variation in the linearised Ricci tensor vanishes. The problem decouples into
two cases:
\begin{itemize}
\item $\delta_i = 0, q_i \neq 0$:\\
  These terms introduce cross products $g_{\varphi i}$ into the metric at
  linear order, and thus contribute to frame-dragging effects. Reversing the
  sign of $\varphi$ will reverse the direction of these effects, hence they are
  called \emph{axial} perturbations.
\item $\delta_i \neq 0, q_i = 0$:\\
  These terms induce no frame dragging and are unaffected by reversal of
  $\varphi$. These perturbations are referred to as \emph{polar}.
\end{itemize}
The terminology polar and axial terminology is misleading as, in a spherically symmetric system, the names suggest that by a rotation one could be transformed into the other. This is certainly not the case!

\paragraph{}
In order to obtain the perturbations we define
\begin{align}
\nu(r,t) &= \nu^{0}_r(r) + \delta_t(r,t)\\
\mu_i(r,\theta,t)  &= \mu_i^0(r,\theta) + \delta_i(r,\theta,t).
\end{align}
We could take the metric, calculate the Ricci tensor directly and then discard higher order terms. Instead I shall find the Ricci tensor in an orthonormal frame and then discard higher order terms. The metric \eqref{eq:general_axisymmetric} suggests the following orthonormal covariant basis:
\begin{align}
{e_{(0)}}_i &= \left(-e^\nu,0,0,0\right)_i\\
{e_{(1)}}_i &= \left(0,e^{\mu_r},0,0\right)_i\\
{e_{(2)}}_i &= \left(0,0,e^{\mu_\theta},0\right)_i\\
{e_{(3)}}_i &= \left(-q_t,-q_r,-q_\theta,1\right)_ie^{\mu_\phi}
\end{align}
The corresponding contravariant basis vectors are given by
\begin{align}
\et^i &= \left(e^{-\nu},0,0,q_t e^{-\nu}\right)^i\\
\er^i &= \left(0,e^{-\mu_r},0,q_r e^{-\mu_r}\right)^i\\
\eth^i &= \left(0,0,e^{-\mu_\theta},q_\theta e^{-\mu_\theta}\right)^i\\
\ephi^i &= \left(0,0,0,e^{-\mu_\phi}\right)^i.
\end{align} 
It is easy enough to check that these do form an orthonormal basis. The definition of a general tetrad is (\cite{book:Wald} chapter 3):
\begin{align*}
e_{(\mu)}^a {e_{(\nu)}}_a &= \eta_{\mu\nu}\\
\eta^{\mu\nu} e_{(\mu)}^a {e_{(\nu)}}_b &= \delta^a_b
\end{align*}
where $\eta_{\mu\nu}$ is a constant matrix and $\eta^{\mu\nu}$ is its
\emph{matrix} inverse. In the case of an orthonormal basis $\eta$ is the
special relativity metric.  To construct the Ricci tensor we require the
\emph{Ricci rotation coefficients} defined by
\begin{equation}
\omega_{\mu\nu a} = {e_{(\mu)}}^b\nabla_a {e_{(\nu)}}_b
\end{equation}
Notice that the first two indices are simply labels, and the 24 \emph{Ricci rotation coefficients} are really 6 one-forms $\omega_{\mu\nu}$. The notation here differs from Wald \cite{book:Wald}, but his notation tends to hide the geometric nature of the rotation coefficients. It should be noted that $\omega_{\mu\nu a}$ is antisymmetric in the first two indices. We rewrite the coefficients in terms of directions of the tetrad, rather than $t$, $r$, $\theta$ and $\varphi$:
\begin{equation}
\omega_{\mu\nu \hat{a}} = \omega_{\mu\nu \sigma} {e^{(\sigma)}}_a.
\end{equation}
We can then evaluate the spin coefficients by combinations of Lie brackets:
\begin{equation}
\omega_{\mu\nu\hat{a}} = \frac{1}{2}\left({e_{(a)}}_b \left[ e_{(\mu)},e_{(\nu)}\right]^b + {e_{(\mu)}}_b \left[ e_{(a)},e_{(\nu)}\right]^b + {e_{(\nu)}}_b \left[ e_{(\mu)},e_{(a)}\right]^b\right).
\end{equation}

\paragraph{}
In a torsion free theory, the Lie bracket is constructed using partial derivatives and the tetrad basis vectors. For example
\begin{align}
\left[e_{(0)},e_{(1)}\right]^a &= {e_{(0)}}^b \partial_b e_{(1)}^a -{e_{(1)}}^b \partial_b e_{(0)}^a\\
&= \left(e^{-\nu}\partial_t  + q_t e^{-\nu}\partial_{\phi}\right) \left(0,e^{-\mu_r(r,t)},0,q_r(r,\theta,t)e^{-\mu_r(r,t)}\right)^a \nonumber\\
&\quad\quad - \left(e^{-\mu_r}\partial_r + q_r e^{-\mu}\partial_{\phi}\right) \,\left(e^{-\nu(r,t)},0,0,q_t(r,\theta,t)e^{-\nu(r,t)}\right)^a\\
&= \exp(-\mu_r-\nu)\left(\nu_{,r}\, ,-\mu_{,t}\, ,0\, ,2q_{[r,t]} + q_t\nu_{,r}\right)^a.
\end{align}
In evaluating the Lie brackets, it is worthwhile noting that the tetrad vectors are independent of $\varphi$, due to the fact that the spacetime is axisymmetric. Any time derivative is automatically at least first order, as the background is independent of $t$. The other Lie brackets are given in table \ref{table:Lie}.

\begin{table}[hbt]
\begin{center}
\begin{tabular}{|lcl|}
\hline
$[e_{(0)},e_{(1)}]$&=&$\exp(-\mu-\nu)\left(\nu_{,r}\, ,-\mu_{,t}\, ,0\, ,2q_{[r,t]} + q_t\nu_{,r} - \mu_{,t}\right)$\\
$[e_{(0)},e_{(2)}]$&=&$\exp(-\nu-\mu_\theta)\left(0\, ,0\, ,-\mu_{\theta,t}\, ,2q_{[\theta,t]} - q_\theta \mu_{\theta,t}\right)$\\
$[e_{(0)},e_{(3)}]$&=&$\exp(-\mu_\varphi - \nu)\left(0\, ,0\, ,0\, ,-\mu_{\varphi,t}\right)$\\
$[e_{(1)},e_{(2)}]$&=&$\exp(-\mu_r-\mu_\theta)\left(0\, ,0\, ,-\mu_{\theta,r}\,,2q_{[\theta,r]}-\mu_{\theta,r}q_\theta\right)$\\
$[e_{(1)},e_{(3)}]$&=&$\exp(-\mu_r-\mu_\phi)\left(0\, ,0\, ,0\, ,-\mu_{\phi,r}\right)$\\
$[e_{(2)},e_{(3)}]$&=&$\exp(-\mu_\theta-\mu_\phi)\left(0\, ,0\, ,0\, ,-\mu_{\varphi,\theta}\right)$\\
\hline
\end{tabular}
\caption{The six independent Lie derivatives of the tetrad basis}\label{table:Lie}
\end{center}
\end{table}

\begin{table}
\begin{center}
\begin{tabular}{|l@{=}l|}
\hline
$\omega_{01  \hat{0}}$& $-\nu_{,r}\exp(-\mu)$\\
$\omega_{01  \hat{1}}$& $-\mu_{,r}\exp(-\nu)$\\
$\omega_{01  \hat{3}}$& $\exp(\mu_\varphi - \nu - \mu_r)q_{[r,t]}$\\
\hline
$\omega_{02  \hat{2}}$& $-\mu_{\theta,t}\exp(-\nu)$\\
$\omega_{02  \hat{3}}$& $\exp(\mu_\varphi - \nu - \mu_\theta)q_{[\theta,t]}$\\
\hline
$\omega_{03  \hat{1}}$& $\exp(\mu_\varphi-\nu - \mu_r)q_{[r,t]}$ = $\omega_{01 \hat{3}}$\\
$\omega_{03  \hat{2}}$& $\exp(\mu_\varphi-\nu-\mu_\theta)q_{[\theta,t]}$=
$\omega_{02\hat{3}}$\\
$\omega_{03  \hat{3}}$& $-\mu_{\varphi,t}\exp(-\nu)$\\
\hline
$\omega_{12  \hat{2}}$& $-\mu_{\theta,r}\exp(-\mu_r)$\\
$\omega_{12  \hat{3}}$& $\exp(\mu_\varphi - \mu_\theta -\mu_r)q_{[\theta,r]}$\\
\hline
$\omega_{13  \hat{0}}$& $-\exp(\mu_\varphi-\nu-\mu_r)q_{[r,t]}$ = $-\omega_{03\hat{1}}$\\
$\omega_{13  \hat{2}}$& $-\exp(\mu_\varphi-\mu_r-\mu_\theta)q_{[r,\theta]}$ =$+\omega_{12\hat{3}}$\\
$\omega_{13  \hat{3}}$& $-\mu_{\varphi,r}\exp(-\mu_r)$\\
\hline
$\omega_{23  \hat{0}}$& $\exp(\mu_\varphi-\nu-\mu_\theta)q_{[t,\theta]}$ = $-\omega_{03 \hat{2}}$\\
$\omega_{23  \hat{1}}$&$\exp(\mu_\varphi-\mu_r - \mu_\theta)q_{[r,\theta]}$=$-\omega_{12 \hat{3}}$\\
$\omega_{23  \hat{3}}$&$-\mu_{\varphi,\theta}\exp(-\mu_\theta)$\\
\hline
\end{tabular}
\caption{Spin coefficients for the given tetrad basis. They have been grouped by the first two parameters (which are antisymmetric) and the last index is the direction of the one-form. All other components are zero.}\label{table:spin_coeff}
\end{center}
\end{table}

\paragraph{}
The procedure of getting the spin coefficients is now straightforward, if somewhat tedious. The result of doing this calculation is summarised in table \ref{table:spin_coeff}. We can now use\\
\begin{equation}
R_{\hat{\rho}\hat{\sigma}\hat{\mu}\hat{\nu}} = 2{e_{[(\rho)|}}^a\partial_a \omega_{\mu\nu|\hat{\sigma}]} - 2\eta^{\alpha\beta}\left(\omega_{\beta\mu[\hat{\rho}}\omega_{|\alpha\nu|\hat{\sigma}]} + \omega_{\beta[\sigma\hat{\rho}]}\omega_{\mu\nu\hat{\alpha}}\right)
\end{equation}
or in less mathematical notation
\begin{align*}
R_{\hat{\rho}\hat{\sigma}\hat{\mu}\hat{\nu}} &= {e_{[(\rho)|}}^a\partial_a \omega_{\mu\nu|\hat{\sigma}]} - \eta^{\alpha\beta}\left(\omega_{\beta\mu[\hat{\rho}}\omega_{|\alpha\nu|\hat{\sigma}]} + \omega_{\beta[\sigma\hat{\rho}]}\omega_{\mu\nu\hat{\alpha}}\right)\\
&\quad - (\text{everything on the first line, but }\sigma\leftrightarrow\rho).
\end{align*}
The orthonormal Ricci tensor is then obtained by contracting over the first and third indices. 

\paragraph{}
Notice that up until this point everything has been exact; no use of the fact that $q_i$ and $\delta_i$ are of first order has been made. Let us denote the Ricci tensor by $R_{\hat{a}\hat{b}}^{0}$. Then the change in the Ricci tensor must be at least second order:
\begin{equation}
\delta R_{\hat{a}\hat{b}} = R_{\hat{a}\hat{b}} - R^{0}_{\hat{a}\hat{b}} = O(\left\{q_i,\delta_i\right\}^2)
\end{equation}
The linearised perturbations fall into two separate categories:
\begin{itemize}
%\item  Two components are \emph{trivial}:
%\begin{align}
%\delta R_{\hat{0}\hat{1}} = \delta R_{\hat{0}\hat{2}} = O(\{q_i,\delta_i\}^2).
%\end{align}
\item Three are \emph{axial}: these terms vanish in the background \emph{and} all the corrections are linear or higher in $q_i$. Hence any term containing $\delta_i$ is of second order.
\begin{subequations}\label{eq_gp:axial_eqns}
\begin{align}
\delta R_{\hat{0}\hat{3}}&=\exp(-2\mu_\varphi^0-\mu_\theta^0)\bigg\{\exp(-\mu^0_r)\left(q_{[t,r]}\exp(3\mu_\varphi^0-\nu^0-\mu^0_r+\mu^0_\theta)\right)_{,r}\nonumber\\
&\quad\quad\quad\quad\quad + \exp(-\mu^0_\theta-\nu^0)\left(q_{[t,\theta]}\exp(3\mu_\varphi^0)\right)_{,\theta}\bigg\}\\
\delta R_{\hat{1}\hat{3}}&=\exp(-2\mu_\varphi^0-\mu^0_\theta-\nu^0)\left(q_{[r,\theta]}\exp(3\mu_\varphi^0-\mu_r^0-\mu_\theta^0+\nu^0)\right)_{,\theta}\nonumber\\
&\quad\quad\quad\quad\quad +\exp(\mu^0_\varphi-2\nu^0-\mu_r^0)q_{[t,r],t}\label{eq:axial_perturb1}\\
\delta R_{\hat{2}\hat{3}}&=-\exp(-2\mu_\varphi^0-\nu^0-\mu^0_r)\left(q_{[r,\theta]}\exp(3\mu_\varphi^0-\mu_r^0-\mu_\theta^0+\nu^0)\right)_{,r}\nonumber\\
&\quad\quad\quad\quad\quad + \exp(-2\nu^0+\mu_\varphi^0-\mu_\theta^0)q_{[t,\theta],t}\label{eq:axial_perturb2}
\end{align}
\end{subequations}
\item The remaining seven are \emph{axial}. These terms do not vanish in the
  background spacetime. The $q_i$ only enter at second order, so are irrelevant
  for calculating the linearised perturbations. The $\delta_i$ do enter at
  first order. At linear order the $q_i$ and $\delta_i$ are \emph{decoupled},
  in the sense that the axial perturbations rely only on $q_i$ and the polar
  perturbations rely only on $\delta_i$. As the $q_i$'s can be ignored for the
  polar perturbations, this problem is spherically symmetric. This symmetry
  reduces the number of \emph{independent} equations from seven to five.\\
  \indent In writing down the axial equations the values of $\mu^0_{\theta}$
  and $\mu^0_{\varphi}$ for the background spacetime have been substituted:
\begin{subequations}\label{eqn_gp:polar_eqns}
\begin{align}
\delta R_{\hat{0}\hat{2}} &= \frac{\exp(-\nu^0)}{r}\frac{\d}{\d t}\left[(\delta_{r}+\delta_{\varphi})_{,\theta} + \left(\delta_{\varphi} - \delta_{\theta}\right)\cot\theta\right]\\
\delta R_{\hat{0}\hat{1}} &= \exp(-\nu^0-\mu_r^0)\frac{\d}{\d t}\bigg[\frac{2}{r}\delta_{r}  - (\delta_{\theta} + \delta_{\varphi})_{,r}\nonumber\\
& \quad\quad +\left(\nu_{,r}^0-\frac{1}{r}\right)(\delta_{\theta}+\delta_{\varphi})\bigg]\\
\delta R_{\hat{1}\hat{2}} &= \frac{\exp(-\mu_r^0)}{r}\bigg[(\delta_{t}+\delta_{\varphi})_{,r\theta} + (\delta_{\varphi}-\delta_\theta)_{,r}\cot\theta\nonumber\\&\quad\quad\quad\quad\quad\quad\quad -\left(\nu_{,r}+\frac{1}{r}\right)\delta_{r,\theta} + (\left(\nu_{,r}-\frac{1}{r}\right)\delta_{t,\theta}\bigg]
\end{align}
\begin{align}
\delta G_{\hat{1}\hat{1}} &= \exp(-2\nu^0)\left(\delta_{\varphi} + \delta_{\theta}\right)_{,tt} -\frac{1}{r^2}\big[(\delta_{t}+\delta_{\varphi})_{,\theta\theta} + 2\delta_{\theta} \nonumber\\
&\quad\quad\quad\quad+(2\delta_{\varphi} + \delta_t - \delta_\theta)_{,\theta}\cot\theta \big]
+ e^{-2\mu_r^0}\bigg[\frac{1}{r}(4\delta_r \nu^0_{,r}-2\delta_{t,r})\nonumber\\
&\quad\quad\quad\quad -\left(\frac{1}{r} + \nu^0_{,r}\right)(\delta_{\theta}+\delta_{\varphi})_{,r}+\frac{2}{r^2}\delta_r\bigg]\\
%
%
%\delta R_{\hat{1}\hat{1}} &= \exp(-2\nu^0)\delta_{r,tt} + \exp(-2\mu^0_r)\bigg(
%\delta_{t,rr} - \delta_{r,r}\nu^0_{,r} + \frac{2}{r}(\delta_{\theta} +\delta_{\varphi})_{,r}+ \delta_{\varphi,rr}\nonumber\\
%&\quad - \frac{2}{r}\delta_{r,r} 
%   -\mu_{r,r}^0 (\delta_{\varphi} + \delta_{\theta}+\nu^0)_{,r} 
%  + \delta_{\theta,rr} +2\nu^0_{,r}\delta_{t,r}\nonumber\\
%&\quad + 2\delta_{r}\mu_{r,r}\nu_{,r}-2\delta_r(\nu_{,r})^2+\frac{4}{r}\mu_{r,r}^0\delta_r-2\delta_r\nu^0_{,rr}\bigg)
%
%
%\delta R_{\hat{2}\hat{2}} &= -\exp(-2\nu^0)\delta_{\theta,tt} + \frac{1}{r^2}\left(\delta_{\varphi,\theta\theta} + 2\delta_{\theta} + (2\delta_{\varphi}-\delta_{\theta})_{,\theta}\cot\theta + (\delta_t+\delta_{r})_{,\theta\theta}\right) \nonumber\\
%&+  \exp(2\mu^0_r)\bigg\{(\nu^0-\mu^0_{r})_{,r}\delta_{\theta,r} - \frac{\delta_{r,r}}{r} + \delta_{\theta,rr} 
%\nonumber\\
%&\quad\quad\quad\quad+\frac{1}{r}\left(\delta_{t} + \delta_{\varphi} + 3\delta_{\theta}\right)_{,r} - \frac{2\delta_r}{r}(\frac{1}{r} - \mu^0_{,r} + \nu^0_{,r})\bigg\}\\
%
%
\delta R_{\hat{3}\hat{3}} &= 
-e^{-2\nu^0}\delta_{\varphi,tt} + \frac{1}{r^2}\left(\delta_{\varphi,\theta\theta} + 2\delta_\theta + (\delta_{r} - \delta_{\theta} + 2\delta_{\varphi} + \delta_t)_{,\theta}\cot\theta\right) \nonumber\\
&\quad\quad + e^{-2\mu^0_r}\bigg(\delta_{\varphi,rr}+ \left[\nu^0_{,r} - \mu^0_{,r} + \frac{2}{r}\right]\delta_{\varphi,r}+\frac{2}{r}(\mu_{r,r}^0-\nu^0_{,r}-\frac{1}{r})\delta_{r}\nonumber\\
&\quad\quad + \frac{1}{r}\left[\delta_{t}-\delta_{r}+\delta_{\theta}+\delta_{\varphi}\right]_{,r}\bigg).
\end{align}
\end{subequations}
\end{itemize}

\subsubsection{Axial perturbations}
Let us now restrict our attention to the axial perturbations. To simplify the equations, we define the variable $Q$ by
\begin{equation}
Q(r,\theta,t) = q_{[r,\theta]}\exp(3\mu^0_\varphi-\mu^0_r-\mu^0_\theta+\nu^0).
\end{equation}
Replacing the $\mu_i^0$ and $\nu^0$ with their values from the background spacetime \eqref{eq:dbh_metric},
\begin{align}
Q &= r^2\sin^3(\theta)\left(1-\frac{b(r)}{r}\right)e^{-\Phi(r)}q_{[r,\theta]}.
\end{align}
We can now rewrite the equations governing the perturbations \eqref{eq:axial_perturb1} and \eqref{eq:axial_perturb2} in the following form:
\begin{align}
Q_{,r} &= \frac{r^2\sin^{3}(\theta)}{e^{-\Phi}(1-b/r)}q_{[t,\theta],t}\\ 
Q_{,\theta} &= -r^4\sin^3(\theta)e^{\Phi}q_{[t,r],t}
\end{align}
We now assume that the perturbations are harmonic i.e. the time dependence is
of the form $\exp(i\omega t)$. The perturbations are then governed by
\begin{align}
\frac{2e^{-\mu_\vartheta}(1-b/r)}{r^2\sin^3(\theta)}Q_{,r} &= 2i\omega q_{[t,\theta]} = i\omega q_{t,\theta} + \omega^2 q_{\theta}\\
\frac{2e^{-\mu_\vartheta}}{r^4\sin^3(\theta)} Q_{,\theta} &= -2i\omega q_{[t,r]} = -i\omega q_{t,r} - \omega^2 q_r.
\end{align}
The $q_t$ term can be eliminated by virtue of the fact that
partial derivatives commute:
\begin{align}
%\left(\frac{2e^{-\psi}(1-b/r)}{r^2\sin^3(\theta)}Q_{,r}\right)_{,r} &= i\omega q_{t,\theta r} + \omega^2 q_{\theta,r}\\
%\left(\frac{2e^{-\psi}}{r^4\sin^3(\theta)} Q_{,\theta}\right)_{\theta} &= -i\omega q_{t,r\theta} - \omega^2 q_{r,\theta}\\
\left(\frac{2e^{-\mu_\vartheta}(1-b/r)}{r^2\sin^3(\theta)}Q_{,r}\right)_{,r} +& \left(\frac{2e^{-\Phi}}{r^4\sin^3(\theta)} Q_{,\theta}\right)_{,\theta} = \omega^2(q_{\theta,r}-q_{r,\theta})\\
&= -2 \frac{\omega^2}{r^2\sin^3(\theta)(1-b/r)e^{-\Phi}} Q. 
\end{align}
This leads to an equation in separable form:
\begin{align}
r^4 e^{\Phi}\left(\frac{e^{-\Phi}(1-b/r)}{r^2}Q_{,r}\right)_{,r} &+ \sin^3(\theta)\left(\frac{1}{\sin^3(\theta)} Q_{,\theta}\right)_{,\theta} + r^2 e^{2\Phi}\frac{\omega^2}{(1-b/r)} Q =0.\label{eq:axial_presep}
\end{align}
We take as our ansatz:
\begin{equation}
Q(r,\theta,t) = rR(r)\vartheta(\theta)\exp(i\omega t).
\end{equation}
which separates out \eqref{eq:axial_presep}:
\begin{align}
r^4 e^{\Phi}\left(\frac{e^{-\Phi}(1-b/r)}{r^2}(rR)_{,r}\right)_{,r} + r^2 e^{2\Phi}\frac{\omega^2}{(1-b/r)} rR - \gamma^2 rR &= 0\label{eq:grav_axial_radial}\\
\sin^3(\theta)\left(\frac{1}{\sin^3(\theta)} \vartheta_{,\theta}\right)_{,\theta} + \gamma^2\vartheta &= 0 \label{eq:grav_axial_angular}
\end{align}
where the separation constant has been denoted $\gamma^2$. The angular equation \eqref{eq:grav_axial_angular} is solved in terms of Legendre polynomials $P^{\ell}_m$ and $Q^{\ell}_m$:
\begin{equation}
\vartheta(\theta) = \sin^2\theta\left[A P^{\frac{\sqrt{9+4\gamma^2}-1}{2}}_{2}(\cos\theta) + B Q^{\frac{\sqrt{9+4\gamma^2}-1}{2}}_{2}(\cos\theta)\right].
\end{equation}
To ensure the solutions are real, we restrict the values of $\ell$ to integers. This requires that $\gamma$ is related to $\ell$ by
\begin{equation}
\gamma^2=\ell^2 + \ell - 2 = (\ell + 2)(\ell - 1).
\end{equation}

\paragraph{}
The aim is to rewrite the radial equation \eqref{eq:grav_axial_radial} in terms of a wave equation over the real line. Changing variables to the tortoise coordinate (cf \eqref{eq:tortoise})
\begin{equation}
\frac{\d r_\star}{\d r} = \frac{1}{\exp(-\Phi)(1-b/r)}
\end{equation}
the radial equation becomes:
\begin{align}
%r^4 e^{\Phi}\left(\frac{\d r}{\d r_\star}(\frac{R}{r^2} + \frac{R_{,r}}{r})\right)_{,r} + r^3 e^{2\Phi}\frac{\omega^2}{(1-b/r)} R - \gamma^2 rR &= 0\\
%r^4 e^\Phi\left(\frac{\d r}{\d r_\star}\frac{R}{r^2} + \frac{R_{,r_\star}}{r}\right)_{,r} + r^3 e^{\Phi}\frac{\d r_\star}{\d r} \omega^2 R - \gamma^2 rR &= 0\\
%r^3 e^\Phi\frac{\d r_\star}{\d r}\left(\frac{\d r}{\d r_\star} \frac{\d\phantom{r}}{\d r}\left(\frac{\d r}{\d r_\star}\right)\frac{R}{r} - 2\left(\frac{\d r}{\d r_\star}\right)^2\frac{R}{r^2} +  R_{,r_\star r_\star}\right) &+  r^3 e^{\Phi}\frac{\d r_\star}{\d r} \omega^2 R - \gamma^2 rR = 0\\
R_{,r_\star r_\star} + \frac{\d r}{\d r_\star} \frac{\d\phantom{r}}{\d r}\left(\frac{\d r}{\d r_\star}\right)\frac{R}{r} - 2\left(\frac{\d r}{\d r_\star}\right)^2\frac{R}{r^2} + \omega^2 R &- \frac{\gamma^2 R}{r^2}\exp(-\Phi)\frac{\d r}{\d r_\star} = 0.
\end{align}

\paragraph{}
This can be written in ``Schr\"{o}dinger form''
\begin{align}
R_{,r_\star r_\star} + \omega^2 R = V(r)R
\end{align}
with the ``potential'' $V(r)$ defined as 
\begin{align}
V(r) &\equiv \frac{\d r}{\d r_\star}\left(- \frac{\d\phantom{r}}{\d r}\left(\frac{\d r}{\d r_\star}\right)\frac{1}{r} + 2\left(\frac{\d r}{\d r_\star}\right)\frac{1}{r^2} + \frac{\gamma^2}{r^2}\exp(-\Phi)\right)\\
&=e^{-\Phi}\left(1-\frac{b}{r}\right)\left(\frac{\Phi_{,r}}{r}\left(1-\frac{b}{r}\right) + \frac{b_{,r}}{r^2} + \frac{\ell(\ell + 1)}{r^2} - \frac{3b}{r^3}\right)
\end{align}
where in the second term we have substituted back \eqref{eq:tortoise} and written $\gamma^2$ in terms of the angular momentum $\ell$.
 
\paragraph{}
So far we have not used the $\delta R_{\hat{0}\hat{3}}$ equation. This equation can be used to complete the solution -- that is, solved to find $q_{t}$ and $q_{(r,\theta)}$. However the form of $q_t$ is not of direct interest to us and it is assumed that all of these equations can be simultaneously satisfied.
%In addition to scalar fields we can also see how the gravitational field is
%perturbed by simply perturbing the metric. Note that no claim is made about
%general spin two fields, these are strictly only perturbations due to gravity.
%Many papers dealing with QNM assume that the generalisations to other spins is
%straightforward, but as observed by Larsen and Cveti\v{c}
%\cite{spin-papersI,spin-papers2} the so-called natural generalisations fails to
%include the Bianci identities when the spin two equations are written down
%\emph{except in the vacuum equations}. It is thus the opinion of (one of the
%authors/ the authors)\footnote{delete as appropriate} that the wave equation
%obtained around dirty black holes will depend on the type of field being
%discussed, not just its spin.

\subsubsection{Polar perturbations}
The equations governing polar perturbations \eqref{eqn_gp:polar_eqns} can be simplified by assuming an $\exp(i\omega t)$ time dependence for all $\delta_i$.  In this case we can simplify \eqref{eqn_gp:polar_eqns}:
\begin{subequations}
\begin{align}
0 &= (\delta_{r}+\delta_{\varphi})_{,\theta} + \left(\delta_{\varphi} - \delta_{\theta}\right)\cot\theta\label{eq:harmonic_polar_1}\\
0 &= \frac{2}{r}\delta_{r}  - (\delta_{\theta} + \delta_{\varphi})_{,r} +\left(\nu_{,r}^0-\frac{1}{r}\right)(\delta_{\theta}+\delta_{\varphi})\label{eq:harmonic_polar_2}\\
0 &= (\delta_{t}+\delta_{\varphi})_{,r\theta} + (\delta_{\varphi}-\delta_\theta)_{,r}\cot\theta -\left(\nu_{,r}+\frac{1}{r}\right)\delta_{r,\theta} + \left(\nu_{,r}-\frac{1}{r}\right)\delta_{t,\theta}\label{eq:harmonic_polar_3}\\
0 &= -\omega^2\exp(-2\nu^0)\left(\delta_{\varphi} + \delta_{\theta}\right) -\frac{1}{r^2}\big[(\delta_{t}+\delta_{\varphi})_{,\theta\theta} + 2\delta_{\theta} \nonumber\\
&\quad\quad\quad\quad+(2\delta_{\varphi} + \delta_t - \delta_\theta)_{,\theta}\cot\theta \big]
+ e^{-2\mu_r^0}\bigg[\frac{1}{r}(4\delta_r \nu^0_{,r}-2\delta_{t,r})\nonumber\\
&\quad\quad\quad\quad -\left(\frac{1}{r} + \nu^0_{,r}\right)(\delta_{\theta}+\delta_{\varphi})_{,r}+\frac{2}{r^2}\delta_r\bigg]\label{eq:harmonic_polar_4}\\
%
%0 &= \omega^2\exp(-2\nu^0)\delta_{\theta} + \frac{1}{r^2}\left(\delta_{\varphi,\theta\theta} + 2\delta_{\theta} + (2\delta_{\varphi}-\delta_{\theta})_{,\theta}\cot\theta + (\delta_t+\delta_{r})_{,\theta\theta}\right) \nonumber\\
%&+  \exp(2\mu^0_r)\bigg\{(\nu^0-\mu^0_{r})_{,r}\delta_{\theta,r} - \frac{\delta_{r,r}}{r} + \delta_{\theta,rr} 
%\nonumber\\
%&\quad\quad\quad\quad+\frac{1}{r}\left(\delta_{t} + \delta_{\varphi} + 3\delta_{\theta}\right)_{,r} - \frac{2\delta_r}{r}(\frac{1}{r} - \mu^0_{,r} + \nu^0_{,r})\bigg\}\label{eq:harmonic_polar_5}
%
0 &= 
\omega^2e^{-2\nu^0}\delta_{\varphi} + \frac{1}{r^2}\left(\delta_{\varphi,\theta\theta} + 2\delta_\theta + (\delta_{r} - \delta_{\theta} + 2\delta_{\varphi} + \delta_t)_{,\theta}\cot\theta\right) \nonumber\\
&\quad\quad + e^{-2\mu^0_r}\bigg(\delta_{\varphi,rr}+ \left[\nu^0_{,r} - \mu^0_{,r} + \frac{2}{r}\right]\delta_{\varphi,r}+\frac{2}{r}(\mu_{r,r}^0-\nu^0_{,r}-\frac{1}{r})\delta_{r}\nonumber\\
&\quad\quad + \frac{1}{r}\left[\delta_{t}-\delta_{r}+\delta_{\theta}+\delta_{\varphi}\right]_{,r}\bigg).\label{eq:harmonic_polar_5}
\end{align}
\end{subequations}
In the vacuum (Schwarzschild) case, the above equations are considerably
simplified by $\mu^0_r = -\nu^0$. In this case, the above equations are identical to equations (31)--(35) in chapter 4 of Chandrasekhar \cite{Book:ChandrasekharI}. Let us introduce the \emph{Ansatz} (due to Friedman \cite{FriedmanAxial}) to separate the equations:
\begin{align}
\delta_t(r,\theta) &= N(r) P^{\ell}(\cos\theta)\\
\delta_r(r,\theta) &= L(r) P^{\ell}(\cos\theta )\\
\delta_\theta(r,\theta) &= T(r)P^{\ell}(\cos\theta) + A(r)\frac{\d^2 P^{\ell}(\cos\theta)}{\d \theta^2}\\
\delta_\varphi(r,\theta) &= T(r)P^{\ell}(\cos\theta) + A(r)\frac{\d P^{\ell}(\cos\theta)}{\d \theta}\,\cot\theta.
\end{align}
Here $P^{\ell}(\cos\theta)$ is a Legendre polynomial, which will be abbreviated to $P^{\ell}$ in future. The first two equations are standard separations, whereas the last pair have been chosen separate out the radial and angular parts. 

\paragraph{}
The $\delta R_{\hat{0}\hat{2}}$ \eqref{eq:harmonic_polar_1} equation shows that not all the radial functions are independent:
\begin{align}
0 &= \delta_{r,\theta} + \delta_{\varphi,\theta} + (\delta_\varphi - \delta_\theta)_{,\theta} \cot\theta\\
&=  (L(r) P^\ell)_{,\theta} + (T(r) - A(r))P^\ell_{,\theta}\\
&= (T + L - A)P^{\ell}_{,\theta}.
\end{align}
Following Chandrasekhar we choose to eliminate $T(r)$ in favour of $L(r)$ and $A(r)$.

\paragraph{}
Turning attention now to \eqref{eq:harmonic_polar_2} we have
\begin{align}
0 &= \frac{2}{r}L P^{\ell} - [2(A - L) - \ell(\ell + 1)A]_{,r}P^\ell \nonumber\\&\quad\quad\quad+\left(\nu_{,r}^0-\frac{1}{r}\right)[2(A-L)-\ell(\ell+1)A]P^\ell\\
&= \left(\frac{\d }{\d r} -\nu_{,r}^0 + \frac{1}{r}\right)[2L + (\ell +2)(\ell- 1)A] + \frac{2}{r}L.
\end{align} 
Equation \eqref{eq:harmonic_polar_3} becomes:
\begin{align}
0 &= \left([N-L]P^\ell+A[P^\ell + P^\ell_{,\theta}\cot\theta]\right)_{,r\theta} + A_{,r}\left(P^\ell_{,\theta}\cot\theta - P^{\ell}_{,\theta\theta}\right)\cot\theta\nonumber\\
&\quad\quad -\left(\nu_{,r}+\frac{1}{r}\right)L P^\ell_{,\theta} + \left(\nu_{,r}-\frac{1}{r}\right)N P^\ell_{,\theta}\\
&= [N-L]_{,r}P^\ell_{,\theta}+A_{,r}[P^\ell_{,\theta}(1 + \cot^2\theta) + (P^\ell_{,\theta}\cot\theta)_{,\theta}  - P^{\ell}_{,\theta\theta}\cot\theta)\nonumber\\
&\quad\quad -\left(\nu_{,r}+\frac{1}{r}\right)L P^\ell_{,\theta} + \left(\nu_{,r}-\frac{1}{r}\right)N P^\ell_{,\theta}\\
&= \left[\left(\frac{\d }{\d r} + \nu_{,r}-\frac{1}{r}\right)N 
- \left(\frac{\d }{\d r} + \nu_{,r}+\frac{1}{r}\right)L 
+A_{,r}\left[\frac{1}{\sin^2\theta} + (\cot\theta)_{,\theta}\right]\right]P^\ell_{,\theta}
\end{align}
which implies that
\begin{align}
\left(\frac{\d }{\d r} + \nu_{,r}-\frac{1}{r}\right)N = 
\left(\frac{\d }{\d r} + \nu_{,r}+\frac{1}{r}\right)L .
\end{align}

\paragraph{}
Up to this point the equations derived agree exactly with Chandrasekhar's treatment of the Schwarzschild black hole. This is not true of the final two equations, where the dirt surrounding the black hole makes a non-trivial difference. Taking \eqref{eq:harmonic_polar_4}:
\begin{align}
0 &= -\omega^2\exp(-2\nu^0)\left(2T -\ell(\ell + 1)A\right)P^\ell -\frac{1}{r^2}\big[N(P^{\ell}_{,\theta\theta}+P^\ell_{,\theta}\cot\theta)\nonumber\\
& \quad\quad\quad\quad -(\ell+2)(\ell-1)TP^\ell  \big]
+ e^{-2\mu_r^0}\bigg[\frac{2}{r}(2L \nu^0_{,r}- N_{,r})\nonumber\\
&\quad\quad\quad\quad -\left(\frac{1}{r} + \nu^0_{,r}\right)(2T - \ell(\ell+1)A)_{,r}+\frac{2}{r^2}L\bigg]P^\ell.
\end{align}
By using the  defining differential equation
\begin{align}
P^{\ell}_{,\theta\theta}+P^\ell_{,\theta}\cot\theta = -\ell(\ell + 1) P^\ell.\label{eq:legendre_poly_relation}
\end{align}
for Legendre polynomial we find 
\begin{align}
&\left(\omega^2e^{-2\nu^0} + e^{-2\mu_r^0}\left[\frac{1}{r} + \nu_{,r}^0\right]\frac{\d}{\d r}\right)\left(2L+(\ell -1)(\ell +2)A\right) +\frac{1}{r^2}\big[N\ell(\ell + 1)\nonumber\\
& +(\ell+2)(\ell-1)(A-L)  \big]
+ e^{-2\mu_r^0}\bigg[\frac{2}{r}\left( 2\nu^0_{,r} + \frac{1}{r}\right)L - \frac{2}{r}N_{,r} \bigg] = 0.
\end{align}

\paragraph{}
The final equation \eqref{eq:harmonic_polar_5} is also the most complicated:
\begin{align}
0 &= 
\omega^2 e^{-2\nu^0}\delta_{\varphi} + \frac{1}{r^2}\left(\delta_{\varphi,\theta\theta} + 2\delta_\theta + (\delta_{r} - \delta_{\theta} + 2\delta_{\varphi} + \delta_t)_{,\theta}\cot\theta\right) \nonumber\\
&\quad\quad + e^{-2\mu^0_r}\bigg(\delta_{\varphi,rr}+ \left[\nu^0_{,r} - \mu^0_{,r} + \frac{2}{r}\right]\delta_{\varphi,r}+\frac{2}{r}(\mu_{r,r}^0-\nu^0_{,r}-\frac{1}{r})\delta_{r}\nonumber\\
&\quad\quad + \frac{1}{r}\left[\delta_{t}-\delta_{r}+\delta_{\theta}+\delta_{\varphi}\right]_{,r}\bigg)
\end{align}

\begin{align}
0 &= 
\omega^2 e^{-2\nu^0}[TP^\ell + AP^\ell_{,\theta}\cot\theta] + \frac{1}{r^2}\left(-(\ell+2)(\ell+1)TP^\ell + (N+L)P^\ell_{,\theta}\cot\theta\right) \nonumber\\
&\quad\quad + e^{-2\mu^0_r}\bigg(\left[ \left[\nu^0_{,r} - \mu^0_{,r} + \frac{2}{r}\right] + \frac{\d}{\d r}\right](T_{,r}P^\ell + A_{,r}P^\ell_{,\theta}\cot\theta)\nonumber\\
&\quad\quad +\frac{2}{r}\left(\mu_{r,r}^0-\nu^0_{,r}-\frac{1}{r}\right)LP^\ell%\nonumber\\
 + \frac{1}{r}\left[N-L+2T-\ell(\ell+1)A\right]_{,r}P^\ell\bigg).
\end{align}
We now collect terms proportional to $P^\ell$ and $P^\ell_{,\theta}$ separately:
\begin{align}
0 &= 
\Big\{ \omega^2 e^{-2\nu^0}T-(\ell+2)(\ell+1)T + \frac{2}{r}e^{-2\mu^0_r}\left(\mu_{r,r}^0-\nu^0_{,r}-\frac{1}{r}\right)L + \nonumber\\
&\quad\quad\frac{e^{-2\mu^0_r}}{r}\left[N-L+2T-\ell(\ell+1)A\right]_{,r} + e^{-2\mu^0_r}\left[ \left[\nu^0_{,r} - \mu^0_{,r} + \frac{2}{r}\right] + \frac{\d}{\d r}\right]T_{,r}\Big\}P^\ell\nonumber\\
&\quad\Big\{\omega^2 e^{-2\nu^0}A + \frac{1}{r^2}(N+L) + e^{-2\mu^0_r}\left[ \left[\nu^0_{,r} - \mu^0_{,r} + \frac{2}{r}\right] + \frac{\d}{\d r}\right]A_{,r}\Big\}P^\ell_{,\theta}\cot\theta\\
 &= 
\Big\{ \omega^2 e^{-2\nu^0}(A-L)-(\ell+2)(\ell+1)(A-L) \nonumber\\
&\quad\quad + \frac{2}{r}e^{-2\mu^0_r}\left(\mu_{r,r}^0-\nu^0_{,r}-\frac{1}{r}\right)L + 
\frac{e^{-2\mu^0_r}}{r}\left[N-3L-(\ell + 2)(\ell-1)A\right]_{,r}\nonumber\\
&\quad\quad + e^{-2\mu^0_r}\left[ \left[\nu^0_{,r} - \mu^0_{,r} + \frac{2}{r}\right] + \frac{\d}{\d r}\right](A-L)_{,r}\Big\}P^\ell\nonumber\\
&\quad\Big\{\omega^2 e^{-2\nu^0}A + \frac{1}{r^2}(N+L) + e^{-2\mu^0_r}\left[ \left[\nu^0_{,r} - \mu^0_{,r} + \frac{2}{r}\right] + \frac{\d}{\d r}\right]A_{,r}\Big\}P^\ell_{,\theta}\cot\theta
\label{eq:separate_long}.
\end{align}
The final expression must vanish for all $\theta$. But the only $\theta$ dependence is in the linearly independent factors $P^{\ell}$ and $P^{\ell}_{,\theta}\cot\theta$. For \eqref{eq:separate_long} to be satisfied for all $\theta$ we require
\begin{subequations}
\begin{align}
&\omega^2 e^{-2\nu^0}(A-L)-(\ell+2)(\ell+1)(A-L) + \frac{2}{r}e^{-2\mu^0_r}\left(\mu_{r,r}^0-\nu^0_{,r}-\frac{1}{r}\right)L + \nonumber\\
&\quad\quad\frac{e^{-2\mu^0_r}}{r}\left[N-3L-(\ell + 2)(\ell-1)A\right]_{,r} + e^{-2\mu^0_r}\left[ \left[\nu^0_{,r} - \mu^0_{,r} + \frac{2}{r}\right] + \frac{\d}{\d r}\right](A-L)_{,r} = 0,
\end{align}
and 
\begin{align}
\omega^2 e^{-2\nu^0}A + \frac{1}{r^2}(N+L) + e^{-2\mu^0_r}\left[ \left[\nu^0_{,r} - \mu^0_{,r} + \frac{2}{r}\right] + \frac{\d}{\d r}\right]A_{,r} &= 0.
\end{align}
\end{subequations}

\subsubsection{Deriving the Zerilli equation}
In the previous section we have taken the equations for polar perturbations, although finding the actual equations we need to satisfy gets lost in the working. The equations governing polar perturbations \eqref{eq:harmonic_polar_1}--\eqref{eq:harmonic_polar_5} and rewritten with the variable $\beta = (\ell + 2)(\ell -1)/2$: 
\begin{subequations}
\begin{align}
&2\left(\frac{\d \phantom{r}}{\d r} - \nu^0_{,r} + \frac{1}{r}\right)\left[L + \beta A\right] + \frac{2}{r}L = 0\label{eq:Z1}\\
&\left(\frac{\d \phantom{r}}{\d r} + \nu^0_{,r} - \frac{1}{r}\right)N = \left(\frac{\d \phantom{r}}{\d r} + \nu^0_{,r} + \frac{1}{r}\right)L\label{eq:Z2}\\
&\left(\omega^2e^{-2\nu^0} + e^{-2\mu_r^0}\left[\frac{1}{r} + \nu_{,r}^0\right]\frac{\d}{\d r}\right)2\left(L+\beta A\right) +\frac{1}{r^2}\big[N\ell(\ell + 1)\nonumber\\
& +2\beta(A-L)  \big]
+ e^{-2\mu_r^0}\bigg[\frac{2}{r}\left( 2\nu^0_{,r} + \frac{1}{r}\right)L - \frac{2}{r}N_{,r} \bigg] = 0\label{eq:Z3}\\
&\omega^2 e^{-2\nu^0}(A-L)-2\beta(A-L) + \frac{2}{r}e^{-2\mu^0_r}\left(\mu_{r,r}^0-\nu^0_{,r}-\frac{1}{r}\right)L + \nonumber\\
&\quad\quad\frac{e^{-2\mu^0_r}}{r}\left[N-3L-2\beta A\right]_{,r} + e^{-2\mu^0_r}\left[ \left[\nu^0_{,r} - \mu^0_{,r} + \frac{2}{r}\right] + \frac{\d}{\d r}\right](A-L)_{,r} = 0\label{eq:Z4}\\
&\omega^2 e^{-2\nu^0}A + \frac{1}{r^2}(N+L) + e^{-2\mu^0_r}\left[ \left[\nu^0_{,r} - \mu^0_{,r} + \frac{2}{r}\right] + \frac{\d}{\d r}\right]A_{,r} = 0.\label{eq:Z5}
\end{align}
\end{subequations}
This is five equations in three variables, namely $N$, $L$ and $A$. The first three equations \eqref{eq:Z1}--\eqref{eq:Z3} form three first order equations in the three variables. As such, we can solve for the first derivatives of $N$, $L$ and $A$ in terms of these same three functions.  The simplest way of solving these equations is to introduce functions $\tilde{a}(r)$, $\tilde{b}(r)$ and $\tilde{c}(r)$ such that
\begin{subequations}
\begin{equation}
N_{,r} = \tilde{a}N + \tilde{b}L + \tilde{c} \beta A, \quad\quad \beta \equiv \frac{(\ell + 2)(\ell - 1)}{2}.\label{eq:zerilli_subN}
\end{equation}
Using equations \eqref{eq:Z2} and \eqref{eq:Z3} easily yield 
\begin{align}
L_{,r} &= \left(\tilde{a} + \nu^0_{,r} - \frac{1}{r}\right) N + \left(\tilde{b} - \nu^0_{,r} - \frac{1}{r}\right) L
  + \tilde{c} \beta A\label{eq:zerilli_subL}\\
 \beta A_{,r} &= -\left(\tilde{a} + \nu^0_{,r} - \frac{1}{r}\right) N + \left(2 \nu^0_{,r} - \tilde{b} - \frac{1}{r}\right) L
    + \left(\nu^0_{,r} - \frac{1}{r} - \tilde{c} \right)\beta A\label{eq:zerilli_subA}.
\end{align}
\end{subequations}
By substituting these expressions into the longer and more messy \eqref{eq:Z3} we can then find these three functions. The results are
\begin{subequations}
\begin{align}
\tilde{a} &= \frac{\ell(\ell + 1)}{2 r} \exp(2\mu^0) = \frac{\beta + 1}{r-b}\\
\tilde{b} &= -\frac{1}{r} - \frac{\beta}{r-b} + \frac{r^3\omega^2}{(r-b)^2}e^{2\Phi} + (\nu^0_{,r})^2 r + \nu^0_{,r}\\
\tilde{c} &= -\frac{1}{r} + \frac{1}{r-b} + \frac{r^3\omega^2}{(r-b)^2}e^{2\Phi} + (\nu^0_{,r})^2 r,
\end{align}
\end{subequations}
where the values for $\nu^0$ and $\mu^0$ have been substituted back into the dirty black hole functions $\Phi(r)$ and $m(r)$ via \eqref{eq:QNM_intro_coord_numu}. The derivative $\nu^0_{,r}$ is complicated, and the expressions are simpler not expanding it.

\paragraph{}
The task now is to find a variable that can be expanded into a one dimensional wave equation. We try the function 
\begin{equation}
Z(r) = f(r) \beta A(r) + g(r) [L(r) + \beta A(r)].
\end{equation}
Intuitively the three functions $N$, $L$ and $A$ span a three dimensional vector space, and the above combination has no $N$ ``component''. If we can express the Zerilli equation as a one-dimensional wave equation then I require to have $Z_{,r_\star r_\star}$ pointing in the same direction, so in particular the $N$ component must cancel. The $L$ and $A$ components have been chosen as candidates for a one dimensional wave equation as \eqref{eq:Z1} tells us that at least $(L + \beta A)_{,r}$ transforms into a component independent of $N$. We have
\begin{align}
\frac{\d Z}{\d r} &= f_{,r} \beta A + f \beta A_{,r} + g_{,r}(L + \beta A) + g(L+\beta A)_{,r}\\
&= f_{,r} \beta A + f\beta A_{,r} +  \left[ g_{,r} + g\left(\nu^0_{,r} - \frac{1}{r}\right)\right](L + \beta A) - \frac{g}{r}L\\
&= f_{,r} \beta A + f \left[
      \left(\frac{1}{r} - \tilde{a} -\nu^0_{,r} \right) N  
   + \left(2\nu_{,r}^0 - \tilde{b} - \frac{1}{r}\right) L 
   + \left( \nu^0_{,r} - \frac{1}{r} - \tilde{c}\right)\beta A
 \right]\nonumber\\
 &\quad +  \left[ g_{,r} + g\left(\nu^0_{,r} - \frac{1}{r}\right)\right](L + \beta A) - \frac{g}{r}L.
 \end{align}
 Before taking the second derivative, we regroup the expression into pieces with $(L+\beta A)$
 \begin{align}
 Z_{,r} &= f \left[
      \left(\frac{1}{r} - \tilde{a} -\nu^0_{,r} \right) N  
   + \left(\nu_{,r}^0  - \frac{1}{r} - \tilde{c}\right)(L+\beta A)+(\tilde{a}+2\nu^0_{,r})L 
 \right]\nonumber\\
 &\quad +  \left[ g_{,r} + g\left(\nu^0_{,r} - \frac{1}{r}\right)\right](L + \beta A) - \frac{g}{r}L + f_{,r}\beta A\\
 &= f \left[
      \left(\frac{1}{r} - \tilde{a} -\nu^0_{,r} \right) N  
   +(\tilde{a}+2\nu^0_{,r})L
 \right] - \frac{g}{r}L + f_{,r}\beta A \nonumber\\
 &\quad +  \left[ 
   g_{,r} + g\left(\nu^0_{,r} - \frac{1}{r}\right)
   + f \left(\nu_{,r}^0  - \frac{1}{r} - \tilde{c}\right) 
 \right](L + \beta A).
 \end{align}
 The \emph{procedure} to obtain the Zerilli equation is now clear: we differentiate this expression again and use the fact that
 \begin{align}
 Z_{,r_\star r_\star}&= e^{2\nu^0 + \Phi}\partial_r\left(e^{2\nu^0 + \Phi} Z_{,r}\right)\\
  &= e^{4\nu^0 + 2\Phi}\left(\left[2\nu^0_{,r} + \Phi_{,r}\right] Z_{,r} + Z_{,rr}\right).\label{eq:zerilli_zrsrs}
  \end{align}
  By eliminating the term in $N$ we will get one differential equation in $f$ and $g$. Another equation is obtained by setting the ratio of the $f$ and $g$ terms to be the same as the ratio of the coefficients of $\beta A$ and $(L + \beta A)$ in $Z_{,r_\star r_\star}$. This is two equations in two unknowns and hence in principle we can determine $f$ and $g$. If the solution turns out to be a linear equation then we trivially have a wave equation for polar perturbations. With some regret, I must report that I have not done this yet. The difficulties are purely mechanical, for example the second derivative is given by
  \begin{align}
  Z_{,rr} &= \left[ \left(\frac{1}{r} - \tilde{a} - \nu^0_{,r}\right)f_{,r} + \left(\tilde{a}\frac{1-b_{,r}}{r-b} - \frac{1}{r^2} - \nu^0_{,rr}\right) f \right] N \nonumber\\
  &\quad + \Bigg[ \left( \tilde{a} - \tilde{c} + 3\nu^0_{,r} - \frac{1}{r}\right) f_{,r}
   + \bigg(2 \nu^0_{,rr} - \tilde{a}\frac{1-b_{,r}}{r-b} - \frac{3\nu^0_{,r}}{r} + \frac{3}{r^2} + \frac{\tilde{c}}{r} + \nu^0_{,rr} \nonumber\\
   &\quad  - \tilde{c}_{,r} + [\nu^0_{,r}]^2 - \tilde{c}(\nu^0_{,r} - 1/r)\bigg)f
   + \left(\frac{3}{r^2} - \frac{3\nu^0_{,r}}{r}  + \nu^0_{,rr} + \frac{1}{r^2} + [\nu^0_{,r}]^2 \right)g\nonumber\\
   &\quad \left(2\nu^0_{,r} - \frac{4}{r}\right)g_{,r} + g_{,rr}
   \Bigg](L+\beta A) + \Bigg[\frac{2}{r}g_{,r} + \frac{g}{r}\left(\nu^0_{,r} - \frac{2}{r}\right) + f_{,rr}\nonumber
   \end{align}
  \begin{align} 
   & - (\tilde{a} + 2\nu^0_{,r})f_{,r} + \left(\frac{\nu^0_{,r}}{r} - \frac{1}{r^2} - \frac{\tilde{c}}{r} + \frac{1-b_{,r}}{r-b}\tilde{a} - 2\nu^0_{,rr}\right)f\Bigg]\beta A \nonumber\\
   &+ \left[\frac{1}{r} - \tilde{a} -\nu^0_{,r}\right]f N_{,r}
 + \left[f(\tilde{a}+2\nu^0_{,r})-\frac{g}{r}\right]L_{,r} + f_{,r}\beta A.
 \end{align}
 We would need to expand the derivatives on the last line and substitute using \eqref{eq:zerilli_subN}--\eqref{eq:zerilli_subA}, then collect terms and place into \eqref{eq:zerilli_zrsrs}. This has not yet been completed.
 
 \subsubsection*{The Zerilli equation for a Schwarzschild black hole}
 I shall not derive it, but I quote the significant results for the special case of a Schwarzschild black hole from Chandrasekhar's book \cite{book:Chandra}. In this case the one dimensional variable that obeys the wave equation is given by
 \begin{equation}
 Z =rA - \frac{r^2}{\beta r + 3M}(L + \beta A).
 \end{equation} 
 The Zerilli equation is given by
 \begin{equation}
 \frac{\d ^2 Z}{\d r_\star^2} + \omega^2 Z = V_{z} Z
 \end{equation}
 where the effective potential $V_{z}$ is given by
 \begin{equation}
 V_z = \frac{2r(r-2M)}{r^5(\beta r + 3M)^2}\left[\beta^2(\beta + 1)r^3 + 3M\beta^2r^2 + 9M^2\beta r + 9 M^3\right].
 \end{equation}
 At some point I would like to find the dirty black hole equivalent and show that it reduces when $b = 2M$ and $\Phi = 0$.

\subsection{Isospectrality and the NP formalism}
Even though the solution has not been obtained for the Zerilli equation, it is at least clear how we could obtain it given enough time and patience. The method does seem devoid of any sort of intuitive interpretation, other than we have realised that there are two decoupled modes. The rest is just going through rather tedious algebra. The NP formalism seems to allow for a cleaner separation of variables, and makes various appeals to the fact that stationary black hole spacetimes are of Petrov type D to considerably simplify matters. Even after finding the principal null directions in static dirty black hole spacetimes the spin coefficients did not separate into two similar sets of equations. To some degree, one would expect that the dirtiness of the spacetimes would complicate matters somewhat, as in the standard approach in vacuum spacetimes the NP formalism a natural way of seeing that the spectra are isospectral. As we know of examples where the spectra are not isospectral (namely AdS) then it is possible that the NP formalism is not of great utility. In any case, I have investigated the possibility and have found the non-zero $\Phi_0$, $\Phi_1$ and $\Phi_2$ have prevented me from finding any obvious separation of variables.

\section{Conclusions}
This chapter has hopefully served as a brief introduction to the idea of QNM and how they are defined, as well as some of their analytic properties. The QNM conjecture has also been outlined, which has served as the motivation for looking at the highly damped QNMs. After all, they are not likely to contribute in any significant manner to any astrophysically relevant process!  The method to obtain the wave equation for a QNM is outlined, and the equations are found explicitly for axial gravitational fields and for scalar fields. In the next chapter methods of probing the asymptotic structure of the QNM spectra are discussed.

\paragraph{}
It has also been pointed out that finding the polar perturbations is non-trivial, and while it has been indicated how one could carry it out in principle the task still remains incomplete. But a more serious problem has developed: while the QNM conjecture looked promising it was ultimately based on a numerical coincidence. As such, it is a promising direction for research and investigation, but we should not argue vehemently that it is true until we have solid arguments leading us to the QNM conjecture from well established basic principles. If one adopts this philosophy, a recent paper by Domagala and Lewandowski \cite{Domagala:2004} has shown the QNM conjecture to be false. Recall that we anticipate being able to set the Barbero--Immirzi parameter as follows
\begin{equation}
\gamma = \frac{\ln 3}{2\sqrt{2} \pi} \approx 0.123637\ldots 
\end{equation}
More generally, if we allowed the black hole to be dominated by links with spin $j$ we found that to get the correct relationship between area and entropy required
\begin{equation}
\gamma_j = \frac{\ln(2 j + 1)}{2\pi \sqrt{j(j+1)}}. \label{eq:BI_general_j}
\end{equation}
The choice $j=1$ is made by the connection to QNM. If the QNM conjecture is not assumed then there is no reason to assume $j=1$, any half-integer value in \eqref{eq:BI_general_j} will do. The Lewandowski and Domagala paper shows that in the case of $j = 1/2$ that \eqref{eq:BI_general_j} does not hold. It is shown that to reproduce the area of the horizon you need $\gamma_0$ to be the solution of
\begin{equation}
\sum_{m=1/2,1,3/2.\ldots}^\infty \exp(2\pi\gamma_{1/2}\sqrt{m(m+1)}) = \frac{1}{2}
\end{equation}
which leads to 
\begin{equation}
\gamma_{j=1/2} = 0.2375329\ldots
\end{equation}
whereas the solution predicted by \eqref{eq:BI_general_j} is
\begin{equation}
\gamma_{j=1/2} = \frac{\ln 2}{\pi\sqrt{3}} \approx 0.127384\ldots
\end{equation}
This is because in considering \eqref{eq:BI_general_j} we did not take into account various constraints, such as the sum of all the spin links intersecting the horizon being zero (as we are considering a Schwarzschild black hole). The value $\gamma_{j=1/2} = 0.2375\ldots$ works over dirty black holes, dilaton black holes, rotating and charged black holes, and hence it seems that it is robust. However, a similar result will be true for the $j=1$ case as well, as the same constraints are still there. Thus $\gamma$ will be different from what is predicted in \eqref{eq:BI_general_j} and we will no longer be able to make a correspondence with the QNMs \emph{while retaining the semi-classical relationship between area and entropy}.

\paragraph{}
Is this too dim and dark a conclusion? Is it not possible to fix up the QNM conjecture in some manner to overcome this difficulty? Not an obvious one that I can see. Of course one can try and repair it, but such a task seems fruitless. The QNM conjecture had a nice intuitive appeal on the basis of particles and some [admittedly odd] use of a correspondence principle. The only evidence for it, however, was not a solid argument but an apparent numerical coincidence. Now even that has gone. As Feynman remarked about proving the mass of the photon is zero, you can only prove that it is less massive than some given upper mass limit. Thus, in principle, if the photon is truly massless you could spend forever getting better and better bounds on its mass. At some point we just accept it is massless until we have a reason to believe otherwise. With the QNM conjecture a similar principle should hold: until we get so desperate for ideas on how to explain aspects of quantum gravity, the QNM conjecture should be discarded.

\chapter{Finding the QNM}\label{chap:qnm_find}

In the previous chapter the definition of QNM was given, and the problem of finding QNMs was related to finding solutions of the Helmholtz equation
\begin{equation}
\frac{\d^2 R(r_\star)}{\d r_\star^2} + \omega^2 R(r_\star) = V(r_{\star}) R(r_\star)\label{eq:QNM_diff_eq}
\end{equation}
with the boundary conditions that the wave is outgoing at both the horizon and infinity, and that the time dependence of the perturbation is of the form $\exp(-i\omega t)$. These boundary conditions can be rewritten explicitly in terms of $r_\star$ as:
\begin{equation}
R(r_\star)\exp(-i\omega t) \rightarrow \left\{\begin{array}{ll}
\exp(-i \omega [r_\star + t])&\text{as }r_\star \rightarrow -\infty\\
\exp(+i \omega [r_\star - t])&\text{as }r_\star \rightarrow +\infty
\end{array}
\right.\label{eq:QNM_boundary_condition}
\end{equation} 
if we take $\Im(\omega) < 0$. Solutions with $\Im(\omega)$ positive are not allowed as they would correspond to solutions that grow exponentially in time, whereas QNMs are the decaying modes of a spacetime.\footnote{In the language of \S \ref{sec:def_QNM} the exponential solutions correspond to $f_+$ and $f_-$, which are the well behaved solutions to the homogeneous equation when either infinity is approached.} It is easily seen that these boundary conditions ensure that the solutions are outgoing at both the horizon and infinity.

\paragraph{}
The previous chapter also showed how to find $V(r_\star)$ for specific fields. Together with information of the initial perturbation we can then find how various fields are perturbed in the spacetime. The goal of this chapter is to outline methods that may be used to calculate the quasinormal frequencies. The problem with solving \eqref{eq:QNM_diff_eq} subject to the boundary conditions are:
\begin{itemize}
\item The generic problem of solving a partial differential equation.
\item While $V(r)$ is a rational function of $r$, there is no
  (simple) relation in terms of $r_\star(r)$. For example, in the Schwarzschild geometry we
  need the \emph{Lambert W function}. Hence $V(r_\star)$ is generally not a
  simple function.
\item The boundary condition is numerically difficult. In general we have a
  superposition of an exponentially growing term and an exponentially decaying
  term as $r_\star \rightarrow \infty$ at fixed $t$. The QNM boundary condition
  \eqref{eq:QNM_boundary_condition} requires that the exponentially
  \emph{decaying} piece vanishes. In a practical numerical solution, the
  exponentially decaying piece would be lost in the numerics, as the
  exponentially growing term would dominate.
\end{itemize}
To find the quasinormal modes we look at four different methods:
\begin{itemize}
\item Analytic approximation methods: By replacing the function $V(r)$ by an
  approximate function $\mathcal{V}(r)$ we are able to solve the new Helmholtz
  equation with the required boundary condition
  \eqref{eq:QNM_boundary_condition} for the frequencies $\tilde{\omega}_n$
  analytically. The underlying assumption is that if $V$ is well approximated
  by $\mathcal{V}$, then $\omega_n$ will be well approximated by
  $\tilde{\omega}_n$.
\item Series expansion methods: By making a series \emph{Antsatz} for the form
  of the solution a recurrence relation or continued fraction method can be
  found for the QNM. Also known as the Leaver method \cite{Leaver:1985ax}, and
  also used by Motl \cite{Motl}.
\item WKB or Monodromy methods: The Helmholtz equation \eqref{eq:QNM_diff_eq}
  is formally analogous to the Schr\"odinger equation. By treating the QNM
  problem as a pure scattering problem, an analogue of the standard
  Bohr--Sommerfeld rule is found. This technique was used by Motl and Neitzke
  in their recent paper \cite{MotlNeitzke}.
\item Born series methods: By treating the QNM problem as a scattering problem,
  it is hoped that the Born series would give accurate results; especially for
  highly damped modes.
\end{itemize}
With the exception of Leaver's method, all of the above methods make an
approximation at the level of the differential equation, rather than finding an
exact solution to the exact differential equation. The above list is only a
survey of possible methods and is by no means exhaustive. In the review article
by Kokkotas and Schmidt \cite{Kokkotas}, the last section outlines several
techniques for finding QNMs. In particular, they mention techniques such as
resonance methods, direct integration methods and variation principles. For
details, see \cite{Kokkotas} and references therein.

\paragraph{}
It is hoped that we could approximate the QNMs of any spherically symmetric
static black hole. This would include the special cases of Schwarzschild,
Schwarzschild--de Sitter, Reissner--Nordstr\"om, as well as the more generic
cases where the black hole was surrounded by spherically symmetric matter
fields.\footnote{The reader is reminded that while axisymmetric matter fields
  are more astrophysically relevant, we are not performing astrophysically relevant
  measurements here. Rather, the goal is to look at highly damped modes and
  investigate the robustness of the QNM conjecture, as highly damped QNM are
  not of astrophysical relevance in any case.} Some methods (notably Leaver's)
do not require spherical symmetry and can be applied even more generally, but
the others apply to \emph{one dimensional} scattering processes and it is
difficult to see how to generalise them.

\paragraph{}
Even within the restrictions of a spherically symmetric black hole things are
still unclear. If we allow arbitrary matter fields, then there is a possibility
of a cosmological horizon or even a ``multi-horizon'' structure. It is argued
that QNMs are a property of the spacetime between two horizons, as this is the only
region accessible to a given observer, under the additional assumption that neither of the horizons are extremal. It is tacitly assumed
that the horizons are boundaries of an inward trapped and outward trapped
region respectively, so that we may apply outward going boundary conditions at
each horizon. By introducing a generalised tortoise coordinate $r_\star$ (as was
done in Chapter \ref{chap:QNM_intro}) that maps the region between the horizons
to the entire real line, the boundary conditions of the differential equation
\eqref{eq:QNM_diff_eq} remain unchanged.

\paragraph{}
Much more subtle is the possibility that the space becomes asymptotically
\emph{anti-de Sitter} (AdS), in which case the space is closed (in the absence
of topological obstructions). Thus there is no hope that the second boundary is
an outward trapped region and the above prescription fails. A QNM prescription is still needed, as energy in the exterior region can propagate through the horizon and become lost, making the use of normal modes inapplicable. To define the QNMs precisely we need to fix a boundary condition at the outer boundary of the spacetime. Conventionally this boundary condition is taken to be the vanishing of fields at the outer boundary that are damped in time. Using this convention the QNMs still correspond to poles in the Green's function, as there is no incoming radiation. A different convention (not adopted here) is that the outer boundary is perfectly reflecting. For a discussion on boundary conditions defining QNMs in AdS spacetime see \S 3.3.1 and references therein of Cardoso's thesis\cite{Cardoso:2004pj} and the paper by Horowitz \cite{Horowitz:1999jd}.

\paragraph{}
Finally the reader is warned that the QNMs found in this thesis are obtained by assuming that the topology is the simplest one compatible with the local geometry. While topological effects may make a significant
difference to the QNM structure, I have assumed throughout that the local
geometry suffices to specify the topology.

\section{Born series method}
The Born series method is the least reliable of all the different ways of finding QNMs and yields very little information. However, it is very simple to apply to wide classes of black holes. The Born approximation is described in detail by many books on quantum mechanics, for example \cite{book:QM_LandauLifshitz,QM:Griffiths}. A very brief discussion is given here.

\paragraph{}
The Born approximation splits the function $R(r_\star)$ into a series
\begin{equation}
R(r_\star) = R^{(0)}(r_\star) + R^{(1)}(r_\star) + R^{(2)}(r_\star) + \ldots
\end{equation}
As we are interested in the limit of highly damped modes, we are really looking at the modes for which 
 $|\omega_n| \gg \sup |V(r)|$, provided that $\sup |V(r)|$ is finite. In this limit, the potential can be ignored in the lowest order (see \eqref{eq:chap2_omega_defined} to recall the definition of $\omega_n$). Hence the lowest order solution $R^{(0)}$ is a plane wave solution. From $R^{(0)}$, the $n^{th}$ Born approximation can be found via
\begin{equation}
\frac{\d^2 R^{(n)}}{\d r_\star^2} + \omega^2 R^{(n)} = V(r_\star) R^{(n-1)}.
\end{equation}
By finding the Green's function $G(r)$ for the Helmholtz equation we convert this into
\begin{equation}
R^{(n)}(r_\star) = \int_{-\infty}^\infty G(r_\star - \tilde{r_{\star}}) V(\tilde{r_\star}) R^{(n-1)}\,\d \tilde{r_{\star}}.
\end{equation}
But for the one dimensional Helmholtz equation the Green's function is well known
\begin{equation}
G(r_\star) = i\exp(i \omega r_\star)
\end{equation}
and hence the Born series becomes
\begin{equation}
R^{(n)}(r_\star) = i\exp(i\omega r_\star)\int_{-\infty}^\infty \exp(-i\omega \tilde{r_{\star}}) V(\tilde{r_\star}) R^{(n-1)}(\tilde{r_\star})\,\d \tilde{r_{\star}}.
\end{equation}

\subsection{Finding the QNM}
We start by utilising the first order Born approximation
\begin{equation}
R^{(0)} + R^{(1)}  = \exp(-i\omega r_\star) - i\exp(i\omega r_\star) \int_{-\infty}^{\infty}\exp(-i\omega \tilde{r_{\star}}) V(\tilde{r_\star}) \exp(-i \omega \tilde{r_\star})\,\d \tilde{r_{\star}}.
\end{equation}
The transmitted part is then in the first term, while the second term describes particles that are scattered back. We want to concentrate on when the back-scattering blows up, so that we can have a case with no incoming particles and only outgoing ones -- the boundary conditions of a QNM!

\paragraph{}
To do this, let us take the second part and label it $a(\omega)$. This is essentially a Fourier transform of the scattered part. Then we have
\begin{align}
a(\omega) &\propto  \int_{-\infty}^{\infty}\exp(-2i\omega \tilde{r_{\star}}) V(\tilde{r_\star})\,\d \tilde{r_{\star}}.
\end{align}
Undergo a change of variables back to $r$. This means that the lower limit is $r_b$, the black hole horizon and the upper limit of integration is $r_c$, the cosmological horizon. If a cosmological horizon does not exist then we formally take $r_c$ to infinity. 
\begin{align}
a(\omega) &\propto  \int_{r_b}^{r_c}\exp(-2i\omega \tilde{r_{\star}}(r)) V(r) \frac{\d r_\star}{\d r}\,\d r\\
&\approx  \int_{r_b}^{r_c}\exp(-2i\omega \tilde{r_{\star}}(r)) V(r) \left[\frac{1}{2\kappa_b(r-r_b)} + \frac{1}{2\kappa_c (r-r_c)} + \alpha \right]\,\d r.
\end{align}
Here we have approximated 
\begin{equation}
\frac{\d r_\star}{\d r} = \frac{1}{2\kappa_b (r-r_b)} + \frac{1}{2\kappa_c (r-r_c)} + \alpha,
\end{equation}
where $\kappa_b$ and $\kappa_c$ are the surface gravities of the black hole and cosmological horizons respectively. The fact that the expression takes this form near a horizon follows from earlier comments about the exponential decay of $r_\star$ as a horizon is approached.\footnote{Or, as a horizon is a place where $2m(r) = r$, we have a pole in the defining expression for the tortoise coordinate. If the horizon is non-extremal by definition this pole is simple and the result follows.} If a cosmological horizon does not exist, the requirement of taking $r_c$ to infinity already kills off the term that contains it. Here we have assumed that the horizon is non-extremal so that the sum of the two terms is a good approximation to the derivative of the tortoise coordinate, at least near the two horizons. The other large assumption is that it is the near horizon geometry that is important. 

\paragraph{}
Let us assume that $r_b \ll r_c$. Then the poles that come about from  the near black hole horizon geometry $\omega_b$ are those for which
\begin{align}
a(\omega_b) &\propto  \int_{r_b}^{r_c}\exp(-2i\omega_b \tilde{r_{\star}}(r)) V(r) \left[\frac{1}{2\kappa_b(r-r_b)} + \frac{1}{2\kappa_c (r-r_c)} + \alpha \right]\,\d r\\
&\approx \int_{r_b}^\infty \exp(-2i\omega_b \tilde{r_\star}(r)) V(r) \left(\frac{1}{2\kappa_b(r-r_b)} + \alpha\right)\,\d r\label{eq:born_integral_I}.
\end{align}
But near the black hole horizon we also have
\begin{equation}
r_\star \approx \alpha(r-r_b) + \frac{1}{2\kappa_b}\ln\left[\frac{r-r_b}{r_b}\right] + \mathcal{O}([r-r_b]^2)
\end{equation}
and hence
\begin{align}
\exp(-2i\omega r_\star) &\approx \exp(-2i\omega\alpha(r-r_b))\left[\frac{r-r_b}{r_b}\right]^{-i\omega/\kappa_b} + \mathcal{O}\left(\exp(2i\omega[r-r_b]^2/\kappa_b) \right)\\
&\approx \exp(+ 2i\omega\alpha r_b)) e^{-z} \left[\frac{r}{r_b}-1\right]^{-i\omega/\kappa_b} 
\end{align}
where $z = -2i\omega\alpha (r-r_b)$.\footnote{This corrects an error in our paper, where we claimed that $z=-2i\omega\alpha r$.} Placing all of this back into the integral \eqref{eq:born_integral_I} we obtain (neglecting factors of proportionality) and making $z$ the variable of integration:
\begin{align}
a(\omega) \approx \int_{0}^\infty e^{-z} z^{-i\omega/\kappa_b-1} V\left[\frac{iz}{2\omega\alpha} + r_b\right] \left(\frac{1}{2\kappa_b } + \alpha  z\right)\,\d z .
\end{align}
We can then expand $V(r_b + iz/2\omega\alpha)$ as a power series in $z$:
\begin{equation}
V\left(r_b + \frac{iz}{2\omega\alpha}\right) = \sum_n a_n z^n
\end{equation}
Hence we have
\begin{align}
a(\omega) &\propto \sum_n a_n \int_{0}^\infty e^{-z} z^{-i\omega/\kappa_b-1+n}\left(\frac{1}{2\kappa_b } + \alpha  z\right)\,\d z \\
&\propto \sum_n b_n \int_{0}^\infty e^{-z} z^{-i\omega/\kappa_b-1+n}\,\d z \\
&\propto \sum_n b_n \int_{0}^\infty e^{-z} z^{-i\omega/\kappa_b-1+n}\,\d z \\
&\propto \sum_n b_n \Gamma\left(-\frac{i\omega}{\kappa_b} + n\right).
\end{align}
Now the Gamma function blows up at all negative integers. The argument of one of the Gamma functions will be a negative integer provided that
\begin{equation}
\omega = -\kappa_b m i,
\end{equation}
for some integer $m$. Note that my sign convention is opposite to the one used in our paper (appendix \ref{app:DBHQNM}). Similarly, by looking at well separated cosmological and black hole horizons we find analogously 
\begin{equation}
\omega = -\kappa_c m i.
\end{equation}
In the case where the horizons are close, or more precisely $(r_c - r_b)/r_b$ is of order one or lower, then treating the two horizons as individual supports for the first Born approximation fails. See appendix \ref{app:squeezed} for details on this case.

\section{Analytical approximation schemes}

Analytical approximation schemes are based around the premise that a ``small''
change in the potential will result in a small change in the QNM frequencies.
It is not clear that this is always the case, or how one should meaningfully
quantify ``small''. One candidate for a measure of smallness is the integrated difference in
the absolute value of the actual potential $V(r_\star)$ and the analytic approximation $\mathcal{V}(r_\star)$
\begin{equation}
\int_{-\infty}^{\infty}| \mathcal{V}(r_\star) - V(r_\star)|\,\d r.
\end{equation}
One possible way of addressing this question is by looking at generic one--dimensional potentials and bounding the reflection and transmission
coefficients, as is done by Visser \cite{Visser:1998} and Fr\"oman and Fr\"oman
\cite{Froman}.

\paragraph{}
So far analytic approximation schemes have been applied to four cases in the
literature. The first paper by Ferrari and Mashhoon \cite{Ferrari:1984} covered
the Schwarzschild, Reissner--Nordstr\"om and slowly rotating Kerr black holes.
Much later the Schwarzschild--de Sitter black hole was investigated by Suneeta
\cite{Suneeta:2003}. If one considers only the near horizon geometry to be
important in dictating the QNM structure (and there is evidence in the Born
series method which suggests this may be the case) Suneeta's results may be
extended to generic dirty black holes as well. 

\subsection{Schwarzschild black holes}
\subsubsection{Matching the potentials}
The potential of choice for modelling black holes has been the Poschl-Teller
potential
\begin{equation}
\mathcal{V}(r_\star) = \frac{V_0}{\cosh^2(\alpha (r_\star - r_{\star}^{0}))}\label{eq:Poschl-Teller}
\end{equation}
where $V_0$ is the value of the potential at the peak $r_\star = r_{\star}^{0}$. The ``matching conditions'' taken by Ferrari and Mashhoon \cite{Ferrari:1984} were
\begin{subequations}\label{eq:QNM_match}
\begin{itemize}
\item That the maxima agreed:
\begin{equation}
V( (r_{\star})_0 ) =\mathcal{V}( (r_{\star})_0 ) .
\end{equation}
\item $\alpha$ was determined by matching the second derivative at the peak:
\begin{equation}
\left.\frac{\d^2 V}{\d r_\star^2}\right|_{(r_{\star})_0} =
\left.\frac{\d^2 \mathcal{V}}{\d r_\star^2}\right|_{(r_{\star})_0}.
\label{eq:match_PT2F}
\end{equation}
\end{itemize}
These conditions emphasise the role of the peak of the potential. Instead of \eqref{eq:match_PT2F}, I adopt a convention that is easier to fit into multi-horizon spacetimes:
\begin{itemize}
\item That the rate of decay at the horizons are equal:
\begin{equation}
\lim_{r_\star \rightarrow -\infty} \frac{\ln(V)}{r_\star} = 
\lim_{r_\star \rightarrow -\infty} \frac{\ln(\mathcal{V})}{r_\star}\label{eq:match_suneeta}.
\end{equation}
This condition determines $\alpha$.
\end{itemize}
\end{subequations}
Equation \eqref{eq:match_suneeta} is inspired by Suneeta's approach to a SdS spacetime. At the end of this section my variation on the theme of the ``Suneeta'' approach will be compared with the results of Ferrari and Mashhoon, and with the numerical results.
 
\begin{figure}[th]
\begin{center}
\begin{minipage}{0.25\textwidth}
\includegraphics[width=\textwidth]{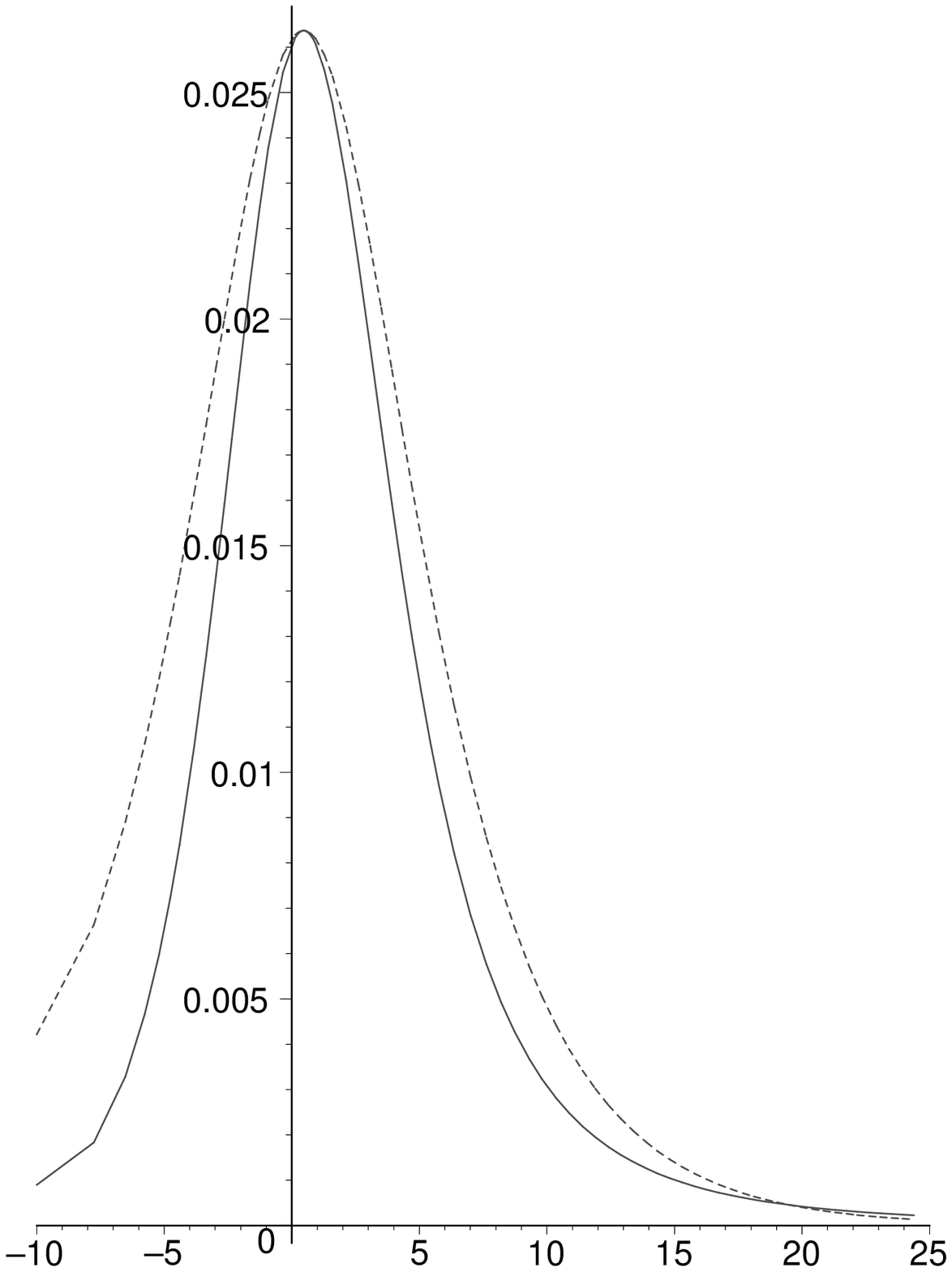}
\end{minipage}
\begin{minipage}{0.25\textwidth}
\includegraphics[width=\textwidth]{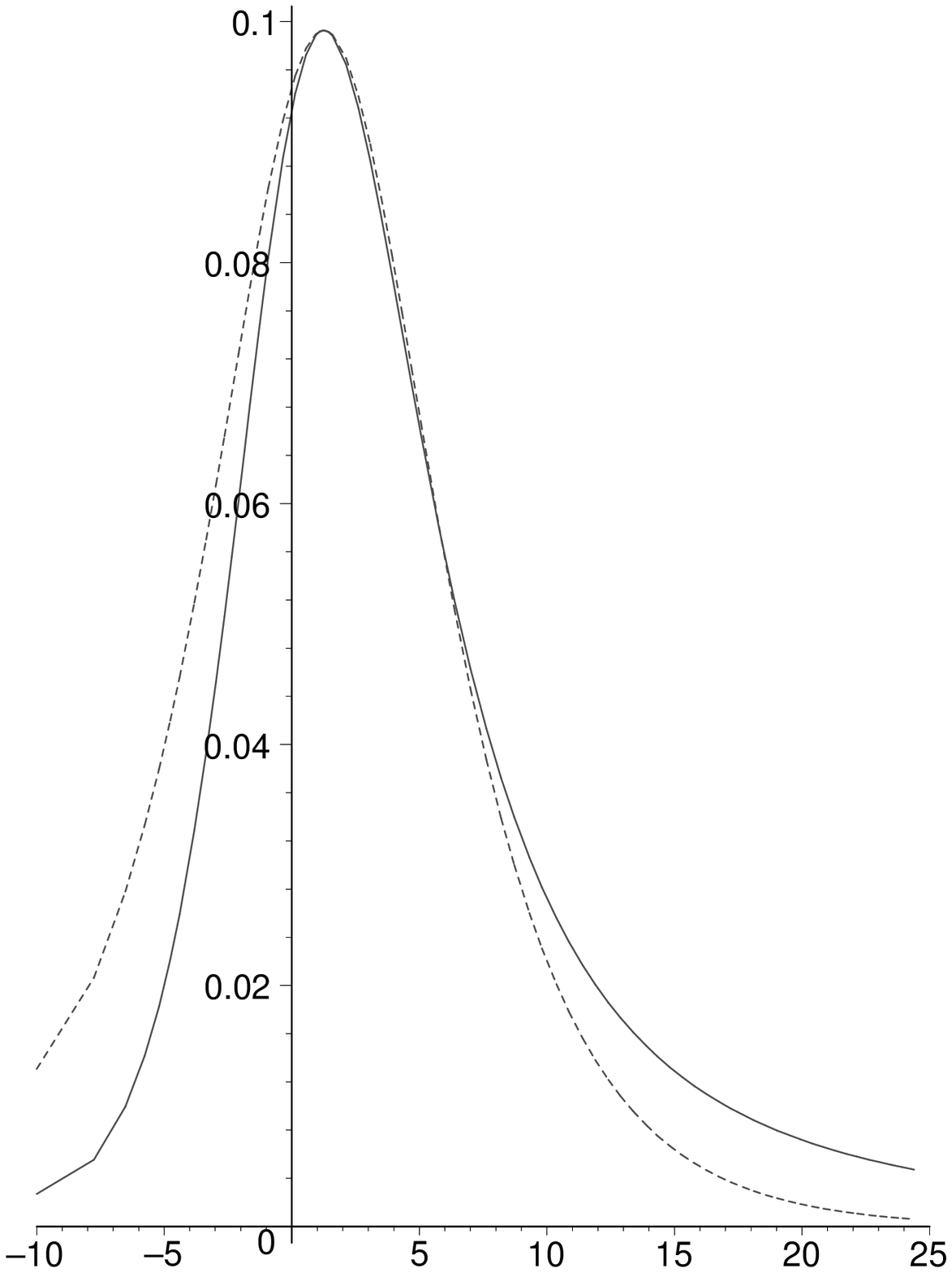}
\end{minipage}
\begin{minipage}{0.25\textwidth}
\includegraphics[width=\textwidth]{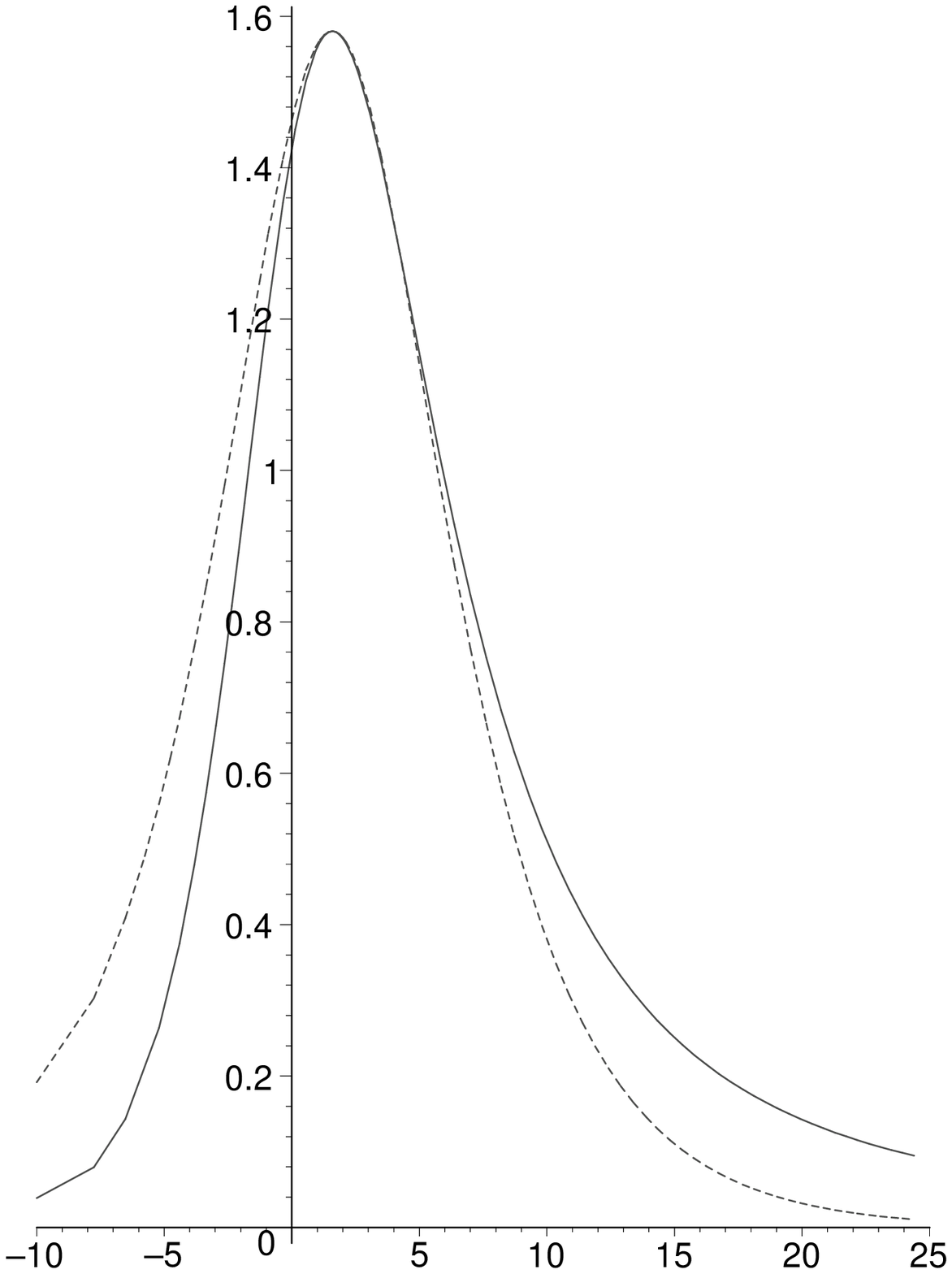}
\end{minipage}
\end{center}
\caption{A comparison of the Regge-Wheeler potential (solid) and the Poschl-Teller potential (dotted) when the matching conditions \eqref{eq:QNM_match} are used. These graphs are, from the left, for $\ell=0$, $1$ and $6$ respectively.}\label{fig:compare_RWPT}
\end{figure}

\paragraph{}
Before starting a discussion on surface gravity, we find some concrete results for the Schwarzschild black hole. The potential for a Schwarzschild black hole is (for a scalar field, cf \eqref{eq:scalar_potential})
\begin{equation}
V(r) = \left(1-\frac{2M}{r}\right)\left(\frac{2M}{r^3} + \frac{\ell(\ell +1)}{r^2}\right).\label{eq:RWpotential}
\end{equation}
We find the location of the maximum:
\begin{align}
\frac{\d V}{\d r} &= -\frac{2}{r^5}(\ell(\ell+1)r^2 + 3M(1-\ell(\ell+1))r -8M^2)=0\\
r_0 &= \frac{3[\ell(\ell+1)-1] + \sqrt{9[\ell(\ell+1) - 1]^2 + 32\ell(\ell+1)}}{2\ell(\ell+1)}M.
\end{align}
It is now simple to find $V_0 = V(r_0)$, where $r_0$ is the location of the
maximum; but such an expression is not illuminating.

\paragraph{}
To impose the second condition we need to investigate the near horizon behaviour of the potential. Introducing the tortoise coordinate (cf \eqref{eq:tortoise}):
\begin{align}
\frac{\d r_\star}{\d r} &= \frac{1}{1-2M/r}\\
r_\star &= r - 2M + 2M\ln(r-2M) = x + 2M\ln(x), \quad\quad x \equiv r-2M.
\end{align}
Near the horizon, $x$ is small. As $x \rightarrow 0$ ( or $r_\star \rightarrow -\infty$)
\begin{equation}
 x \sim \exp(r_\star/2M).
\end{equation}
From equation \eqref{eq:RWpotential} the near horizon form of the potential is found:
\begin{equation}
V(x) = \frac{x}{8M^3}\left(1-\frac{x}{2M}\right)\bigg(1-\frac{3x}{2M} + \ell(\ell +1)\left[1-\frac{x}{M}\right]\bigg) + O\left(\frac{x^4}{M^6}\right),
\end{equation}
or in terms of $r_\star$:
\begin{equation}
V(r_\star) = \frac{\exp(r_\star/2M)}{8M^3}\bigg(\ell^2 + \ell + 1 \bigg) + O\left(\frac{\exp(r_\star/M)}{M^3}\right).
\end{equation}
Hence the near horizon decay rate is $1/2M$, or more formally
\begin{equation}
\lim_{r_\star \rightarrow -\infty} \frac{\ln(V(r_\star))}{r_\star} = \frac{1}{2M}.
\end{equation}

\paragraph{}
The Poschl-Teller potential has exponential decay as one heads toward the horizon as well:
\begin{align}
\lim_{r_\star \rightarrow -\infty} \mathcal{V}(r_\star) &=
  \frac{V_0}{\exp(-2\alpha[r_\star - r_{\star}^{0}])/2}\\
& = \left[2V_0\exp(-2\alpha r_{\star}^{0})\right] \exp(2\alpha r_\star).
\end{align}
The decay rates will agree near the horizon if and only if
\begin{equation}
\alpha = \frac{1}{4M} = \kappa_h
\end{equation}
where $\kappa_h$ is the \emph{surface gravity} of a Schwarzschild black
hole.\footnote{To find the surface gravity for any stationary black hole see
  (for example) Wald \cite{book:Wald}.}

\subsubsection{Finding the QNM}
 Our attention is now turned to finding the QNM frequencies of the approximate Poschl-Teller potential. That is, we wish to find the QNM given
\begin{equation}
\frac{\d^2 R}{\d r_\star^2} + \omega^2 R = \frac{V_0}{\cosh^2\left(\alpha r_\star\right)}R,
\end{equation}
where the $r_\star$ axis has been translated so that $r_{\star}^{0}=0$. The exact solution is given in terms of hypergeometric functions:
\begin{align}
R(r) &=  C_1(2\cosh(2\alpha r_\star)+2)^{1/4+G/\alpha}\nonumber\\
&\quad\times {}_{2}F_{1}\left(\frac{-2i\omega + \alpha + 4G }{4\alpha},\frac{\alpha+2i\omega+4G}{4\alpha};1 + \frac{4G}{2\alpha};\frac{1}{2}\cosh(2\alpha r_\star) + \frac{1}{2}\right)\nonumber\\
&\quad+C_2(2\cosh(2\alpha r_\star)+2)^{1/4-G/\alpha}\nonumber\\
&\quad\times{}_{2}F_{1}\left(\frac{2i\omega + \alpha - 4G }{4\alpha},\frac{\alpha-2i\omega-4G}{4\alpha};1 - \frac{4G}{2\alpha};\frac{1}{2}\cosh(2\alpha r_\star) + \frac{1}{2}\right),\label{eq:QNM_PTsol}\\
G &= \frac{1}{4}\sqrt{\alpha^2 - 4V_0}.
\end{align}
We are wanting to look at the behaviour as $r_\star$ goes to infinity. The argument of the hypergeometric is dominated by the exponentially growing piece:
\begin{align}
&\lim_{r_\star\rightarrow \pm\infty}R(r) = \nonumber\\
& C_1\exp([\alpha+4G]|r_\star|/2) {}_{2}F_{1}\left(\frac{\alpha-2i\omega + 4G }{4\alpha},\frac{\alpha+2i\omega+4G}{4\alpha};1 + \frac{4G}{2\alpha};\frac{1}{4}e^{2\alpha |r_\star|} \right)\nonumber\\
&+C_2\exp([\alpha-4G]|r_\star|/2){}_{2}F_{1}\left(\frac{\alpha+  2i\omega  -4G }{4\alpha},\frac{\alpha-2i\omega-4G}{4\alpha}];1 - \frac{4G}{2\alpha};\frac{1}{4}e^{2\alpha |r_\star|}\right).
\end{align}
To express this in terms of exponential functions (and hence ingoing and outgoing waves) near infinity the following result (see equation (e.6) of \cite{book:QM_LandauLifshitz})
\begin{align}
{}_{2}F_{1}(a,b;c;z) &= \frac{\Gamma(c)\Gamma(b-a)}{\Gamma(b)\Gamma(c-a)}(-z)^{-a}{}_2F_{1}\left(a,a+1-c;a+1-b;\frac{1}{z}\right) \nonumber\\
& + \frac{\Gamma(c)\Gamma(a-b)}{\Gamma(a)\Gamma(c-b)}(-z)^{-b}{}_2F_{1}\left(b,b+1-c;b+1-a;\frac{1}{z}\right).
\end{align}
To find the asymptotic form at infinity the series expansion for the hypergeometric function is employed:
\begin{equation}
{}_2F_{1}(a,b;c;z) = 1 + \frac{ab}{c}\frac{z}{1!} + \frac{a(a+1)b(b+1)}{c(c+1)}\frac{z^2}{2!} + \ldots
\end{equation}
This allows us to replace the hypergeometric near infinity with unity.
\vfill
\pagebreak
The asymptotic form of \eqref{eq:QNM_PTsol} is then (up to an irrelevant constant factor):
\begin{align}
&\lim_{r\rightarrow \pm \infty}R(r) \propto C_1 \Gamma\left(1+\frac{\sqrt{\alpha^2-4V_0}}{2\alpha}\right)\nonumber\\
&\times\Bigg[\frac{\Gamma(i\omega/\alpha)}{\Gamma([\alpha + 2i\omega + \sqrt{\alpha^2-4V_0}]/4\alpha)\Gamma([3\alpha + \sqrt{\kappa^2 - 4V_0} + 2i\omega]/4\alpha)}e^{ i\omega |r_\star|}\nonumber\\
&+\frac{\Gamma(-i\omega/\alpha)}%
{\Gamma([\alpha + \sqrt{\alpha^2 - 4V_0} - 2i\omega]/4\alpha)
\Gamma([3\alpha + \sqrt{\alpha^2 - 4V_0} - 2i\omega]/4\alpha)}e^{-i\omega |r_\star|}\Bigg]\nonumber\\
& + C_2\Gamma\left(1-\frac{\sqrt{\alpha^2-4V_0}}{2\alpha}\right)\nonumber\\
&\times\Bigg[\frac{\Gamma(-i\omega/\alpha)}{\Gamma([\alpha - 2i\omega - \sqrt{\kappa^2 -4V_0}]/4\alpha)\Gamma([3\alpha - 2i\omega-\sqrt{\alpha^2 - 4V_0}]/4\alpha)}e^{-i\omega|r_\star|}\nonumber\\
&+\frac{\Gamma(i\omega/\alpha)}%
{\Gamma([\alpha + 2i\omega - \sqrt{\alpha^2 - 4V_0}]/4\alpha)\Gamma([3\alpha+2i\omega - \sqrt{\alpha^2 - 4V_0}]/4\alpha)}e^{i\omega|r_\star|}\Bigg].
\end{align}
The QNM boundary condition means that the incoming waves (the part proportional to $\exp(-i\omega|r_\star|)$) must vanish. This implies that
\begin{align}
&0=\Gamma\left(-\frac{i\omega}{\alpha}\right)\Bigg\{C_1 \Gamma\left(1+\frac{\sqrt{\alpha^2-4V_0}}{2\alpha}\right)\nonumber\\
&\quad\times\left[\Gamma\left(\frac{\alpha - 2i\omega + \sqrt{\alpha^2-4V_0}}{4\alpha}\right)\Gamma\left(\frac{3\alpha + \sqrt{\kappa^2 - 4V_0} - 2i\omega}{4\alpha}\right)\right]^{-1}\nonumber\\
&\quad+C_2\Gamma\left(1-\frac{\sqrt{\alpha^2-4V_0}}{2\alpha}\right)\times\nonumber\\
&\quad\left[
\Gamma\left(\frac{\alpha -2i\omega - \sqrt{\alpha^2 - 4V_0}}{4\alpha}\right)
\Gamma\left(\frac{3\alpha-2i\omega - \sqrt{\alpha^2 - 4V_0}}{4\alpha}\right)
\right]^{-1}\Bigg\}e^{- i\omega|r_\star|}.
\end{align}
It is well known that the gamma function has no zeros in the complex plane. Candidates for QNMs then come from the fact that the Gamma function diverges at the negative integers. Specifically, letting $n$ be a \emph{positive} integer, candidate QNM modes occur at
\begin{itemize}
\item $C_1 = 0$: Then we have one of the two remaining Gamma functions encountering a pole, i.e.:
\begin{subequations}
\begin{equation}
\alpha - 2i\omega - \sqrt{\alpha^2 -4V_0} = -4n\alpha\label{eq:PT_solve_QNM}
\end{equation}
or
\begin{equation}
3\alpha - 2i\omega - \sqrt{\alpha^2 - 4V_0} = -4n\alpha\label{eq:PT_solve_QNM_redundent}
\end{equation}
which implies that 
\begin{equation}
\omega_n = \sqrt{V_0 - \left(\frac{\alpha}{2}\right)^2} - i\left(n+\frac{1}{2}\right)\alpha.
\end{equation}
\item $C_2 = 0$: The only change is the sign of the square root changes.
\end{subequations}
\end{itemize} 
We can finally summarise the QNM for the Poschl-Teller potential:
\begin{equation}
\omega_n = (\pm)\sqrt{V_0 - \left(\frac{\alpha}{2}\right)^2} + i\left(n + \frac{1}{2}\right)\alpha\label{eq:actual_PT_QNM}.
\end{equation}

\subsection{Comparison of results}
In comparison, the results of Ferrari and Mashhoon \cite{Ferrari:1984} are given by ($\ell \gg 1$)
\begin{align}
\omega_n &\approx (\pm)\frac{\ell + 1/2}{3\sqrt{3} M} + i\frac{n+1/2}{3\sqrt{3}M}\\
&= \frac{\alpha}{3\sqrt{3}}\left(\pm2(2\ell + 1) + i\left(n+\frac{1}{2}\right)\right).
\end{align}
These results are presumably given for gravitational waves, although no such statement is made by the authors (electromagnetic QNM are independent of $\ell$, and the paper focuses on gravitational and electromagnetic QNM).

\paragraph{}
It should be noted that the type of field and the angular momentum $\ell$ determine the potential completely. In particular, they determine $V_0$, and it follows that the \emph{real} part of $\omega_n$ will depend on $\ell$, irrespective of the matching condition chosen.

\paragraph{}
The values of $n$ occur from the imposition of the boundary conditions on the Poschl--Teller potential. Hence the imaginary part will be evenly spaced if we match to the Poschl--Teller potential \emph{irrespective of the details of the matching conditions}. The value of the spacing between the levels does depend on the matching condition. More specifically
\begin{itemize}
\item Matching decay rate:
Real part depends on $\ell$ via $V_0$. The matching condition gives
\begin{equation}
\alpha = \kappa_h
\end{equation}
which is independent of $\ell$, hence the imaginary spacing is constant.
\item Ferrari and Mashhoon:
Now $\alpha$ depends on $\ell$ as well:
\begin{equation}
\alpha = \left(1-\frac{2M}{r_0}\right)\frac{1}{\sqrt{2V_0}}\sqrt{\left.\frac{\d^2 V}{\d r}\right|_{r=r_0}}.
\end{equation}
This condition is $\ell$ dependent, so the imaginary spacing would rely on both $n$ and $\ell$. The real part has $\ell$ dependence from both $V_0$ and $\alpha$.
\end{itemize}
It is worth emphasising again that the imaginary spacing of the real QNMs for a given $\ell$ does not have to be constant; this is a property of the approximate Poschl--Teller potential.

%Contains a table of computed QNM, and numerically found QNM
\begin{sidewaystable}[p]
\begin{center}
{\large $M\omega_n$ for a Schwarzschild black hole:}\\
 Numerical and analytic approximations
\begin{tabular}{|r|cccccc|c|}
\hline
\backslashbox{$n$}{$\ell$} 
&1&2&3&4&5&6&$\Im(M\omega_s)$\\
\hline
0 & ?+?$i$ & 0.374 + 0.089$i$ & 0.599+0.093$i$ & 0.809 + 0.094$i$ &?&?&0\\
1 & ?+?$i$ & 0.347 + 0.274$i$ & 0.583+0.281$i$ & 0.797 + 0.283$i$ &?&?&0.375\\
2 & ?+?$i$ & 0.301 + 0.478$i$ & 0.552+0.479$i$ & 0.772 + 0.480$i$ &?&?&0.625\\
3 & ?+?$i$ & 0.252 + 0.705$i$ & 0.512+0.690$i$ & 0.740 + 0.684$i$ &?&?&0.875\\
4 & ?+?$i$ & 0.208 + 0.947$i$ & 0.470+0.916$i$ & 0.702 + 0.898$i$ &?&?&1.125\\
5 & ?+?$i$ & 0.169 + 1.196$i$ & ?+?$i$ & ? &?&?&1.375\\
200 & ?+?$i$ & 0.061 + 49.860$i$ & 0.078 +  49.843$i$ & 
      0.085 + 49.837$i$&?& 0.241 + 49.397$i$&50.125\\
400 & ?+?$i$ & 0.056 + 99.864$i$ & 0.069 + 99.852$i$ & 
      0.083 + 99.837$i$ &?& 0.129 + 99.678$i$ &100.125\\
600 & ?+?$i$ & 0.054 + 149.866$i$ & 0.065 + 149.856$i$ &
      0.078 + 149.843$i$ &?&0.0360 + 149.548$i$&150.125\\
800 & ?+?$i$ & 0.052 + 199.867$i$ & 0.062 + 199.858$i$ & 
      0.074 + 199.847$i$ &?& 0.080 + 199.817$$&200.125\\
1000 & ?+?$i$ &0.051 + 249.866$i$ & 0.060 + 249.860$i$ & 
      0.072 + 249.850$i$ &?&0.086 + 249.824$i$ &250.125\\
3000 & ?+?$i$ &0.048 + 749.871$i$ & 0.053 + 749.866$i$ & 
      0.060 + 749.860$i$ &?&0.077 + 749.844$i$   &750.125\\
6000 & ?+?$i$ &0.0469 + 1499.872$i$ & 0.051 + 1499.868$i$ & 
      0.056 + 1499.864$i$ &?& 0.068 + 1499.852$i$&1500.125\\
10000  & ?+?$i$ &0.0462 + 2499.873$i$ &0.049+ 2499.870$i$ & 
      0.053 + 2499.867$i$ &?&0.063 + 2499.8571&2500.125\\
\hline
$\Re(M\omega_s)$& 0.060 & 0.368 &0.597 & 0.807 &1.011&1.211&\\
$\Re(M\omega_f)$& 0.108 & 0.378 & 0.602 & 0.811 &1.014&1.214&  \\
\hline
\end{tabular}
%\end{tiny}
\caption{The modes $n=0,1,2,3,4,5$ were obtained from Anderson and Glampedakis \cite{Glampedakis:2003dn}. The higher modes were taken from work kindly provided by Emanuele Berti (private communication). Here $\omega_s$ are the frequencies by imposing \eqref{eq:match_suneeta} (the ``Suneeta'' condition), while $\omega_f$ comes from imposing \eqref{eq:match_PT2F} (the ``Ferrari'' condition). Question marks denote frequencies that I have not found numerical results for. The real parts of both Poschl-Teller approximations are independent of $n$. The imaginary part of $\omega_s$ is independent of $\ell$ as it is related to the surface gravity; but as $\omega_f$ relies on the curvature of the ($\ell$ dependent) potential the $\Im(\omega_f)$ changes as $\ell$ does.}
\end{center}
\end{sidewaystable}

\begin{sidewaystable}[p]
\begin{center}
  {\large $M\omega_n$ for a Schwarzchild black hole:}\\
  Using the Mashhoon \& Ferrari matching condition
\begin{small}
\begin{tabular}{|r|cccccccc|}
\hline
\backslashbox{$\Im(\omega_{n,\ell})i$}{$\ell$} 
&1&2&3&4&5&6&100&500\\
\hline
$n=0$& 0.054$i$&0.189$i$&0.301$i$&0.406$i$&
  0.507$i$&0.607$i$&9.669$i$&48.160$i$\\
1& 0.163$i$&0.567$i$&0.904$i$&1.217$i$&
  1.521$i$&1.820$i$&29.008$i$&144.481$i$\\
2& 0.271$i$&0.946$i$&1.506$i$&2.029$i$&
  2.535$i$&3.034$i$&48.347$i$&240.802$i$\\ 
3& 0.379$i$&1.324$i$&2.108$i$&2.840$i$&
  3.549$i$&4.247$i$&67.686$i$&337.123\\ 
4& 0.488$i$&1.702$i$&2.711$i$&3.651$i$&
  4.564$i$&5.461$i$&87.025$i$&433.444$i$\\ 
5& 0.596$i$&2.081$i$&3.313$i$&4.463$i$&
  5.578$i$&6.674$i$&106.364$i$&529.764$i$\\
200& 21.738$i$&75.844$i$&120.785$i$&162.692$i$&
  203.330$i$&243.315$i$&3877$i$&19312$i$\\
400& 43.422$i$&151.498$i$&241.268$i$&324.978$i$&
  406.154$i$&486.023$i$&7745$i$&38576$i$\\ 
600& 65.105$i$&227.153$i$&361.751$i$&487.264$i$&
  608.977$i$&728.731$i$&11612$i$&57841$i$\\ 
800& 86.789$i$&302.808$i$&482.235$i$&649.551$i$&
  811.801$i$&971.440$i$&15481$i$&77105$i$\\ 
1000& 108.473$i$&378.462$i$&602.718$i$&811.837$i$&1014.624$i$&1214.148$i$&19349$i$&96369$i$\\ 
3000& 325.311$i$&1135.008$i$&1807.552$i$&2434.699$i$&3042.858$i$&3641.230$i$ &58026$i$&289011$i$\\ 
6000& 650.568$i$&2269.827$i$&3614.803$i$&4868.992$i$&6085.209$i$&7281.853$i$&116042$i$&577973$i$\\ 
10000& 1084.244$i$&3782.919$i$&6024.471$i$&8114.716$i$&10141.677$i$&12136.017$i$&193398$i$&963256$i$\\ 
\hline
$\Re(\omega_{n,\ell})$&0.108&0.378&0.602&0.811&1.014&1.214&19.339&96.321\\
\hline
\end{tabular}
\end{small}
\caption{This table shows the QNM calculated using Ferrari and Mashhoon's method. The real part depends only on $\ell$, and is displayed at the bottom of the table to prevent needless duplication. At constant $\ell$ the spacing between adjacent $n$ levels is constant.}
\end{center}
\end{sidewaystable}

\paragraph{}
The tabulation of the QNM mode frequencies show that, at least for low values of $\ell$, the decay rate of the potential allows much more accurate modelling than matching at the peak. It should be noted that the decay toward infinity is actually polynomial, as 
\begin{equation}
\lim_{r \rightarrow \infty}V(r) = \frac{\ell(\ell+1)}{r^2}
\end{equation}
independent of the ``spin''. At large $r$ there is little difference between $r$ and $r_\star$.

%\subsection{Schwarzschild--de Sitter black holes}

%\section{Monodromy method}

%\section{Leaver's method}

%\section{Conclusions}

\chapter{Near horizon conformal field theory}
\label{chap:conformal}

%\section{Introduction}
When entropy was first introduced into classical physics, it was as a state
function and bore no relation to ``microscopic degrees of freedom'' until the
pioneering work of Boltzmann. It is then perhaps not completely surprising that the
explanation and calculation for black hole entropy yields an answer but does
not enlighten us as to the \emph{microscopic} origins of black hole
entropy. What \emph{is} surprising is that the answer is definite; in classical
physics the entropy was determined only up to an additive constant, whereas the
prescription of the logarithm of the number of microstates yields an exact
value. However the calculations of black hole entropy do not yield any clue to
the nature, location or even existence of black hole microstates.\footnote{It
may be argued that the existence of entropy is proof of the number of
microstates. It is possible that black holes are fundamentally different from
other systems and possess entropy without microstates but this is a fairly
radical and desperate idea.}

\paragraph{}
The black hole temperature and entropy calculations rely on very basic physics,
and can easily be extended into other dimensions. This has led to the
hypothesis of \emph{universality}:\footnote{Universality is overused in the
context of black hole entropy. Carlip \cite{Carlip:1999db} uses universality to
mean that many different theories of quantum gravity reproduce these results
hence they are universal. Fursaev \cite{Fursaev:2004qz} calls a method
universal if it gives the correct result for any spacetime.} any explanation
of black hole entropy must also apply in any number of spacetime
dimensions. This may seem puzzling at first -- we only live in four spacetime
dimensions so the fact that the calculation may not work in other dimensions
seems largely irrelevant. Having the argument work in only four dimensions
could actually be seen as a positive sign if there was some fundamental reason
for it, as it may be an ``explanation'' for why the world is four
dimensional. The problem is that general relativity is written in the language
of differential geometry, a language that is very flexible about the
dimensionality and type of manifold that it is placed on. Consequently we can
take our semi-classical entropy and temperature calculations and apply them to
mathematically legitimate (but presumably unphysical) black holes. Given that
these \emph{same} semi-classical calculations provide the only justification
for us to look for microstates of black holes it would be odd if the unphysical spacetimes did not posses similar microstates; hence the universality
principle. It is only a hypothesis, and it may just be (however unlikely) that
quantum gravity links quantum field theory and general relativity in such a way
that the only consistent theory is in four dimensions;
hence in other dimensions the classical (and hence semi-classical)
approximations break down. Until evidence of such a claim can be provided (or
indeed suggested) by \emph{any} calculation, the universality principle seems
like a good guiding principle.

\paragraph{}
We also have a large number of quantum gravity candidates that are able to
reproduce the correct temperature and area of a black hole such as string
theory, D-branes, loop quantum gravity (LQG) and Sakharov induced gravity
models. In black holes with negative cosmological constant (anti-de Sitter
space or AdS) there is a correspondence with conformal field theories (CFTs)
that gives the correct entropy and temperature as well. An excellent and recent
review of calculations of black hole entropy that shall be used extensively
throughout this chapter is Fursaev \cite{Fursaev:2004qz}. None of these models
is entirely satisfactory. The LQG calculations were outlined in chapter
\ref{chap:QNM_intro}, where the microstates were \emph{spin-links} that
penetrated the horizon. String theory calculations are often done in a
spacetime that is dual to the original spacetime. As such the microstates are
not even located in the right space! It is not clear that the string theory
calculations can be carried out on a generic black hole; see Fursaev
\cite{Fursaev:2004qz} for details. The fact that the
entropy and temperature seems generic across many different theories seems to
suggest one of the three possible alternatives:
\begin{itemize}
\item It is \emph{just} a numerical coincidence. Any quantum gravity
calculation that does not give the Hawking temperature or Bekenstein-Hawking
entropy maybe discarded. It is just an odd curiosity that so many
theories remain.
\item These theories are not really different. They are the same underlying
theory written in a different way, and they are formally equivalent once
correct identifications are made.
\item Something fundamental is being assumed in all of the models. One such
suggestion is that a symmetry in the classical theory is being preserved under all
these quantisations making the temperature and entropy calculations generic.
\end{itemize}
To some degree the first idea must be correct, as it is possible to concoct
theories which do not give rise to the correct entropy or temperature relations
(e.g. the completely indistinguishable spin-links in Polychronakos
\cite{Polychronakos:2003}). To claim that there is no reason for the existence
of so many completely different theories that reproduce these results seems
highly unlikely. A literal interpretation of the second idea is demonstrably
false; among string theories alone many are known to be inequivalent!
In the spirit of M-theory it is possible that many of the
theories are different ways of writing one theory in different languages and
in different limits.

\paragraph{}
This chapter is written as part of an investigation into the third possibility,
that a \emph{near horizon classical conformal symmetry} is preserved on
quantisation and fixes the density of states and hence the entropy. This idea
started when Brown and Henneaux \cite{Brown:1986nw} noted in \emph{classical}
AdS$_3$ that the generators of the group of diffeomorphisms at large $r$ is
becomes (two copies of) the generators of CFTs, known as Virasoro
algebras. Brown and Henneaux where able to conclude that any theory of quantum
gravity on AdS$_3$ that was diffeomorphism invariant must be a conformal field
theory! In two dimensions the group of conformal transformations is infinite
dimensional (any homomorphic function is a conformal transformation) and the
conformal symmetry is a powerful constraint; it determines the density of
states up to a constant known as the central charge. In classical theories
\emph{without boundaries} the central charge vanishes, and hence it is normally
associated with quantum theories. However Brown and Henneaux
\cite{Brown:1986nw} showed that even classical field theories could acquire a
central change if they had a boundary. In the AdS$_3$ calculation this boundary
was taken to be the $t-\varphi$ plane at infinite radial distance. Strominger
\cite{Strominger:1998eq} showed that it was possible to obtain the
Hawking-Bekenstein entropy of a negative cosmological constant black hole from
the conformal field theories. In higher dimensional spaces the CFTs are finite
dimensional and do not fix the density of states, and it was unclear how to
perform direct generalisations. Even in three dimensions the CFT was unable to
attribute individual black holes with an entropy as the CFT correspondence only
applied at large $r$. I think that this problem is somewhat overstated as
entropy cannot be localised; entangled particles can fly arbitrarily far apart
in a ``pure state'' without increasing the entropy. When I allow one particle
to interact with a heat bath I change the entropy of the entangled
pair. Naturally one would like to be able to associate entropy with
(essentially) non-interacting subsystems of the universe, but if black holes
are essentially non-interacting then the asymptotic symmetry should be
``essentially'' obtained far enough away from each black hole that each one
could be assigned an entropy. It is not a forgone conclusion that it is
possible to associate entropy with individual black holes if they are strongly
interacting (such as in a binary orbit). A more serious objection is that the
method of using a CFT at infinity cannot distinguish between a spacetime that
contains a star and a spacetime that contains a black hole.

\paragraph{}
To have a CFT account for the black hole microstates in a generic spacetime
there must be a boundary, and a 2D surface on which the generators of
diffeomorphisms look like (multiple copies of) Virasoro algebras. In generic
black hole spacetimes one such candidate for the boundary is a
horizon\cite{Carlip:1998qw}. As the horizon has a dimensionality one lower than
the spacetime any boundary conditions specified on the horizon eliminate gauge
freedom there. The question of what type of horizon needs to be used is not
resolved, but for the stationary case where a Killing field is available a
Killing horizon is appropriate. In the stationary case the two plane spanned by
the normal to the (spacelike section of the) horizon and the Killing field
possess a 2D CFT as the horizon is
approached\cite{Carlip:1998qw,Solodukhin:1998tc,Carlip:1999db,Carlip:2002be}. As
the CFT is on the horizon, the microstates can be associated with an individual
black hole. It should be emphasised that the black hole horizon is not being
treated as a physical boundary, but as a convenient place to set boundary
conditions that ensure the existence of a black hole (and hence in the
stationary case a 2D CFT)\cite{Carlip:1998qw}. As a consequence the ``near
horizon'' CFTs can distinguish between stationary black holes (which posses 2D
CFTs) and stationary stars (which generically do not).

%\section{Computations of black hole entropy}
%An excellent review of different models of black hole entropy and different ways of looking at it has recently been given by Fursaev \cite{Fursaev:2004qz}
%\subsection{Modelling the microstates}
%Include 2D models and make like particle models. follows Fursaev \S 2.1

%\subsection{String theory computations}\label{sec:string}
%Fursaev \S 2.3

%\section{Conformal field theories}\label{sec:cft}
%\subsection{Introdunction to conformal field theories}
%\subsection{Central charge and Cardy's formula}
%\subsection{CFT and relation to black hole entropy}

\chapter{Near horizon stress tensors}\label{chap:nhst}

%\section{Types of stress tensor}

\section{Stationary metrics}\label{sec:stationary_metrics}
In this chapter we investigate \emph{stationary} metrics, meaning that there is an isometry of the spacetime, the orbits of which are timelike curves. We have two possibilities for a spacetime containing a black hole:
\begin{itemize}
\item The spacetime is \emph{static}; that is the timelike (Killing) vector
field is orthogonal to a family of spacelike surfaces of ``constant
time''. This can be shown to be equivalent to the statement that there exist coordinates
in which all mixed space/time metric
coefficients of the form $g_{t i}$ vanish ($i=1,2,3$). See for example Wald,
page 119 \cite{book:Wald}.
\item The spacetime is \emph{stationary} and axisymmetric. An axisymmetric spacetime is a spacetime with an isometry that has closed spacelike curves. For a spacetime to be both stationary and axisymmetric the action of the isometries must also commute.
\end{itemize}
There is actually a third case that is less often discussed:
\begin{itemize}
\item The spacetime is static and non-axisymmetric. For this to occur, an ergosphere would have to exist and it could not intersect the horizon
\end{itemize}
There are no known solutions of the third type, and good reasons to suppose that they do not exist. Hawking and Ellis \cite{Hawk74} point out that a ``Penrose process'' allows the existence of negative energy particles (as measured by an observer at infinity) as long as there is an ergosphere.\footnote{To see this, remember that if the four momentum of a particle is $k^a$ and the asymptotically timelike Killing vector is $\zeta^a$ then the energy of the particle measured by an observer at infinity is $E = -k^a \zeta_a$. When $\zeta^a$ becomes spacelike, $E$ may be less than zero. But $\zeta^a$ spacelike defines the position of the ergosphere, hence negative energy particles can only exist in the ergosphere.} Hawking and Ellis then argue that if the horizon and the ergosphere have no point in common then the negative energy particles cannot pass through the event horizon but must remain in the ergosphere. If it was possible to continue this process indefinitely we could mine an infinite amount of energy from the black hole. The ergosphere is constrained from spontaneously vanishing, as we require a place for the negative energy particles. The other possibility is that the event horizon and the ergosphere eventually touch, allowing particles to escape from the ergosphere into the singularity. Once there are no negative energy particles the ergosphere can be reduced to zero, as any negative energy particles can be forced into the black hole. 

\paragraph{}
The above arguments come from \S 9.3 of Hawking and Ellis, but do not completely preclude the idea that the ergosphere is disconnected from the event horizon. For example, the Penrose process could be limited by the gravitational effects of the numerous test particles fired in to extract energy! There is a stronger statement by Hajicek \cite{Hajicek77} that insists that the ergosphere and event horizon must coincide. Hajicek's result requires the mild technical assumption that the orbits of the timelike Killing field are complete.

\paragraph{}
Once we have discarded the possibility that the ergosphere and the event horizon do not intersect, we have to then show that a non-axisymmetric metric does not exist.

\section{The ADM decomposition}

\label{sec:ADM}
The \emph{Arnowitt--Deser--Misner (ADM) decomposition} \cite{ADM_orig} is the process of breaking a four--dimensional spacetime into a time dependent three--dimensional geometry. To do this the spacetime (or a region of the spacetime) is foliated with three dimensional spacelike surfaces, labelled by a continuous parameter $t$. The idea is that the surface $\Sigma_t$ represents the geometry of space at ``time'' $t$. There is no natural way of doing this even in special relativity, as even if one observer sees two spatially separated events as simultaneous other observers do not necessarily agree! The decomposition of spacetime into space and time (or more precisely, a series of spacelike surfaces labelled by a parameter $t$) is thus highly non-unique. For obvious reasons the ADM decomposition is also referred to as a $(3+1)$ decomposition. While there are many references available on the ADM decomposition, such as the living review article \cite{living_review_initial_data} or \cite{ADM_orig}, I have found an elementary treatment lacking. Thus the work in this section is meant to be clarifying and does not contain any new results.\footnote{One of the examiners (David Wiltshire) has pointed out some elementary resources that I had not found. He suggests the references \cite{constrained_hamiltonians, constrained_dynamics, Wiltshire_intro_quantum_cosmology}.}

\begin{figure}
\begin{center}
\includegraphics[width=0.7\textwidth]{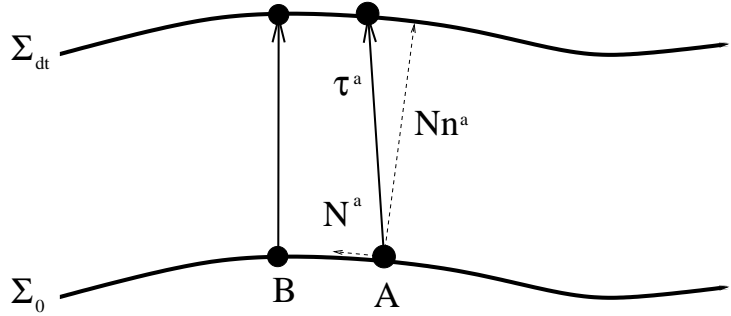}
\caption{The decomposition of $\tau^a$ into the shift and lapse. Picture inspired by Wald \cite{book:Wald}.}\label{fig:ADM_shiftlapse}
\end{center}
\end{figure}

\paragraph{}
What do points on different surfaces $\Sigma_{t_1}$ and $\Sigma_{t_2}$ have to do with one another? By introducing a timelike vector field $\tau^a$ representing the ``flow of time'' we are able to identify points on $\Sigma_{t_1}$ with $\Sigma_{t_2}$ by insisting that if two points are on the same orbit as $\tau^{a}$ then they are the same point (see figure \ref{fig:ADM_shiftlapse}). Instead of thinking of spacetime as a foliation of spacelike hypersurfaces we can instead consider space to be the surface $\Sigma_0$ with a time dependent metric. Notice that the vector field $\tau^a$ is highly non-unique, and different choices lead to different decompositions.

\paragraph{}
To actually carry out the ADM decomposition we need to make three (non-canonical) choices:
\begin{itemize}
\item An initial surface $\Sigma_0$, representing time zero.
\item A \emph{time function} $t: M \rightarrow \Re$ that assigns a time to each point in the spacetime. To be consistent with the interpretation of $\Sigma_0$ we require that $\forall p \in \Sigma_0: t(p) = 0$.
\item A flow of time $\tau^a$.
\end{itemize}
A change in any of these will lead to a different decomposition of the same spacetime. Notice that the choices are not all independent or arbitrary. For example, the time function $t$ is constrained so that the surfaces of constant $t$ (i.e. $\Sigma_t$) are spacelike. A function $t$ that satisfies this constraint is called a \emph{valid time function}. Normally the normalisation condition 
\begin{equation}
\tau^a \nabla_a t = 1 \label{eq:normalise_flow}
\end{equation}
is insisted upon. I shall not insist on it, as $\tau^a$ becomes null as the horizon is approached.

\begin{figure}
\begin{center}
\begin{minipage}{0.4\textwidth}
\includegraphics[width=\textwidth]{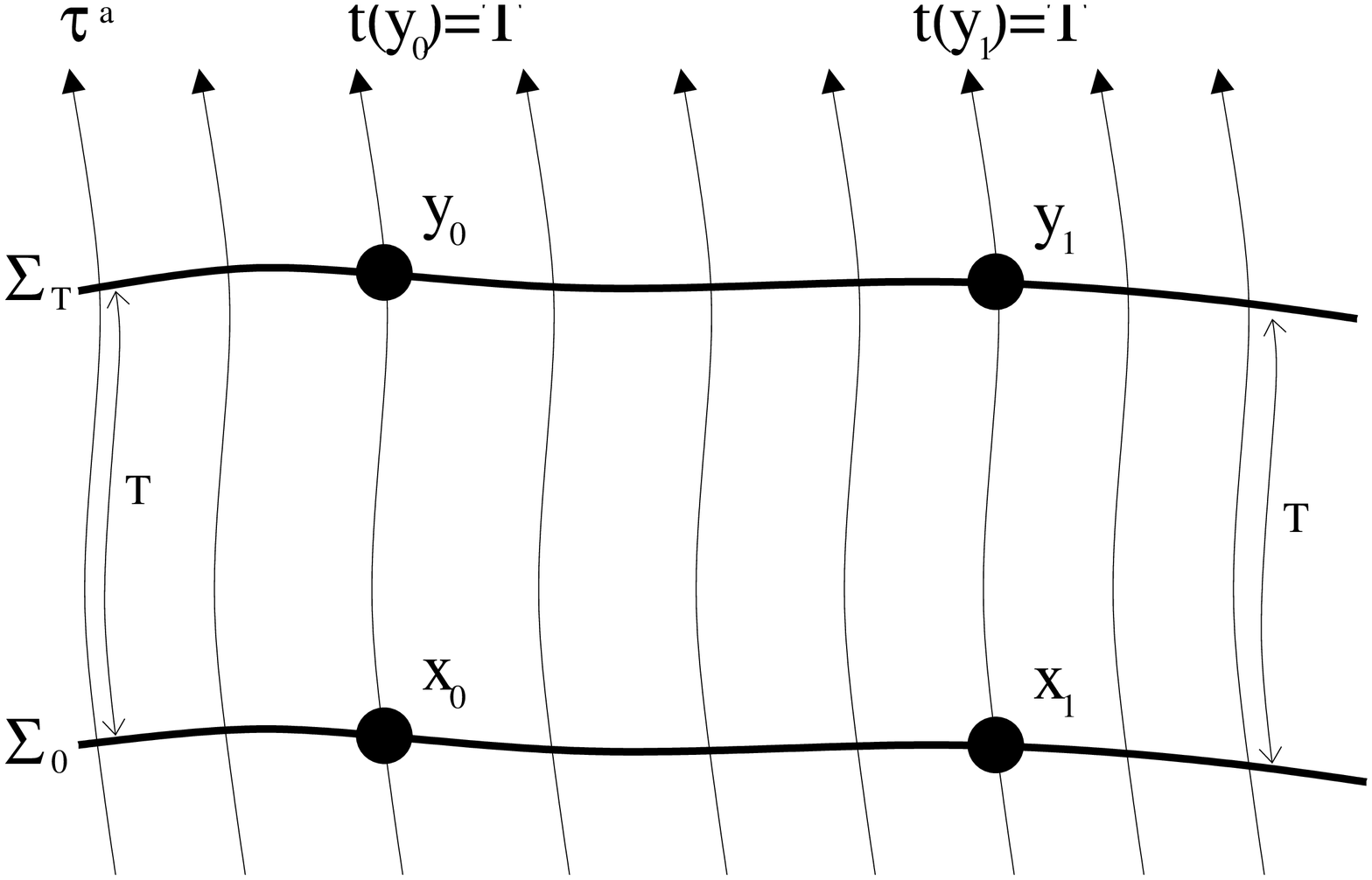}
\end{minipage}
\begin{minipage}{0.4\textwidth}
\includegraphics[width=\textwidth]{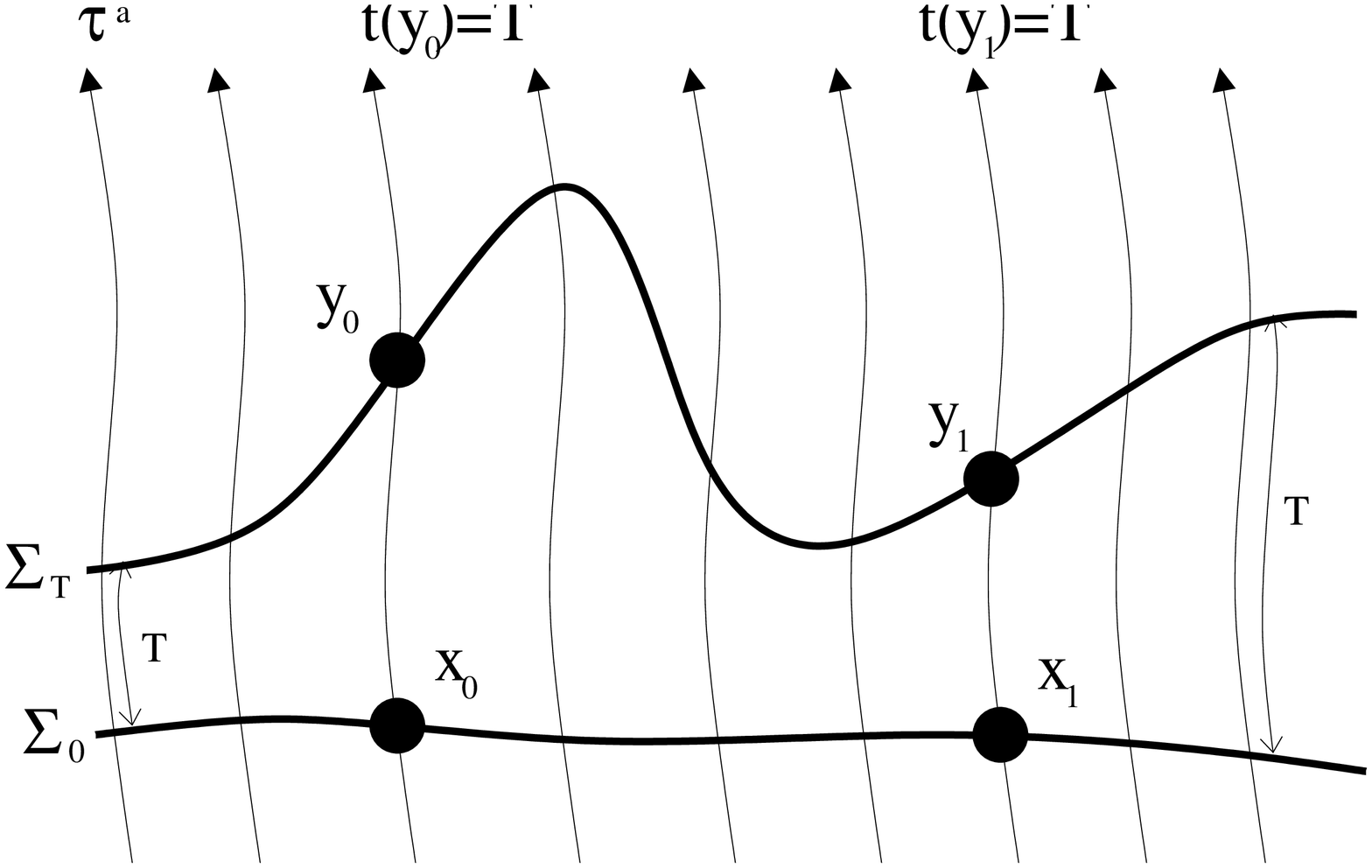}
\end{minipage}
\caption[ADM decomposition]{Two figures showing the dragging of the points of $\Sigma_0$ along integral curves of the flow $\tau^a$. The time function $t$ is different on the LHS and RHS. The level surface $t=T$ (i.e. $\Sigma_T$) on the LHS is spacelike and well-behaved, if this occurs for all $T$ then $t$ is a valid time function. On the RHS the surface $\Sigma_T$ is not spacelike so $t$ is not a valid time function.}\label{fig:ADM}
\end{center}
\end{figure}

\paragraph{}
Once we have defined the time function $t$ we have a future pointing timelike vector field, $\nabla t$. Note that there is no \emph{a priori} reason to pick $\tau^a$ so that it points in the same direction as $\nabla t$. In general, we can decompose the flow $\tau^a$ into a timelike normal part and a spacelike part. First, we introduce a timelike normal
\begin{equation}
n^a = g^{ab} \frac{\nabla_b t}{\sqrt{\left|g^{cd}(\nabla_c t)(\nabla_d t)\right|}}
\end{equation}
and then decompose the flow of time in the following manner:
\begin{equation}
\tau^a = N n^a + N^a \label{eq:decompose_flow}
\end{equation}
where $N$ is a scalar function called the \emph{lapse} and $N^a$ is a spacelike vector called the shift. In an intuitive sense $N^a$ lies in $\Sigma_t$. Of course, this cannot be taken literally as $n^a$ and $N^a$ both lie in the tangent space of a point in $M$, not in $M$ itself which is where $\Sigma_t$ is defined! However, this intuitive idea is captured by the orthogonality requirement of the normal and the shift:
\begin{equation}
N^a n_a = 0.
\end{equation}
Figure \ref{fig:ADM_shiftlapse} shows this decomposition. Getting the lapse is easy, as one only needs to find the component of $\tau^a$ along $n^a$:
\begin{equation}
N = - \tau^a n_a,
\end{equation}
or by using the normalisation condition \eqref{eq:normalise_flow}
\begin{equation}
N =  +\frac{1}{\sqrt{\left|g^{cd}(\nabla_c t)(\nabla_d t)\right|}}\label{eq:interpret_lapse}.
\end{equation}
The shift can be obtained by a rearrangement of \eqref{eq:decompose_flow}. However, it is nicer to introduce a projector that will allow us to find the component of an arbitrary vector $s^a$ orthogonal to $n^a$. First we note that the component of $s^a$ parallel to $n^a$ is
\begin{subequations}
\begin{equation}
s^a_{\textrm{along n}} = -n_a s^a
\end{equation}
and hence the orthogonal component is
\begin{equation}
s^a_{\bot \textrm{to n}} = s^a - s^a_{\textrm{along n}} n^a = s^a + n_b s^b n^a.
\end{equation}
Finally we note the identity
\begin{equation}
s^a = g^{a}{}_{b} s^b
\end{equation}
and hence
\begin{equation}
s^a_{\bot \textrm{to n}} = \left(g^{a}{}_b + n^a n_b \right) s^b.
\end{equation}
The quantity in brackets projects an arbitrary vector to the spacelike surface orthogonal to $n^a$, so we define
\begin{equation}
h^{a}{}_b = g^a{}_b +n^a n_b.
\end{equation}
\end{subequations}
In particular, we can apply the projector to the flow vector field to obtain the shift:
\begin{equation}
N^a = h^{ab} \tau_b.
\end{equation}

\paragraph{}
Now we give some insight into the meaning of the shift, lapse and projector. Let us consider a timelike observer moving along an integral curve of $n^a$. The distance measured by such an observer is given by 
\begin{equation}
\d s^2 = -N^2 \d t^2.
\end{equation}
Using $\lambda$ as the proper time of such an observer we have
\begin{equation}
\d t = \frac{\d \lambda}{N}.
\end{equation}
Thus a small lapse means that a observers travelling along integral curves of $n^a$ (``normal observers'') experience a small ``lapse'' in time between two constant time surfaces. For example, if $t$ is the Killing time coordinate it takes an infinite amount of Killing time to cross an event horizon, while the proper time of a ``stationary'' observer slows as the horizon is approached\footnote{I really want an apparent horizon here, rather than an event horizon. But as I have used a Killing time I have assumed the spacetime is stationary and hence the apparent horizon and event horizons coincide.}. When using Killing time as the time function we then when an event horizon exists we have $N=0$. 

\paragraph{}
\begin{minipage}{0.9\textwidth}
Note: \textit{There is an error in the paper reproduced in Appendix \ref{app:conformal2} and the published version. We strongly imply that the statements ``the horizon'', ``$N=0$'' and ``infinite red-shift surface'' can be used interchangeably. This is not the case: the ergosphere is an infinite redshift surface (with respect to Killing time) but has no special causal properties.}
\end{minipage}

\paragraph{}
What is the interpretation for the shift? Let $(t=x^0,x^1,x^2,x^3)$ be the coordinates in a patch of the four dimensional manifold $M$. Let $(y^1,y^2,y^3)$ be the three coordinates on a patch of $\Sigma_t$. Now consider $\Sigma_t$ and $\Sigma_{t + \d t}$. By construction, we are identifying the point $y\in \Sigma_t$ with the point in $\Sigma_{t + \d t}$ if they lie on the same integral curve of $\tau^a$. 
In fact, we can see the stronger statement
\begin{equation}
\d y^a = \d x^a - N^a\d t.
\end{equation}
With a coordinate change, we can find the metric in terms of the coordinates $(t = y^0,y^1,y^2,y^3)$. We have
\begin{equation}
\d s^2 = g_{ab}\, \d x^a\,\d x^b = g_{ab}\, \d y^a \,\d y^b
\end{equation}
or, in terms of the $y$ coordinate system
\begin{equation}
\d s^2  = g_{tt}\, \d t^2 + 2 g_{ti} \d t\,(\d y^i + N^i \d t) + g_{ij} (\d y^i + N^i \d t)(\d y^j + N^j \d t)
\end{equation}
where $i$ and $j$ run over spacelike indices only. But the second term is zero, as the $t$ direction is orthogonal to the spacelike directions by construction. The last term has sums over $\d y^a$ and $N^a$, both of which are orthogonal to $n^a$. Thus there is no different between using $g_{ij}$ or $h_{ij}$ and we finally obtain
\begin{align}
\d s^2 &= g_{tt} \,\d t^2 + h_{ij} (\d y^i + N^i \d t)(\d y^j + N^j \d t)\\
&= \left(g_{tt} + h_{ij} N^i N^j\right) \d t^2 + 2 (h_{ij} N^i)\d y^i\;\d t + h_{ij} \d y^i \;\d y^j. 
\end{align}
Thus we see that the projector $h_{ab}$ is really the time dependent metric on the three dimensional geometry $\Sigma_t$. To identify the lapse we look again at the normal observers
\begin{equation}
\d s^2 = -N^2 \d t^2 = \left(g_{tt} + h_{ij} N^i N^j\right) \d t^2
\end{equation}
or that 
\begin{equation}
g_{tt} = -(N^2 + h_{ij}N^iN^j).
\end{equation}
Hence we can write the metric in the terms of the shift and lapse by
\begin{equation}
g_{ab} = \left(\begin{array}{c|ccc}
-(N^2 + h_{ij}N^i N^j) & &N_i& \\
\hline
&&&\\
N_j& & h_{ij} & \\
&&&\\ 
\end{array}
\right)\label{eq:ADM_metric_form}.
\end{equation}

\subsection{Modified ADM for stationary spacetimes with horizons}
A glance at eq. (G.10) would lead one to think naively that the lapse is given by $N$ and the shift by $\omega\; \d \varphi$, based on a simple comparison with the form of the metric in  \eqref{eq:ADM_metric_form}. It is true, but quite difficult to prove, that $N=0$ is the horizon. The difficulty stems from showing that the geodesics are trapped, rather than just having a perverse coordinate system that ``shifts'' close to the shifts at the speed of light. 

\paragraph{}
To see this, we note that we cannot use the Killing vector $\zeta^a$ to generate time translations as $\zeta$ is spacelike in the vicinity of a rotating black holes horizon. In fact, this is the definition of the ergosphere! As the vector field $\zeta^a$ is not timelike this also implies that I cannot use the Killing time $t$ as the parameter labelling different surfaces. To distinguish the Killing coordinate $t$ and the time function I shall denote the latter by $\mathcal{T}$. There is a Killing vector field that is null on the horizon and timelike in the vicinity of the horizon, formed by a linear combination of the time translation Killing vector field $\zeta^a$ and the axisymmetric Killing field $\psi^a$:
\begin{equation}
\chi^a = \zeta^a + \Omega_h \psi^a.
\end{equation}
As a linear combination of Killing vector fields, $\chi^a$ is also a Killing vector field. It is this Killing field that generates the horizon. The constant $\Omega_h$ is the rotational rate of the black hole, defined by taking the locally measured rotation
\begin{equation}
\omega = -\frac{g_{t\varphi}}{g_{\varphi\varphi}}
\end{equation}
in the limit as the horizon is approached. The result that the rate of rotation is constant along the horizon is the \emph{rigidity theorem} due to Carter \cite{Carter_cent_survey}. 

\paragraph{}
We could use $\chi^a$ as our flow of time vector field, but that is not timelike at large distances. Instead, we use 
\begin{equation}
\tau^a = \zeta^a + \omega\;\psi^a.
\end{equation}
Notice that $\tau^a$ is \emph{not} a Killing vector field. As $\omega$ is the locally measured rate of rotation,  it does have the physically nice property of identifying a point on one time slice with a point on another time slice that has ``drifted'' along with the space. We also choose the time function so that $\tau = \nabla\mathcal{T}$. Hence the flow and the normal are pointing in the same direction and the shift vanishes. The lapse is given by
\begin{align}
-N^2 &= \tau^a \tau_a = \zeta^a \zeta_a + \omega\;(\zeta^a \psi_a + \psi^a \zeta_a) + \omega^2 \psi^a \psi_a\\
&= g_{tt} + 2\omega\;g_{t\varphi} + \omega^2 g_{\varphi\varphi}\\
&= g_{tt} - 2\frac{g_{t\varphi}^2}{g_{\varphi\varphi}} + \frac{g_{t\varphi}^2}{g_{\varphi\varphi}}= g_{tt} - \frac{g_{t\varphi}^2}{g_{\varphi\varphi}}\\
&= g_{tt} - \omega^2\;g_{\varphi\varphi}.
\end{align}

\section{Near horizon geometry for stationary Killing horizons}
\subsection{Metric for stationary non-static Killing horizons}\label{sec:time_reversal}
In this section we finally tie together the loose threads developed in previous sections. First we use the assumption of an stationary (but not static) metric with a horizon; then from \S \ref{sec:stationary_metrics} we find the spacetime is also axisymmetric. Hence we may introduce coordinates $(t,\varphi,x_1,x_2)$ so that the metric functions $g_{\mu\nu}$ are independent of $t$ and $\varphi$. Then the condition that the spacetime is \emph{time reversible} is imposed; that is the spacetime is invariant under the simultaneous changes
\begin{equation}
t \rightarrow -t, \quad\quad\quad \varphi \rightarrow -\varphi.
\end{equation}
As the metric components themselves are independent of $t$ and $\varphi$, but the \emph{intervals} are not, distances will be preserved iff
\begin{equation}
g_{x_{i}t} = g_{x_{i}\varphi} = 0, \quad\quad i = 1,2.
\end{equation}
Conditions that ensure the existence of a time reversal symmetry are discussed in \S \ref{sec:time_reversal}. Equivalently, time reversal symmetry can be assumed as a ``physically reasonable condition''. Together these conditions ensure that the metric can be taken locally as 
\begin{equation}
g_{\mu\nu} = \left(\begin{array}{cc|cc}
g_{tt}(x_1,x_2)& g_{t\varphi}(x_1,x_2)& 0 &0\\
g_{\varphi t}(x_1,x_2)&g_{\varphi\varphi}(x_1,x_2)& 0&0\\\hline
0&0& g_{x_1 x_1}(x_1,x_2)& g_{x_1 x_2}(x_1,x_2)\\
0&0& g_{x_2 x_1}(x_1,x_2)& g_{x_2 x_2}(x_1,x_2)
\end{array}\right)_{\mu\nu} \label{eq:2by2_decomposed_metric}.
\end{equation}

\paragraph{}
As the spacetime decomposes into the direct sum of two 2-$d$ manifolds, an index convention that respects this is now adopted. Greek indices $\alpha,\beta,\ldots$ will represent a general spacetime index; that is it may take values from 0 to 3 representing the coordinates $(t,\varphi,x_1,x_2)$. Upper case latin indices $A,B,\ldots$ take the values 0 or 1 and represent the $t-\varphi$ submanifold. Lower case latin indices $a,b,\ldots$ take the value 2 or 3 and represent the coordinates $x_1$ or $x_2$ respectively. This convention is \emph{not} used in earlier parts of this thesis. Such a convention allows us to write \eqref{eq:2by2_decomposed_metric} in the much more compact form
\begin{equation}
g_{\mu\nu} = \left(\begin{array}{c|c}
g_{AB}(x_1,x_2)&0\\\hline
0&g_{ab}(x_1,x_2)
\end{array}\right)_{\mu\nu}.
\end{equation}

\paragraph{}
The ADM-like decomposition described at the end of \S\ref{sec:ADM} is performed on the $t-\varphi$ submanifold. That is, 
\begin{align}
g_{tt} &= -\left(N^2 - \frac{g_{t\varphi}^2}{g_{\varphi\varphi}}\right) = -\left(N^2 - \omega^2 g_{\varphi\varphi}\right)\\
\omega &\equiv -\frac{g_{t \varphi}}{g_{\varphi\varphi}}
\end{align}
where the horizon is located at $N=0$. The parameter $\omega$ defines how much the horizon ``rotates'' by as one travels along the orbits of $\chi^a$, the vector field that is null on the horizon. We have the earlier result that $\omega$ on the horizon is $\Omega_h$.

\paragraph{}
So far no use has been made of the coordinate freedom in $x_1$ and $x_2$. Near the horizon we can introduce a \emph{normal coordinate system}. The coordinate $n$ measures the distance normal from the horizon, with the horizon occurring at $n=0$. The coordinate $z$ is a spacelike vector that lies on the horizon. The decomposition is then
\begin{equation}
g_{ab} = \left(\begin{array}{cc}
1&0\\
0&g_{zz}
\end{array}\right)_{ab}.
\end{equation}
Notice that the (now overused) word normal in this context means a spacelike vector that lies in the $x_1$--$x_2$ plane that is orthogonal to the horizon. For example, in the Schwarzschild geometry the radial coordinate $r$ would be a normal coordinate in this sense of the word. The full metric can now be given
\begin{equation}
g_{\mu\nu} = \left(\begin{array}{cc|cc}
-(N^2 - \omega^2 g_{\varphi\varphi})& -\omega g_{\varphi\varphi}&0&0\\
-\omega g_{\varphi \varphi} & g_{\varphi\varphi}&0&0\\\hline
0&0&1&0\\
0&0&0&g_{zz}
\end{array}\right)_{\mu\nu}.\label{eq:final_metric_form}
\end{equation}
Note that this coordinate system only exists locally, it will fail as soon as two of the geodesics leaving the horizon in the $n$ direction intersect. As we are only interested in the near horizon results, the local construction is good enough.

\subsection{Static metrics}
If the metric is static then we no longer have the result that it is also axisymmetric. Physically this is because there is no rotation, and hence no tidal friction. As we know there that the Killing vector field is hypersurface orthogonal we have the result
\begin{equation}
g_{\mu\nu} = \left(\begin{array}{c|ccc}
-N^2 & 0 & 0 & 0\\
\hline 
0& & & \\
0& &h_{ij} & \\
0& & & 
\end{array}\right)_{\mu \nu}.
\end{equation}
In the neighbourhood of any point on the horizon we may decompose the coordinates into an on horizon part $(\varphi,z)$ and a normal coordinate $n$. The metric then takes the form
\begin{align}
&
\hspace{1.3cm}\begin{array}{c@{\hspace{0.6cm}}c@{\hspace{0.6cm}}c@{\hspace{0.6cm}}c}
t&n& \varphi & z\\
\downarrow&\downarrow&\downarrow&\downarrow
\end{array} \nonumber\\
g_{\mu\nu} &= \left(
\begin{array}{cc|cc}
-N^2 & 0 & 0 & 0\\
0 & 1 & 0 & 0\\ \hline
0 & 0 & g_{\varphi\varphi} & g_{z\varphi}\\
0 & 0 & g_{\varphi z} & g_{zz}
\end{array}\right)_{\mu\nu}
\begin{array}{l}
\leftarrow t\\
\leftarrow n\\
\leftarrow \varphi\\
\leftarrow z.\label{eq:static_metric_final form}
\end{array}
\end{align}
This exhausts the four coordinate choices we have available (choosing Killing time $t$, picking the normal $n$, and choosing $z$ and $\varphi$ so that $g_{nz} = g_{n\varphi} = 0$). Without the introduction of more symmetries no further reduction is possible.

\section{Out, damned singularity!}

Once we have adopted appropriate coordinates (either for a static or stationary non-static spacetime) near the horizon, we can insist that the curvature scalars remain finite as the horizon is approached. Specifically, we require that
\begin{itemize}
\item $R^2$
\item $R^{ab}R_{ab}$
\item and $R_{abcd}R^{abcd}$
\end{itemize}
all remain finite as the horizon is approached. (Any linear combination with constant coefficients will also suffice, as a computational matter it is sometimes more convenient to use the traceless versions of the Ricci tensor and Riemann tensor). If any of these scalars becomes infinite as the horizon is approached then a genuine curvature singularity exists, as a scalar must be independent of the coordinates used. It is possible to have a directional curvature singularity, where any tetrad will have directions distorted as the singularity is approached. However, it can happen that the curvature in two or more independent directions can contribute opposite amounts to the curvature scalars, rendering the limit finite. Our finiteness condition is thus necessary but it is not sufficient.

\paragraph{}
In order to investigate the near-horizon geometry we take $n=0$ as the horizon and expand the lapse $N$ and the rotation rate $\omega$:
\begin{align}
N(n,z) &= \kappa_h(z) + \kappa_1(z) n +\frac{1}{2!}\kappa_2(z) n^2 +\ldots\\
\omega(n,z) &= \Omega_h + \omega_1(z) + \frac{1}{2!}\omega_2(z) n^2 + \ldots
\end{align}
Naturally, $\omega$ is trivial in the case of static black holes! We have used the additional information that $n=0$, the horizon, corresponds to $N=0$ and that the rigidity theorem implies that $\omega(n=0)=\Omega_h$. 

\paragraph{}
The rest of the details are simply a case of calculating the Ricci scalar, Ricci tensor squared and the Weyl tensor squared. I summarise the case for the stationary non-static case below. The static case is similar, and in both cases more details can be found in the appendices

\subsection{The curvature invariants for stationary non-static horizons}

\subsubsection{The Ricci Scalar}
The Ricci scalar for a stationary non-static black hole at the horizon is given
by
\begin{equation}
R =
\frac{g_{\varphi\varphi}(0,z)}{2\kappa_h(z)^2}\frac{
\omega_1(z)^2 }{n^2} + \mathcal{O}\left(\frac{1}{n}\right).
\end{equation}
Assume:
\begin{itemize}
\item Non-extremal black hole: $\kappa_h(z) \neq 0$.
\item The horizon geometry is regular: $g_{\varphi\varphi} \neq 0$ and is
finite
\end{itemize}
Hence we must have $\omega_1(z) \equiv 0$ so that $R$ remains finite as
$n\rightarrow 0$. 

\subsubsection{The traceless Ricci tensor}

After using the fact that $\omega_1(z) = 0$ we find:
\begin{align}
R_{ab}R^{ab} - \frac{1}{4}R^2 &= \bigg\{\left(\left.\frac{\d \ln g_{zz}}{\d
n}\right|_{n=0}\right)^2 + \left(\left.\frac{\d \ln g_{\varphi\varphi}}{\d
n}\right|_{n=0}\right)^2\nonumber\\
&\hspace{-4cm} + \left(\frac{4 \kappa_2(z)}{\kappa_h(z)}\right)^2 +
 \frac{8}{g_{zz}(0,z)\kappa_h(z)}\left(\left.\frac{\d \kappa_h}{\d
z}\right|_{n=0}\right)^2\bigg\}\frac{1}{n^2}+\mathcal{O}\left(\frac{1}{n}\right).
\end{align}
As this is a sum of squares, we may conclude
\begin{itemize}
\item An alternative derivation of the \emph{zeroth} law: the surface
$\kappa_h$ is constant.
\item The next term in the expansion of the lapse, $\kappa_2(z)$, vanishes.
\item Both $\partial_n g_{zz}$ and $\partial_n g_{\varphi\varphi}$ vanish on
 the horizon.
\end{itemize}

\subsubsection{The other constraints}

We have yet to apply the constraint that the Weyl tensor squared yields a
finite result, or use the $n^{-1}$ terms in the Ricci scalar and square of the
traceless Ricci tensor. The six constraints found above render the square of
the Weyl tensor finite at the horizon upon substitution, and also ensure the
$n^{-1}$ pieces of the Ricci scalar and traceless Ricci tensor vanish as well.

\section{Near horizon geometry}
In all the stationary black hole cases, when we construct a tetrad from $(\chi, n, \varphi,\zeta)$ we have the result that the Einstein block diagonalises
\begin{equation}
G_{\hat{a}\hat{b}} = \left(\begin{array}{cc|c}
G_{\hat{\chi}\hat{\chi}} & 0 & 0 \\
0& -G_{\hat{\chi}\hat{\chi}}&0 \\
\hline
0 & 0 & G_{\bot}
\end{array}\right)
\begin{array}{l}
\leftarrow \chi\\
\leftarrow n\\
\leftarrow (\varphi,z)
\end{array}
\end{equation}
The exact form of the different components depends if you are looking at the static case or the stationary non-static case. However, the general form of the stress tensor remains in tact. It is at this point, and this point only, that we can invoke the Einstein equations to tell us that a Killing horizon requires that
\begin{equation}
\rho = - p_n
\end{equation}
if $\rho$ is the energy density in the timelike direction $\hat{\chi}$. In particular, this conclusion is reached without any consideration for the energy conditions.

\chapter{Conclusions}

With a range of problems so diverse, it is hard to think of a concise way of summarising all of them. Indeed, to a large extent I already have summarised them at the end of each section. Instead of explaining the problems again and how far through completion they are, I would like to leave with a list of things that I believe are easily do-able but have not yet been done. I am afraid that to the reader that this may read as a checklist of things to do, and that is indeed how I intend to use it. But after providing a 250+ page thesis, I hope that the reader will allow me this indulgence.

\subsection*{Algorithmic constructions}
\begin{itemize}
\item I believe that looking at spherically symmetric but anisotropic stars is going to be a hard problem, and likely one of little physical relevance. While I believe that very compact objects such as quark stars or neutron stars are probably going to be \emph{very} anisotropic, especially at the surface where crusts are expected to form, the assumption that they are static will almost certainly be false. Even a very slowly spinning star will have a large angular momentum, resulting in a very quickly rotating neutron star that has formed as the result of gravitational collapse.
\item Finding a bound on the compactness for isotropic stars that also obeys the DEC looks like a simple problem. The goal would be to find a family of stars that get arbitrarily close to saturating the $8/9$ths bound, or to improve the bound and saturate that.
\item Improving the variational changes method in the gravity profile to allow for arbitrarily large changes in the profile, so that more comparison bounds can be done. As can be seen from appendix \ref{app:bounds_paper}, many bounds were obtained by comparison with the Schwarzschild solution.
\item The problems of ensuring the positivity of integrals, such as the expression for $\rho(r)$, seem too difficult. A more modest problem of finding large classes of functions $g(r)$ that ensure the energy density is positive everywhere may, however, be achievable.
\end{itemize}

\subsection*{Quasinormal modes}
This section is somewhat depressing for two reasons. The first is that the motivation for looking at highly damped QNMs does not exist anymore, as the recent paper by Domagala and Lewandowski showed that the QNM conjecture was almost certainly false \cite{Domagala:2004}. The other problem is that everything is very messy; one may consider that the adjective ``dirty'' could refer to finding the potentials as well as or better than to the spacetime! Finding an approach that is more intuitive or concrete would make me feel better than just going through screeds of algebra. This may, of course, be impossible. After all, the nice geometric results tend to span classes of spacetimes and tend to be existence theorems [for example, singularity theorems]. Here we really are interested in numerical answers for QNM, not existence answers to existence questions such as ``are there an infinite number of QNMs for every black hole spacetime''? And as a numerical question, there is some expectation that the answer will come about from arduous working for different spacetimes. There is one avenue that seems worthwhile pursuing, namely
\begin{itemize}
\item Can we express the problem in a clean and separable way in the Newman-Penrose formalism? If so, can we assign a similar interpretation as in the Schwarzschild case?
\end{itemize}
The NP formalism allows separation of  Reissner--Nordstr\"om, Kerr, Kerr-Newman and of course Schwarzschild. It is easier to see the relationship between the spectrum of the axial and polar cases when in the NP formalism, and indeed it provides a clear way to show that the two cases are isospectral in these examples. As we know that general dirty black hole spacetimes are not isospectral, we can ask the related question
\begin{itemize}
\item Is the utility of the Newman--Penrose formalism only significant if the spectra under consideration are isospectral?
\end{itemize}
A slightly different, but related question, would be
\begin{itemize}
\item How useful is the NP formalism in a situation where the standard massless fields are isospectral (as, excluding the scalar field, massless fields only possess two polarisations) at describing a \emph{massive} field? Do all polarisations occur independently? Are all polarisations isospectral?
\end{itemize}
It would also be interesting and feasible, although algebraically tedious, to extend the potentials for the standard fields to the stationary dirty black holes used in the conformal symmetry section rather than just static and dirty black holes.

\paragraph{}
Although it is not something that I plan to spend much time on, I would be interested in at least seeing if the claims about the real part of highly damped QNM changes now that the QNM conjecture is almost certainly false. Do these analytical proofs hold up to closer scrutiny, or are physicists just as prone to the bandwagon effect as sociologists\footnote{ Physicist Alan Sokal managed to get a paper entitled ``Transgressing the boundaries: Toward a Transformative Hermeneutics of Quantum Gravity'' published  in \emph{Social Text} \#46/47 pp. 217--252 (1996). It is still available online from \emph{http://www.physics.nyu.edu/faculty/sokal}.}?

\subsection*{Conformal symmetry}
There are two things that I would like to work through in more detail when it comes to the work that we have done on conformal transformations. The first is to see if I can find a different coordinate system in which both the central charge $c$ and the eigenvalue of $L_0$ remain finite as the horizon is approached. This would suggest that it is possible to have a well behaved CFT on the horizon, rather than having a theory which one is only allowed to answer certain questions for which well defined limits exist. To do this I would want to have a better understanding of the role that a central charge plays. Steve Carlip has suggested looking at a $(1+1)$ Dilaton black hole and looking at null sheets to see if a well defined central charge can be obtained in this toy model.

\paragraph{}
The other thing that would be interesting does not relate to the CFT, but rather simply the form of the geometry near the horizon. We have a null surface, and by hypothesis of stationarity the fields on that surface cannot evolve in time. One can imagine that even if this null surface was not a Killing horizon that the stress tensor would be constrained -- after all massive \emph{particles} cannot travel along a null surface! So while intuitively we may expect that on a null surface only ``zero mass'' fields would be able to exist, as massive ``quanta'' would be forced to travel at the speed of light to remain on this surface, I face two difficulties. The first is that it is difficult to express in general terms what is meant by a field with zero mass quanta in terms of allowed stress energies. At the moment I am not playing with a Lagrangian, only the locally defined stress energy tensor. The second point is slightly more problematic: the energy conditions have not been imposed anywhere! A timelike particle is still required to have a non-null four momentum, and so it is still possible that the only way to make such quanta have null momenta is to have singularities. This is an interesting problem, as it indicates that the restrictions on the stress energy we are seeing may have no bearing on the question of the existence of a CFT near a Killing horizon! Still, as an idea it is still too vague for me to use to prove anything with. 

\paragraph{}
Finally, as the near-horizon question is more one of existence I would hope that a nice clean argument could be presented that gives the same result. A candidate for such a result is the bifurcate Killing horizon argument outlined in appendix \ref{app:conformal2}.

\addcontentsline{toc}{chapter}{Bibliography}
\nocite{*}
\bibliographystyle{eprint}
\bibliography{chap1,chap2,chap3,chap4}

\appendix

% Algorithmic construction
%------------------------------------------------
% Algorithmic construction of static perfect fluid spheres
% Damien Martin and Matt Visser
% 24 June 2003
% V2 March 2004
% Uses revtex4
%------------------------------------------------
%------------------------------------------------
%------------------------------------------------
%------------------------------------------------
%------------------------------------------------
%------------------------------------------------
\chapter{Algorithmic construction of static perfect fluid spheres}\label{app:algorithm}
\markboth{Appendix \ref{app:algorithm}: Algorithmic construction...}{Phys. Rev. {\bf D 69}: 104028, 2004, (gr-qc/030609)} 
%------------------------------------------------
%\author{Damien Martin}
%\affiliation{School of Mathematical and Computing Sciences, 
%Victoria University of Wellington, PO Box 600, Wellington, New Zealand\\}
%------------------------------------------------
%\author{Matt Visser}
%\email{matt.visser@vuw.ac.nz}
%\homepage{http://www.mcs.vuw.ac.nz/~visser}
%\affiliation{School of Mathematical and Computing Sciences, 
%Victoria University of Wellington, PO Box 600, Wellington, New Zealand\\}
%------------------------------------------------
%\author{\ } %to get some semi-decent spacing
%------------------------------------------------
\paperauthor{Damien Martin and Matt Visser}
%\date{24 June 2003; 9 March 2004; 
%\LaTeX-ed \today}
%-----------------------------------------------
\begin{abstract}
%------------------------------------------------

  Perfect fluid spheres, either Newtonian or relativistic, are the
  first step in developing realistic stellar models (or models for
  fluid planets). Despite the importance of these models, explicit and
  fully general solutions of the perfect fluid constraint in general
  relativity have only very recently been developed.  In this Brief
  Report we present a variant of Lake's algorithm wherein: (1) we
  re-cast the algorithm in terms of variables with a clear physical
  meaning --- the average density and the locally measured
  acceleration due to gravity, (2) we present explicit and fully
  general formulae for the mass profile and pressure profile, and (3)
  we present an explicit closed-form expression for the central
  pressure. Furthermore we can then use the formalism to easily
  understand the pattern of inter-relationships among many of the
  previously known exact solutions, and generate several new exact
  solutions.

%\vskip 0.50cm
%\noindent
%  File {\sf algorithm2.tex}; Version 1.00
\centerline{{\sf damien.martin@mcs.vuw.ac.nz}, 
            {\sf matt.visser@mcs.vuw.ac.nz}}
%------------------------------------------------
\end{abstract}
%------------------------------------------------
\publishnotice{Published as Physical Review D {\bf 69} (2004) 104028.}
%-----------------------------------------------------------------
\pacs{04.20.-q, 04.40.Dg, 95.30.Sf }
%-----------------------------------------------------------------
\keywords{Fluid spheres; gr-qc/0306109}
%-----------------------------------------------------------------

%------------------------------------------------
%------------------------------------------------

% Bounds
%-----------------------------------------------------------------------
% Bounds on the interior geometry and pressure profile 
% of static fluid spheres
% Damien Martin and Matt Visser
% 10 June 2003
% Uses iopart.cls setstack.sty
%----------------------------------------------------------------------

%------------------------------------------------
\chapter{Bounds on the interior geometry and pressure profile of static fluid spheres}\label{app:bounds_paper}
\markboth{Appendix \ref{app:bounds_paper}: Bounds on perfect fluid spheres}{Class. Quant. Grav. {\bf 20}:3699-3716,2003 
 (gr-qc/0306038)}
%------------------------------------------------
\paperauthor{Damien Martin and Matt Visser}
%------------------------------------------------
%\address{School of Mathematical and Computing Sciences, 
%Victoria University of Wellington, PO Box 600, Wellington, New Zealand}
%------------------------------------------------

%------------------------------------------------
\begin{abstract}
%------------------------------------------------

  It is a famous result of relativistic stellar structure that (under
  mild technical conditions) a static fluid sphere satisfies the
  Buchdahl--Bondi bound $2M/R\leq8/9$; the surprise here being that
  the bound is not $2M/R\leq1$.  In this article we provide further
  generalizations of this bound by placing a number of constraints on
  the interior geometry (the metric components), on the local
  acceleration due to gravity, on various combinations of the internal
  density and pressure profiles, and on the internal compactness
  $2m(r)/r$ of static fluid spheres.  We do this by adapting the
  standard tool of comparing the generic fluid sphere with a
  Schwarzschild interior geometry of the same mass and radius --- in
  particular we obtain several results for the pressure profile (not
  merely the central pressure) that are considerably more subtle than
  might first be expected.

\centerline{{\sf damien.martin@mcs.vuw.ac.nz}, 
            {\sf matt.visser@mcs.vuw.ac.nz}}
%------------------------------------------------
\end{abstract}
%------------------------------------------------
\publishnotice{Published as Classical and Quantum Gravity 20:3699-3716,2003}
\indent gr-qc/0306038

% QNM I
%----------------------------------------------------------------------------
% Dirty black holes: Quasinormal modes.
% A.J.M. Medved
% Damien Martin
% Matt Visser
% Uses iopart.cls and setstack.cls
% October 2003
%---------------------------------------------------------------------------
% V2: one additional reference added
%----------------------------------------------------------------------------
% V3: minor changes due to referee feedback
%----------------------------------------------------------------------
%------------------------------------------------
%------------------------------------------------
%------------------------------------------------
\chapter{Dirty black holes: Quasinormal modes}\label{app:DBHQNM}
\markboth{Appendix \ref{app:DBHQNM}: Dirty black holes: QNM}{Class. Quant. Grav. {\bf 21}:1393-1406, 2004, (gr-qc/0310009)} 
%------------------------------------------------
\paperauthor{A.J.M.~Medved, Damien Martin and Matt Visser}
%------------------------------------------------
%\address{School of Mathematical and Computing Sciences,
%Victoria University of Wellington, PO Box 600, Wellington, New Zealand}
%------------------------------------------------
%------------------------------------------------
\begin{abstract}
%------------------------------------------------
 
  In this paper, we investigate the asymptotic nature of the
  quasinormal modes for ``dirty'' black holes --- \emph{generic}
  static and spherically symmetric  spacetimes for which a central
  black hole is surrounded by arbitrary ``matter'' fields.  We
  demonstrate that, to the leading asymptotic order, the [imaginary]
  spacing between modes is precisely equal to the surface gravity,
  independent of the specifics of the black hole system.
  
  Our analytical method is based on locating the complex poles in the
  first Born approximation for the scattering amplitude. We first
  verify that our formalism agrees, asymptotically, with previous studies
  on the Schwarzschild black hole.  The analysis is then generalized
  to more exotic black hole geometries.  We also extend considerations
  to spacetimes with two horizons and briefly discuss the
  degenerate-horizon scenario.

\centerline{{\sf damien.martin@mcs.vuw.ac.nz}, 
            {\sf joey.medved@mcs.vuw.ac.nz}}
\centerline{ {\sf matt.visser@mcs.vuw.ac.nz}}
%-----------------------------------------------------------------
\end{abstract}
%-----------------------------------------------------------------
\publishnotice{Published as Classical and Quantum Gravity 21:1393-1406, 2004}
%-----------------------------------------------------------------
\pacs{04.70.Dy, 03.65.Nk} 
%-----------------------------------------------------------------
\keywords{quasinormal modes, black holes, scattering amplitude, gr-qc/0310009} 
%-----------------------------------------------------------------

%-----------------------------------------------------------------
%\maketitle
%-----------------------------------------------------------------

% Squeezed horizons
%----------------------------------------------------------------------------
% Dirty black holes: Quasinormal modes for ``squeezed'' horizons
% A.J.M. Medved
% Damien Martin
% Matt Visser
% Uses iopart.cls and setstack.cls
% October 2003
%----------------------------------------------------------------------------
%----------------------------------------------------------------------
%
%------------------------------------------------
%------------------------------------------------
%------------------------------------------------
%------------------------------------------------
\chapter[Dirty black holes: Quasinormal modes for ``squeezed''\\
horizons]{Dirty black holes: Quasinormal modes for ``squeezed''
horizons}\label{app:squeezed}
\markboth{Appendix \ref{app:squeezed}: DBH: QNM for ``squeezed'' horizons}{Class. Quant. Grav. {\bf 21}: 2393--2405, 2004(gr-qc/0310097)}
%------------------------------------------------
\paperauthor{A.J.M.~Medved, Damien Martin and Matt Visser}
%------------------------------------------------
%\address{School of Mathematical and Computing Sciences,
%Victoria University of Wellington, PO Box 600, Wellington, New Zealand}
%------------------------------------------------

%------------------------------------------------
\begin{abstract}
%------------------------------------------------
 
  We consider the quasinormal modes for a class of black hole
  spacetimes that, informally speaking, contain a closely ``squeezed''
  pair of horizons.  (This scenario, where the relevant observer is
  presumed to be ``trapped'' between the horizons, is operationally
  distinct from near-extremal black holes with an external observer.)
  It is shown, by analytical means, that the spacing of the
  quasinormal frequencies equals the surface gravity at the squeezed
  horizons.  Moreover, we can calculate the real part of these
  frequencies provided that the horizons are sufficiently close
  together (but not necessarily degenerate or even ``nearly
  degenerate'').  The novelty of our analysis (which extends a
  model-specific treatment by Cardoso and Lemos) is that we consider
  ``dirty'' black holes; that is, the observable portion of the
  (static and spherically symmetric) spacetime is allowed to contain
  an arbitrary distribution of matter.

\centerline{{\sf damien.martin@mcs.vuw.ac.nz}, 
            {\sf joey.medved@mcs.vuw.ac.nz}}
\centerline{ {\sf matt.visser@mcs.vuw.ac.nz}}
%-----------------------------------------------------------------
\end{abstract}
%-----------------------------------------------------------------
\publishnotice{Published as Classical and Quantum Gravity 21:2393--2405,2004}
\pacs{04.70.-s, 04.30-w} 
%-----------------------------------------------------------------
\keywords{quasinormal modes, black holes, gr-qc/0310097}
%-----------------------------------------------------------------

% Tale of two treatments
%\documentstyle[prd,preprint,aps]{revtex}
%%%%%My Macros%%%%%%%%%%%%%%%%%%%%%%%
%\newcommand{\half}{{1\over2}}
%%% Macros for equations
%\newcommand{\be}{\begin{equation}}
%\newcommand{\ee}{\end{equation}}
\newcommand{\bea}{\begin{eqnarray}}
\newcommand{\eea}{\end{eqnarray}}
\newcommand{\none}{\nonumber \\}
\newcommand{\req}[1]{Eq.(\ref{#1})}
\newcommand{\reqs}[1]{Eqs.(\ref{#1})}
%%%%%%%%%%%%%%%%%%%%%%%%%%%%%%%%%%%%%

\newcommand{\I} {{\cal I}}
\newcommand{\re} {(r_{+})_{ext}}
\newcommand{\M} {{\tilde M}}
\newcommand{\V} {{\cal V}_n}
\newcommand{\Y} {{\overline \psi}}

%\begin{document}

\reversemarginpar
%\tighten

\chapter{A note on quasinormal modes: A tale of two treatments}\label{app:tale} 
\markboth{Appendix \ref{app:tale}: A tale of two treatments}{Gen.~Rel.~Grav. {\bf 37}: 1529--1539, 2005, (gr-qc/0311086)}
\paperauthor{A.J.M.~Medved and Damien Martin}
%\address{
%School of Mathematical and Computing Sciences\\
%Victoria University of Wellington\\
%PO Box 600, Wellington, New Zealand \\}

%\maketitle

\begin{abstract}
  
  There is an apparent discrepancy in the literature with regard to the
  quasinormal mode frequencies of Schwarzschild--de Sitter black holes in the
  degenerate-horizon limit.  On the one hand, a Poschl--Teller-inspired method
  predicts that the real part of the frequencies will depend strongly on the
  orbital angular momentum of the perturbation field whereas, on the other
  hand, the degenerate limit of a monodromy-based calculation suggests there
  should be no such dependence (at least, for the highly damped modes). In the
  current paper, we provide a possible resolution by critically re-assessing
  the limiting procedure used in the monodromy analysis.
  
\centerline{{\sf damien.martin@mcs.vuw.ac.nz}, 
            {\sf joey.medved@mcs.vuw.ac.nz}}

\end{abstract}

\publishnotice{Published as General Relativity and Gravitation {\bf 37} (2005) 1529--1539}

\keywords{QNMs, gr-qc/0311086}

% Static conformal symmetries
%%%%%%%%%%%%%%%%%%%%%%%%%%%%%%%%%%%%%%%%%%%%%%%%%%%%%%%%%%%%%%%%%%%%%%
% Dirty black holes: Spacetime geometry and near-horizon symmetries
%   Joey Medved, Damien Martin, and Matt Visser
% 14 February 2004
%%%%%%%%%%%%%%%%%%%%%%%%%%%%%%%%%%%%%%%%%%%%%%%%%%%%%%%%%%%%%%%%%%%%%%
% Attention: uses plain LaTeX
%--------------------------------------------------------------------------
%--------------------------------------------------------------------------
%---------------------------------------------------------------------------
\chapter{Dirty black holes: Spacetime geometry and near-horizon symmetries}
\label{app:conformalI}
\markboth{Appendix \ref{app:conformalI}: DBH: Spacetime geometry and near horizon \ldots}{Class. Quant. Grav. {\bf 21}: 3111--3126, 2005, (gr-qc/0402069)}
%---------------------------------------------------------------------------
\paperauthor{A J M Medved, Damien Martin and Matt Visser}
%------------------------------------------------
%------------------------------------------------
%---------------------------------------------------------------------------
\date{22 February 2004}
%---------------------------------------------------------------------------
%---------------------------------------------------------------------------
%---------------------------------------------------------------------------
%\maketitle
%---------------------------------------------------------------------------
%---------------------------------------------------------------------------

%---------------------------------------------------------------------------
%---------------------------------------------------------------------------
\begin{abstract}
  
  We consider the spacetime geometry of a static but otherwise generic
  black hole (that is, the horizon geometry and topology are not
  necessarily spherically symmetric).  It is demonstrated, by purely
  geometrical techniques, that the curvature tensors, and the Einstein
  tensor in particular, exhibit a very high degree of symmetry as the
  horizon is approached.  Consequently, the stress-energy tensor will
  be highly constrained near any static Killing horizon.  More
  specifically, it is shown that --- at the horizon --- the
  stress-energy tensor block-diagonalizes into ``transverse'' and
  ``parallel'' blocks, the transverse components of this tensor are
  proportional to the transverse metric, and these properties remain
  invariant under static conformal deformations.  Moreover, we
  speculate that this geometric symmetry underlies Carlip's notion of
  an asymptotic near-horizon conformal symmetry controlling the
  entropy of a black hole.

\centerline{{\sf damien.martin@mcs.vuw.ac.nz}, 
            {\sf joey.medved@mcs.vuw.ac.nz}}
\centerline{ {\sf matt.visser@mcs.vuw.ac.nz}}
\end{abstract}
%----------------------------------------------------------------------------
\publishnotice{Published as Classical and Quantum Gravity {\bf 21} (2004) 3111--3126}
\indent gr-qc/0402069
%\enlargethispage{250pt}

% Stationary conformal symmetries
%%%%%%%%%%%%%%%%%%%%%%%%%%%%%%%%%%%%%%%%%%%%%%%%%%%%%%%%%%%%%%%%%%%%%%
% Dirty black holes: Symmetries at stationary non-static horizons
%   Joey Medved, Damien Martin, and Matt Visser
% 5 March 2004
%%%%%%%%%%%%%%%%%%%%%%%%%%%%%%%%%%%%%%%%%%%%%%%%%%%%%%%%%%%%%%%%%%%%%%
% Attention: uses plain LaTeX
%--------------------------------------------------------------------------
%--------------------------------------------------------------------------
%---------------------------------------------------------------------------
\chapter[Dirty black holes: Symmetries at stationary non-static\\ horizons]{Dirty
black holes: Symmetries at stationary non-static horizons}
\label{app:conformal2}
\markboth{Appendix \ref{app:conformal2}: DBH: Symmetries at stationary horizons}{Phys. Rev. {\bf D 70}: 024009, 2004, (gr-qc/0403026)}
%---------------------------------------------------------------------------
\paperauthor{A J M Medved, Damien Martin and Matt Visser}
%------------------------------------------------
%------------------------------------------------
%---------------------------------------------------------------------------
\date{5 March 2004}
%---------------------------------------------------------------------------
%---------------------------------------------------------------------------
%---------------------------------------------------------------------------
%\maketitle
%---------------------------------------------------------------------------
%---------------------------------------------------------------------------

%---------------------------------------------------------------------------
%---------------------------------------------------------------------------
\begin{abstract}
   
  We establish that the Einstein tensor takes on a highly symmetric
  form near the Killing horizon of any stationary but non-static (and
  non-extremal) black hole spacetime. [This follows up on a recent
  article by the current authors, gr-qc/0402069, which considered
  static black holes.]  Specifically, at any such Killing horizon ---
  irrespective of the horizon geometry --- the Einstein tensor
  block-diagonalizes into ``transverse'' and ``parallel'' blocks, and
  its transverse components are proportional to the transverse metric.
  Our findings are supported by two independent procedures; one based
  on the regularity of the on-horizon geometry and another that
  directly utilizes the elegant nature of a bifurcate Killing horizon.
  It is then argued that geometrical symmetries will severely
  constrain the matter near any Killing horizon.  We also speculate on
  how this may be relevant to certain calculations of the black hole
  entropy.

\centerline{{\sf damien.martin@mcs.vuw.ac.nz}, 
            {\sf joey.medved@mcs.vuw.ac.nz}}
\centerline{ {\sf matt.visser@mcs.vuw.ac.nz}}
\end{abstract}
%----------------------------------------------------------------------------
\publishnotice{Published as Physical Review D {\bf 70 } (2004) 024009}
\indent {gr-qc/0403026}
%\enlargethispage{250pt}

%%%%%%%%%%%%%%%%%%%%%%%%%%%%%%%%%%%%%%%%%%%%%%%%%%%%%%%
% and of course book style knows about backmatter
\backmatter
% and of course report style doesn't
%%%%%%%%%%%%%%%%%%%%%%%%%%%%%%%%%%%%%%%%%%%%%%%%%%%%%%%

\end{document}